\documentclass[a4paper,11pt]{article}
\pdfoutput=1 % if your are submitting a pdflatex (i.e. if you have
\usepackage{jcappub}

\usepackage[utf8]{inputenc}
\usepackage[dvipsnames]{xcolor}
\usepackage{amsmath,mathtools,slashed}
\usepackage{dcolumn}% Align table columns on decimal point
\usepackage{bm}
\usepackage{comment}
\usepackage[normalem]{ulem} % to strike through
\usepackage{enumerate}
\usepackage{amssymb,amsmath,amsfonts,mathrsfs}
\usepackage[dvipsnames]{xcolor}
\usepackage{graphicx}
\usepackage{longtable}
\usepackage{verbatim}
\usepackage{color}
\usepackage[mathscr]{euscript}
\usepackage{framed}
\usepackage{hyperref}
\usepackage{xcolor}
\hypersetup{colorlinks, citecolor=blue, linkcolor=red, urlcolor=blue}
\usepackage[margin=1in,footskip=0.25in]{geometry}
\usepackage{ulem}
\usepackage{tabularx}
\usepackage{pifont}
\usepackage{tikz}

\usepackage{hyperref}
\hypersetup{
    colorlinks=true,
    linkcolor=blue,
    filecolor=magenta,      
    urlcolor=blue,
    citecolor=red,
}
\def\bea{\begin{eqnarray}}
\def\eea{\end{eqnarray}}
\def\be{\begin{equation}}
\def\ee{\end{equation}}
\newcommand{\Mpl}{M_{\rm{Pl}}}
\newcommand{\Mp}{M_{\rm{Pl}}}
\def\d{\mathrm{d}}
\newcommand{\bpm}{\begin{pmatrix}}
\newcommand{\epm}{\end{pmatrix}}
\newcommand{\Msun}{M_\odot}
\newcommand{\lp}{\left(}
\newcommand{\rp}{\right)}
\newcommand{\lb}{\left[}
\newcommand{\rb}{\right]}

\newcommand{\llp}{\left [}
\newcommand{\rrp}{\right ]}

\def\npbh{\bar n_\text{PBH}}

\def\mpbh{m_\text{PBH}}

\def\fpbh{f_\text{PBH}}
\def\rhopbh{ \rho_\text{PBH}}

\newcommand{\cmark}{\textcolor{Green}{\ding{51}}}%
\newcommand{\xmark}{\textcolor{Red}{\ding{55}}}%

\def\PBH{\text{PBH}}

\def\vk{{\vec{k}}}

\def\hr{\hat{r}}

\def\d{{\rm d}}
\def\vx{{\vec{x}}}

\def\PBH{\text{PBH}}

\newcommand{\OGW}{\Omega_\text{GW}}

\definecolor{grey}{rgb}{0.4,0.4,0.4}
\definecolor{dullmagenta}{rgb}{0.4,0,0.4}
\definecolor{darkblue}{rgb}{0,0,0.4}
\definecolor{midblue}{rgb}{0,0,0.5}
\definecolor{midred}{rgb}{0.5,0,0}
\definecolor{orange}{rgb}{1,0.5,0}
\definecolor{lightbrown}{rgb}{0.75,0.5,0.25}
\definecolor{tan}{cmyk}{0.14,0.42,0.56,0}
\definecolor{djunglegreen}{cmyk}{0.99,0,0.52,0}
\definecolor{lightgreen}{rgb}{0,1,0}
\definecolor{olivegreen}{cmyk}{0.64,0,0.95,0.40}
\definecolor{midgreen}{rgb}{0.0,0.675,0.0}
\definecolor{darkgreen}{rgb}{0,0.5,0}

\title{\hspace{12cm}\vspace{-2cm}\href{https://lisa.pages.in2p3.fr/consortium-userguide/wg_cosmo.html}{\includegraphics[height=2cm]{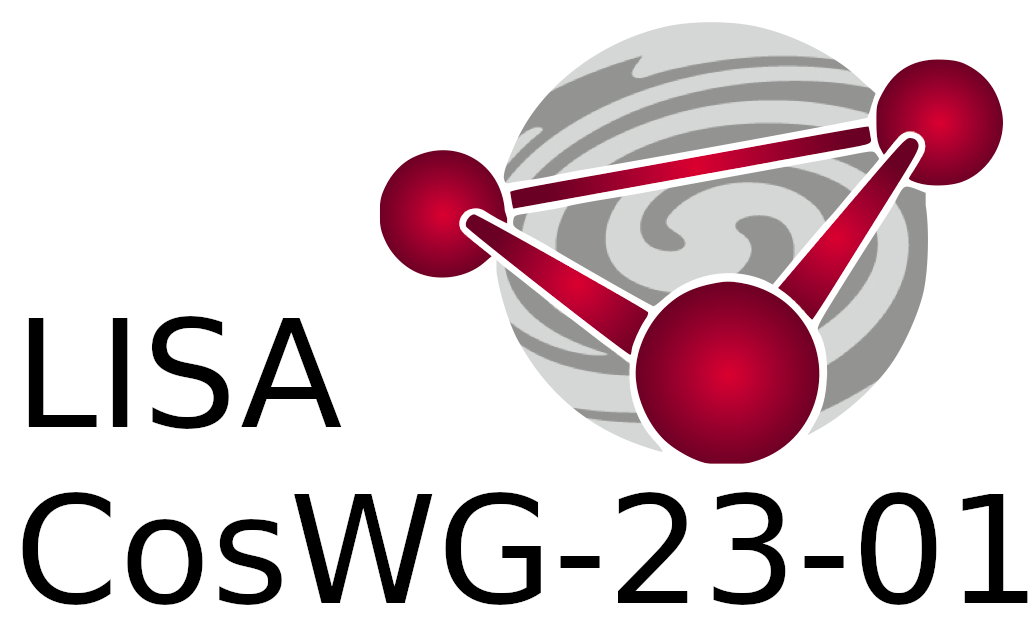}} \\[2cm] Primordial black holes \\
and their gravitational-wave signatures
}

\author[a]{Eleni Bagui}
\author[a]{\!\!, Sebastien Clesse\footnote{Project coordinator: sebastien.clesse@ulb.be}}
\author[b]{\!\!, Valerio De Luca}
\author[c]{\!\!, Jose Mar\'ia Ezquiaga}
\author[d]{\!\!, Gabriele Franciolini}
\author[e]{\!\!, Juan Garc\'ia-Bellido\footnote{Project coordinator: juan.garciabellido@uam.es}}
\author[f,g]{\!\!, Cristian Joana}  
\author[h]{\!\!, Rajeev Kumar Jain}
\author[e,i]{\!\!, Sachiko Kuroyanagi}
\author[j,k]{\!\!, Ilia Musco}
\author[l,m,n]{\!\!, Theodoros Papanikolaou}
\author[o,p,q,d]{\!\!, Alvise Raccanelli}
\author[r]{\!\!, S\'ebastien~Renaux-Petel} 
\author[s]{\!\!, Antonio~Riotto}
\author[e,t]{\!\!, Ester Ruiz Morales}
\author[u]{\!\!, Marco Scalisi}
\author[v,w,x,y]{\!\!, Olga Sergijenko}
\author[z,a2,b2]{\!\!, Caner Unal}
\author[c2]{\!\!, Vincent Vennin}
\author[d2]{\!\!, David Wands}
\author[]{\\ \centering \texttt{(For the LISA Cosmology Working Group)}}

\affiliation[a]{Service de Physique Th\'eorique, Univesrit\'e Libre de Bruxelles (ULB), Boulevard du Triomphe, CP225, B-1050 Brussels, Belgium}
\affiliation[b]{Center for Particle Cosmology, Department of Physics and Astronomy, University of Pennsylvania 209 S. 33rd St., Philadelphia, PA 19104, USA}
\affiliation[c]{Niels Bohr International Academy, Niels Bohr Institute, Blegdamsvej 17, DK-2100 Copenhagen, Denmark}
\affiliation[d]{Theoretical Physics Department, CERN, 1 Esplanade des Particules, 1211 Geneva 23, Switzerland}
\affiliation[e]{Instituto de F\'isica Te\'orica UAM/CSIC, Universidad Aut\'onoma de Madrid, Cantoblanco, 28049 Madrid, Spain}
\affiliation[f]{CAS Key Laboratory of Theoretical Physics, Institute of Theoretical Physics, Chinese Academy of Sciences, Beijing 100190, China}
\affiliation[g]{Cosmology, Universe and Relativity at Louvain (CURL), Institute of Mathematics and Physics, University of Louvain,	Chemin du Cyclotron 2,	1348 Louvain-la-Neuve,	Belgium}
\affiliation[h]{Department of Physics, Indian Institute of Science, C. V. Raman Road, Bangalore 560012, India}
\affiliation[i]{Department of Physics, Nagoya University, Furo-cho Chikusa-ku, Nagoya 464-8602, Japan}
\affiliation[j]{Dipartimento di Fisica, Sapienza Università di Roma, Piazzale Aldo Moro 5, 00185, Roma, Italy} 
\affiliation[k]{INFN, Sezione di Roma, Piazzale Aldo Moro 2, 00185, Roma, Italy}
\affiliation[l]{Scuola Superiore Meridionale, Largo S. Marcellino 10, I-80138, Napoli, Italy}
\affiliation[m]{Laboratoire Astroparticule et Cosmologie, CNRS Universit\'e de Paris, 75013 Paris, France}
\affiliation[n]{National Observatory of Athens, Lofos Nymfon, 11852, Athens, Greece}
\affiliation[o]{Dipartimento di Fisica Galileo Galilei, Universit\` a di Padova, I-35131 Padova, Italy}
\affiliation[p]{INFN Sezione di Padova, I-35131 Padova, Italy}
\affiliation[q]{INAF-Osservatorio Astronomico di Padova, Italy}
\affiliation[r]{Institut d’Astrophysique de Paris, GReCO, UMR 7095 du CNRS et de Sorbonne Universit\'e, 98bis boulevard Arago, 75014 Paris, France}
\affiliation[s]{D\'epartement de Physique Th\'eorique and Gravitational Wave Science Center (GWSC), Universit\'e de Gen\`eve, CH-1211 Geneva, Switzerland CH-1211 Geneva, Switzerland}
\affiliation[t]{Department of Physics-ETSIDI, Universidad Polit\'ecnica de Madrid, 28012 Madrid, Spain}
\affiliation[u]{Max-Planck-Institut f\"ur Physik (Werner-Heisenberg-Institut), F\"ohringer Ring 6, 80805, M\"unchen, Germany}
\affiliation[v]{Main Astronomical Observatory of the National Academy of Sciences of Ukraine, Zabolotnoho str., 27, 03143, Kyiv, Ukraine}
\affiliation[w]{AGH University of Krakow, Aleja Mickiewicza, 30, 30-059, Krakow, Poland}
\affiliation[x]{Faculty of Natural Sciences, National University "Kyiv Mohyla Academy", Skovorody str., 2, 04070, Kyiv, Ukraine}
\affiliation[y]{Astronomical Observatory of Taras Shevchenko National University of Kyiv, Observatorna str., 3, 04053, Kyiv, Ukraine}
\affiliation[z]{Department of Physics, Ben-Gurion University of the Negev, Beer Sheva 84105, Israel}
\affiliation[a2]{CEICO, FZU – Institute of Physics of the Czech Academy of Sciences, Na Slovance 2, 182 21, Prague, Czech Republic}
\affiliation[b2]{Feza Gursey Institute, Bogazici University, Kandilli, Istanbul, Turkey}
\affiliation[c2]{Laboratoire de Physique de l'\'Ecole Normale Sup\'erieure, ENS, CNRS, Universit\'e PSL, Sorbonne Universit\'e,  Universit\'e Paris Cit\'e, F-75005, Paris, France}
\affiliation[d2]{Institute of Cosmology and Gravitation, University of Portsmouth, Dennis Sciama Building, Burnaby Road, Portsmouth, PO1 3FX, United Kingdom}

\abstract{In the recent years, primordial black holes (PBHs) have emerged as one of the most interesting and hotly debated topics in cosmology. Among other possibilities, PBHs could explain both some of the signals from binary black hole mergers observed in gravitational wave detectors and an important component of the dark matter in the Universe.  Significant progress has been achieved both on the theory side and from the point of view of observations, including new models and more accurate calculations of PBH formation, evolution, clustering, merger rates, as well as new astrophysical and cosmological probes.  In this work, we review, analyse and combine the latest developments in order to perform end-to-end calculations of the various gravitational wave signatures of PBHs. Different ways to distinguish PBHs from stellar black holes are emphasized. Finally, we discuss their detectability with LISA, the first planned gravitational-wave observatory in space.}
\begin{document}

\maketitle
%\newpage

%\tableofcontents

\newpage

\section{Introduction}

\subsection{A brief history of primordial black holes (PBHs)}

The idea that black holes could have been formed in the early Universe dates back to the late 1960's with the pioneering work of Zel'dovich and Novikov~\cite{Zeldovich:1967lct} and to the early 1970's with the work of 
% to be conistent, I think we should not use initials - DW
Hawking \cite{Hawking:1971ei}. This triggered Hawking's famous discovery of black hole evaporation~\cite{Hawking:1974sw}. Subsequently, Carr, Hawking's PhD student at the time, continued to investigate PBHs \cite{Carr:1974nx,Carr:1975qj}.  Already in \cite{Hawking:1971ei,Carr:1974nx,Chapline:1975ojl}, it was suggested that such primordial black holes (PBHs) could contribute to the suspected Dark Matter (DM) in the Universe or to the seeds of supermassive black holes.  The first formation scenarios in the context of inflation were proposed in the 1990's \cite{Dolgov:1992pu,Carr:1993aq,Carr:1994ar,Ivanov:1994pa,GarciaBellido:1996qt,Kim:1996hr,Kawasaki:1997ju,Green:1997sz,Ivanov:1997ia,Yokoyama:1998qw,Kotok:1998rp}, but these usually led to (evaporating) PBHs of very small mass, except for~\cite{GarciaBellido:1996qt} who predicted solar mass PBHs that could account for the MACHO microlensing events observed towards the Magellanic clouds~\cite{Aubourg:1993wb,Alcock:1996yv}.  
Other mechanisms were also proposed, e.g., based on phase transitions~\cite{Jedamzik:1996mr,Niemeyer:1997mt}, an early matter era~\cite{Khlopov:1980mg,Polnarev:1986bi,Green:1997pr}, scalar field instabilities~\cite{Khlopov:1985jw}, collapse of topological defects~\cite{Polnarev:1988dh}, modified gravity~\cite{Barrow:1996jk,Kawai:2021edk,Zhang:2021rqs}, string theory~\cite{Cicoli:2018asa,Nanopoulos:2020nnh,Mavromatos:2022yql}, bouncing cosmological scenarios~\cite{Copeland:1998gj,Carr:2011hv,Quintin:2016qro,Chen:2016kjx,Banerjee:2022xft} as well as quantum gravity setups~\cite{Papanikolaou:2023crz}. In the late 1990’s stellar-mass PBHs were seriously considered as a dark matter candidate, following the possible detection in the MACHO survey~\cite{Aubourg:1993wb,Alcock:1996yv} of several microlensing events towards the Magellanic clouds. However, the EROS~\cite{EROS-2:2006ryy} and OGLE~\cite{Wyrzykowski:2010bh,Wyrzykowski:2010mh,Wyrzykowski:2011tr,Novati:2013fxa} surveys later set more stringent limits on the PBH abundance, and at the same time, very stringent constraints from cosmic microwave background (CMB) observations were claimed in~\cite{Ricotti:2007au}.

Despite some pioneering numerical studies~\cite{Nadezhin,Bicknell,Polnarev}, only more recently it has become possible to fully understand the mechanism of PBH formation with detailed spherically-symmetric numerical simulations~\cite{Shibata:1999zs,Musco:2004ak,Polnarev:2006aa}, showing that superhorizon cosmological perturbations would collapse to PBHs after re-entering the cosmological horizon, if their amplitude $\delta$ is larger than a certain threshold value $\delta_c$. This quantity, measured at horizon crossing, was initially estimated with a simplified Jeans length argument in Newtonian gravity~\cite{Carr:1975qj}, giving $\delta_c \sim c_s^2$, where $c_s^2=1/3$ is the sound speed of the cosmological radiation fluid measured in units of the speed of light. More recently this argument has been generalized within general relativity using a three-zone model~\cite{Harada:2013epa,Papanikolaou:2022cvo} which gives $\delta_c\simeq0.4$ for a radiation-dominated Universe. 
Numerical simulations showed that the mechanism of critical collapse characterizing perfect fluids~\cite{Neilsen:1998qc} arises 
naturally in the context of PBH formation~\cite{Yokoyama:1998qw,Niemeyer:1997mt,Jedamzik:1999am,Musco:2004ak}.
In particular, the collapse is characterized by a relativistic wind which progressively separates the collapsing perturbation from the expanding background, approaching a self-similar critical solution characterizing the threshold~\cite{Musco:2008hv,Musco:2012au}.   %Important achievements of these methods include the calculation of the overdensity threshold leading to the gravitational collapse of pre-existing, super-Hubble inhomogeneities and the PBH mass in the critical collapse regime.  
Finally, a very recent detailed study has found a clear relation between the value of the threshold, $\delta_c$, and the initial configuration of the curvature (or energy density) profile used to set the initial condition of the numerical simulations, with $0.4\leq \delta_c \leq 2/3$, where the shape is identified by a single parameter~\cite{Musco:2018rwt,Escriva:2019phb}. This range is reduced to $0.4\leq \delta_c \lesssim 0.6$ when the initial perturbations are computed, using peak theory~\cite{Bardeen:1985tr}, from the primordial power spectrum of cosmological perturbations~\cite{Musco:2020jjb}. 

Since 2016, the real game-changer that has rekindled the idea that PBHs may exist and constitute a significant fraction to the total dark matter~\cite{Bird:2016dcv,Clesse:2016vqa,Sasaki:2016jop} has been the first gravitational-wave (GW) detection from a black hole merger by LIGO/Virgo~\cite{Abbott:2016blz}.  Initially, the merger rate of early PBH binaries seemed to only allow a small fraction of the dark matter to be made of PBHs~\cite{Sasaki:2016jop}, but N-body simulations later showed that these rates are in fact significantly suppressed~\cite{Raidal:2018bbj,Trashorras:2020mwn}.   Nowadays, the importance of the different PBH binary formation channels, the possible abundance of PBHs, their viable mass function, are a subject of intense activity and are hotly debated (for recent reviews, see e.g.~\cite{Carr:2020xqk,Carr:2020gox}).   Recent analysis based on the rate and mass~\cite{Clesse:2017bsw,Raidal:2018bbj,Carr:2019kxo,Hall:2020daa,Jedamzik:2020ypm,Jedamzik:2020omx,Clesse:2020ghq,Hutsi:2020sol,Escriva:2022bwe}, or spin~\cite{Fernandez:2019kyb,DeLuca:2020bjf, Garcia-Bellido:2020pwq}, or both \cite{DeLuca:2020qqa,Wong:2020yig,DeLuca:2021wjr, Franciolini:2021tla,Franciolini:2021nvv,Franciolini:2021xbq,Franciolini:2022iaa,Franciolini:2022tfm} distributions of compact binary coalescences observed in the first, second~\cite{LIGOScientific:2018mvr} 
and third~\cite{Abbott:2020niy,LIGOScientific:2021djp}
%and first half of the third~\cite{Abbott:2020niy} 
observing runs of {the LIGO/Virgo/Kagra collaboration (LVK)}, suggest that some black holes may be primordial.  Some are based on a Bayesian approach~\cite{Clesse:2017bsw,Raidal:2018bbj,Hall:2020daa,Hutsi:2020sol,Wong:2020yig,DeLuca:2020sae,Kritos:2020wcl,DeLuca:2021wjr, Franciolini:2021tla,Franciolini:2021nvv,Franciolini:2022tfm,Fernandez:2019kyb,Garcia-Bellido:2020pwq} that will ultimately allow us to compare statistically the different possible origins and scenarios.  However due to the broad variety of PBHs and astrophysical models as well as the large theoretical and observational uncertainties, it is still premature to firmly affirm (or deny) a PBH origin of GW observations.  In this context, the next generation of GW detectors like Einstein Telescope~\cite{Maggiore:2019uih}, Cosmic Explorer~\cite{Reitze:2019iox} and the Laser Interferometer Space Antenna (LISA)~\cite{Audley:2017drz} will play a crucial role.
Also, the 
% great 
development of ultra-high frequency GW detectors may additionally complement the search for GW signatures of asteroid-mass PBHs \cite{Herman:2022fau,Franciolini:2022htd} (see also e.g.  
\cite{Aggarwal:2020olq,Ringwald:2020ist,Berlin:2021txa,Domcke:2022rgu,Berlin:2022hfx}).

Important progress has also been made on the limits on the allowed PBH abundance set by various types of observations.  Shortly after the first GW detection, the CMB limits were re-analysed and found to be much less stringent than initially thought~\cite{Ali-Haimoud:2016mbv}, thereby re-opening the stellar mass region.  At the same time, the degree of validity of the microlensing limits has been questioned~\cite{Hawkins:2015uja,Clesse:2015wea,Garcia-Bellido:2017xvr,Green:2017qoa,Calcino:2018mwh,Carr:2019kxo} for more realistic scenarios than monochromatic and homogeneous distributions of PBHs.   New probes, like ultra-faint-dwarf galaxies~\cite{Brandt:2016aco,Green:2016xgy,Li:2016utv} and X-ray sources towards the galactic center~\cite{Gaggero:2016dpq}, have revived the mass region between $1$ and $100 M_\odot$, 
while the constraining power of other probes like high-cadence microlensing~\cite{Niikura:2017zjd}, neutron stars~\cite{Capela:2013yf} and white dwarves~\cite{Capela:2012jz}, have been reduced, which has reopened the asteroid-mass region~\cite{Smyth:2019whb}.  Finally, a series of recent microlensing events from OGLE~\cite{Niikura:2019kqi,Wyrzykowski:2019jyg}, HSC~\cite{Niikura:2017zjd}, as well as quasars~\cite{Hawkins:1993yud,Hawkins:2020zie,Hawkins:2020rqu,Hawkins:2022vqo}, may hint at the existence of an important population of planetary-mass and solar-mass dark compact objects. The unexplained spatial correlations between infrared and X-ray backgrounds at high redshift~\cite{2005Natur.438...45K} could also be due to an important population of stellar-mass PBHs~\cite{Kashlinsky:2016sdv,2018RvMP...90b5006K,Cappelluti:2021usg}.

Furthermore, since PBHs are formed by the collapse of large density fluctuations, there is an associated stochastic GW background (SGWB)   sourced by these perturbations at second-order. It has been calculated~\cite{Ando:2017veq,Garcia-Bellido:2017aan} that if BHs detected by LVK have a primordial origin, there is an inevitable  accompanying SGWB 
%\footnote{This SGWB is ineludible in the sense that it does not require any further assumption other than general relativity and large density perturbations. It is a standard SGWB formed by anisotropic stress which is quadratic order scalar perturbations.}  
peaking around pulsar-timing-array frequencies. In September 2020, NANOGrav claimed the possible detection of a SGWB at nanohertz frequency using pulsar timing arrays~\cite{Arzoumanian:2020vkk}, which may have been sourced by primordial density perturbations which could be the origin of stellar-mass PBH formation~\cite{Vaskonen:2020lbd,DeLuca:2020agl,Kohri:2020qqd}.  
{Remarkably, such claim 
was confirmed by more recent pulsar timing array data released in 2023 by the NANOGrav~\cite{NG15-SGWB,NG15-pulsars}, EPTA (in combination with InPTA)\,\cite{EPTA2-SGWB,EPTA2-pulsars,EPTA2-SMBHB-NP}, PPTA\,\cite{PPTA3-SGWB,PPTA3-pulsars,PPTA3-SMBHB} and CPTA\,\cite{CPTA-SGWB} collaborations, who found evidence for a Hellings and Downs angular correlation,  typical of an homogeneous spin-2 GW background and consistent with the quadrupolar nature of GWs in GR~\cite{1983ApJ...265L..39H}}

On the theory side, a plethora of new models have been proposed in the last four years.  Most of them are still subject to fine-tuning issues related to the amplitude of fluctuations required to form PBHs with a significant abundance.  Nevertheless, one should note a new class of models that do not require a modification of the primordial power spectrum on the cosmological scale, the PBH formation being related to a modification of the Gaussian tail of the distribution of primordial perturbations~\cite{Ezquiaga:2019ftu,Carr:2019hud}.  In these models, a large population of PBHs could arise much more naturally.   PBH masses of order of the solar mass and around $30 M_\odot$ could also arise naturally due to the thermal history of the Universe and the change in the equation of state at the QCD epoch~\cite{Jedamzik:1996mr,Niemeyer:1997mt,Byrnes:2018clq,Carr:2019kxo,Franciolini:2022tfm,Escriva:2022bwe}, which should strongly boost PBH formation.   Finally, besides explaining the dark matter or a significant fraction of it, PBHs could also be related to baryogenesis and different mechanisms have been proposed~\cite{Dolgov:2000ht,Baumann:2007yr,Garcia-Bellido:2019vlf,Carr:2019hud,Garcia-Bellido:2019tvz,DeLuca:2021oer}.

Any firm detection would open a new window on the physics at play in the very early Universe and a possible way to solve various long-standing astrophysical and cosmological puzzles~\cite{Clesse:2017bsw,Carr:2019kxo}. 
Ultimately, the best way to distinguish PBHs from stellar black holes would be to detect sub-solar-mass black holes.   Recently, four sub-solar black hole triggers have been found in the second observing run of LVK~\cite{Phukon:2021cus,Morras:2023jvb}, {and three in the third observing run~\cite{LIGOScientific:2022hai}.  These are not statistically significant enough to claim a firm detection, but} if confirmed, these would strongly support a PBH origin of some LVK black holes and make a significant contribution to the dark matter in the Universe.

\subsection{Why a(nother) review on PBHs?}

There already exist a few recent and relatively complete review articles dedicated to PBHs, see e.g.~\cite{Sasaki:2018dmp,Garcia-Bellido:2017fdg,Carr:2020xqk,Carr:2020gox,Carr:2023tpt}. Earlier reviews that pre-date the first GW detection also exist, see e.g.~\cite{Khlopov:2008qy,Carr:2009jm}.  All of them are written by only a few authors.  It has therefore been quite challenging for them to include detailed discussions and analysis of the broad range of topics covered by PBHs.  Indeed, these include early Universe cosmology, numerical relativity, dark matter, astrophysics, celestial mechanics, GW astrophysics, etc.  Those reviews therefore summarize some of the results obtained in these topics but do not include model specific A-to-Z calculations of PBH related observables.   Very often, some of the most recent developments like the precise determination of the critical threshold density leading to PBH formation, which is highly model-dependent and varies with the evolution of the equation of state of the Universe, or the rate suppression/boost linked to Poisson clustering, are not always considered when models are compared to observations.  

The goal of this paper is to review these most recent developments and discuss how they impact the GW signatures of PBHs.  Compared to other recent reviews, we aim at integrating together the most recent and accurate models, e.g., of PBH formation, evolution, clustering, merger rates, in order to compute the GW signatures of PBHs, and to explore their detectability.  For this purpose, we have developed in parallel a numerical code that implements all those models and recent scenarios.  The code will be released soon in a separate publication, but it has been already used in this paper in order to produce the key figures. The task of including these increasingly complex models has been eased by the specific expertise of many authors, who have actively contributed to these developments, covering a broad range of topics.  

Finally, another original aspect of this work is the fact that we plan to have a \textit{living} review format, with  bi-yearly updates of the present manuscript, in order to track the rapid progress in this very active field of research and provide at all times an up-to-date reference publication.  In future, this review could be expanded to include new sections on the detectability by PTAs and ground-based GW detectors, or on the cosmological signatures of PBH scenarios.  

All of this encourages us to go beyond the present state, and provide a new kind of review on PBHs.

\subsection{Probing PBHs with LISA}

This project is hosted by the cosmology working group of the LISA consortium.  Therefore our discussion has been focused on the GW signatures of PBHs in the particular context of the LISA mission.

PBHs cover a very wide range of masses and therefore frequency ranges of GWs. In particular, LISA will be sensitive to a broad band around the millihertz GW frequencies~\cite{Garcia-Bellido:2017aan,Cai:2018dig,Bartolo:2018evs,Unal:2018yaa}, which is complementary to the ground-based GW detectors~\cite{Maggiore:2019uih,Reitze:2019iox} and electromagnetic probes~\cite{Ali-Haimoud:2019khd}.   The search for a SGWB from PBH formation in the radiation era can be indirectly used for their discovery or alternatively to constrain their existence in the $10^{-12}~M_\odot$ mass range~\cite{Saito:2008jc,Garcia-Bellido:2017aan,Cai:2018dig,Bartolo:2018evs,Unal:2018yaa}. Moreover, the coalescence of the heavy seeds of supermassive black holes (SMBH) in the late Universe could leave clear signatures of their primordial origin, or even detect individual events at high redshift which could not have arisen from astrophysical mechanisms. 

The great sensitivity of LISA at mHz frequencies~\cite{Baibhav:2019rsa,Amaro-Seoane:2022rxf,LISA:2022kgy,Karnesis:2022vdp} opens the possibility to detect the mergers of $10^3 - 10^4~M_\odot$ PBHs all the way to $z\simeq100$. Furthermore, the isotropic SGWB from the coalescence of PBHs since recombination should have an amplitude and spectral shape that will make it easily detectable by LISA. In this review, we will describe the different features of known PBH scenarios that can be probed with LISA.

{LISA will not only probe the GWs from individual PBH mergers and we will review numerous other types of signatures of PBHs that can be probed with LISA.  Among them, the GW background from second order curvature fluctuations will probe the very interesting asteroid-mass range.  Other backgrounds include the one from ultralight, evaporated PBHs, the ones from early and late PBH binaries as well as from hyperbolic encounters.  GW from individual sources could also be observed by LISA, for instance from intermediate-mass PBH mergers, including at high-redshifts, as above mentioned, or from inspirals with extreme mass ratios which are expected for extended PBH mass distributions.  LISA will also have a key role if data are combined to LVK or other detector observations at higher frequencies, for long-duration signals including mergers with a subsolar component.  Subsolar-mass PBHs could also produce detectable quasi-monochromatic continuous waves in the LISA frequency range.  Finally, a more exotic possibility is to use LISA as a detector of near-Earth asteroid-mass PBHs. }  

\subsection{Outline of this review}

This review is organized as follows.  In Section~\ref{sec:theory}, we present the different classes of theoretical PBH formation scenarios, with a focus on recent inflation models but including a discussion of other production mechanisms such as (p)reheating, curvatons, phase transitions, topological defects and primordial magnetic fields.  Among inflationary scenarios, we especially focus on recent proposals in which the usual Gaussian distribution of primordial curvature fluctuations is modified in a more complex way than can be described in terms of conventional non-Gaussian deviation parameters ($f_{\rm NL}$, etc).   Ultimately this could be one possible way to resolve the long-standing fine-tuning issue for primordial fluctuations as the origin of PBHs.

Section~\ref{sec:PBHformation} is dedicated to PBH formation and the computation of the PBH mass distribution.  We start with the standard formation formalism but then we summarize and include in our calculations the results of the most recent and accurate studies based on simulations in numerical relativity, such as non-linear and non-Gaussian effects, the impact of the shape of the spectrum and the curvature/overdensity profiles, the changes in the equation of state linked to the thermal history, the mass and spin evolution after formation, etc.

Once the PBH mass distribution is known, the next important step is the calculation of PBH merger rates, covered in Section~\ref{sec:rates}.  PBH mergers are due to two dominant channels: early binaries that are formed before matter-radiation equality, and late binaries formed in PBH clusters.  We use the latest prescriptions, e.g., based on N-body simulations, in order to derive merger rates for different mass scales and for some representative models, including for very low mass ratios and intermediate-mass BH mergers. Finally, we consider the case of hyperbolic encounters in PBH clusters that can also lead to a large number of GW burst events.

In Section~\ref{sec:SGWBs}, we review the different sources of stochastic backgrounds related to PBHs. One particularly relevant source of SGWB comes from the curvature fluctuations that source GWs at second order in perturbation theory.  Other SGWBs are due to early or late PBH binaries as well as due to PBH isocurvature fluctuations.

In Section~\ref{sec:GWandLSS}, we review the possible ways to use the cross-correlations between gravitational wave observations of individual mergers or the stochastic background, and large-scale structure (LSS), in order to constrain the existence and abundance of PBHs.

The current limits on the abundance of PBHs from various probes, including GWs, are reviewed and discussed in Section~\ref{sec:limits}.  In particular, we aim to present them in a model-dependent way.   We emphasize and comment on the underlying hypotheses and discuss how these limits change according to the different classes of models.  

Finally, the detectability of various GW signatures from PBHs with LISA is discussed in Section~\ref{sec:LISA}, including individual merger events with extreme mass ratios or intermediate-masses, hyperbolic encounters, stochastic backgrounds, continuous waves, high-redshift signals, signal combination with other ground-based detectors, etc.  

At the end of each section, we provide an augmented discussion of recent results, the current limitations and pave the way for future analyses by listing some interesting perspectives.   Our conclusions are presented in Section~\ref{sec:ccl}.

\section{Theoretical models}  \label{sec:theory}

In this Section, we  provide an overview of the principal mechanisms that can lead to large curvature fluctuations and PBH formation.   We start with inflationary scenarios (see \cite{Ozsoy:2023ryl} for a recent review), either in single-field (quasi-inflection point) or multi-field realisations (hybrid inflation, turns in field space, gauge field interactions, etc.).  Where possible, we provide some general formulae for the calculation of the primordial power spectrum.  Then we consider recent models relying on quantum diffusion of the inflaton or of a stochastic spectator field, generically leading to non-Gaussian distributions of the curvature fluctuations.  The third part of this Section is dedicated to other models of PBH formation, due to curvaton fields, preheating, phase transitions, early matter era, cosmic strings, domain walls and primordial magnetic fields.  

\subsection{Single-field inflationary models}
\label{sec:single-field}

In the following, we provide a review of the main production mechanisms of primordial black holes in the context of single-field inflationary models.
\subsubsection{The basic idea and the slow-roll approximation}

The inflationary epoch provides an ideal setting for producing PBHs in the very early Universe, with single-field inflation being the minimal framework. While a nearly scale-invariant spectrum of primordial curvature perturbations is remarkably consistent with the observations of CMB temperature anisotropies at large scales, the spectrum on smaller scales remains largely unconstrained. When small-scale perturbations with sufficiently large overdensities re-enter the expanding horizon during the radiation-dominated epoch, their gradients induce a gravitational collapse that cannot be overcome even by the radiation pressure of the expanding plasma, thereby 
producing PBHs with a mass of the order of the horizon mass at the time of re-entry of a given wavemode, $k$. The abundance of PBHs crucially depends on the collapse process and the nature of primordial curvature perturbations, which are discussed later.  If PBHs start evaporating due to Hawking radiation right after their production, one finds that PBHs with mass $m_{\rm PBH} \gtrsim 10^{-18}\, M_{\odot}$ will survive until today and dominate the expansion of the Universe at matter-radiation equality unless their initial abundance is strongly suppressed. 
% {\bf SK: This may be not the case for subdominant PBHs (and I guess we want to include such possibilities in later sections). DW: now addressed I hope
%JGB: There is another issue, that extremely light PBH evaporate until their temperature drops below that of the surrounding thermal bath, which in the early universe can be very high. - DW: commented out. Left for future discussion!

Since single-field slow-roll inflationary models typically predict a nearly scale-invariant power spectrum of curvature perturbations on all scales, most slow-roll single-field inflation models do not produce PBHs, because a significant abundance of PBHs requires a substantial growth of the power spectrum on small scales. Nevertheless, 
a large positive running of the scalar primordial power spectrum can lead to PBH formation~\cite{Carr:1994ar}.  Another possibility is to have a transition in the slope of the potential, with the scalar field slowly rolling towards a region of almost constant potential, as one gets for instance in the original hybrid inflation model along the valley.  But the difficulty is then to end inflation in a limited number of e-folds while, in order to produce a large power, the potential is extremely flat.  Another possibility for producing PBHs in single-field inflation is through the violation of the usual slow-roll conditions~\cite{Ivanov:1994pa,Garcia-Bellido:2017mdw,Motohashi:2017kbs,Inomata:2021uqj,Inomata:2021tpx,Inomata:2022yte}, although one should then pay a special attention to  perturbative control \cite{Inomata:2022yte}. 
%{\bf JGB: One can imagine single-field models with running positive tilt that reach large amplitudes on small scales, w/o the need to violate SR conditions. RKJ: Yes, with large enough positive running, one can also obtain the desired amplitude of power spectrum at small scales. However, one needs to be sure that the Taylor expansion of the power spectrum about the pivot point makes sense. } 
Moreover, not all single-field inflationary models can produce the relevant abundance of PBHs as it crucially depends on the shape of the potential and various model parameters.  This typically leads to a high degree of fine-tuning of model parameters.

The power spectrum of curvature perturbations in single-field models is given at first order and in the slow-roll approximation, by
%\be
%\mathcal P_{\zeta}(k) = \frac{H_*^2}{ \pi m_{\rm pl}^2 \epsilon_{1*}}, \hspace{1cm} {\rm where \hspace{1cm}} \epsilon_1 \equiv - \frac{\dot H}{H^2}
%\ee 
\be
\mathcal P_{\zeta}(k) = \frac{H_*^2}{ \pi m_{\rm Pl}^2 \epsilon_{1*}} \left[ 1 - 2(C+2) \epsilon_{1*} + C \epsilon_{2*}-(2 \epsilon_{1*} + \epsilon_{2*}) \ln \left( \frac{k}{k_*} \right) \right]~,
\ee
{where $C = 0.578 +2 \ln 2- 2$, $\epsilon_1 \equiv - \dot H /H^2$ is the first slow-roll {Hubble-flow} parameter, $\epsilon_2 \equiv {\rm d} \ln \epsilon_1 / {\rm d} N$ {is the second Hubble-flow parameter}, $m_{\rm Pl}$ is the Planck mass {(and $M_{\rm Pl} = \sqrt{8 \pi}~ m_{\rm Pl }$ denotes the reduced Planck mass)} and a star denotes a quantity evaluated at the time the comoving wavenumber, $k_*$, crosses the Hubble radius during inflation when $k_*=a_*H_*$.  Therefore, one can naively expect that an enhancement of $\mathcal P_{\zeta}$ can occur if there is a dynamical phase during the inflationary expansion in which $\epsilon_1$ becomes much smaller than unity. Such an epoch is often called an \textit{ultra slow-roll} (USR) phase which can be achieved by means of a plateau region in the inflationary potential, which slows down the inflaton field more rapidly than in slow roll,
before ending inflation. An efficient and interesting way to produce this plateau is with the introduction of an inflection point in the potential. However, one must ensure that the inflaton does not stay too long at the inflection point otherwise all the inflationary fluctuations that had successfully imprinted the metric perturbations on large scales to explain the CMB anisotropies will be expanded away. A schematic plot of the potential with an inflection point is shown in Figure~\ref{potential}. 
\begin{figure}[t]
\begin{center}
\includegraphics[width=7.7cm, height=6.0cm]{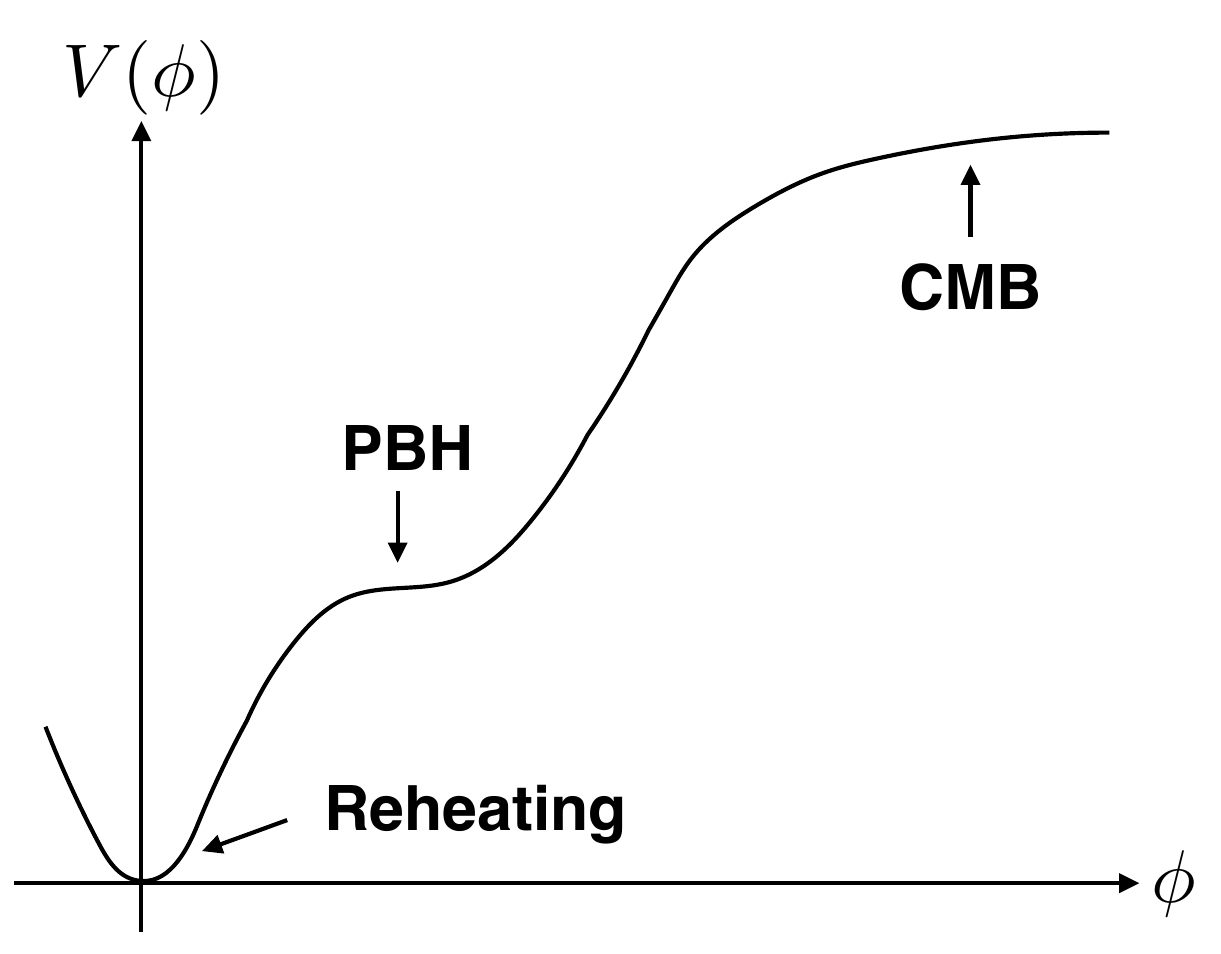}
\caption{A schematic plot of the inflationary potential with an inflection point. Adapted from Ref.~\cite{Garcia-Bellido:2017fdg}. }
\label{potential}
\end{center}
\end{figure}

\subsubsection{Inflection point potentials}

The simplest way to achieve an inflection or a near-inflection point in the potential is to use a suitable polynomial potential or ratio of polynomials so as to obtain the desired nearly scale invariant spectrum on CMB scales~\cite{Garcia-Bellido:2017mdw}. A very generic feature of such a potential is that although the inflaton starts in a slow-roll regime, it dynamically enters into an USR stage close to the inflection point for a short duration of e-folds that induces an enhancement in the spectrum. Nevertheless, it is possible to evade the USR regime in quasi-inflection point potentials~\cite{Ezquiaga:2017fvi}. However, one must always be careful that this new feature in the potential does not affect the large-scale dynamics which is already strongly constrained by CMB and large-scale structure (LSS). These types of potentials arise naturally in models of Higgs inflation \cite{Bezrukov:2007ep,Ezquiaga:2017fvi}, MSSM inflation \cite{Allahverdi:2006iq, Allahverdi:2006we}, accidental inflation \cite{Linde:2007jn}, 
string inflation \cite{Cicoli:2018asa,Ozsoy:2018flq}, $\alpha$-attractor models~\cite{Dalianis:2018frf,Iacconi:2021ltm,Kallosh:2022vha}, 
punctuated inflation \cite{Jain:2008dw, Jain:2009pm}, non-canonical inflation~\cite{Kamenshchik:2018sig,Lin:2020goi,Papanikolaou:2022did}, as an effective description of multifield inflation \cite{Geller:2022nkr} and in more general inflection point inflation \cite{Hotchkiss:2011am}. However, to produce sufficient PBHs in these models, one crucial difference is that, rather than imposing the initial conditions on the inflaton evolution close to the inflection point, one must assume that the field starts in the slow-roll regime at larger values of the potential well above the inflection point, and then slow-rolls down towards the minimum of the potential, crossing the inflection point, typically $30$ to $40$ e-folds before the end of inflation. Such a dynamical evolution must ensure that large-scale dynamics are consistent with the CMB constraints, PBHs are produced at some intermediate smaller scales during the USR regime and inflation naturally ends when the field finally rolls down towards the true minimum of the potential.
In this context, note that there has been an intense discussion recently on loop corrections induced on CMB scales by enhanced scalar perturbations on PBH scales \cite{Kristiano:2022maq,Riotto:2023hoz,Kristiano:2023scm,Riotto:2023gpm,Firouzjahi:2023aum,Firouzjahi:2023ahg,Franciolini:2023lgy,Tasinato:2023ukp,Cheng:2023ikq}.
Hints for significant one-loop contribution, to the point of undermining the consistency of such set-ups, have appeared in computations including a subset of vertices as derived from the interaction Hamiltonian. However, it has been recently shown that these effects disappear due to cancellations once all relevant interaction terms, at least at cubic order, are taken into account \cite{Fumagalli:2023hpa}, and this subject will likely deserve further studies. Note also that enhanced fluctuations can truly lead, through loop effects, to an increase of power on (intermediate) larger scales due to infrared rescattering \cite{Fumagalli:2023loc}, in single-field models and beyond, with potentially important consequences for the PBH population.

\subsubsection{Example model: critical Higgs inflation}

Critical Higgs Inflation (CHI) is probably among the best physically-motivated single-field models of inflation which can give rise to a quasi-inflection point \cite{Ezquiaga:2017fvi} and a high plateau in the matter power spectrum. The non-minimally coupled Higgs action is given by
\begin{equation}
    S\!=\!\!\int \!{\rm d}^{4}x\sqrt{g} \lb \lp\frac{M_{\rm Pl}^2}{2}
+\frac{\xi(\phi)}{2}\phi^{2}\rp\!R -\frac{1}{2}(\partial\phi)^{2}-\frac{1}{4}\lambda(\phi)\phi^{4}\rb
\end{equation}
 with the running of the couplings  is parametrized by
\begin{eqnarray}
\lambda(\phi)&=&\lambda_0+b_\lambda \ln^{2}\lp\phi/\mu\rp\,, \label{eq:RGEl} \\[2mm] \label{eq:RGExi} 
\xi(\phi)&=&\xi_0+b_\xi \ln\lp\phi/\mu\rp\,,
\end{eqnarray}
around the critical point $\phi=\mu$. %Here $\kappa^2\equiv8\pi G$.

After standard metric and scalar field redefinitions, 
% \begin{eqnarray}
%  g_{\mu\nu} &\to& \left(1+\xi(\phi)\phi^2\right)\,g_{\mu\nu}\,,\\[2mm]
% \phi &\to & \varphi=\hspace{-2pt}\int\hspace{-2pt} {\rm d}\phi \frac{\sqrt{1+\xi(\phi)\phi^2(1+6(\xi(\phi)+\phi\xi'(\phi)/2)^2)}}
% {1+\xi(\phi)\phi^2}\,,
% \end{eqnarray}
% \color{blue}
\begin{eqnarray}
 g_{\mu\nu} &\to& \left(1+\Mpl^{-2}\xi(\phi)\phi^2\right)\,g_{\mu\nu}\,,\\[2mm]
\phi &\to & \varphi=\hspace{-2pt}\int\hspace{-2pt} {\rm d}\phi \frac{\sqrt{1+\Mpl^{-2}{\xi(\phi)\phi^2}\left(1+6\left(\xi(\phi)+\frac12{\phi\xi'(\phi)}\right)^2\right)}}
{1+\Mpl^{-2} {\xi(\phi)\phi^2}}\,,
\end{eqnarray}
% {CJ: $\phi$ is dimensionless  in eq. 2.5/2.6, but not in 2.2} \\
% \color{black}
the effective inflationary potential becomes
\begin{equation}
    V(x)=\frac{V_0\,(1+a\,\ln^{2}x)\,x^{4}}{(1+c\,(1+b\,\ln x)\,x^{2})^{2}}\,,
\label{eq:potential}
\end{equation}
with $V_0 = \lambda_0\mu^4/4$, $a=b_\lambda/\lambda_0$, 
$b=b_\xi/\xi_0$ and $c=\xi_0\,\Mpl^{-2}\mu^2$. The potential has a flat plateau 
at large values of the field $x=\phi/\mu$, see Fig.~\ref{potential}, where
$V(x\gg x_{\rm c}) \simeq V_0 \,\frac{a}{(b\,c)^2} = \frac{\Mpl^4}{4}\,\frac{b_\lambda}{b_\xi^2} \ll M_{\rm Pl}^4\,.$
The potential also has a short secondary plateau around the critical point, $\phi_{\rm c} = \mu$, where the inflaton-Higgs slows down and induces a large peak in the curvature power spectrum.  This second plateau is induced by a quasi-inflection point at $x=x_{\rm c}$, where $V'(x_{\rm c}) \simeq 0, \, V''(x_{\rm c}) \simeq 0$. As a consequence, the number of $e$-folds has a sharp jump, $\Delta N$, at that point, plus a slow rise towards larger field values, corresponding to CMB scales.
This behavior is very similar to the one discussed in Ref.~\cite{Garcia-Bellido:2017mdw}.

In order to have large PBH production and good agreement with the CMB constraints, the allowed range of CHI parameters can be found to be $\lambda_0 \approx (0.01–8)\times 10^{-7}$, $\xi_0\approx(0.5-15)$, $x^2 \mu^2\approx(0.05–1.2)$, $b_\lambda \approx(0.008–4)\times 10^{-6}$ and $b_\xi \approx(1–18)$, for {the USR phase to occur} $N \in (30, 35)$ {e-folds before the end of inflation}. The question arises whether these values, corresponding to the model parameters at the critical scale $\mu$, are consistent with the values of the SM parameters at the EW scale. Given the latest values of {the top quark mass} $m_{\rm top}$ and {strong coupling} $\alpha_s$~\cite{CMS:2019esx,ATLAS:2019guf}, the values of $\lambda_0$ and $b_\lambda$ that we consider for the Higgs quartic coupling, are consistent, within $2\sigma$, with the existence of a critical point.

\begin{figure}[t]
\begin{center}
\includegraphics[width=10cm, height=7.0cm]{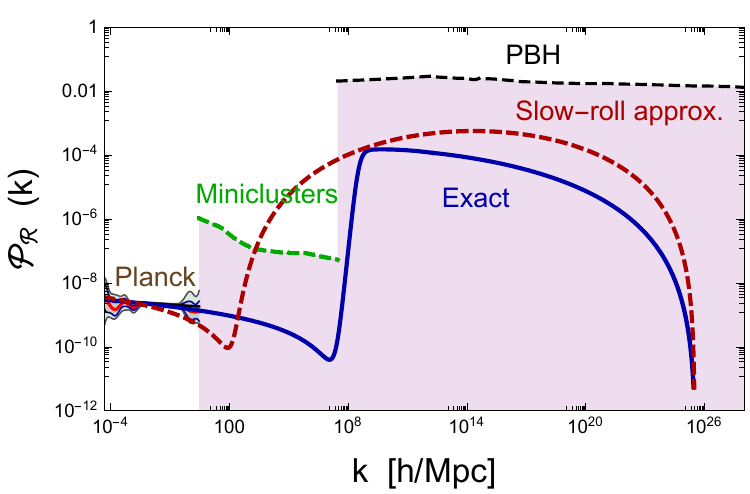}
\caption{Power spectrum $\mathcal{P}_{\mathcal{R}=\zeta} (k) $ for the single-field Critical Higgs Inflation model, with an inflection point at $N\approx 36$ satisfying the Planck 2018 constraints. Adapted from Ref.~\cite{Garcia-Bellido:2017mdw}.}
\label{PowerSpectrum}
\end{center}
\end{figure}

Taking into account the running of both the Higgs self-coupling $\lambda$ and the non-minimal coupling to gravity $\xi$, there are regions of parameter space allowed by the Standard Model for which the inflaton-Higgs potential acquires a second plateau at smaller scales, around the critical point $\lambda(\mu) \simeq \beta_\lambda(\mu) = 0$. This plateau gives a super-slow-roll evolution of the Higgs, inducing a high peak in the curvature power spectrum which is very broad. When those fluctuations reenter the horizon during the radiation era, they collapse to form PBHs with masses in the range $0.1$ to $100\,\Msun$.

In single-field models with a near-inflection point, the inflaton slows down right before the end of inflation, creating a stronger backreaction and a quick growth in curvature fluctuations, giving rise to a significant increase in the power spectrum at scales much smaller than those probed by CMB and LSS observations. 
%The exact expression of the power spectrum is given by
%$$P(k) =\frac{\kappa^2 H^2(\phi)}{8 \pi^2 \epsilon(\phi)},$$
%where $\epsilon(\phi)$ is the slow-roll parameter during inflation.
Since the spectral amplitude is essentially inversely proportional to the parameter $\epsilon_1$, whose exact evolution can deviate significantly from the slow-roll approximation \cite{Garcia-Bellido:2017mdw}, the power spectrum has to be calculated by integrating numerically
the evolution of the inflaton field $\phi$ following the exact (beyond slow-roll) equations.
In Fig.~\ref{PowerSpectrum}, we show a typical $\mathcal{P}(k)$ produced in quasi-inflection-point models like CHI~\cite{Ezquiaga:2017fvi}.  One can notice the difference between the exact power spectrum and that obtained in the slow-roll approximation.  The power spectrum can be parametrized as a double step with different amplitudes and different tilts:
\begin{equation}
{\mathcal P}_\zeta(k) = \left\{
   \begin{array}{cc}
     A_1 \left(\frac{k}{k_1}\right)^{n_{{\rm s}1}-1} \hspace{5mm} &{\rm for} \ \ k < k_{\rm c}
     \\[2mm]
     A_2\left(\frac{k}{k_2}\right)^{n_{{\rm s}2}-1}  \hspace{5mm} &{\rm for} \ \ k > k_{\rm c}
\end{array}
\right.
\end{equation}
where $A_i$, $n_{{\rm s}i}$, {$k_i$ are two different amplitudes, spectral tilts and arbitrary pivot scales} for the power spectrum, before and after the cut at $k_{\rm c}$. It is also possible to introduce some small amount of running of the tilt.
This power spectrum neglects quantum diffusion, which induces non-Gaussian contributions \cite{Biagetti:2018pjj,Ezquiaga:2018gbw} and exponential tails~\cite{Ezquiaga:2019ftu}, which can be responsible for a large probability of PBH formation even with a relatively small power spectrum amplitude, as will be discussed in subsequent sections.

\begin{figure}[!t]
% \vspace*{-3cm}
\centering 
% \hspace*{-1cm}
\includegraphics[width = 0.8\textwidth]{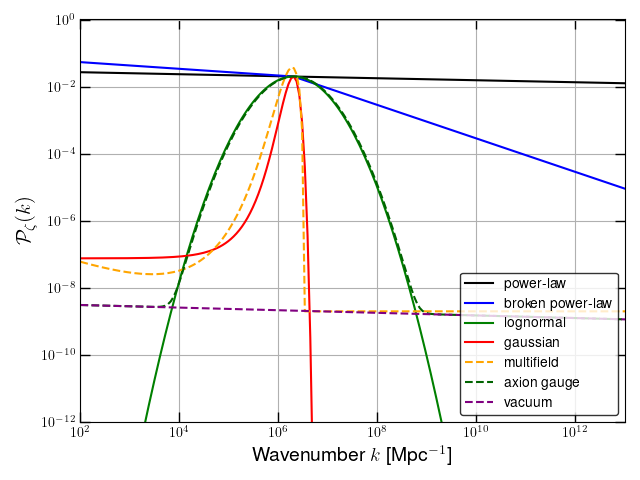}
% \vspace*{-3cm}
\caption{Examples of primordial power spectra of curvature fluctuations, leading to (stellar-mass) PBH formation:  power-law, broken power-law, Gaussian and log-normal models (solid lines), and particular examples of multifield and axion-gauge models (dashed lines).  
%{GF: this plot should be adjusted. FIRAS data force the spectrum to be ${\cal P}_\zeta <10^{-4}$ for $k \lesssim 10^{4}$ Mpc$^{-1}$.} {CU: Axion gauge curve, broken-law and multifield here are not correct. Axion gauge is a log-normal like bump or growing to smaller scales not and decrease. Broken-Law curve produces so much PBH at large masses. Gaussian curve is also not correct since the scales less than $10^4$ produces SMBHs too many! }
 % {CJ:Updated lines for gaussian, axion-gauge and multifield, now consistent with FIRAS.}
}
\label{fig:Pzetas}
\end{figure}

\subsubsection{Reverse engineering approach}

In single-field inflationary models, the commonly adopted strategy is to start from some scalar potential that features an approximate stationary inflection point, as shown in Fig.~\ref{potential}; 
the latter is controlled by various
free parameters, which are fine-tuned to 
guarantee the desired enhancement 
of the power spectrum of curvature perturbations \cite{Inomata:2016rbd,Garcia-Bellido:2017mdw,Ballesteros:2017fsr,Hertzberg:2017dkh,Kannike:2017bxn,Dalianis:2018frf,Inomata:2018cht,Cheong:2019vzl,Ballesteros:2020qam,Iacconi:2021ltm,Kawai:2021edk}.  
Following Ref.~\cite{Franciolini:2022pav}, 
a different, and much more powerful, approach can be entertained. 
In a reverse engineering procedure, 
the model building can be moved away from the 
scalar potential---that from this perspective is the output of the analysis instead of the starting point---and focused on the underlying inflationary dynamics (see refs.\,\cite{Ragavendra:2020sop,Tasinato:2020vdk,Ng:2021hll,Karam:2022nym} for a similar viewpoint).

The inflationary background dynamics can be captured by specifying a minimal set of inputs. 
Consider the Hubble parameters $\epsilon_1$ and 
$\eta \equiv  \epsilon_1 - \epsilon_2/2$,
and the number of e-folds $N$, defined by $\mathrm{d}N = H\mathrm{d}t$. 
Ref.~\cite{Franciolini:2022pav}
devised the following analytical {\it ansatz} 
for the time-evolution of $\eta$
\begin{align}
2 \eta(N) = 
& \left[
\eta_{\rm I} - \eta_{\rm II} + (\eta_{\rm II}-\eta_{\rm I})\tanh\left(\frac{N - N_{\rm I}}{\delta N_{\rm I}}\right)
\right] + 
% \nonumber \\ &
\left[
\eta_{\rm II} + \eta_{\rm III} + (\eta_{\rm III}-\eta_{\rm II})\tanh\left(\frac{N - N_{\rm II}}{\delta N_{\rm II}}\right)
\right] + \nonumber \\
&  \left[
\eta_{\rm IV}-\eta_{\rm III} + (\eta_{\rm IV}-\eta_{\rm III})\tanh\left(\frac{N - N_{\rm III}}{\delta N_{\rm III}}\right)
\right]\,.\label{eq:MainEqEta}
\end{align}
The inflationary dynamics is therefore divided in different stages: 
{\it i)} a first stage where $\eta_{\rm I}$ is constant and negative (necessarily taken to be small to reproduce the  conventional slow-roll dynamics) 
so that %$\epsilon$ 
{$\epsilon_1$} increases relatively gently; 
{\it ii)}
a USR phase characterized by 
%$\eta_{\rm II} > (3+\epsilon)/2 \simeq 3/2$
{$\eta_{\rm II} > (3+\epsilon_1)/2 \simeq 3/2$}
, where negative friction 
makes the parameter 
%$\epsilon$ 
{$\epsilon_1$} decrease abruptly
{\it iii)} an optional intermediate stage 
where 
%$\epsilon$ 
{$\epsilon_1$}
remains constant, 
achieved if $\eta_{\rm III}\simeq 0$, thus generating a second slow roll phase 
(i.e. an enhanced plateau in the curvature power spectrum \cite{DeLuca:2020agl,Franciolini:2022pav});
{\it iv)} a final phase characterized by 
$\eta_{\rm IV} < 0$, that brings 
%$\epsilon$ 
{$\epsilon_1$}
back to 
$O(1)$ values for inflation to end.  
The parameters $\delta N_{\rm I,II,III}$ 
control the sharpness of the transitions.

As shown in Refs.~\cite{Franciolini:2022pav,Franciolini:2022tfm}, it is possible to choose the parameters controlling the ansatz {of Eq.}~\eqref{eq:MainEqEta} in such a way that the power spectrum of curvature perturbations is compatible with CMB constraints at large scales as well as featuring an enhanced peak or plateau at scales beyond the FIRAS constraints 
(i.e. for $k\gtrsim  10^{4} / {\rm Mpc}$), where a sizeable PBH abundance can be obtained,
corresponding to the possible formation
of masses smaller than ${\cal O}(10^4)M_\odot$.

Once the Hubble parameters are known, one can compute the inflationary potential by means of 
\begin{align}
\frac{V(N)}{V(N_{\rm ref})} & =\exp\left\{
   -2\int_{N_{\rm ref}}^{N}dN^{\prime}\left[\frac{\epsilon_1(3-\eta)}{3-\epsilon_1}\right]
   \right\}\,,
   ~~~~~~~
\phi(N)  = \phi(N_{\rm ref}) \pm \int_{N_{\rm ref}}^N dN^{\prime}\sqrt{2\epsilon_1}\,,   \label{eq:recPot1}
% \label{eq:recPot2}
\end{align}
{where $N_{\rm ref}$ denotes an arbitrary reference e-fold time, and where} in the second equation we consider the minus sign having in mind a large-field model in which the field value decreases as inflation proceeds. 
Combining $V(N)$ and $\phi(N)$, we reconstruct the 
profile $V(\phi)$ of the inflationary potential in field space\,\cite{Byrnes:2018txb}. 
Eq.\,\eqref{eq:recPot1} shows the convenience of modelling the inflationary dynamics directly at the level of $\eta$ instead of $V(\phi)$. 
This is because the Hubble parameters enter in the exponent in the definition of $V(N)$, 
and thus allow for a much finer control on power spectral features 
when performing the reverse engineering procedure. 
This approach facilitates model building, 
even though it does not provide
an interpretation of the reconstructed potential 
in terms of high-energy theories of inflation.

\subsection{Multi-field inflationary models}
As seen above, a common challenge of PBH formation within single-field inflationary models is keeping control over the two 
% perturbation 
phases that lead respectively to the seeds for the CMB at large cosmic scales and for PBHs at smaller scales.

Multi-field inflationary scenarios can help separate and better control these two phases of perturbation generation, while also being very natural candidates from the UV perspective. Furthermore, violation of the standard slow-roll conditions can be efficiently realized in a context with more degrees of freedom~\cite{Clesse:2015wea}.

\subsubsection{Hybrid inflation}

In a setup with two scalar fields, the most studied scenario is perhaps {\it hybrid inflation}~\cite{Linde:1993cn,Copeland:1994vg}.  In this class of models, inflation takes place along an almost flat valley in the direction of an initial slowly rolling scalar field, $\phi$.  Inflation ends with a waterfall instability when $\phi$ crosses a critical point, $\phi_{\rm c}$, due to the presence of a second field, $\psi$.   The original hybrid potential can be written as 
\be
V(\phi,\psi) = \Lambda^4 \left[ \left(1 - \frac{\psi^2}{M^2} + \frac{\phi^2}{\mu^2} + 2 \frac{\phi^2 \psi^2}{\phi_{\rm c}^2 M^2}\right)  \right].
\ee
The general shape of the hybrid potential is shown in Fig.~\ref{HybridModel}.  

\begin{figure}[b!]
\begin{center}
\includegraphics[width=8cm]{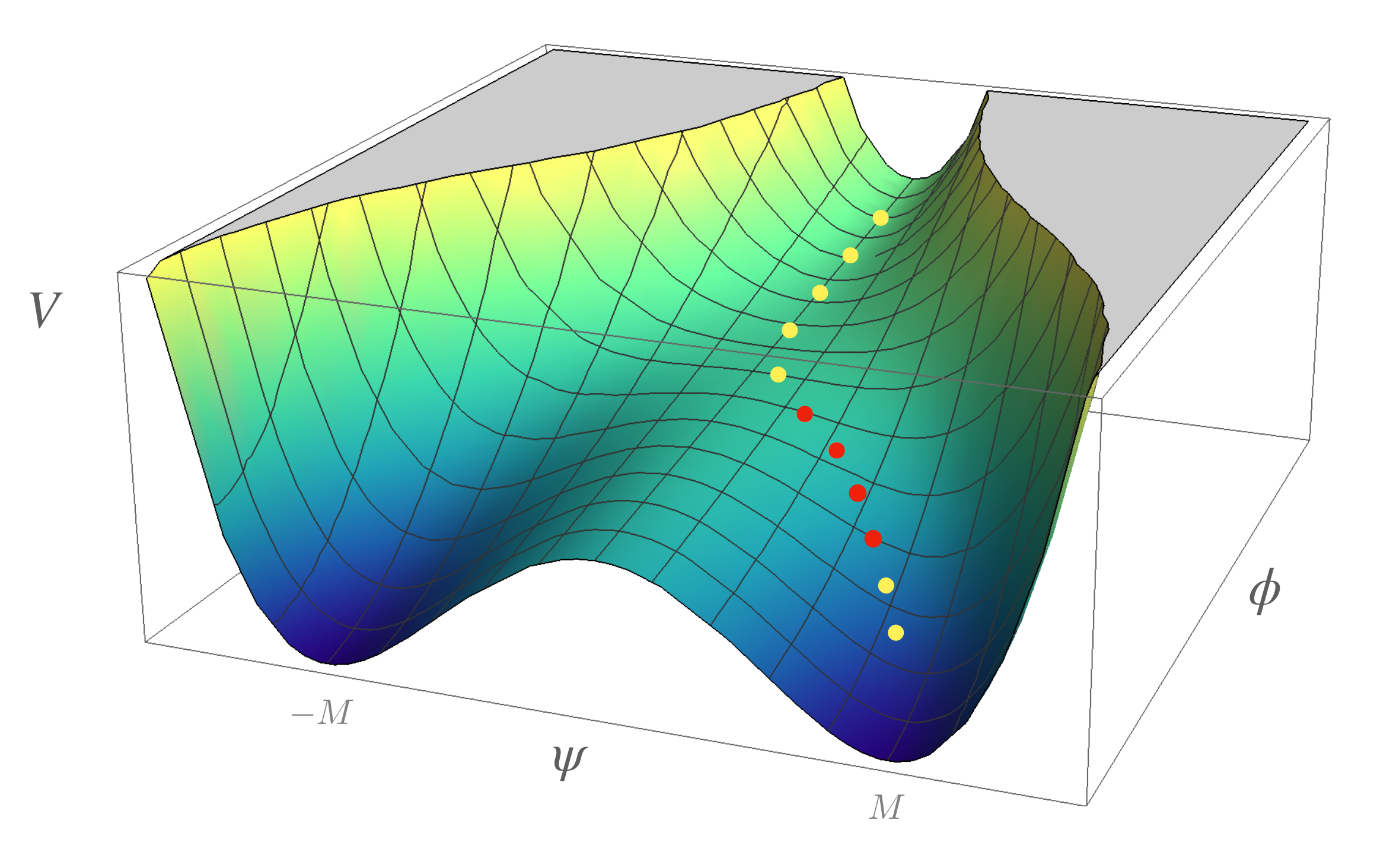}
\caption{Representation of a typical hybrid inflation potential, $V$, with a possible trajectory in two-field space (dotted line). CMB perturbations are created along the valley at $\psi=0$, during a first phase of inflation. Curvature perturbations suitable for production of PBHs are generated in a second flat part of the potential (red dotted line), {when} the mass-squared of the field $\psi$ changes sign.}
\label{HybridModel}
\end{center}
\end{figure}

When $\mu M \ll M_{\rm Pl}^2$, the waterfall phase transition is nearly instantaneous and lasts less than one e-fold.  The global minima of the potential are located at $\phi = 0$ and $\psi = \pm M$.  Inflation can take place along the valley $\psi = 0$ with the effective potential
\be
V(\phi,\psi) = \Lambda^4 \left( 1 + \frac{\phi^2}{\mu^2} \right)
\ee
that most often produces a \textit{blue-tilted} power spectrum for the original potential (a \textit{red-tilted} spectrum can nevertheless be obtained for some specific parameter values~\cite{Clesse:2015wea}).  Other hybrid realisations, like F-term or D-term inflation, can produce other shapes for the effective potential, leading to a \textit{red-tilted} spectrum, compatible with the Planck observations (see e.g. \cite{Braglia:2022phb}, which  realises this scenario via hybrid $\alpha$-attractors models with scalar potentials with no tachyonic directions and, therefore, bounded from below). PBHs of tiny mass can be produced during a fast waterfall phase~\cite{GarciaBellido:1996qt}, which would have evaporated today, eventually leading to the reheating of the Universe.

In the other regime $\mu M \gg M_{\rm Pl}^2$, the waterfall phase is slow, lasting $N \gg 60$ e-folds~\cite{Clesse:2010iz}, and it is then crucial to consider the exact two-field dynamics to calculate the shape of the primordial power spectrum, which is however {not} compatible with Planck CMB measurements~\cite{Clesse:2013jra}.
 
Hybrid inflation becomes interesting for the production of massive PBHs in the transient regime, 
when $\mu M \sim M_{\rm Pl}^2$, producing a significant but not too large number of e-folds, $10 \lesssim N \lesssim 40$.   
During the mild waterfall phase, the multi-field perturbation dynamics leads to a strong, exponential power spectrum enhancement (that can be sharper than the typical maximal $\propto k^4$ increase in single-field models~\cite{Byrnes:2018txb}) followed by a relatively sharp decrease, when going from large to small scales.   
The position and amplitude of the resulting broad peak is driven by a single combination of the hybrid potential parameters, $\Pi \equiv M \sqrt{\phi_{\rm c} \mu_1} / M_{\rm Pl}^2$ where $\mu_1$ comes from a linear expansion of the potential around the critical point $V\propto (\phi - \phi_{\rm c}) / \mu_1$.  
Numerical calculations of the primordial power spectrum of curvature fluctuations using either the exact multi-field perturbation dynamics or the $\delta N$ formalism, were first done in Ref.~\cite{Clesse:2015wea}. The resulting power spectrum is well approximated by a log-normal in mass spectrum, 
\be
\mathcal P_\zeta (k) \approx P_{\rm p} \exp \left[ - \frac {(N_k - N_{\rm p})^2 }{2 \sigma_{\rm p}^2} \right] \,,
\ee
of maximal amplitude $P_{\rm p}$ and width $\sigma_{\rm p} $. Some examples of power spectra are shown in Fig.~\ref{fig:Pzetahybrid}.   Finally, let us emphasize that the exact amplitude of the power spectrum also depends on the quantum stochastic dynamics of the auxiliary field $\psi$ close to the critical instability point.  In principle, this should lead to slightly different power spectrum amplitudes in different regions of the Universe, and therefore different PBH abundances.  A natural outcome of the hybrid inflation scenario would thus be that PBHs are formed in clusters whose mass and size depends on these quantum stochastic dynamics.  This can be also somehow used to reduce the fine-tuning linked to the required power spectrum amplitude, since many realisations would have been generically produced within the Universe.

\begin{figure}[t]
\begin{center}
\includegraphics[width=10cm, height=7.0cm]{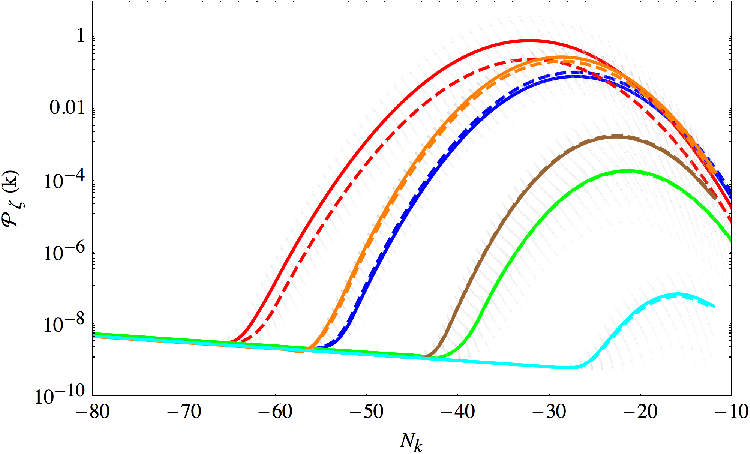}
\caption{Power spectrum of curvature perturbations for hybrid inflation parameters values $M = 0.1 \Mpl$, $\mu_1 = 3 \times 10^5 \Mpl$ and $\phi_{\rm c} = 0.125 \Mpl $ (red)$, 0.1 \Mpl $ (blue), $0.075 \Mpl $ (green), and $0.05 \Mpl $ (cyan).  Those parameters correspond respectively to $\Pi^2 = 375 / 300 / 225/150$.  The power spectrum is degenerate for lower values of $M,\phi$ and larger values of $\mu_1$, keeping the combination $\Pi^2$ constant.  For larger values of $M, \phi_{\rm c}$ the degeneracy is broken: power spectra in orange and brown are obtained respectively for $ M = \phi_{\rm c} = \Mpl$ and $ \mu_1 = 300 \Mpl / 225 \Mpl$.  Dashed lines assume $\psi_c = \psi_0$ whereas solid lines are obtained after averaging over 200 power spectra obtained from initial conditions on $\psi_c$ distributed according to a Gaussian of width $\psi_0$.  The power spectra corresponding to these realizations are plotted in dashed light gray for illustration.  The $\Lambda$ parameter has been fixed so that the spectrum amplitude on CMB anisotropy scales is in agreement with Planck data.  The parameter $\mu_2 = 10 \Mpl$ so that the scalar spectral index on those scales is given by $n_{\rm s} = 0.96$. 
Figure from Ref.~\cite{Clesse:2015wea}.
}
\label{fig:Pzetahybrid}
\end{center}
\end{figure}

\subsubsection{Turns in multi-field inflation}

Embeddings of inflation in high energy physics motivate scenarios in which multiple scalar fields participate in the inflationary dynamics. An ubiquitous phenomenon in this context is the one of turns in field space, corresponding to a bending of the inflationary trajectory, or more accurately to its deviation from a geodesic of the field space,\footnote{In the context of string theory compactifications, it has recently been shown \cite{Freigang:2023ogu} that deviations from geodesic trajectories must be negligible near the boundary of the moduli space. Large turns can therefore be engineered just in the moduli space bulk.} quantified by a dimensionless parameter $\eta_\perp(t)$ measuring the acceleration of the trajectory perpendicular to its direction \cite{GrootNibbelink:2000vx,GrootNibbelink:2001qt}. When such a turn is sufficiently strong, with $\eta_\perp \gtrsim 1$ (see \cite{Renaux-Petel:2021yxh} for a short overview), it leads to a transient tachyonic instability of the scalar field fluctuations before Hubble crossing \cite{Cremonini:2010ua,Brown:2017osf,
Garcia-Saenz:2018ifx}, similar in essence to the one arising in axion gauge-field inflation (see section \ref{sec:axion-gauge}). 
This gives rise to an exponential enhancement of the power spectrum on those scales, compared to the ones that cross the Hubble radius before and after the turn, and thus constitutes a natural mechanism to seed PBH formation \cite{Palma:2020ejf,Fumagalli:2020adf,Braglia:2020eai,Iacconi:2021ltm}. Interestingly, owing to its inherently multi-field origin, the growth rate of the power spectrum can overcome the bound deduced in single-field setups
\cite{Byrnes:2018txb,Carrilho:2019oqg,Ozsoy:2019lyy}, i.e., the peak of the power spectrum can be sharper. An important parameter is the duration of the turn in \textit{e}-folds $\delta\equiv \delta N$, which leads to qualitatively different behaviours for broad ($\delta \gtrsim 1$) and sharp turns ($\delta \ll 1$). While broad turns lead to a featureless bump of the power spectrum, sharp turns lead to a localised peak of the envelope of the power spectrum, modulated by order-one rapid oscillations, as described in simple cases by the analytical power spectrum (assuming a constant value of $\eta_\perp$ during the turn) \cite{Palma:2020ejf,Fumagalli:2020nvq}
\begin{equation}
\label{eq:P-analytic-xi}
\frac{\mathcal{P}_\zeta(k)}{{\cal P}_0} =   \, \frac{e^{2 \sqrt{(2-\kappa)\kappa} \, \eta_\perp \delta} }{2 (2-\kappa) \kappa}
\times \sin^2\left[e^{-\delta/2}\kappa \eta_\perp+\arctan\left(\frac{\kappa}{\sqrt{(2-\kappa)\kappa}}\right) \right]\,.
\end{equation}
Here, ${\cal P}_0$ denotes the amplitude of the power spectrum in the absence of the transient instability, $\kappa=k/k_\textrm{f}$ is the dimensionless wavenumber in units of the maximally enhanced one, with $k_\textrm{f}$ the scale crossing the Hubble radius at the time of the sharp turn, and the formula is valid for $\kappa \leq 2$ (see \cite{Fumagalli:2020nvq} for generalisations). The oscillations there, which are to a very good approximation periodic in $k$, are characteristic of a sharp feature, and go hand in hand with a boosted power spectrum generated by an event localised in time \cite{Fumagalli:2020nvq}. 
The mass function of PBHs is highly sensitive to the tail of the probability density function (PDF) of the smoothed density contrast, which is not yet known for the class of models described here. However, tentative results for the mass function indicate that the oscillatory patterns are washed out when assuming Gaussian statistics of the primordial curvature fluctuations \cite{Fumagalli:2020adf}, and hence further investigation is required to assess the observational consequences for PBHs of the oscillatory features. What is known, however, is that the SGWB generated by scalar fluctuations at horizon re-entry does keep an imprint of the oscillations of the scalar power spectrum: they manifest as oscillations in the frequency profile of the SGWB, providing a robust probe of small-scale features that would be inaccessible otherwise \cite{Fumagalli:2020nvq,Braglia:2020taf,Fumagalli:2021cel,Witkowski:2021raz,Fumagalli:2021dtd}.

If one uses the standard criterion that the power spectrum has to reach values ${\cal P}_\zeta \sim 10^{-2}$ to generate a substantial amount of PBHs, it is interesting that the required enhancement from the baseline value ${\cal P}_0 \sim 10^{-9}$ necessitates
a total angle $\Delta\theta= \eta_\perp \delta$ swept by the trajectory during the turn exceeding the value $\pi$ obtainable in flat field space, and hence requires the inflationary field space to be curved \cite{Palma:2020ejf}.
Such large values of the power spectrum are not without consequences however, and considerations of backreaction and perturbative control are important to take into account to delineate the space of theoretically viable models \cite{Fumagalli:2020nvq}. Indeed, the criterion for perturbative control parametrically  depends on the large parameter $\eta_\perp$ in a way that imposes the bound $\eta_\perp^6 {\cal P}_\zeta \lesssim 1$ (or $\eta_\perp^4 {\cal P}_\zeta \lesssim 1$) for broad \cite{Fumagalli:2019noh,Bjorkmo:2019qno} (or sharp) turns \cite{Fumagalli:2020nvq}. 
Thus, an important open question is whether the inevitably present non-Gaussianities of the curvature fluctuations in these models, enhanced in flattened configurations \cite{Garcia-Saenz:2018vqf,Fumagalli:2019noh,Ferreira:2020qkf}, are such that a substantial amount of PBHs can be generated with a value of the power spectrum less than $10^{-2}$ while ensuring perturbative control. 

\subsubsection{Two-stage models}

{There are several concrete examples of PBH formation resulting from a turn in the field space during inflation that separates two stages of inflation.  The first stage produces the primordial power spectrum on CMB scales while the second stage is responsible for the enhancement of power on smaller scales, leading to PBH formation.  Oscillations are expected at the transition, as explained in the previous Section.   Pi et al.~\cite{Pi:2017gih} have proposed a model combining Starobinsky inflation with a scalaron field that becomes massive at the end of the first stage of inflation and a curvature field that plays the role of the inflaton in a second stage.  Kawasaki et al~\cite{Kawasaki:2016pql} proposed a scenario with a second phase of double inflation in a supergravity framework.  One should notice that non-Gaussianity is typically small in these models, since most of the field trajectories are effectively single-field.  Other advantages of these models are that one does not especially need to tune the position of a peak in the power spectrum, since it can be nearly scale invariant, as generically predicted by slow-roll inflation, and that the transition between the two regimes can be very sharp.   Due to the features at the transition in the power spectrum, such models can also produce multi-modal distributions of black holes, as proposed in~\cite{Carr:2018poi}.  }

\subsubsection{Axion-gauge scenario} \label{sec:axion-gauge}

PBHs can also arise in models in which rolling axions interact with gauge fields, enhancing primordial density perturbations \cite{Linde:2012bt,Bugaev:2013fya,Domcke:2017fix,Garcia-Bellido:2016dkw,Garcia-Bellido:2017aan,Ozsoy:2020kat,Ozsoy:2020ccy}. In natural inflation \cite{Freese:1990rb,Adams:1992bn}, the couplings of the inflaton to matter respect a shift symmetry, $\phi \to \phi + {\rm constant}$. Therefore, they do not provide by themselves quantum corrections to the inflaton potential (that arise from a breaking of the shift symmetry, as for instance from instanton effects), which can therefore be technically naturally small and flat. Also, UV-completed theories, such as string theory or supergravity models of inflation, contain a large number of pseudo-scalar particles that could be identified with the axion inflaton, or some other dynamically relevant axion. The shift-symmetric coupling between the axion and the gauge field is 
\begin{equation}
    {\cal L}  \supset \frac{a}{f} F {\tilde F} ,
\label{aFFt}
\end{equation}
where $a$ is the pseudoscalar axion, $F$ is the gauge field strength, ${\tilde F}$ its dual, and $f$ is the axion decay constant. Shift-symmetry is satisfied due to the fact that $F {\tilde F}$ is a total derivative and the constant term becomes a topological invariant.
This interaction amplifies one helicity of the vector field around  horizon crossing. The amplified gauge quanta might affect the evolution of the axion field \cite{Anber:2009ua} or, even if produced in a smaller amount, source scalar density perturbations \cite{Barnaby:2011vw} and parity violating tensor modes \cite{Sorbo:2011rz}. These effects are exponentially sensitive to the parameter $\xi \equiv \frac{\dot{a}}{2 f H}$, and visible effects are obtained for $f \simeq 10^{-2} M_{\rm Pl}$, namely for an axion decay constant parametrically equal to the string scale. On one hand, this enhances the possibility that models of string cosmology have a rich phenomenology. On the other hand, no effect is expected for models of a single axion inflaton, that are required to have a super-Planckian excursion, unless some specific mechanism allows excursions much greater than the scale of inflation, such is the case for monodromy \cite{Silverstein:2008sg} or for aligned axion inflation \cite{Kim:2004rp}. 

Scalar perturbations are produced via inverse decay $A + A \to a$ (where $A$ denotes the gauge field), and therefore, they obey a non-Gaussian $\chi^2$ distribution. The amplified gauge fields also produce GWs via an analogous $2 \to 1$ process. We note that these primordial GWs, which are produced during inflation, form a different stochastic GW background with respect to the secondary GWs that are sourced by density perturbations in the process of horizon re-entry in radiation or matter-dominated eras. 

The power spectrum of density perturbations can be parametrized as \cite{Namba:2015gja,Peloso:2016gqs,Garcia-Bellido:2016dkw}
\begin{equation}
    \mathcal{P}_\zeta(k) = \mathcal{P}_{\rm vacuum}(k) + \mathcal{P}_{\rm sourced}(k) \; ,
\end{equation}
where $\mathcal{P}_{\rm vacuum} \simeq 2 \times 10^{-9} \, \left(k/k_{\rm cmb} \right)^{n_{\rm s}-1}$ is the standard nearly scale invariant vacuum signal (with $k_{\rm cmb}$ being some pivot wavenumber at CMB scales). As already mentioned, the sourced signal is exponentially sensitive to the parameter $\xi \equiv \dot{a}/{2 f H}$, and it is therefore highly sensitive to the speed of the axion, and, ultimately, to its potential. For axion inflation with a single monomial potential the signal is typically blue, and one simply needs to require that the production is sufficiently small so that no stable PBH is produced via this mechanism \cite{Linde:2012bt}. This does not need to be the case for a more complicated potential, or in the case in which the axion is not identified with the inflaton \cite{Garcia-Bellido:2016dkw,Garcia-Bellido:2017aan}. In this second case, one can imagine a situation in which the axion rolls significantly only for a limited number of e-folds \cite{Namba:2015gja,Peloso:2016gqs} and the sourced spectrum can be described as a nearly log-normal bump, peaked at the scales that were exiting the horizon while $a$ was rolling. The infra-red part of the spectrum scales as $\mathcal{P} \propto k^3$ due to causality, while, close to the peak, the spectrum is well parametrized by \cite{Namba:2015gja} 
\begin{equation}
   \mathcal{P}_{\rm sourced}(k)= {\cal A}(\xi_*, {\cal W}) \, {\rm exp} \left[ - \frac{\ln^2(k/k_*)}{2\sigma^2(\xi_*,{\cal W})} \right] \;,
\label{Ps-bump}
\end{equation}
 where the amplitude ${\cal A}$, the central position $k_*$, and the width $\sigma$ of the peak are functions of the maximum value of the particle production parameter $\xi$ and of the amount of e-folds ${\cal W}$ during which the axion is significantly rolling. An analogous parametrization also holds for the amount of sourced GWs \cite{Namba:2015gja}.

\begin{figure}[t!]
\begin{center}
\includegraphics[width=0.75\textwidth]{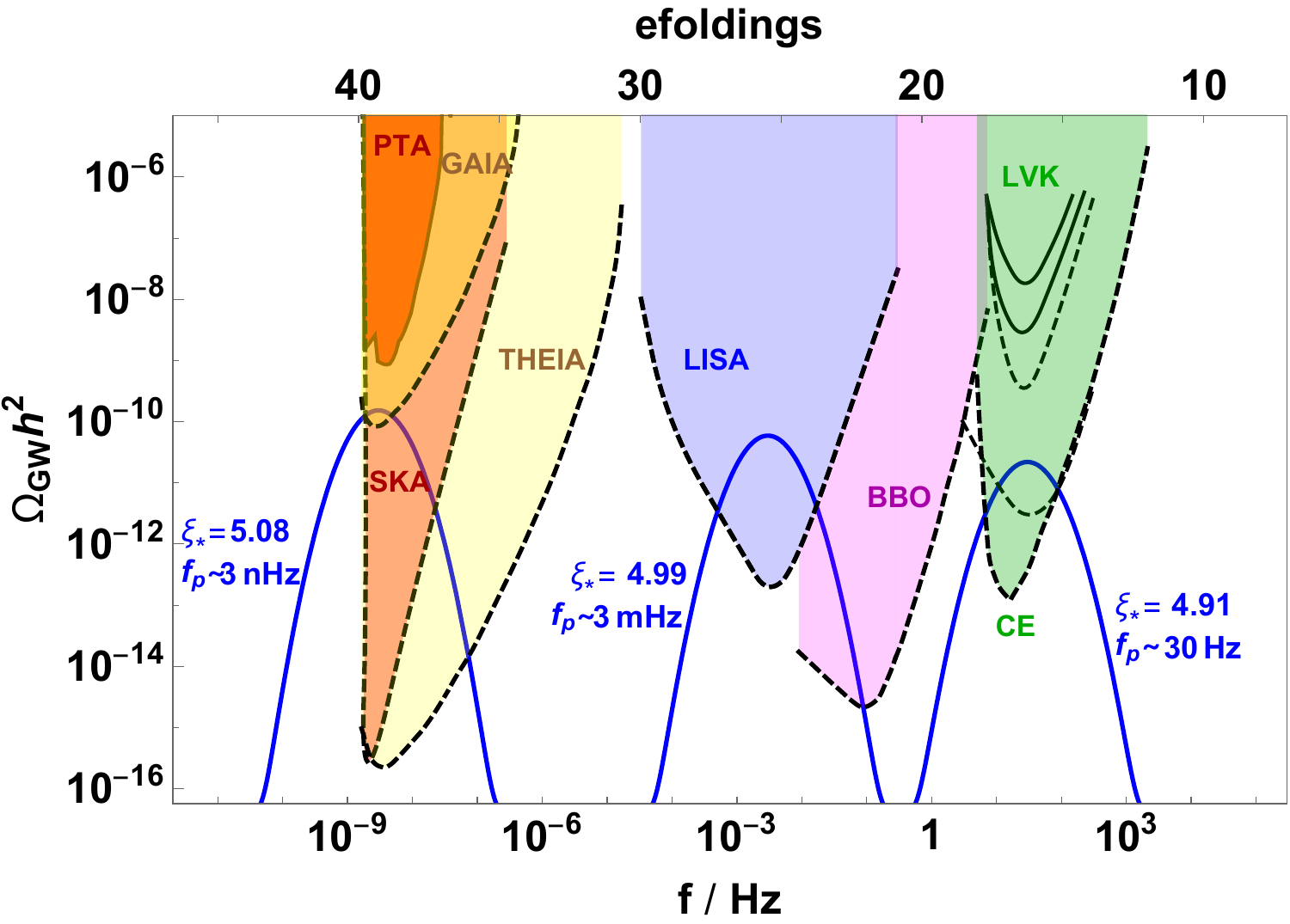}
\caption{The stochastic GWs produced by axion inflation at four main scales of interferometers {for which we show the approximate sensitivities}, nHz (PTAs and SKA), $\mu$Hz (Gaia and Theia), mHz {(LISA and BBO) and Hz (LVK, Einstein Telescope, and Cosmic Explorer)}. The enhanced density perturbations may produce PBHs which are a significant fraction of dark matter for $1-100 M_{\odot}$ and the totality in the $10^{-14}-10^{-11}M_{\odot}$ mass range and, remarkably, the enhanced perturbations leave inevitable GW backgrounds at the most sensitive regimes of GW detectors: for the first range of BHs this corresponds to PTA frequencies (and possibly future SKA) and LISA frequencies for the second range. The PBHs formed from fluctuations peaking at Hz scales will be so light that they are expected to be part of thermal history through Hawking radiation. The Figure is updated (with slight modifications) from Refs.~\cite{Garcia-Bellido:2017aan,Unal:2018yaa,Garcia-Bellido:2021zgu}}
\label{fig:Axioninflationpnhinterferometers}
\end{center}
\end{figure}

The sourced scalar perturbations are highly non-Gaussian 
\cite{Barnaby:2011vw}. For a nearly scale-invariant signal (as is the case if the axion is the inflaton), their bispectrum has a shape which is very close to the equilateral one. For the case of short-duration production leading to {Eq.~}(\ref{Ps-bump}), the bispectrum is a maximum when the magnitude of the three momenta is parametrically equal to $k_*$, and its scale-dependence can also be described accurately via a log-normal shape \cite{Namba:2015gja}. The fact that the sourced scalar perturbations obey a $\chi^2$ statistics makes PBH production much more significant compared to Gaussian perturbations with the same power. Therefore, an equal amount of PBHs is produced with a much smaller power spectrum than in the Gaussian case. This results in a significantly smaller amount of stochastic GWs produced at second order by the scalar perturbations, with a different prospect for detection at LISA and PTA-SKA, see for example Fig.~\ref{fig:Axioninflationpnhinterferometers}. This is discussed in more detail in Subsection~\ref{sec:secondordergw}.
 
\subsubsection{False-vacuum models}

{
It is also possible to form PBHs in both single and multi-field scenarios if some spatial regions end up in a false-vacuum.  
These regions can collapse into black holes before the field can pass the potential barrier through quantum tunneling and evolve towards the true vacuum.  Several examples have been proposed.   }

{
For instance, the Higgs field is suspected to be metastable.  If it is a spectator field during inflation, some regions may have ended in this region of the potential, thereby leading to the formation of black holes~\cite{Passaglia:2019ueo}.  Nevertheless, their abundance cannot be significant enough to contribute to the dark matter.   Another possibility is a tiny step in the inflaton potential that can basically be overcome by most of field trajectories, but not all~\cite{Cai:2021zsp}.  This can radically change the tail of the Gaussian fluctuations, even if the conventional non-linearity parameters ($f_{\rm NL}, g_{\rm NL}$,...) remain small.  The regions with large curvature fluctuations later form PBHs.}

{
It has also been proposed that a light spectator field during inflation, having an asymmetric polynomial potential, experiences  stochastic fluctuations that displace the field from the global minimum of its potential and populate a false vacuum state~\cite{Maeso:2021xvl}. Lattice simulations have been used to show how when the Hubble radius reaches the false-vacuum bubble size, the potential energy is transferred to the kinetic and gradient energies of the bubble wall, such that it begins to contract and possibly collapse into a black hole.   Finally, multi-field realisations have been proposed, for instance in~\cite{Garriga:2015fdk,Deng:2017uwc}.
}

\subsection{Non-Gaussian models}
\label{ssection:nG}

\subsubsection{Motivations}

In Subsection~\ref{sec:single-field} we discussed how PBHs could be generated in single-field models due to an enhancement of the power spectrum caused, for example, by a second plateau in the potential. 
However, as we will discuss in more detail in Subsection~\ref{sec:PBHformation}, what really matters to determine the PBH formation is the full profile of the PDF of the primordial curvature perturbations, in particular the probability to be above a given threshold. This implies that PBH abundance is largely controlled by the tails of the PDF, which at the same time are highly sensitive to any non-Gaussian corrections.

Non-Gaussian corrections can either enhance or suppress the production of PBHs depending on their effect on the tail of the PDF. 
One way of accounting for these effects is to compute the different statistical moments of the distribution. Gaussian profiles are fully characterized by the second moment (the power spectrum), while general PDFs can have contributions from any of the higher moments. For illustrative purposes, we plot in Fig.~\ref{fig:PDF} the effect of the third and fourth moment as determined by the skewness and kurtosis, respectively. 
As it can be clearly seen, a positive kurtosis has a large impact in the tail of the PDF. 
Interestingly, one can analytically compute the effect of non-zero higher order moments in the abundance of PBHs \cite{Franciolini:2018vbk,Matsubara:2022nbr}. 

Non-Gaussian corrections are known to be relevant in multi-field models of inflation. 
For instance, models of axion inflation in which the gauge field sources the curvature perturbations~\cite{Garcia-Bellido:2016dkw} display $\chi^2$ statistics for the curvature perturbation. This has implications both for the PBH abundance and for the GW signature, as studied in detail in~\cite{Garcia-Bellido:2017aan}. 
In particular, a strong increase of the PBH abundance associated with a greater tail of the PDF does not correspond to a strong increase of the associated GW production. Therefore, for equal PBH abundance, a smaller GW signal is produced with respect to the Gaussian case. 

Moreover, these non-Gaussian corrections can also enhance the abundance of PBHs in single-field models due to the dynamics of the ultra-slow roll phase \cite{Biagetti:2021eep,Cai:2022erk} and the effect of quantum diffusion beyond slow-roll \cite{Firouzjahi:2018vet,Biagetti:2018pjj,Ezquiaga:2018gbw,Pattison:2021oen}\footnote{A violation of slow-roll is necessary for non-Gaussian effects to be sufficiently large to change the PBH abundance in single-field models~\cite{Passaglia:2018ixg}.}. 
In fact, in general terms, quantum diffusion introduces an exponential tail in the PDF of curvature perturbations \cite{Pattison:2017mbe,Ezquiaga:2019ftu}. For this reason, we discuss next the effect of quantum diffusion in the production of PBHs.

% FIG. 
\begin{figure}[!t]
\centering
\includegraphics[width = 0.478\textwidth]{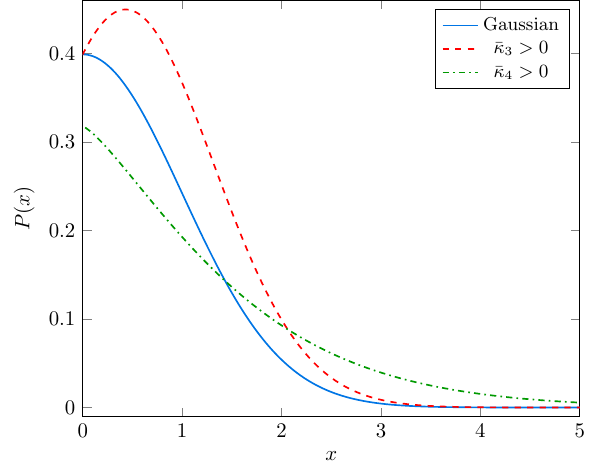}
\includegraphics[width = 0.49\textwidth]{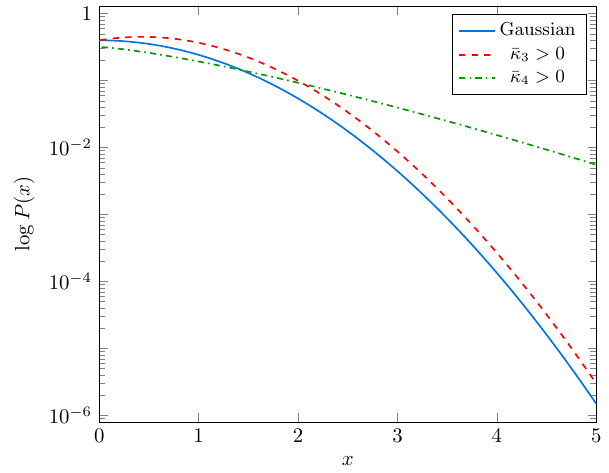}
\vspace{-5pt}
\caption{Illustration of the effect of a positive skewness, $\bar{\kappa}_3$, and a positive (excess) kurtosis, $\bar{\kappa}_4$, for the probability density function $P(x)$ (left) and its logarithm (right) in comparison with a Gaussian distribution ($\bar{\kappa}_{n>2}=0$). The dominant effect in the tail is given by a positive kurtosis. Plots reproduced from \cite{Ezquiaga:2018gbw}.}
\label{fig:PDF}
\end{figure}

\subsubsection{Quantum diffusion}

\begin{figure}[t!]
\centering 
\includegraphics[width=.8\textwidth]{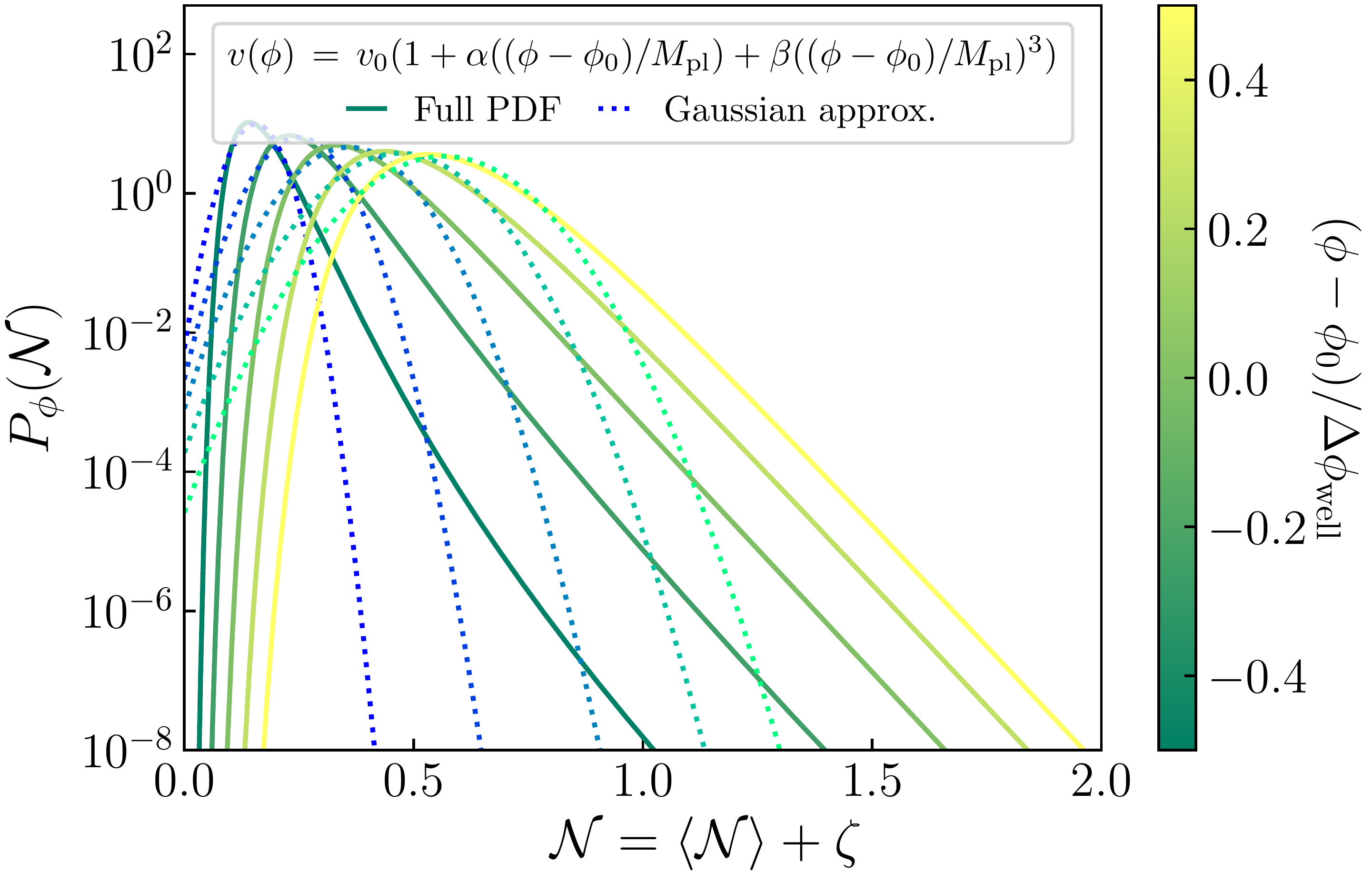}
 \caption{Probability density functions of the curvature perturbations generated by an inflection-point inflationary potential (see inset). Solid lines correspond to the full distribution functions computed by mean of the stochastic-$\delta N$ formalism, where different colours correspond to different locations in the potential $v(\phi)$ where the scale under consideration emerges from the Hubble radius and where $\Delta \phi_{\rm well} $ denotes the field range for which stochastic effects dominate over the classical dynamics. The dotted lines correspond to the standard result, which provides a good Gaussian approximation for the maximum of the distribution but that however fails to describe the exponential tails, where PBHs are nonetheless produced. This figure is adapted from Ref.~\cite{Ezquiaga:2019ftu}.}
 \label{fig:quantum_diffusion:full_vs_gaussian}
\end{figure}

Primordial curvature perturbations are expected to be well-described by a quasi-Gaussian distribution when they are small and close to the maximum of their probability distribution. This can be modelled using conventional cosmological perturbation theory where the free field fluctuations during inflation describe a Gaussian distribution and higher-order interactions lead to non-Gaussian corrections. However, PBHs result from the collapse of large density perturbations, which may be far from the peak of the distribution. In this regime, we require a non-perturbative approach, such as the stochastic-$\delta N$ formalism \cite{Enqvist:2008kt, Fujita:2013cna, Vennin:2015hra}. 

Stochastic inflation \cite{Starobinsky:1982ee} splits the inflaton field into small-scale quantum fluctuations and a large-scale effectively classical field above a coarse-graining scale (typically around the Hubble length). The large-scale field follows the non-linear evolution of a (locally) homogeneous and isotropic cosmology~\cite{Salopek:1990jq,Wands:2000dp,Rigopoulos:2003ak}, except that small-scale modes are constantly swept up from the quantum vacuum state, crossing into the coarse-grained field, leading to quantum diffusion $\langle \Delta \varphi^2 \rangle \approx (H/2\pi)^2$ per Hubble volume per Hubble time. In regimes where PBHs are formed, the inflaton must undergo large quantum fluctuations (for instance as the result of travelling through a very flat potential), which implies that stochastic corrections are important and must be taken into account~\cite{Pattison:2017mbe, Ezquiaga:2018gbw, Biagetti:2018pjj, Ezquiaga:2019ftu}. 

In order to derive the predicted abundance of PBHs in stochastic inflation, we make use of the $\delta N$ formalism~\cite{Starobinsky:1982ee,Starobinsky:1986fxa,Sasaki:1995aw,Wands:2000dp,Lyth:2004gb}, in which the primordial curvature perturbation is identified with the perturbation in the number of e-folds $N=\int H\, \mathrm{d}t$ with respect to the average, $\zeta={\cal N}-\langle{\cal N}\rangle$ (where $\mathcal{N}$ denotes the duration of inflation starting from an initial field configuration, measured in number of $e$-folds $N$). One thus has to extract the distribution $P_\phi(\mathcal{N})$ of the number of e-folds of expansion $\mathcal{N}$ from a given inflaton field value $\phi$ to the end of inflation. This is the program of the stochastic $\delta N$ formalism, in which it can be shown~\cite{Vennin:2015hra,Pattison:2017mbe,Ezquiaga:2019ftu, Ando:2020fjm} that the Fokker-Planck equation driving the stochastic evolution of the coarse-grained field with respect to the integrated expansion during inflation, $\partial P_N(\phi)/\partial N = \mathcal{L}_{\mathrm{FP}}(\phi)\cdot P_N(\phi)$, can be recast in terms of the adjoint Fokker-Planck operator acting on the duration distribution $P_\phi(\mathcal{N})$, such that $\partial P_\phi(\mathcal{N})/\partial \mathcal{N} = \mathcal{L}_{\mathrm{FP}}^\dagger(\phi) \cdot P_\phi(\mathcal{N})$.

By solving this equation in different setups, one can show that quantum diffusion leads unavoidably to non-Gaussian tails \cite{Pattison:2017mbe,Ezquiaga:2019ftu}. In particular, if the classical evolution gives a finite number of e-folds $N_{\rm c}$ from a given field value to the end of inflation, then the large-${\cal N}$ limit (where ${\cal N}\gg N_{\rm c}$) is inevitably dominated by quantum diffusion since it requires many quantum kicks to evade the classical drift towards the end of inflation. Quite generally, the probability distribution approaches an exponential tail, $P_\phi({\cal N})=\sum_n a_n e^{-\Lambda_n \mathcal{N}}\propto e^{-\Lambda_0{\cal N}}$ as ${\cal N}\to\infty$ \cite{Pattison:2017mbe,Ezquiaga:2019ftu,Jackson:2022unc}, which cannot be described by a quasi-Gaussian distribution (see Fig.~\ref{fig:quantum_diffusion:full_vs_gaussian}). Thus predictions for the abundance of PBHs, which will be described by the integrated probability distribution function above a given threshold, may differ by several orders of magnitude from results obtained assuming a Gaussian distribution \cite{Pattison:2017mbe, Ezquiaga:2019ftu, Biagetti:2021eep,Kitajima:2021fpq,DeLuca:2022rfz,Gow:2022jfb}. 

Let us also stress that these exponential tails cannot be properly described by usual, perturbative parametrisations of non-Gaussian statistics (such as those based on computing the few first moments of the distribution and the non-linearity parameters $f_{\mathrm{NL}}$, $g_{\mathrm{NL}}$, etc.), which can only account for polynomial modulations of Gaussian tails. A non-perturbative approach, such as the one presented here, is therefore necessary. 

It is also worth mentioning that the details of the tail (such as the value of $a_0$ and $\Lambda_0$) are determined by the specifics of the inflationary model under consideration. For instance, in the simplest setup where inflation is realised with a single scalar field, if the potential energy $V$ is constant across a field region of size $\Delta\phi$, then \cite{Pattison:2017mbe} $\Lambda_0=V/(96\Mp^2\Delta\phi^2)$ and $\Lambda_n = (2n+1)^2\Lambda_0$. Wider ``quantum wells'' (i.e. larger values of  $\Delta\phi$) thus lead to smaller $\Lambda_0$ and to wider tails. This result is valid if the initial velocity of the inflaton into the quantum well is vanishing. Otherwise, a transient period of ``ultra slow roll'' takes place until the inflaton relaxes to the slow-roll attractor, and on top of the eigenvalues $\Lambda_n$ given above, a second set of decay rates $\Lambda_n'=\Lambda_n+3$ also arises \cite{Pattison:2021oen}. When the initial velocity is large compared to the velocity that would be required to cross the well classically, the amplitudes $a_n'$ associated with those eigenvalues are larger than those for the rates $\Lambda_n$, hence the tails are more suppressed. Another way to suppress the tails is to add a tilt in the inflationary potential, $V=24\pi^2\Mp^4 v_0(1+\alpha \phi/\Mp)$, for which the decay rates are all shifted by a constant value~\cite{Ezquiaga:2019ftu}, i.e. $\Lambda_n'=\Lambda_n + \alpha^2/(4 v_0)$. Unless $\alpha^2$ is fine tuned to values less than the potential energy in Planckian units, the exponential tails are heavily suppressed. As a last example, let us mention the case where the potential features an inflection point, close to which one can approximate $V=24\pi^2\Mp^4 v_0[1+\beta(\phi/{\Mp})^{3}]$. In such models, if slow roll is not violated when approaching the inflection point (which depends on the details of the inflationary potential away from that point), one finds~\cite{Ezquiaga:2019ftu} $\Lambda_n = (3/2)^{2/3}\pi^2 (v_0\alpha)^{1/3} (n+1/2)^2$, such that primordial black holes are always overproduced. 

These examples show that the abundance of primordial black holes is ultimately related to the details of the inflationary potential where quantum diffusion plays an important role. PBHs would thus provide access to regions of the inflationary potential that can otherwise not be probed with other cosmological measurements such as the cosmic microwave background.

\subsubsection{Stochastic spectator field}

Another recent scenario~\cite{Carr:2019hud} takes the advantage of the dynamics of a stochastic spectator field to produce PBHs, while the primordial power spectrum remains at the CMB level almost everywhere in our Universe, except in subdominant regions where PBHs can form.  If PBHs are linked to the generation of the baryon asymmetry of the Universe, thereby connecting 1) the PBHs and baryon abundance, 2) the baryon-to-photon ratio to the PBH density at the time of their formation, the stochastic and quantum nature of the spectator field fluctuations allows one to invoke an anthropic selection mechanism to resolve the long-standing parameter fine-tuning issues related to PBH formation~\cite{Garcia-Bellido:2019vlf,Carr:2019hud,Garcia-Bellido:2019tvz}.

In this scenario, the stochastic spectator field experiences quantum fluctuations during inflation that lead it to explore a wide range of the potential, including its slow-roll part, in different Hubble patches. But the field does not have any impact on the inflationary dynamics, thus being a \textit{spectator} field.  Because the number of patches is huge, there necessarily exist some in which the spectator field, at the time of Hubble exit, has the required mean value for subsequent quantum fluctuations (but sub-Hubble at that time) to later induce large curvature fluctuations.  Our observable Universe would correspond to one of these patches.  

After inflation, the spectator field and its fluctuations within these patches remain frozen during the radiation era until its potential energy starts to dominate the density of the Universe, well after inflation but also possibly well before the time of PBH formation.  At this point, in the regions where the field is in the slow-roll part of the potential, a very short second inflationary phase (for at most a few e-folds) takes place, whereas in the rest of the patch the field quickly rolls down its potential without inflating.  Such an extra expansion corresponds to the production of local  non-linear curvature fluctuations that are still super-horizon at that time.  Only later they re-enter the horizon and collapse into PBHs.  But in the rest of the Universe the curvature fluctuations are statistically Gaussian and behave as expected in standard slow-roll inflation, unaffected by the spectator field.

The probability to form a PBH in a given region (that can be connected to the function $\beta$ introduced in the next Section) has been calculated as~\cite{Carr:2019hud}
\be
P_{\rm PBH} = \sqrt{\frac 2 \pi} \frac{\Delta \psi^{\rm sr}}{\sqrt{H_N^2 / 4 \pi^2 +\langle \delta \psi^2 \rangle _{N-1} }} \times \exp \left[ -\frac{\langle \psi \rangle ^2}{2 (H_N^2 /4 \pi^2 + \langle \delta \psi^2 \rangle _{N-1})}\right],
\ee
where $\Delta \psi^{\rm sr}$ is the size of the slow-roll region where the field can induce $\mathcal O(1)$ curvature fluctuations.  The subscript $N$ is used to indicate that a quantity is evaluated at the e-fold time at which the scale exited the horizon during inflation.  It can be noticed that the mathematical form of $P_{\rm PBH}$  is quite similar to the one of $\beta$ (see next Section) in more standard models of Gaussian inflationary perturbations.  As a consequence, the shape of the PBH mass function is also very similar and driven by the evolution of $H$ during inflation.  There is also no need to precisely fix the potential parameters in this model, the only condition being that the potential is of plateau or small-field type and that it can dominate the density of the Universe prior the QCD epoch (if one is interested to get stellar-mass or heavier PBHs).

\subsection{Curvaton}
A different possibility to generate PBHs, while keeping the generation mechanisms for curvature perturbations at different scales decoupled, is represented by a simple modification of the original {\it curvaton scenario} (in the original curvaton model \cite{Lyth:2001nq,Moroi:2001ct}, one field, the inflaton, is responsible only for sustaining the inflationary background evolution while the other field, the curvaton, produces the primordial curvature perturbations). In this setup with two fields, the primordial curvature perturbations on CMB scales are produced by the inflaton, which acts very similarly to the standard single-field scenario, while the curvaton field becomes responsible for perturbations on smaller scales and thus formation of PBHs.

The power spectrum of curvature perturbations will be given as the sum of the contributions of the inflaton and of the curvaton
\begin{equation}
    \mathcal{P}(k)=\mathcal{P}_{\text{infl}}(k) + \mathcal{P}_{\text{curv}}(k)\,.
\end{equation}
where $\mathcal{P}_{\text{curv}}(k)$ can be described by a lognormal function such as in Eq.~\ref{Ps-bump}.

At large scales (reasonably $k\lesssim 1$ Mpc$^{-1}$), the first term will be dominating, with its magnitude being determined by CMB normalization, that is $\mathcal{P}_{\rm infl}(k)\simeq 2\times10^{-9}$. Moving to smaller scales (larger $k$), the second contribution, given by the curvaton field, increases and becomes dominant. 

So, as long as the curvature perturbation on large cosmological scales is dominated by inflaton fluctuations with an almost scale-invariant spectrum, the curvaton field is not required to be light during inflation and its fluctuations may have a steep blue spectrum. Curvaton fluctuations could then give the dominant contribution to the primordial density perturbations on small scales after the curvaton decays. The curvaton may have a steep blue spectrum either due to interactions with the inflaton or other fields evolving during inflation~\cite{Yokoyama:1995ex},
a non-trivial kinetic term~\cite{Pi:2021dft}
or in an axion-like model, where the curvaton is identified with the phase of a complex field whose modulus decreases rapidly during inflation~\cite{Kawasaki:2012wr}.

An important characteristic of any curvaton model is that the fluctuations originate from non-adiabatic field fluctuations during inflation, a fact which gives rise to a non-Gaussian distribution in the primordial curvature perturbations, including a finite non-Gaussianity in the squeezed limit. Although this local-type non-Gaussianity is typically small ($f_{\rm NL}\sim1$), it can nonetheless have a significant effect on the abundance of PBHs, as discussed in Subsection~\ref{ssection:nG}.

\subsection{Preheating}

Another way through which one can produce  PBHs is the preheating instability. During the preheating period after inflation, when the inflaton oscillates coherently at its ground state and decays to other degrees of freedom,  it is widely argued that resonant amplification of quantum field fluctuations, responsible for particle production, take place ~\cite{Kofman:1994rk,Kofman:1997yn}. These amplified quantum fluctuations should be accompanied by a resonant amplification of the scalar metric fluctuations (usually quoted as metric preheating ~\cite{Finelli:1998bu,Bassett:1999mt,Jedamzik:1999um,Bassett:1999cg}), responsible for gravitational fluctuations in the curvature, since the two are coupled through Einstein’s equations.  One then expects that the amplified metric perturbations give rise to large curvature/matter perturbations, which in their turn collapse and form PBHs.

Historically, PBHs emanating from the preheating instability were speculated in the context of multi-field inflation and in particular in the context of two-field chaotic inflation~\cite{Bassett:1998wg, Green:2000he,Bassett:2000ha,Suyama:2004mz} and more recently in~\cite{Torres-Lomas:2012tna,Torres-Lomas:2014bua}, since in this case the parametric amplification of entropy/isocurvature fluctuations can source the parametric amplification of the adiabatic/curvature fluctuations in the regime of broad resonance. Thus, these amplified curvature fluctuations could break the scale-invariance of the primordial power spectrum and give birth to PBHs through non-linear gravitational fluctuations inducing large density contrasts \cite{Kou:2019bbc,Joana:2022uwc}.

It is predicted as well~\cite{Jedamzik:2010dq} (see also~\cite{Easther:2010mr}) that in the context of single-field inflation, there is a pronounced resonant instability structure in the narrow regime where amplified metric perturbations can induce the production of PBHs as recently studied in~\cite{Martin:2019nuw,Auclair:2020csm}. At this point, one should point out that the narrow resonant structure of metric preheating was shown in~\cite{Martin:2020fgl} to be immune to the decay of the inflaton into a radiation fluid, ensuring in this way the transition to the Hot Big Bang phase of the Universe.

\paragraph{The case of metric preheating}

Most of the studies in which PBHs are produced during preheating treat them in the context of multi-field and especially in the context of two-field inflation. This makes it difficult to predict in an analytic way the matter power spectrum responsible for the PBH formation, which is constructed consequently numerically. On the contrary, in the context of single-field inflation, one can in principle extract in a direct way the matter power spectrum responsible for the PBH production. In what follows we emphasize on a specific scenario studied in~\cite{Martin:2019nuw,Auclair:2020csm}, in which PBHs are produced during the preheating phase due to the emergence of a resonant instability structure concerning the equation of motion of the scalar metric perturbations that are enhanced on small scales (metric preheating), which exit the Hubble radius close to the end of inflation. Specifically, the modes concerned satisfy the relation
\begin{equation}\label{enhanced modes}
aH< k < a\sqrt{3Hm},
\end{equation}
where $k$ is the comoving scale concerned, $a$ is the scale factor, $H$ is the Hubble parameter and $m$ is the mass parameter of the inflationary potential.  For these modes, as shown in~\cite{Martin:2019nuw}, the density contrast scales linearly with the scale factor, i.e. $\delta_k \sim a$, manifesting in this way an effective ``dust" behavior during the instability phase in which PBHs can be easily produced. 

Considering only the modes entering the instability band from above, namely the left hand side of the inequality (\ref{enhanced modes}), the matter power spectrum responsible for the PBH formation at the end of the preheating instability phase can be written analytically as~\cite{Martin:2019nuw}
\begin{equation}
\label{matter power spectrum - preheating}
\mathcal{P}_{\delta}\left( t_\mathrm{\Gamma}, \frac{k}{k_\mathrm{end}}\right)=\left(\frac{k}{k_\mathrm{end}}\right)^{4}\left(\frac{\rho_\mathrm{inf}}{\rho_\Gamma}\right)^{2/3} \left(\frac{6}{5}\right)^2 \mathcal{P}_{\zeta,\mathrm{end}}\left(\frac{k}{k_\mathrm{end}}\right) ,
\end{equation}
where $k_\mathrm{end}$ is the comoving scale exiting the Hubble radius at the end of inflation, $\rho_\mathrm{inf}$ is the energy scale at the end of inflation, $\rho_\Gamma$ is the energy scale at the end of the preheating instability phase and $\mathcal{P}_{\zeta,\mathrm{end}}\left(\frac{k}{k_\mathrm{end}}\right)$ is the curvature power spectrum at the end of inflation which in the slow-roll approximation reads as ~\cite{Schwarz:2001vv, Gong:2001he}
 \begin{equation}\label{eq:Pzeta:end:SR}
\mathcal{P}_{\zeta,\mathrm{end}}\left(\frac{k}{k_\mathrm{end}}\right) =
  \frac{H_*^2\left(k\right)}{8\pi^2 \Mp^2 \epsilon_{1*}
    \left(k\right)}\left[1+\left(\frac{k}{k_\mathrm{end}}\right)^2\right]
  \left[1-2\left(C+1\right)\epsilon_{1*}\left(k\right)
    -C\epsilon_{2*}\left(k\right)\right]\ \text{for}\ k<k_\mathrm{end}.
\end{equation}
The functions $H_*(k)$, $\epsilon_{1*}(k)$ and $\epsilon_{2*}(k)$ 
denote respectively the values of $H$, $\epsilon_1$ and $\epsilon_2$ at the time when the mode $k$ exits the Hubble radius during inflation. The parameter $C\simeq-0.7296$ is a numerical constant. 

The relevant range of the density contrast at the end of the preheating phase is
\begin{align}
\label{eq:cond:delta}
\left(\frac{\rho_\mathrm{inf}}{\rho_\mathrm{\Gamma}}\right)^{1/3}\left(\frac{k}{k_\mathrm{end}}\right)^{2} \left(\frac{3\pi}{2}\right)^{2/3}
\left[\left(\frac{k}{k_\mathrm{end}}\right)^3 \sqrt{\frac{\rho_\mathrm{inf}}{\rho_\Gamma}}
  -1\right]^{-2/3}<\delta_{k}(t_\mathrm{\Gamma})< 1,
\end{align}
while the relevant parameter space $(\rho_\mathrm{\Gamma},\rho_\mathrm{inf})$ so that we do not have PBH overproduction during preheating reads as
\begin{align}
\label{eq:reheating:via:PBH:evap:criterion}
\frac{\rho_\Gamma}{\Mp^4}<\frac{4}{125\sqrt{3}\pi^{5}}
\left(\frac{\rho_\mathrm{inf}}{\Mp^4}\right)^{5/2}\ .
\end{align}
Below, in Fig.~\ref{fig:Matter Power Spectrum - Preheating} we see the behavior of $\mathcal{P}_{\delta}\left(t_\mathrm{\Gamma}, \frac{k}{k_\mathrm{end}} \right)$ [see Eq.~\ref{matter power spectrum - preheating}] by fixing $\rho^{1/4}_\mathrm{inf} = 10^{14}\mathrm{GeV}$ and $\rho^{1/4}_\mathrm{\Gamma} = 10^{7}\mathrm{GeV}$. At this point, it is important to stress out that exactly at $k=k_\mathrm{end}$, Eq.~\eqref{eq:Pzeta:end:SR} is not trustful since there the slow-roll approximation breaks down.

\begin{figure}[t!]
\begin{center}
\includegraphics[width=0.796\textwidth, clip=true]               {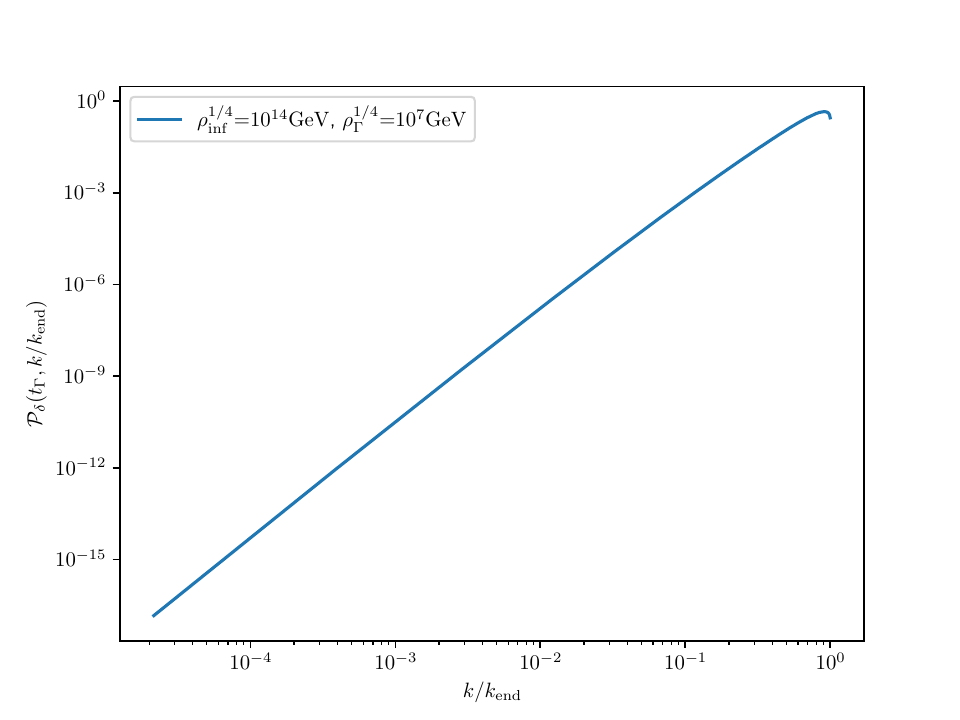}
\caption{The matter power spectrum at the end of the preheating instability phase. The energy scales at the end of inflation and at the end of the preheating instability phase are fixed as $\rho^{1/4}_\mathrm{inf} = 10^{14}\mathrm{GeV}$ and $\rho^{1/4}_\mathrm{\Gamma} = 10^{7}\mathrm{GeV}$.}
\label{fig:Matter Power Spectrum - Preheating}
\end{center}
\end{figure}

The corresponding mass of the PBHs produced is related to the mass inside the Hubble radius at the time when a scale $k$ re-enters the horizon, namely
\begin{equation}\label{PBH mass - Preheating}
m_\mathrm{PBH} = \gamma M_\mathrm{H}  =
\gamma \frac{\left(3M^2_\mathrm{pl}\right)^{3/2}}{ \sqrt{\rho_\mathrm{inf}}}
\left(\frac{k}{k_\mathrm{end}}\right)^{-3},
\end{equation}
where $\gamma$ is an efficiency parameter of order one depending on the details of the gravitational collapse. Here, it is important to notice that in \eqref{PBH mass - Preheating}, one clearly sees the dependence of the PBH mass on the energy scale at the end of inflation and as a consequence on the underlying inflationary model.

Regarding the characteristic PBH mass produced during preheating one can choose $k/k_\mathrm{end}=0.1$, and using a relative high energy scale at the end of inflation $\rho^{1/4}_\mathrm{inf} = 10^{14}\mathrm{GeV}$, from Eq.~\eqref{PBH mass - Preheating} one obtains $m_\mathrm{PBH} \simeq 10^{-27}M_\odot$. This mass, around which the matter power spectrum peaks, is very small compared to the mass range LISA can detect through the GW portal of PBH mergers. However, as pointed out recently in \cite{Auclair:2020csm}, in which the PBH mass function is computed more accurately with the use of the excursion-set formalism, the mass of PBHs produced from the metric preheating instability, at which the PBH mass function peaks, is higher than the PBH mass quoted above and can be as high as the solar mass, lying in this way within the mass range probed by LISA.

At this point, one should also stress that ultralight PBHs, such as the ones produced from the preheating instability, can be potentially detected by LISA through the stochastic background of  GWs induced at second order from PBH isocurvature perturbations, as argued in \cite{Papanikolaou:2020qtd}. For these GWs, the peak frequency could lie within the LISA band despite the smallness of the PBH mass.

\subsection{Early matter era}  

An early prolonged period of matter domination -- after the end of inflation -- could arise due to the domination of moduli fields which are a natural prediction of any supergravity or superstring-inspired theory of the early Universe. Since the curvature threshold for the PBH collapse depends on the equation of state of the Universe, it might be easier to obtain the same abundance of PBH during the moduli dominated epoch from a peak in the inflationary power spectrum less strong than the typical $O(10^{-2})$ required in the standard scenario of radiation domination. Several mechanisms have been proposed to lead to such a production during an early matter dominated era, see e.g.~\cite{Khlopov:1980mg,Khlopov:1985jw,Green:1997pr,Nayak:2009pk,Harada:2016mhb,Carr:2017edp,Carr:2018nkm,Kokubu:2018fxy,Ballesteros:2019hus,deJong:2021bbo,DeLuca:2021pls}.  In such scenarios, the mass and fraction of PBHs produced from a given inflationary potential depend on whether the moduli dominated epoch is immediate or delayed after the end of inflation. Recently, it has also been pointed out that even if PBHs are formed during radiation domination, the presence of a reheating epoch can still crucially affect the PBH abundance~\cite{Cai:2018rqf, Bhaumik:2019tvl}.  
%calculated at the matter-radiation equality 
%{\bf DW: Why ``at matter-radiation equality''? Ambiguous whether this means the usual matter-radiation equality, or at the end of reheating.}

% DW: comment out this paragraph as it is now replaced by the preceding one
%The curvature threshold for the PBH collapse depends on the equation of state of the Universe.  It goes to zero in a matter dominated era, and so it is possible that PBH have formed from relatively small amplitude primordial curvature fluctuations if the early Universes has undergone a transient matter-dominated era.  Several mechanisms have been proposed to be at the origin of such early matter era, see e.g.~\cite{Khlopov:1980mg,Khlopov:1985jw,Green:1997pr,Nayak:2009pk,Ballesteros:2019hus}.  

In models of inflation based on moduli stabilisation in string theory compactifications, for example, the inflaton may be identified with a modulus field which is only gravitationally coupled to the visible sector, leading to a slow-reheating phase, where the inflaton decay is via Planck-suppressed operators, and hence a prolonged duration of matter-domination after inflation~\cite{Green:1997pr,Ballesteros:2019hus}. Ref.~\cite{Ballesteros:2019hus} finds that PBH formation is most effective for reheating temperatures $\lesssim10^6$~GeV, in which case PBHs can form a significant fraction of the dark matter density today when primordial curvature fluctuations from inflation are of order $10^{-4}$ in a matter-dominated era, rather than $\sim10^{-2}$ usually required in a radiation-dominated era.

\subsection{Phase transitions}  

The formation and abundance of PBHs can also be affected by phase transitions in the early Universe.
During the radiation era, the formation of PBHs may be facilitated if pre-existing super-horizon density fluctuations enter the cosmological horizon during a first order phase transition which proceeds in approximate equilibrium~\cite{Khlopov:1985jw,Jedamzik:1999am}. During this phase transition the speed of sound tends to zero and the pressure response of the fluid vanishes and does not counter-balance the gravitational collapse of horizon-sized primordial over-densities.  Detailed numerical simulations using general-relativistic hydrodynamics show that the overdensity threshold required for PBH formation by fluctuations that enter the cosmological horizon around the phase transition is smaller than the ordinary value for PBH formation during the radiation epoch. This would produce an increase in the PBH mass spectrum around the corresponding horizon mass in which the phase transition occurs. 

It has been argued that the QCD phase transition could produce an enhancement of the population of PBH's around $1\Msun$~\cite{Jedamzik:1996mr}, but lattice QCD calculations have shown that the QCD phase transition in the Standard Model is not first order but a continuous cross-over~\cite{Bhattacharya:2014ara}. Nevertheless, it is possible to collapse a significant amount of PBHs at the QCD transition if there are large fluctuations and radiation pressure drops as the relativistic quark and gluon degrees of freedom disappear from the plasma~\cite{Byrnes:2018clq,Carr:2019kxo,Jedamzik:2020omx}

Alternative scenarios have been proposed, where the dynamics of the QCD phase transition incorporates extra degrees of freedom, like lepton number violation~\cite{Bodeker:2020stj}
and solitosynthesis~\cite{Garcia-Bellido:2021zgu}.
In other models, the QCD transition could be first order if the number of light quarks were larger than three at the confinement. Models in which the dynamics of a light scalar field have been tailored to suppress quark masses before the QCD transition, while recovering the measured values after the transition, have been developed in~\cite{Davoudiasl:2019ugw}.

Another possibility is that the collision of bubbles formed during a first order phase transition in the early Universe could create a sufficiently large local concentration of the energy stored in the walls to cause gravitational collapse and the formation of a primordial black hole~\cite{Hawking:1982ga, Liu:2021svg,Jung:2021mku}. This process could be affected by spatial clustering properties of bubble generated in high-energy phase transitions \cite{Pirvu:2021roq, DeLuca:2021mlh},
Recent models~\cite{Baker:2021nyl} consider the possibility that there may be particles whose mass increases significantly during the phase transition, suppressing the transmission of the corresponding particles through the advancing bubble walls. This effect can build a sufficiently large overdensity in front of the walls that collapse into a primordial black hole. The primordial black hole density and mass distribution depend on the model parameters.
Other models~\cite{Kawana:2021tde} consider fermion species that gain big masses in the true vacuum, so that the corresponding particles get trapped in the false vacuum as they do not have sufficient energy to penetrate the bubble wall. After the first order phase transition, the fermions are compressed into the false vacuum remnants to form Fermi-balls that finally collapse to PBH due to the Yukawa attractive force.

Higher-order phase transitions periods in the early Universe may also lead to the formation of PBHs. For example,
domain walls can be generated during a non-equilibrium second order phase transition in the vacuum state of a scalar field with a flat direction during inflation~\cite{Rubin:2000dq}. The background de Sitter fluctuations of such an effectively massless scalar field could provide non–equilibrium redefinition of correlation length and give rise to the islands of one vacuum in the sea of another one. After the phase transition takes place in the Friedman– Robertson-Walker (FRW) epoch, the two vacua are separated by a wall. Some of these closed and sufficiently large walls could accumulate enough energy to collapse and form a massive black hole. The mass spectrum of the PBHs which can be created in such a way depends on the details of the potential of the scalar field that has the flat direction during inflation and triggers the phase transition.

\subsection{Domain walls and cosmic strings} 

Topological defects in the early Universe may also lead to the production of PBHs.
S. W. Hawking~\cite{Hawking:1987bn} was the first to show that PBHs could form from the collapse of cosmic string loops. He argued that loops that collapse by a factor $(G\mu)^{-1}$, where $\mu$ is the mass per unit length of the strings, will inevitably collapse to form black holes. 
PBH production can be used to place bounds on the cosmic string network~\cite{Hawking:1990tx,Polnarev:1988dh,Caldwell:1993kv,Garriga:1993gj,Caldwell:1995fu,MacGibbon:1997pu,Helfer:2018qgv,James-Turner:2019ssu} or cosmic strings nucleated during inflation~\cite{Garriga:1992nm}.
More recently it has been argued that cosmic string cusps can collapse gravitationally into PBHs with a mass function that could extend up to stellar-masses~\cite{Jenkins:2020ctp}, although this claim has been disputed in Ref.~\cite{Blanco-Pillado:2021klh}

Moreover, domain walls, which arise in theories with a broken discrete symmetry, can also form PBHs when such domain walls are smaller than a critical radius fall within the cosmological horizon and collapse due to their own tension \cite{Ferrer:2018uiu} or are produced by tunnelling during inflation~\cite{Liu:2019lul}.  In such scenarios, the mass function of PBHs in general has a spikelike structure. Moreover, in QCD axion models, PBHs can form either by the collapse of long-lived string-domain wall networks or from the collapse of closed domain walls \cite{Ferrer:2018uiu,Ge:2019ihf}. The mass and abundance of PBHs formed in these scenarios crucially depends on the QCD axion mass. However, interestingly, various observational constraints on the PBH abundance in turn limit the QCD axion parameter space.

%Topological defects may have lead to the production of PBHs.  Hawking~\cite{Hawking:1987bn} was the first to show that PBHs could form from the collapse of cosmic string loops. He argued that loops that collapse by a factor $(G\mu)^{-1}$, where $\mu$ is the mass per unit length of the strings, will inevitably collapse to form black holes. PBH production can be used to place bounds on the cosmic string network~\cite{Hawking:1990tx,Polnarev:1988dh,Caldwell:1993kv,Garriga:1993gj,Caldwell:1995fu,MacGibbon:1997pu,Helfer:2018qgv,James-Turner:2019ssu} or cosmic strings nucleated during inflation~\cite{Garriga:1992nm}. More recently it has been argued that cosmic string cusps can collapse gravitationally into PBHs with a mass function that could extend up to stellar-masses~\cite{Jenkins:2020ctp}.  The collapse of domain walls, e.g. produced by tunnelling during inflation~\cite{Deng:2016vzb,Liu:2019lul} or in QCD axion models~\cite{Ferrer:2018uiu,Ge:2019ihf} is another class of PBH formation scenarios.  

\subsection{Primordial magnetic fields}  

Primordial magnetic fields generated in the early Universe are considered one of the feasible candidates to provide the required seeds for the observed large-scale intergalactic magnetic fields. In addition, such primordial fields also generally induce an anisotropic stress that can act as a source for the evolution of the super-horizon curvature perturbations, leading to the formation of PBHs with a broad possible range of masses. Since a large abundance of PBHs requires a large amplitude of density perturbations, the strength of these primordial magnetic fields on small scales should also be large enough for this mechanism to work~\cite{Saga:2020ics}.  Based on this idea, strong constraints on the amplitude of primordial magnetic fields have been established, a few hundreds nano-Gauss on scales in the wave-number range $10^2\, {\rm Mpc}^{-1} \le k \le 10^{18}\, {\rm Mpc}^{-1}$.

Alternatively, it has been proposed that PBHs can act as seeds of cosmic magnetic fields due to their accretion disks, either due to a \textit{Biermann battery}~\cite{Safarzadeh:2017mdy,Papanikolaou:2023nkx} mechanism or more exotic mechanisms such as monopoles~\cite{Maldacena:2020skw,Araya:2020tds} or Kerr-Newman PBHs~\cite{Hooper:2022cvr}. In particular, in~\cite{Papanikolaou:2023nkx} it was shown that locally isothermal disks around supermassive PBHs with masses $M>10^{10}M_\odot$,  can generate a seed primordial magnetic field of the order of $10^{-30}\mathrm{G}$, which can be later amplified by various dynamo/instability processes and provide the seed for the present day magnetic field in intergalactic scales. Interestingly enough, this population of magnetised supermassive PBHs can induce a stochastic GW background which can be used to set tight constraints on the supermassive PBH abundances [See~\cite{Papanikolaou:2023cku} for more details].

%It is suspected that the required seeds for extra-galactic magnetic fields have an origin in the early Universe.  These primordial magnetic fields induce an anisotropic stress that can act as a source of large super-Hubble curvature fluctuations, leading to PBH formation with a broad possible range of masses~\cite{Saga:2020ics}. 

\subsection{Summary}

In this Section, we have explored multiple mechanisms described in the literature for generating primordial black holes, from gravitational collapse of primordial fluctuations and inflation to preheating, phase transitions, collisions of bubble walls and cosmic strings. Each of these mechanisms has its own mass spectrum and individual characteristics, which makes it possible to differentiate between them. The most predictive one is that of primordial fluctuations, where specific models exist that can be put to test by present and future observations. But we also emphasized the possible role of non-Gaussian tails in the distribution of curvature fluctuations, which arise in models with an ultra-slow-roll phase of inflation or in models with a stochastic spectator field.  This is an interesting, recent and active research area since this could soften or even resolve the fine-tuning issues associated to the formation of PBHs with significant abundances.  In the next Sections, we will explore the dynamics of gravitational collapse and how it depends on the thermal history of the Universe, with its impact on the PBH mass function, their spin distribution and merging rates.

\section{PBH formation, mass function and clustering} 
\label{sec:PBHformation}

\subsection{From curvature perturbations to density contrast}
PBHs form after cosmological perturbations re-enter the cosmological
horizon. Assuming spherical symmetry on superhorizon scales, the local 
region of the Universe describing such perturbations has the following asymptotic form of the metric
\bea
\d s^2 & = & -\d t^2 + a^2(t) \left[ \frac{\d r^2}{1-K(r)r^2} + r^2\d\Omega^2 \right] = - \d t^2+a^2(t)e^{2\zeta(\hr)}  \left[ \d\hr^2 + \hr^2\d\Omega^2 \right] ,
\label{pert_metric}
\eea
where $a(t)$ is the scale factor, while $K(r)$ and $\zeta(\hr)$ are the
conserved comoving curvature perturbations, defined on a super-Hubble scale,
converging to zero at infinity where the universe is taken to be unperturbed
and spatially flat. The equivalence between the radial and the angular parts
of these two forms of the metric gives
\begin{equation} 
\left\{
\begin{aligned}
& r = \hr e^{\zeta(\hr)} \,,\\ 
& \displaystyle{\frac{\d r}{\sqrt{1-K(r)r^2}}} = e^{\zeta(\hr)} \d\hr \,
\end{aligned}
\right.
\label{K_zeta}
\end{equation}
and the difference between the two Lagrangian coordinates $r$ and $\hr$ is
related to the parametrisation of the comoving coordinate, fixed by the
curvature perturbation chosen in the metric (i.e. $K(r)$ or $\zeta(\hr)$).
From a geometrical point of view, the coordinate $\hr$ considers the
perturbed region as a local FLRW separated universe with the curvature
perturbation $\zeta(\hr)$ modifying the local expansion, while the
curvature profile $K(r)$ is defined with respect to the background FLRW
solution ($K=0$) and measures more directly the spatial geometry of space-time. 
 
On superhorizon scales, where the curvature profile is time independent, these two are related as
\be \label{Kzeta}
K(r)r^2 =  - \hr\zeta'(\hr) \left[ 2+\hr\zeta'(\hr) \right]\,.
\ee
In this regime, we can use the gradient expansion
approach~\cite{Shibata:1999zs,Tomita:1975kj,Salopek:1990jq,Polnarev:2006aa},
based on expanding the time dependent variables, like energy density and
velocity profiles, as power series of a small parameter $\epsilon \ll 1$ 
up to the first non zero order, where $\epsilon$ is conveniently identified
with the ratio between the Hubble radius and the length scale of the 
perturbation. Although this approach reproduces the time evolution of 
the linear perturbation theory, it also allows having non linear curvature
perturbations if the spacetime is sufficiently smooth on the scale of the
perturbation (see~\cite{Lyth:2004gb}). This is equivalent to saying that in
this regime pressure gradients are negligible, not playing an important role
in the perturbation evolution, which grows with the Universe expansion in a self similar way. 

In this approximation, the energy density profile can be written as \footnote{$K'(r)$ denotes 
differentiation with respect to $r$ while $\zeta'(\hr)$ and
$\nabla^2\zeta(\hr)$ denote differentiation with respect to $\hr$.}
\cite{Yoo:2018kvb,Musco:2018rwt} 
\bea
\label{rel} 
\frac{\delta\rho}{\rho_{\rm b}} &\equiv& \frac{\rho(r,t) - \rho_{\rm b}(t)}{\rho_{\rm b}(t)} =  
\frac{1}{a^2H^2} \frac{3(1+w)}{5+3w} \frac{ \left[ K(r)\,r^3 \right]^\prime}{3r^2} \nonumber \\
&=& - \frac{1}{a^2H^2} \frac{4(1+w)}{5+3w} e^{-5\zeta(\hr)/2}\nabla^2 e^{\zeta(\hr)/2} \,,
\eea
where $H(t)=\dot{a}(t)/a(t)$ is the Hubble parameter. The
parameter $w$ is the coefficient of the equation of state $p = w \rho_{\rm b}$,
relating the total (isotropic) pressure $p$ to the total background energy density $\rho_{\rm b}$.

In the linear regime of curvature perturbations, when $\zeta \ll 1$, 
this expression can be written as 
\be 
\label{linear}
\frac{\delta\rho}{\rho_{\rm b}} \simeq - \frac{1}{a^2H^2} \frac{2(1+w)}{5+3w} \nabla^2 \zeta(\hr) = - \frac{k^2}{a^2H^2} \frac{2(1+w)}{5+3w} \zeta_k \,,
\ee
where the second equality is obtained from the Fourier transformation 
$\nabla^2\zeta = k^2 \zeta_k$, showing that in the linear regime there is a simple one to one mapping between the real space, where the perturbations collapses, and the Fourier space where the power spectrum of cosmological perturbations is defined. If the curvature perturbation $\zeta$ is a Gaussian variable, the density contrast $\delta\rho/\rho_{\rm b}$ has also a Gaussian distribution within the linear regime described by \eqref{linear}. 
However, because the amplitude of the threshold $\delta_{\rm c}$ for PBHs is non linear, the linear regime does not give an accurate description of the statistics of the density contrast, and the mapping between the Fourier space and the real space requires a more elaborate approach. As we will see later, this allows to compute the threshold $\delta_{\rm c}$ from
the shape of the power spectrum $\mathcal{P}_\zeta$. 

For this reason, the definition of the threshold in terms of $\zeta_{\rm c}$ computed from the linear approximation given by \eqref{linear},  does not give an accurate description and should be dropped in favour of $\delta_{\rm c}$ which is clearly defined in the following Section. We anticipate that the nonlinear relation between the density contrast and the comoving curvature perturbations causes the former to obey non-Gaussian statistics, even in absence of primordial non-Gaussianity. Computations based on peak theory and on threshold statistics show that this unavoidable non-Gaussianity makes the production of PBHs more difficult. In case of a peaked power spectrum, this effect is compensated by an increase of the amplitude by an order 2-3 factor \cite{DeLuca:2019qsy,Young:2019yug}.

\subsection{Perturbation amplitude}
To define the amplitude of a cosmological perturbation, it is useful to introduce the compaction function~\cite{Shibata:1999zs, Musco:2018rwt}, defined as
\be
\label{a}
\mathcal{C} \equiv 2\frac{\delta M(r,t)}{R(r,t)} \,,
\ee
where $R(r,t)$ is the areal radius and $\delta M(r,t)$ is the difference 
between the Misner-Sharp mass within a sphere of radius $R(r,t)$, and the
background mass \mbox{$M_{\rm b}(r,t)=4\pi \rho_{\rm b}(r,t)R^3(r,t)/3$} within the same
areal radius but calculated with respect to a spatially flat FLRW metric. 
In the superhorizon regime (i.e. $\epsilon \ll 1$) the compaction 
function is time independent, and directly related to the curvature profile as
\be \label{comp}
\mathcal{C}(r) = \frac{3(1+w)}{5+3w} K(r)r^2,
\ee
where this expression can be written in terms of $\zeta(\hr)$ using \eqref{Kzeta}.

The comoving length scale of the perturbation can be conveniently identified
with the distance where the compaction function reaches its peak (which is 
a maximum for a positive perturbation and a minimum for a negative one) when
$r=r_{\rm m}$ \cite{Musco:2018rwt}(i.e. \mbox{$\mathcal{C}'(r_{\rm m}) = 0$}), characterized by a constraint relation in terms of the curvature profile
\bea \label{rm_condition} 
K(r_{\rm m})+\frac{r_{\rm m}}{2}K'(r_{\rm m})=0, \quad \textrm{or} \quad
\zeta'(\hr_{\rm m})+\hr_{\rm m}\zeta''(\hr_{\rm m})=0 \,.
\eea
Given the curvature profile, the parameter $\epsilon$ of the gradient expansion is therefore defined as
\be
\epsilon \equiv \frac{R_{\rm H}(t)}{R_{\rm b}(r_{\rm m},t)} = \frac{1}{aHr_{\rm m}} = \frac{1}{aH\hr_{\rm m} e^{\zeta(\hr_{\rm m})}} \,,
\ee
where $R_{\rm H}=1/H$ is the cosmological horizon and $R_{\rm b}(r,t)=a(t)r$ is 
the background component of the areal radius. With these definitions, the expression written in Eq.~\eqref{rel} is valid with good accuracy for $\epsilon \ll 1$. 

According to this, we can now consistently define the perturbation amplitude
as being the mass excess of the energy density within the scale $r_{\rm m}$ when $\epsilon=1$ ($aHr_{\rm m}=1$). This corresponds to defining the horizon crossing 
time $t_{\rm H}$ with a linear extrapolation from the superhorizon regime. Although this is not very accurate, it provides a well defined criterion to measure and compare consistently the amplitude of different perturbations, understanding how the threshold is varying with respect to the initial curvature profiles.

The amplitude of the perturbation measured at $t_{\rm H}$, which we refer to as 
$\delta_{\rm m} \equiv \delta(r_{\rm m},t_{\rm H})$, is given by the excess of mass averaged over a spherical volume of radius $R_{\rm m}$, defined as
\be
\label{delta}
\delta_{\rm m} = \frac{4\pi}{V_{R_{\rm m}}} \int_0^{R_{\rm m}}   \frac{\delta\rho}{\rho_{\rm b}} \,R^2 \d R\,  = 
\frac{3}{r_{\rm m}^3} \int_0^{r_{\rm m}} \frac{\delta\rho}{\rho_{\rm b}} \, r^2 \d r  \,,
\ee
where $V_{R_{\rm m}} = {4\pi}R_{\rm m}^3/3$. The second equality is obtained by 
neglecting the higher order terms in $\epsilon$, approximating \mbox{$R_{\rm m} \simeq a(t)r_{\rm m}$}, which allows to simply integrate over the comoving volume 
of radius $r_{\rm m}$. Inserting the  expression for $\delta\rho/\rho_{\rm b}$ given by 
\eqref{rel} into \eqref{delta}, one obtains $\delta_{\rm m} = \mathcal{C}(r_{\rm m})$, 
and with simple calculation seen in~\cite{Musco:2018rwt}, one obtains the fundamental relation
\be
\delta_{\rm m} = 3 \frac{\delta\rho}{\rho_{\rm b}} (r_{\rm m},t_{\rm H}),
 \label{delta_m}
\ee
showing that at $r_{\rm m}$ the perturbation amplitude is not affected by the location where this is measured.

\subsection{Threshold for PBH formation}\label{sec:th PBH form}
PBHs form when the perturbation amplitude $\delta_{\rm m} > \delta_{\rm c}$, where the value of the threshold $\delta_{\rm c}$ depends on the shape of the energy density profile and the equation of state. For a radiation-dominated Universe ($p=\frac{1}{3}\rho$), it has been found that $2/5 \leq \delta_{\rm c} \leq 2/3$, with the shape of the cosmological perturbation simply characterised by one dimensionless parameter corresponding to the width of the peak of the compaction function~\cite{Musco:2018rwt,Escriva:2019phb}
\be \label{alpha}
\alpha = - \frac{ \mathcal{C}^{\prime\prime}(r_{\rm m}) r_{\rm m}^2 }{ 4 \mathcal{C}(r_{\rm m}) }\,.
\ee
This allows to compute the threshold $\delta_{\rm c}$ as a function of the shape parameter $\alpha$, with an analytic expression up to a few percent precision~\cite{Escriva:2019phb}
\be \label{delta_c}
\delta_{\rm c} \simeq \frac{4}{15} e^{-\frac{1}{\alpha}} \frac{\alpha^{1-\frac{5}{2\alpha}}}{\Gamma\left(\frac{5}{2\alpha}\right)- \Gamma\left(\frac{5}{2\alpha}, \frac{1}{\alpha} \right)},
\ee
 where $\Gamma$ identifies the special Gamma-functions. This is consistent with the numerical analysis made in~\cite{Musco:2018rwt}, where it was shown that the effects on the threshold of additional parameters are negligible, because the collapse in spherical symmetry is basically characterized by the configuration of the region within the forming apparent horizon ($r\leq r_{\rm m}$), whose initial configuration is fully described by the shape parameter $\alpha$. 

The mass of a PBH is determined by the amplitude $\delta$ of the perturbation with respect to the corresponding threshold $\delta_{\rm c}$, according to the scaling law of critical collapse
\be \label{eq:masspbh2}
m_{\rm PBH} = \mathcal{K}(\delta-\delta_{\rm c})^\eta M_{\rm H}
\ee
where the critical exponent $\eta$ depends only on the equation of state ($\eta\simeq0.36$ for a radiation-dominated Universe), while $\mathcal{K}$, as for $\delta_{\rm c}$, depends on the initial configuration of the energy density profile, roughly varying between $1$ and $10$. For the standard scenario of a radiation-dominated Universe, expression \eqref{eq:masspbh2} describes the mass with good accuracy when $(\delta-\delta_{\rm c})\lesssim10^{-2}$, corresponding to
$m_{\rm PBH}\lesssim M_{\rm H}$.

Although non linear cosmological density perturbations are described by a non Gaussian random field, when the Universe is still radiation-dominated one can compute the 
threshold $\delta_{\rm c}$ from the shape of the Gaussian inflationary power spectrum following a simple prescription. The algorithm, divided into a few simple 
 steps, accounts for both the non linear effects associated with the relation between the Gaussian curvature perturbation and 
 the density contrast as well as for those arising at horizon crossing. While a more refined description of the various steps will be found in \cite{Musco:2020jjb}, we give here a synthetic overview of the prescription one needs to follow.
 
 \begin{enumerate}
 
\item {\bf The power spectrum of the curvature perturbation}: take the primordial power spectrum $\mathcal{P}_\zeta$ of the  
Gaussian curvature perturbation and compute, on superhorizon scales, its convolution with the transfer function $T(k,\eta)$
\begin{equation} 
P_\zeta(k,\eta) = \frac{2 \pi ^2}{k^3}\mathcal{P}_\zeta (k)  T^2(k,\eta).
\end{equation}	

\item {\bf The comoving length scale $\hr_{\rm m}$} of the perturbation is related to the characteristic scale $k_*$ of the 
 power spectrum $P_\zeta$. Compute the value of $k_* \hr_{\rm m}$ by solving the following integral equation
 \begin{equation}
\hspace{-0.45cm}\int  {\rm d}k k^2 \!\left[ ( k^2\hr_{\rm m}^2 - 1 ) \frac{\sin(k\hr_{\rm m})}{k\hr_{\rm m}}  + \cos{(k\hr_{\rm m})} \right] \!P_\zeta(k,\eta) = 0\,. 
 \end{equation}
 
\item {\bf The shape parameter:} compute the corresponding shape parameter $\alpha$ of the collapsing perturbation, 
including the correction from the non linear effects, by solving the following equation
 \begin{align}\label{alpha_sigma_ee}
 F(\alpha) \left[ 1 + F(\alpha) \right] \alpha &= 
 2 \alpha_{\text{\tiny G}} 
 \end{align}
 with 
 \begin{align}
\alpha_\text{\tiny G} &= - \frac{1}{4} \left[ 1 + \hr_{\rm m} \frac{ \int {\rm d}k k^4 \cos{(k\hr_{\rm m})} P_\zeta (k,\eta) }{ \int {\rm d}k k^3 \sin{(k\hr_{\rm m})} 
P_\zeta (k,\eta) } \right],
%  \!- \!\frac{1}{2} \!\left[ 1 \!+\! \hr_m \!\frac{ \int dk k^4 \cos{(k\hr_m)} 
%  P_\zeta (k,\eta) }{ \int dk k^3 \sin{(k\hr_m)} P_\zeta (k,\eta)} \right]
 \nonumber \\
 F(\alpha) & = \sqrt{ 1 - \frac{2}{5} e^{-\frac{1}{\alpha}} \frac{\alpha^{1-\frac{5}{2\alpha}}}{\Gamma\left(\frac{5}{2\alpha}\right) - 
\Gamma\left(\frac{5}{2\alpha},\frac{1}{\alpha}\right)} } \,.    
 \end{align}
A numerical fit of the shape parameter $\alpha$ as a function of $\alpha_\text{\tiny G}$, which represents the solution of Eq.~\eqref{alpha_sigma_ee} 
with a percent accuracy, is given by \cite{Franciolini:2021nvv}
 \begin{equation}\label{alphafitalphaG}
 	\alpha \simeq
 \left\{
\begin{aligned}
 & 1.758 \, \alpha_\text{\tiny G}^{2.335} + 1.912 \, \alpha_\text{\tiny G} 
 \quad\quad &0.1 \lesssim \, \alpha \lesssim 4.5, \\ 
& 4 \,\alpha_\text{\tiny G}^{2} + 3.930 \, \alpha_\text{\tiny G}  
\quad\quad\quad &
\alpha \gtrsim 8. \,\,\,\, 
\end{aligned}
\right. 
\end{equation}
In the intermediate region where $4.5 < \alpha <8$, no simple power law fit with a percent accuracy was found, and one needs to solve Eq.~\eqref{alpha_sigma_ee} numerically.

\item {\bf The threshold $\delta_{\rm c}$:} compute the threshold as function of $\alpha$, fitting the numerical simulations. 
\begin{itemize}
\item At \emph{superhorizon scales} making a linear extrapolation at horizon crossing ($aHr_{\rm m} = 1$).
\begin{equation}
\delta_{\rm c}  \simeq
 \left\{
\begin{aligned} 
& \alpha^{0.047} - 0.50  \quad &0.1\lesssim \, \alpha \lesssim \ 7 \,  \\ 
& \alpha^{0.035}  - 0.475  \quad\quad  &7\lesssim \, \alpha \lesssim 13  \\ 
& \alpha^{0.026} - 0.45  \quad  &13\lesssim \, \alpha \lesssim 30
\end{aligned}
\right.
\end{equation}
\item At \emph{horizon crossing} taking into account also the non linear effects. 
 \begin{equation}
\delta_{\rm c}  \simeq
 \left\{
\begin{aligned}
 &\alpha^{0.125} - 0.05  \quad &0.1 \lesssim \, \alpha \lesssim 3 \\ 
&\alpha^{0.06} + 0.025  \quad\quad  & 3 \lesssim \,\alpha \lesssim 8 \\ 
& \quad \quad 1.15  \quad  &\alpha \gtrsim 8 
\end{aligned}
\right. 
 \end{equation}
 \end{itemize}
 \end{enumerate} 

This prescription is using the fact that, if $\mathcal{P}_\zeta$ follows a Gaussian distribution, $\Phi_{\rm L}\equiv-\hat{r}\zeta^\prime(\hat{r})$ is also a Gaussian variable, and one can write $\delta_{\rm m}$ as
\be \label{Phi}
\delta_{\rm c} = \frac{4}{3} \Phi_{\rm c, L} \left( 1 - \frac{1}{2} \Phi_{\rm c, L} \right)  
\ee
where $\Phi_{\rm m, L} = \Phi_{\rm L}(\hr_{\rm m})$, and $\Phi$ is such that $0.37 \lesssim \Phi_{\rm c,L} \leq 1$.

In Fig.~\ref{fig:deltac_alpha}, we show the threshold density contrast as a function of the shape parameter (left panel). Also, in the right panel, we report the critical collapse parameter $\mathcal{K}$ defined in Eq.~\eqref{eq:masspbh2}. We recall that the critical collapse exponent $\eta$ is independent from the shape of the collapsing overdensity peak (i.e. independent from $\alpha$) and it is fixed to be $\eta = 0.358$ in a perfect radiation fluid. In the following, we will discuss how these parameters are modified when thermal effects during the evolution of the universe, causing a significant departure from the perfect radiation fluid approximation.
\begin{figure}
	\centering
    \includegraphics[width = 0.45\textwidth]{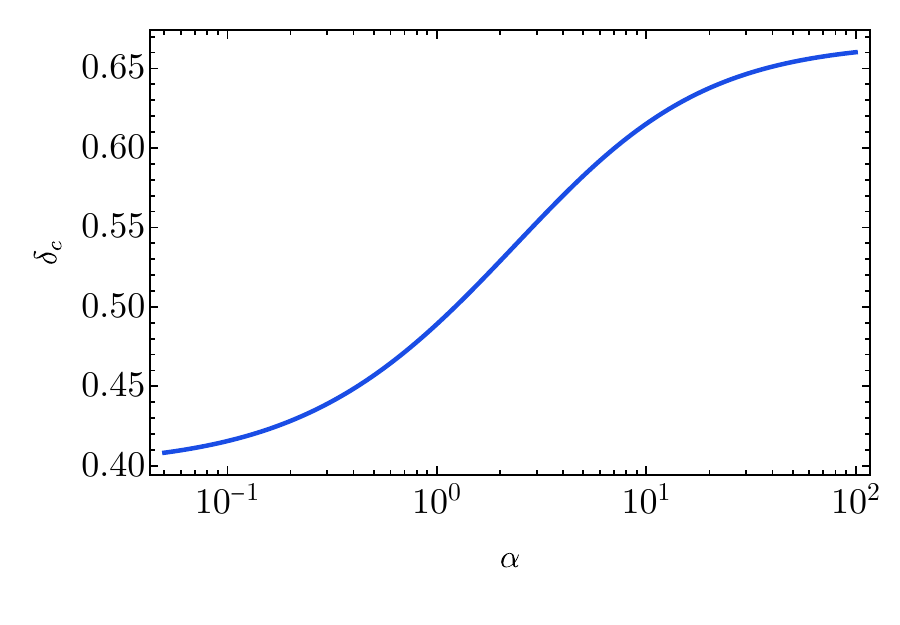}
    \includegraphics[width = 0.45\textwidth]{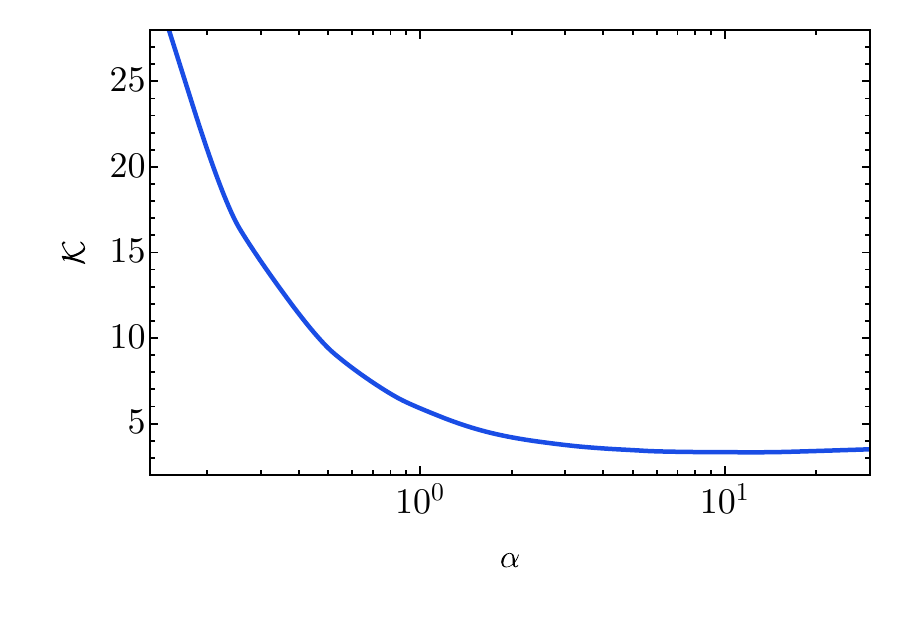}
	\caption{
 Threshold $\delta_c$ and critical collapse parameter ${\cal K}$ as a function of the shape parameter $\alpha$, assuming a universe dominate by perfect radiation (i.e. $w = 1/3$). Figure adapted from Ref.~\cite{Musco:2020jjb}
    }  \label{fig:deltac_alpha}
\end{figure}

Before concluding this Section, we mention two relevant points. First, it has been recently observed that  corrections from the non-linear radiation transfer function and the determination of the true physical horizon crossing tend to decrease the PBH abundance \cite{DeLuca:2023tun}. Secondly, 
a recent study~\cite{Musco:2021sva} which investigated the formation of PBHs with an anisotropic relativistic perfect fluid, using a covariant form of the EoS in terms of pressure and energy density gradients. This indicates that the value of the threshold $\delta_{\rm c}$ is increasing with the amplitude of the anisotropy, if this is not large, while further studies are necessary for a nonlinear modification of the EoS.  

\subsection{Gravitational collapse of primordial curvature perturbations - standard formalism}

%\JME{Need to emphasis the key ingredients discuss later in the subsections: 0) Mass function from PDF 1) $\zeta\to\delta$ effects, 2) threshold for collapse 3) thermal history}
%\JME{Start from PDF($\delta$)}

Although PBHs can be generated by many different mechanisms, we will 
focus here on the collapse of cosmological perturbations. In this case, 
the fraction $\beta_{\rm f}$ of PBHs formed in the early Universe is determined by the probability that a given primordial curvature fluctuation $\zeta$, characterized by an amplitude $\delta$ of the density contrast, is above a certain threshold $\delta_{\rm c}$. The fraction of PBHs at formation is then usually given by
\be \label{eq:pbh_fraction}
\beta_{\rm f}(m_{\rm PBH}) = \int_{\delta_{\rm c}}^{\infty} P(m_{\rm PBH},\delta) 
{\rm d} \delta \,,
\ee
and the abundance of PBHs is sensitive to the PDF $P(m_{\rm PBH},\delta)$ as to the value of the threshold $\delta_{\rm c}$. In the previous section, we have discussed in details how the amplitude $\delta$ is computed from the curvature $\zeta$, understanding how this is related to the shape of the cosmological power spectrum of curvature perturbations $\mathcal{P}_\zeta(k)$. Later we will also  discuss the effects of primordial non-Gaussianity on the abundance.

In Eq. \eqref{eq:pbh_fraction}, we have expressed the PDF in terms of the mass $m_{\rm PBH}$ of the PBHs formed instead of the number of $e$-folds because in first approximation the mass of PBHs is related to the size of the causal horizon collapsing, depending on the time of formation (see e.g. \cite{GarciaBellido:1996qt}).
%\be \label{eq:masspbh}
%M_{_\text{PBH}}\simeq 4\pi\gamma %\frac{M_{\text{pl}}^2}{H_{\text{inf}}}e^{2N}\,,
%\ee
%where $M_{\text{pl}}$ is the reduced Planck mass, $H_{\text{inf}}$ the energy scale of inflation and $\gamma$ efficiency parameter encapsulating the details of the gravitational collapse and the efficiency of reheating, that we fix to be $\gamma\sim0.2$. \IM{This relation is very rough, I would replace it with )
This implies that 
\be 
m_{\rm PBH} \simeq \gamma M_{\rm H}
\ee
where $M_{\rm H}$ is the mass of the cosmological horizon computed at horizon crossing, while $\gamma$, varying roughly between $0.1$ and $1$, is the efficiency parameter encapsulating the details of the gravitational collapse. 

Given the fraction of PBHs $\beta_{\rm f}(M)$, one can compute the contribution of PBHs to the energy density of the Universe and determine the fraction of DM they represent today. In this sense it is convenient to compute this quantity at the time of matter-radiation equality
\be
\Omega_{\rm PBH}^{\rm eq}=\int_{M_\text{ev}}^{M_{\rm eq}}\beta_{\rm eq}(m_{\rm PBH}){\rm d}\ln m_{\rm PBH}
\ee
where $M_\text{ev}$ is the lower bound due to Hawking evaporation, i.e. PBHs with $M < M_\text{ev}$ would already been evaporated by now. We assume that the fraction of PBH at the time of matter-radiation equality has grown because of the cosmic expansion followed after the time of formation, which gives 
\be
\beta_\text{eq}(m_{\rm PBH})=e^{(N_{\rm eq}-N_{\rm f})}\beta_{\rm f}(m_{\rm PBH}) \,.
\ee
The PDF is characterized by the physics in the early Universe while the threshold depends on the conditions at the time of formation. 
The analysis of the gravitational collapse of curvature perturbations to form PBHs and the appropriate threshold condition has been an active line of research in the past years \cite{Musco:2004ak,Polnarev:2006aa,Kopp:2010sh,Harada:2013epa,Young:2014ana}, with recent analysis on the dependence of the threshold on the shape of the power spectrum \cite{Yoo:2018kvb,Germani:2018jgr} and on the non-linear relation between the curvature perturbations and the density contrast \cite{Musco:2018rwt,Musco:2020jjb}, which affects significantly the computation of the abundance with respect to the standard calculation done with the linear Gaussian approximation \cite{Young:2019yug,DeLuca:2019qsy}. 

Note that the threshold is also sensitive to the equation of state (EoS). For example, the QCD phase transition is lowering the ratio between the pressure and the energy density, reducing the threshold and therefore increasing the production of PBHs formed during the QCD phase, characterized by a typical mass of the order of $1\Msun$ \cite{Jedamzik:1996mr,Byrnes:2018clq}. 

Large curvature fluctuations can be produced by very different means in the early Universe. If the curvature perturbations were Gaussian, then the PDF would be described only by its variance $\sigma^2$ %=\PSzeta(k_M)$ 
and the fraction of PBHs measured at the time of formation is usually computed as
\be 
\beta_{\rm f}(m_{\rm PBH}) = 2 \int_{\delta_{\rm c}}^{\infty} \frac{1} {\sqrt{2\pi} \sigma} e^{-\frac{\delta^2}{2\sigma^2}} {\rm d} \delta =  
1 - {\rm erf} \left( \frac{\delta_{\rm c}}{\sqrt{2} \sigma}\right) =
{\rm erfc}\left( \frac{\delta_{\rm c}}{\sqrt{2} \sigma}\right)~,
\ee
where the factor $2$ outside the integral is sometimes introduced when the abundance is computed using the excursion set method (see, for example, \cite{MoradinezhadDizgah:2019wjf,DeLuca:2020ioi}). This expression, however, does not take into account that when $\delta$ is larger then a certain value $\delta_{\rm max}$ (originally estimated $\delta_{\rm max}\sim 1$ by Bernard Carr in 1975 \cite{Carr:1975qj}), the perturbation forms a separate closed universe, topologically disconnected. A more accurate version can nevertheless be obtained re-normalizing the previous expression as
\be \label{eq:beta_pbh_Gaussian}
\beta_{\rm f}(m_{\rm PBH})
= \frac{\displaystyle{\int_{\delta_{\rm c}}^{\delta_{\rm max}} \frac{1} {\sqrt{2\pi \sigma^2}} e^{-\frac{\delta^2}{2\sigma^2}} {\rm d} \delta }} {\displaystyle{\int_0^{\delta_{\rm max}} \frac{1}{\sqrt{2\pi \sigma^2}} e^{-\frac{\delta^2}{2\sigma^2}} {\rm d} \delta }} =
1 - \frac{{\rm erf} \lp \displaystyle{\frac{\delta_{\rm c}}{\sqrt{2}\,\sigma}}\rp}
{{\rm erf}\lp\displaystyle{\frac{\delta_{\rm max}}{\sqrt{2}\,\sigma}}\rp} = 
\frac{ {\rm erfc}\left( \displaystyle{\frac{\delta_{\rm c}}{\sqrt 2 \sigma}} \right) - {\rm erfc}\left( \displaystyle{\frac{\delta_{\rm max}}{\sqrt 2 \sigma}} \right) }{ 1 - {\rm erfc}\left( \displaystyle{\frac{\delta_{\rm max}}{\sqrt 2 \sigma}} \right)}\,.
\ee
PBHs are produced through the tail of the PDF, ie. the area left under the PDF curve where $\delta$ is larger than the critical threshold $\delta_{\rm c}$. Therefore, PBH abundance is exponentially sensitive to the threshold and the statistical properties of the primordial perturbations. Even tiny modifications in the tail of the distribution can change the PBH abundance by many orders of magnitude. 

The variance of the field of density perturbations $\sigma$, according to the Gaussian distribution of $\delta$, consistent with the linear approximation of the relation between the energy density contrast and the curvature perturbation (see next Subsection), is given by
\be \label{eq:beta_pbh_Gaussian2}
\sigma^2 = \langle \delta^2 \rangle = \int\limits_0^\infty \frac{\mathrm{d}k}{k} \mathcal{P}_{\delta}(k,r) = \frac{16}{81}\int\limits_0^\infty \frac{\mathrm{d}k}{k}(kr)^4 \tilde{W}^2(k,r) T^2 (k,r) \mathcal{P}_\zeta(k)\,,
\ee
where ${\cal P}_\delta(k,r)$ and $\mathcal{P}_\zeta(k)$ are the density and the curvature power spectra, while $\tilde{W}(k,r)$ %= 3 (\sin k r - k r \,\cos k r)/(k r)^3$ 
is the Fourier transform of the top-hat smoothing function and $T(k,r)$
is the linear transfer function
\begin{align}
 \tilde{W}(k,r) &
 = 3\left [ \frac{\mathrm{sin}(k r)- k r \mathrm{cos}(k r)}{(k r)^3} \right]
 \label{eqn:window} 
 \\
T(k,r) &= 3 \left [
\displaystyle{ \frac{\mathrm{sin}\left(\frac{k r}{\sqrt{3}}\right)- \frac{k r}{\sqrt{3}} \mathrm{cos}\left(\frac{k r}{\sqrt{3}}\right)}{\left(\frac{k r}{\sqrt{3}}\right)^3} }
\right] \,.
\label{eq:T}
\end{align}
The quantities $\delta_{\rm c}$ and $\sigma$, giving the critical threshold $\nu_{\rm c}\equiv \delta_{\rm c}/\sigma$, are typically computed at super horizon scale, when the curvature perturbation $\zeta$ for adiabatic perturbations is time independent, making a linear extrapolation at horizon crossing time, which is approximately the time when PBHs are formed (we will discuss this later in more details). All this shows that a larger amplitude of the power spectrum, combined with a shape decreasing the corresponding value of $\nu_{\rm c}$, could increase significantly the fraction of PBHs because the abundance of PBHs is exponentially sensitive to the value of $\delta_{\rm c}$. 

The linear transfer function\footnote{{Ref.~\cite{DeLuca:2023tun} has recently investigated the role of non-linear corrections, coming from the non-linear radiation transfer function and the determination of the true physical horizon crossing, on the PBH abundance, showing that, while the critical threshold is larger than what routinely assumed, the
variance is unlikely to be changed with respect to the linear one.}} ensures that the variance would always converge, because the window function by itself, chosen consistently with the top-hat window function in real space used in the definition of $\delta_{\rm c}$, does not guarantee convergence if the power spectrum considered is sufficiently broad. An alternative approach is to use a Gaussian window function, which provides always convergence of the variance, but this is contaminating the value of $\delta_{\rm c}$, introducing an error in the final computation of the abundance \cite{Young:2019osy}.

%The value $k_M$ is the wave-number associated to the scale of PBH forned with a mass $M$. A larger power spectrum would then clearly increase the fraction of PBHs. In terms of the density contrast, for a Gaussian distribution, the relevant quantity is
%\be
%\sigma_\delta^2=\int_0^\infty\frac{dk}{k}\mcP_\delta(k)W^2(k ,R) ,
%\ee
%where $\mcP_\delta(k)$ is the density power spectrum, $W(k\,R)$ is the smoothing window function and $R$ the horizon scale. The fraction is then $\beta(M)={\rm erfc}\lp\delta_c/\sqrt{2}\sigma_\delta\rp/2$ and the reasoning with the power spectrum follows similarly.

\subsection{The inevitable non-Gaussianity of the primordial black hole abundance}

The non-linear relation between the curvature perturbation and the density contrast in \eqref{rel} induces an inescapable non-Gaussianity of the density contrast, even assuming a Gaussian primordial initial condition \cite{Musco:2018rwt,Young:2019yug,DeLuca:2019qsy}. 
This can be captured in the non-linear expression \cite{Young:2019yug}
 \be\label{NL_rel_d}
 \delta_{\rm m}=\delta_{\rm l}-\frac{3}{8}\delta_{\rm l}^2,
 \quad\,\,\,\, \delta_{\rm l}=-\frac{4}{3}r_{\rm m}\zeta'(r_{\rm m}).
 \ee
It highlights two  key points. First of all, the probability of forming PBHs does not depend on the comoving curvature perturbation itself $\zeta$, but on its derivative $\zeta'$. This is expected, given that on superhorizon scales one can always add or subtract to the comoving curvature perturbation a constant by a coordinate transformation and  this  may not influence any physical result. Secondly, thanks to the conservation of the probability
\be
P(\delta_{\rm l})=P[\delta_{\rm m}(\delta_{\rm l})]  \left|\frac{ {\rm d} \delta_{\rm m}}{ {\rm d} \delta_{\rm l}}\right|,
\ee
one can compute the abundance of PBHs simply using the Gaussian probability of $\delta_{\rm l}$, 
integrating it from the critical amplitude for $\delta_{\rm l}$
\be
\delta_{{\rm l},{\rm c}}=\frac{4}{3}\left(1-\sqrt{1-\frac{3}{2}\delta_{\rm c}}\right),
\ee
in terms of the critical amplitude $\delta_{\rm c}$.

\subsection{Collapse of some initial non-Gaussian perturbations}
It has been observed that nearly all models discussed in Sec. \ref{sec:theory} generate a considerable amount of primordial non-Gaussianity. We briefly summarize the $\beta$ function in case of large local non-Gaussianity, which turns the PDF into a $\chi^2$ distribution, explicitly $\delta_{\chi^2}=G^2 - \langle G^2\rangle $; and also in case of large cubic interactions, which turn the PDF into a cubic-Gaussian distribution, explicitly $\delta_{G^3}=G^3$, with $G$ a field obeying Gaussian statistics \cite{Lyth:2012yp,Linde:2012bt,Young:2013oia,Bugaev:2013fya}.
\bea \label{eq:beta_pbh_ng}
&& \beta_{{\rm f}, \, \chi^2}(m_{\rm PBH}) =
{\rm erfc}\left( \sqrt{ \frac{1}{2} + \frac{\delta_{\rm c}}{\sqrt 2 \sigma} } \right)~, \nonumber\\
&& \beta_{{\rm f}, \, G^3}(m_{\rm PBH}) = 
{\rm erfc}\left( \left( \frac{\delta_{\rm c}}{\sqrt \frac{8}{15} \sigma} \right)^{1/3} \right).
\eea
We also note that the threshold density $\delta_{\rm c}$ depends slightly on the non-Gaussianity \cite{Kehagias:2019eil,Kitajima:2021fpq,Escriva:2022pnz}.

\subsection{Non-perturbative abundance from non-Gaussian perturbations}\label{sec:NG_nonpert}

Refs.~\cite{Ferrante:2022mui,Gow:2022jfb} addressed the problem of computing the impact of a general form of local primordial NG on PBH abundance beyond the perturbative approach.  In particular, let us focus on the functional form
\begin{align}
    \zeta(\vec{x}) = F(\zeta_\text{\tiny G}(\vec{x}))\,,\label{eq:MainF}
\end{align}
where $F$ is a generic non-linear function of the Gaussian component $\zeta_\text{\tiny G}$. 
There is a large number of relevant cases in the literature of PBH formation 
where the NGs can be modelled as in Eq.\,(\ref{eq:MainF}).
We summarize few relevant examples in 
Tab.~\ref{tabNGmm} (adapted from Ref.~\cite{Ferrante:2022mui}).

{
\renewcommand{\arraystretch}{2}
\begin{table}[!h]
% \vspace{.1cm}
\begin{tabularx}{1 \columnwidth}{|X|l|}
\hline
\hline
Power-series expansion
\cite{Bugaev:2013vba,Nakama:2016gzw,Byrnes:2012yx,Young:2013oia,Yoo:2018kvb,Kawasaki:2019mbl,Yoo:2019pma,Riccardi:2021rlf,Taoso:2021uvl,Meng:2022ixx,Escriva:2022pnz} &
 $\zeta = \zeta_\text{\tiny G} + \frac{3}{5}f_\text{\tiny NL}\zeta_\text{\tiny G}^2
 + \frac{9}{25}g_{\rm NL}\zeta_\text{\tiny G}^3 + \dots$
\\
\hline 
Curvaton \cite{Sasaki:2006kq,Pi:2021dft} &  $\zeta = \log\left[X(r_{\rm dec},\zeta_\text{\tiny G})\right]$
\\
\hline
USR \cite{Atal:2019cdz} &
$\zeta = -\left(\frac{6}{5}f_\text{\tiny NL}\right)^{-1}\log\left(1-  
    \frac{6}{5}f_\text{\tiny NL}\zeta_\text{\tiny G}\right)$
\\
\hline
USR with an upward step\,\cite{Cai:2022erk} &
 $\zeta = -\frac{2}{|h|}
    \left[\sqrt{1-|h|\zeta_\text{\tiny G}} - 1\right]$
\\
\hline
\hline
\end{tabularx}
\caption{
Some examples of PBH formation scenarios characterised by
local non-Gaussianity of the form defined in Eq.~\eqref{eq:MainF}.
The function $X(r_{\rm dec},\zeta_\text{\tiny G})$ given in the second row is defined in \cite{Sasaki:2006kq}.
}
\label{tabNGmm}
\end{table}
}

It is important to stress here that going beyond the perturbative approach becomes a necessity at least when dealing with broad spectra of curvature perturbations. 
This is because, as shown in Ref.~\cite{Ferrante:2022mui}, 
in these cases the probability distribution function of $\zeta_\text{\tiny G}$ is evaluated beyond the radius of convergence of the series expansion, leading to inaccurate results if the series is truncated at finite order. Notice that in non-attractor models the sign of non-Gaussianity is always positive \cite{Firouzjahi:2023xke}, thus enhancing the PBH abundance. 

Using the relation between the curvature perturbation and the compaction function in the gradient expansion \eqref{rel}, it is possible write
\begin{align}\label{eq:CompactionFull}
\mathcal{C}(r) = 
-2\Phi\,r\,\zeta^{\prime}(r)\left[
1 + \frac{r}{2}\zeta^{\prime}(r)
\right] = 
\mathcal{C}_1(r) - \frac{1}{4\Phi}\mathcal{C}_1(r)^2\,,
~~~~~~~~~~~~~
\mathcal{C}_1(r) \equiv -2\Phi\,r\,\zeta^{\prime}(r)\,,
\end{align} 
where $\mathcal{C}_1(r)$ defines the so-called linear component of the compaction function and we introduced the prefactor $\Phi$ that accounts for potential thermal effects (see more in depth discussion below Eq.~\eqref{eqn:non-linear}).
Substituting in the previous expression the relation $\zeta = F(\zeta_\text{\tiny G})$ (see eq.\,\eqref{eq:MainF}), one obtains that the linear component of the compaction function takes the form
\begin{align}\label{eq:C1expl}
\mathcal{C}_1(r) = -2\Phi\,r\,\zeta_\text{\tiny G}^{\prime}(r)\,
\frac{dF}{d\zeta_\text{\tiny G}} = 
\mathcal{C}_\text{\tiny G}(r)\,
\frac{dF}{d\zeta_\text{\tiny G}}\,
\end{align}
with $\mathcal{C}_\text{\tiny G}(r) \equiv 
-2\Phi\,r\,\zeta_\text{\tiny G}^{\prime}(r)$. Consequently, the compaction function reads
 \begin{align}\label{eq:CCgau}
\mathcal{C}(r) = 
\mathcal{C}_\text{\tiny G}(r)\,
\frac{dF}{d\zeta_\text{\tiny G}} 
 - \frac{1}{4\Phi}
 \mathcal{C}^2_\text{\tiny G}(r)
 \left(\frac{dF}{d\zeta_\text{\tiny G}}
 \right)^2\,.
 \end{align}
Crucially,  $\mathcal{C}(r)$ depends on both the Gaussian linear component $\mathcal{C}_\text{\tiny G}$ and the Gaussian  
 curvature perturbation $\zeta_\text{\tiny G}$. 
 Both these random variables are Gaussian; $\zeta_\text{\tiny G}$ is Gaussian by definition while  
 $\mathcal{C}_\text{\tiny G}$ is defined by means of the derivative of the Gaussian variable $\zeta_\text{\tiny G}$.

Adopting threshold statistics on the compaction function, one can express the non-Gaussian mass fraction as an integration of the joined PDF in the domain which results in overthreshold perturbations, i.e. 
\begin{align}
\beta_{\rm NG} & = \int_{\mathcal{D}}\mathcal{K}(\mathcal{C} - \mathcal{C}_{\rm th})^{\gamma}
\textrm{P}_\text{\tiny G}(\mathcal{C}_\text{\tiny G},\zeta_\text{\tiny G})
\d\mathcal{C}_\text{\tiny G} 
\d\zeta_\text{\tiny G}\,,
\label{eq:CompactionIntegral}
 \\
 \mathcal{D} & = 
\left\{
\mathcal{C}_\text{\tiny G},\,\zeta_\text{\tiny G} \in \mathbb{R}~:~~
\mathcal{C}(\mathcal{C}_\text{\tiny G},\zeta_\text{\tiny G}) > \mathcal{C}_{\rm th}  
~\land~\mathcal{C}_1(\mathcal{C}_\text{\tiny G},\zeta_\text{\tiny G}) < 2\Phi
\right\}\,,\label{eq:RegionD}
\end{align} 
where the multivariate normal distribution of $(\mathcal{C}_\text{\tiny G},\zeta_\text{\tiny G})  $ can be written as
\begin{equation}
    \textrm{P}_\text{\tiny G}
    (\mathcal{C}_\text{\tiny G},\zeta_\text{\tiny G})  
    =
 \frac{1}{(2\pi)\sigma_c\sigma_{r}\sqrt{1-\gamma_{cr}^2}}
 \exp\left(
 -\frac{\zeta_\text{\tiny G}^2}{2\sigma_r^2}
 \right)
 \exp\left[
 -\frac{1}{2(1-\gamma_{cr}^2)}\left(
 \frac{\mathcal{C}_\text{\tiny G}}{\sigma_c} - \frac{\gamma_{cr}\zeta_\text{\tiny G}}{\sigma_r}
 \right)^2
 \right]\,.
\end{equation}
To shorten the notation, we have followed Ref.~\cite{Young:2022phe} and defined the element of the covariance matrix as 
 \begin{align}
 \langle\mathcal{C}_{\rm G}\mathcal{C}_{\rm G}\rangle & = \sigma_c^2 =
  \frac{4\Phi^2}{9}\int_0^{\infty}\frac{dk}{k}
  (kr_m)^4 \tilde W^2(k,r_m) T^2(k,r_m) P_{\zeta}(k)\,,\label{eq:Var1}
   \\
 \langle\mathcal{C}_{\rm G}\zeta_{\rm G}\rangle & = \sigma_{cr}^2 = 
 \frac{2\Phi}{3}\int_0^{\infty}\frac{dk}{k}(kr_m)^2
 \tilde W(k,r_m)
 W_s(k,r_m) T^2(k,r_m) P_{\zeta}(k)\,,
  \\
  \langle\zeta_{\rm G}\zeta_{\rm G}\rangle & = \sigma_r^2 =   \int_0^{\infty}\frac{dk}{k}
  W_s^2(k,r_m) T^2(k,r_m) P_{\zeta}(k)\,,\label{eq:Var3}
 \end{align}
where $W_s(k,r) = \sin(kr)/kr$, $\tilde W(k,R)$ and $T(k,\tau)$ are given in Eqs.\,\eqref{eq:T} and \eqref{eqn:window},
while $\gamma_{cr} \equiv {\sigma_{cr}^2}/{\sigma_c \sigma_r}$.

\subsection{Thermal history}

\subsubsection{The equation of state and the number of relativistic degrees of freedom}

The reheating at the end of inflation should have filled the Universe with radiation. In the absence of extensions beyond the Standard Model (SM) of particle physics, the Universe remains dominated by relativistic particles with an energy density decreasing as the fourth power of the temperature as the Universe expands and cools down. The number of relativistic degrees of freedom remains constant ($g_{*} = 106.75$) until around $200$ GeV, when the temperature of the Universe falls to the mass thresholds of SM particles.

As shown in Fig.~\ref{fig:g-and-w-of-T} (left panel), the first particle to become non-relativistic is the top quark at $T \simeq m_{\rm t} = 172$ GeV, followed by the Higgs boson at $125$ GeV, and the $Z$ and $W$ bosons at $92$ and $81$ GeV, respectively. These particles become non-relativistic at nearly the same time and this induces a significant drop in the number of relativistic degrees of freedom down to $g_{*} = 86.75$. There are further changes at the $b$ and $c$ quark and $\tau$-lepton thresholds but these are too small to appear in Fig.~\ref{fig:g-and-w-of-T}. Thereafter, $g_{*}$ remains approximately constant until the QCD transition at around $200$ MeV, when protons and neutrons condense out of the free light quarks and gluons. The number of relativistic degrees of freedom then falls abruptly to $g_{*} = 17.25$. A little later the pions become non-relativistic and then the muons, giving $g_{*} = 10.75$. After that, $g_{*}$ remains constant until $e^{+}e^{-}$ annihilation and neutrino decoupling at around $1$ MeV, when it finally drops to $g_{*} = 3.36$.

\begin{figure}[!h]
	\centering
    \includegraphics[width = 0.45\textwidth]{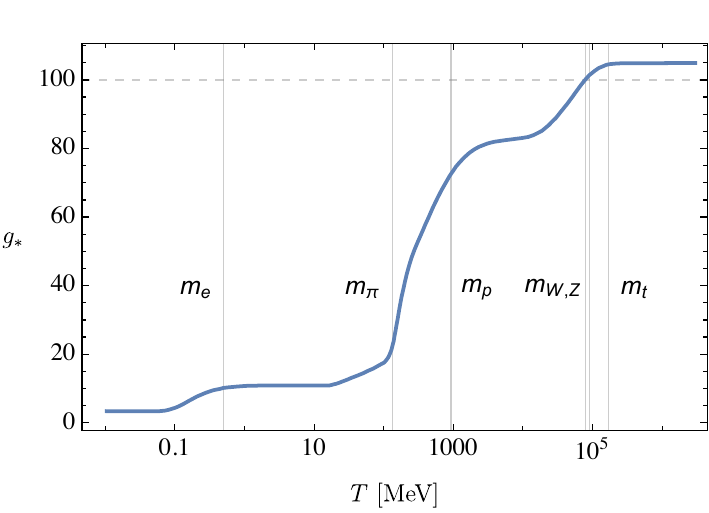}
    \includegraphics[width = 0.45\textwidth]{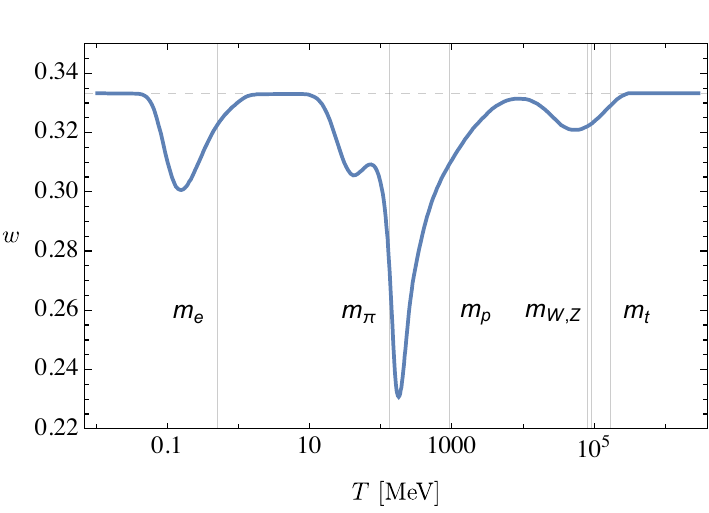}
	\caption{Relativistic degrees of freedom $g_{*}$ ({\it Left panel}) 
			and equation-of-state parameter $w$ ({\it Right panel}), 
			both as a function of temperature $T$ (in MeV).
			The grey vertical lines correspond to the masses of the electron, 
			pion, proton/neutron, $W$, $Z$ bosons and top quark, respectively.
			The grey dashed horizontal lines indicate values of 
			$g_{*} = 100$ and $w = 1 / 3$, respectively.
	\label{fig:g-and-w-of-T}
	}  
%\end{figure}

% %\begin{figure}
%  %   \centering
%     \includegraphics[width=9cm,angle=90]{Figures/fPBH_ns_varying}
%     \caption{Expected PBH mass function for a power-law primordial power spectrum with spectral index $n_{\rm s}= 0.965/0.97/0.975$ (red, blue and green curves, respectively) and an amplitude fixed to get an integrated fraction $f_{\rm PBH} = 1$, including the effects of the thermal history.
%     %\textcolor{red}{The figure contains also an incomplete set of the constraints on the PBH abundance (black lines). To compare it with a more complete and updated version see Fig.~\ref{fig:PBHlimits}. 
%     Figure from~\cite{Carr:2019kxo}.}
%     \label{fig:fPBH_thermal_hist}
\end{figure}

Whenever the number of relativistic degrees of freedom suddenly drops, it changes the effective equation of state parameter $w$. As shown in Fig.~\ref{fig:g-and-w-of-T} (right panel), there are thus three main periods in the thermal history of the Universe when $w$ decreases. After each period, $w$ resumes its relativistic value of $1 / 3$ but each sudden drop modifies the probability of gravitational collapse of any large curvature fluctuations present at that time.

\subsubsection{Thermal effects on PBH formation and abundance}
% and the PBH abundance beyond the threshold}
Because the threshold $\delta_{\rm c}$ { is a measure of the pressure gradients working against the gravitational collapse, it} depends on the equation-of-state, and since the PBH abundance depends exponentially on the value of the threshold, the thermal history should have left imprints in the PBH mass distribution. In particular, one finds that the PBH mass distribution is sufficiently broad, in the form of a high peak at the solar-mass scale, a bump around $30 M_\odot$, which would have implications for {LVK} observations, as well as one bump at planetary-masses from the electroweak scale and a final one for intermediate-mass PBHs~\cite{Byrnes:2018clq,Carr:2019kxo}.  

The expected mass distribution for an almost scale invariant power spectrum Fig.~\ref{fig:fPBH2models} for two representative models of PBH formation from a (nearly) scale invariant power spectrum with spectral index $n_{\rm s} = 0.97$ and $n_{\rm s} = 1$.  These models are later used in this review for computing the PBH merger rates and SGWB. 
%\sout{For getting the} 
{ However to obtain Fig.~\ref{fig:fPBH2models} the effects of the variation of the equation-of-state and of the pressure gradients \textit{during} the PBH formation process were neglected.  }

%\sout{But} 
{ More recently new  numerical simulations of PBH formation at the QCD epoch have been performed~\cite{Escriva:2022bwe,Musco:2023dak}, including the time and radial variations of the equation of state. 
%\sout{and showing that the resulting} 
These modify significantly the PBH mass function characterised by 
%\sout{is significantly impacted, in particular with} 
a lower but broader QCD-induced peak.} The physical reason of this broadening is the time duration of the formation process, during which the changes in the equation-of-state favor the collapse.  As a result, the threshold and the resulting PBH mass distribution do not only depend on the fluctuation mass at horizon crossing, but also on the precise evolution of the equation of state, as well as the details of the curvature fluctuation profile.  

%The resulting mass distributions are shown in Fig.~\ref{fig:fofmPBH_newsims}for different spectral indices and compared to the case where these effects are neglected.  Similar simulations have been used in~\cite{Franciolini:2022tfm,Musco:2023dak} that included additional effects on the PBH mass in the critical regime. We will discuss these results in more detail in the next Section.}

%\textcolor{red}{For reference, in the following section we will adopt the mass distribution derived with the techniques described here, as shown in Fig.~\ref{fig:fPBH2models}.}

 \begin{figure}
      \centering
  	\includegraphics[width = 0.49\textwidth]{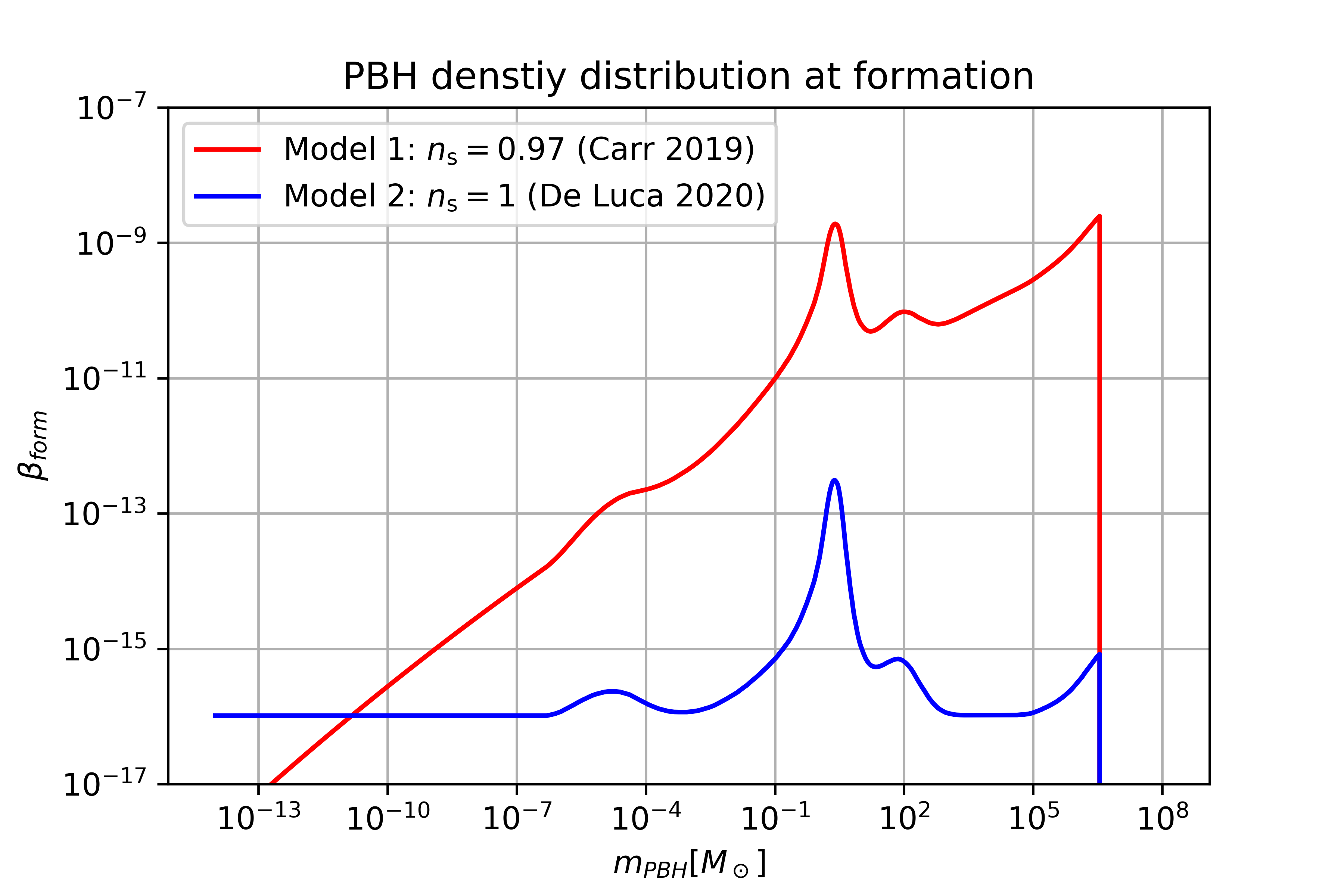}
  	\includegraphics[width = 0.49\textwidth]{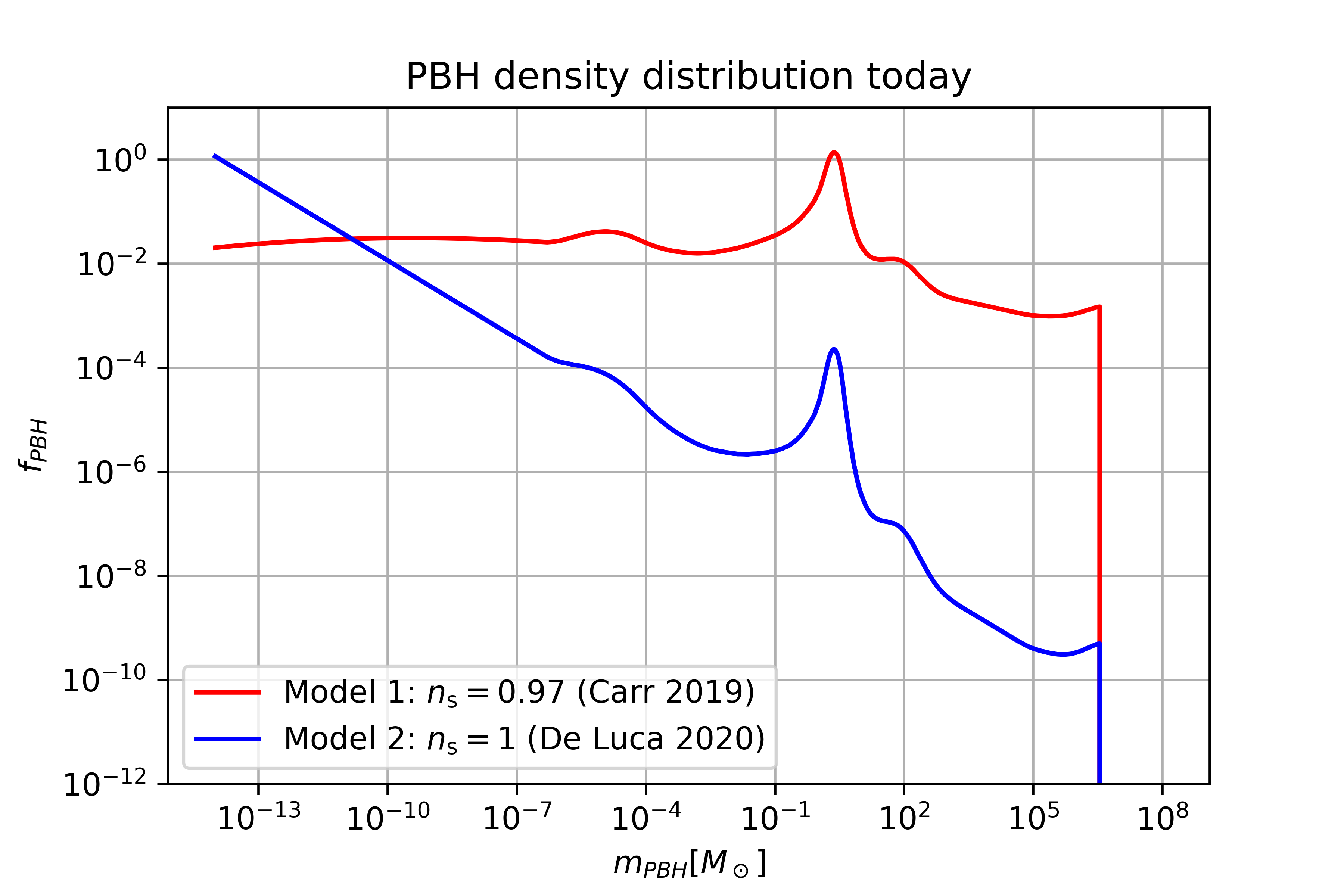}
      \caption{PBH density fraction at formation $\beta^{\rm form}$ (left panel) and the corresponding PBH mass function $f_{\rm PBH}$ today (right panel), neglecting the effects of PBH growth by accretion and hierarchical mergers, for two models with a power-law primordial power spectrum and including the effects of thermal history:  Model 1 from~\cite{Carr:2019kxo,Clesse:2020ghq} with spectral index $n_{\rm s} = 0.97$; Model 2 from~\cite{DeLuca:2020agl,Byrnes:2018clq} with $n_{\rm s} = 1.$ and a cut-off mass of $10^{-14} M_\odot$.  The transition between the large-scale and small-scale power spectrum is fixed at $k=10^3 {\rm Mpc}^{-1}$.  The power spectrum amplitude is normalized such that both models produce an integrated PBH fraction $f_{\rm PBH} =1$, i.e. PBH constitute the totality of Dark Matter.  A value of $\gamma = 0.8$ (ratio between the PBH mass and the Hubble horizon mass at formation) was assumed.  Figure produced for~\cite{LISACosmologyWorkingGroup:2022jok}.
      }
      \label{fig:fPBH2models}
  \end{figure}

%\begin{figure}
%   \centering
%  \includegraphics[width=9cm]{Figures/comparison_mass_function.pdf}
%    \caption{Comparison of the expected PBH mass function for a power-law primordial power spectrum with spectral index $n_{\rm s}= 0.965/0.97/0.975$ (blue, red and green curves, respectively) and an amplitude fixed to get an integrated fraction $f_{\rm PBH} = 1$, including the effects on the threshold from the thermal history with (solid curves) or without (dashed curves) taking the time and radial variations of the equation-of-state in the simulations, during the PBH formation process.  Figure from~\cite{Escriva:2022bwe}.}
%    \label{fig:fofmPBH_newsims}
%\end{figure}
  
%\subsubsection{Thermal effects on the PBH abundance beyond the threshold}

{The thermal history, affecting the equation of state of the fluid dominating the energy density at the time of PBH collapse, impacts many aspects of the PBH formation. 
Following Refs.~\cite{Franciolini:2022tfm,Musco:2023dak}, here we discuss %\sout{two new effects that complement the modulation of threshold introduced in the previous section.} 
{ how the threshold is varying according to the equation of state,}
focusing on 
%\sout{the solar mass range, we will discuss} 
the modification of the relation between the density contrast and the curvature perturbation 
%\sout{induced by a time variation of} 
{ when} the equation of state { is varying with time.} 
%\sout{as well as} 
{This has an impact also on the PBH masses with} a modified { scaling  behaviour} of the critical collapse.}
%\sout{scaling}.

%While the former reduces the impact of the threshold reduction, the latter induces a pile-up effect around the solar mass that enhances the QCD peak. 

When the equation-of-state is not constant, 
Eq.~\eqref{rel}
is modulated by an overall factor $\Phi(t)$, which we can define in the long wavelength approximation
as \cite{Franciolini:2022tfm}
\begin{equation} 
\frac{\delta\rho}{\rho_b}(r,t) = - 
\frac{4}{3} \Phi
\left(\frac{1}{aH}\right)^2 e^{-5\zeta(r)/2} \nabla^2 e^{\zeta(r)/2}.
\label{eqn:non-linear}
\end{equation}
$\Phi(t)$ can be computed by solving the
 equation~\cite{Polnarev:2006aa}
\begin{equation}\label{eq:Phievo}
    \frac{1}{H}\frac{\d \Phi(t) }{\d t} + \frac{5+3 w(t)}{2} \Phi(t)-\frac{3}{2} (1+ w(t)) = 0
\end{equation}
integrated from past infinity to the time when the amplitude of the perturbation is computed.
Assuming the standard models of the very early Universe, 
the initial condition for Eq.~\eqref{eq:Phievo} is derived  assuming a radiation dominated medium, with an equation-of-state $p=w\rho$ and $w=1/3$.
When a constant $w(t) = \bar w$ characterises the fluid dominating the energy budget of the Universe, we have 
${\d \Phi(t) }/{\d t} = 0$ and one obtains 
\begin{equation}\label{solPhi}
\bar \Phi = \frac{3(1+ \bar w)}{(5+3 \bar w)},
\end{equation}
yielding $\bar \Phi=2/3$ for a radiation fluid with $\bar w = 1/3$.
Such a prefactor is standardly reported in Equation~\eqref{rel}.
The behavior of $\Phi$ across the QCD phase transition, obtained by solving Eq.~\eqref{eq:Phievo}, differs from the average $\bar \Phi$, particularly in the region where $w$ and $c_s^2$ are quickly varying with respect to $M_H$ (left panel of Fig.~\ref{fig:EoS}). 
The resulting evolution of $\Phi$ as a function of horizon crossing mass is shown in the middle panel of Fig.~\ref{fig:EoS}.

\begin{figure}[t!]
% \vspace{-1.6cm}
	\centering
    \includegraphics[width=0.49\textwidth]{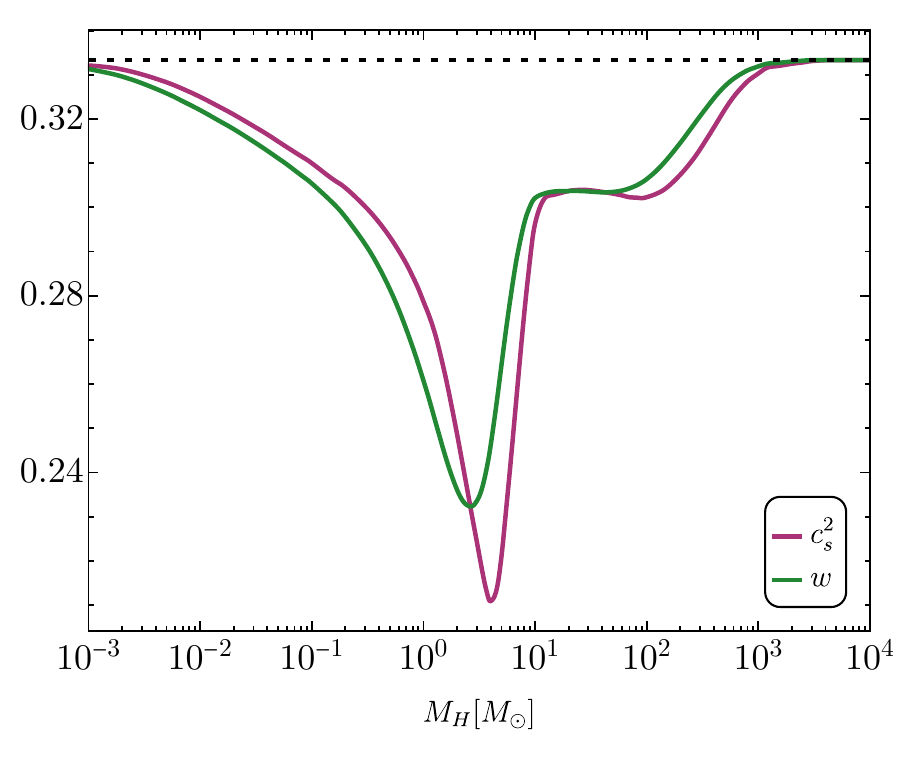}
    \includegraphics[width=0.49\textwidth]{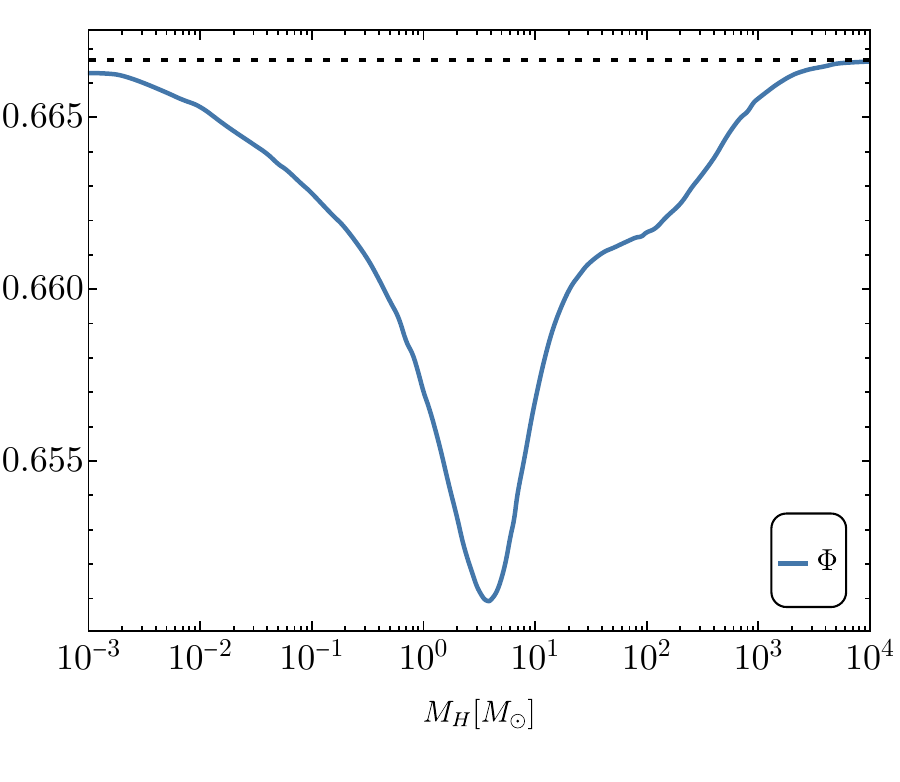}
\caption{
	Figure taken from \cite{Franciolini:2022tfm}.
	{Left panel:} the equation-of-state parameter $w=p/\rho$ (red) and squared speed of sound (blue) as functions of the cosmological horizon mass $M_H$.
	{Right panel:} 
	Evolution of the equation-of-state dependent parameter $\Phi$, relating the density contrast to the curvature perturbation as functions of the cosmological horizon mass $M_H$.
	}
\label{fig:EoS}
\end{figure}
\begin{figure}[t!]
% \vspace{-1.6cm}
	\centering
    \includegraphics[width=0.59\textwidth]{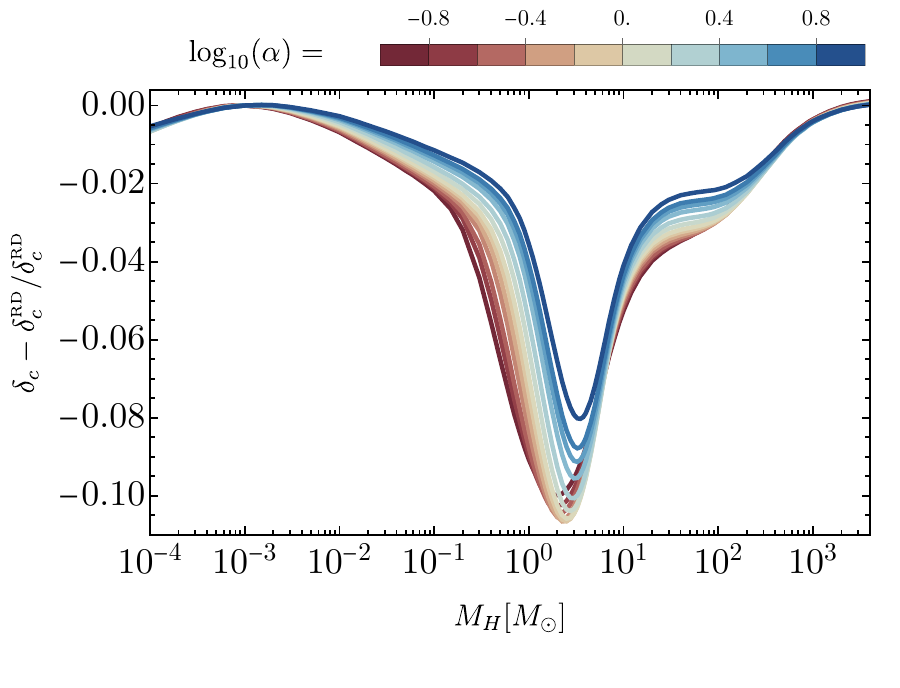}
	\caption{
 Relative variation of the threshold compared to what is obtained assuming perfect radiation as a function of the horizon crossing time (parametrised here with $M_H$) induced by the QCD thermal effects. 
The color code indicates the different values of $\log_{10}(\alpha)$ as indicated by the bar on top of the frame.  Figure adapted from Ref.~\cite{Musco:2023dak}.
	}
\label{fig:Threshold}
\end{figure}

Overall, the QCD phase transition introduces an additional degree of freedom into the problem, which is the characteristic scale of the horizon crossing of the cosmological perturbation. {In Figure~\ref{fig:Threshold}, obtained from the computations done in~\cite{Musco:2023dak}, we summarise the modified behaviour of the threshold $\delta_{\textrm{c}}$ for different values of the shape parameter $\alpha$ defined in \eqref{alpha}. The different lines with a color varying between red, for smaller values of $\alpha$ and blue, for larger values, shows as the threshold is varying during the QCD epoch, as function of the cosmological horizon mass $M_H$. This is in agreement with the results obtained in~\cite{Escriva:2022bwe}, apart from a few percent difference due to the different set up of the initial conditions. }

{In~\cite{Musco:2023dak} the authors has computed also the modified critical scaling behaviour defined in Eq.~\eqref{eq:masspbh2}, which
is modified as
\be \label{eq:masspbh22}
m_{\rm PBH} = \mathcal{K}(M_H)
\llp 
\delta-\delta_{\rm c}(M_H)
\rrp^{\gamma(M_H)}
M_{\rm H},
\ee
where $\delta_c(M_H)$, $\gamma(M_H)$ and $\mathcal{K}(M_H)$ are fitted to numerical simulations performed in Ref.~\cite{Musco:2023dak} as a function of $M_H$, i.e. when the perturbation is crossing the cosmological horizon, defined as 
\begin{equation}
    M_H \simeq  
 17 M_\odot\left(\frac{g_*}{10.75}\right)^{-1/6}
 \left(\frac{k/\kappa}
 {10^6{\rm Mpc}^{-1}}\right)^{-2}, 
 \label{M-k}
\end{equation}
where $g_*$ is the number of degrees of freedom of relativistic particles and $\kappa\equiv r_m k$ relates the spectral wavenumber with the horizon crossing size of perturbations as derived in Sec.~\ref{sec:th PBH form}. }
The different lines shown in Fig.~\ref{fig:MF_critical_collapse_1} with a color varying between red, for smaller values of $M_H$, and blue for larger values, shows how the scaling law is modified by the characteristic scale of the problem. 
Notice that an exact power-law critical behaviour is only obtained close enough to the density threshold $(\delta-\delta_c \lesssim 10^{-5})$, where the PBH masses are significantly smaller than the cosmological horizon mass, not able to affect significantly the collapse, while for larger values the equation-of-state during the QCD epoch induces further modifications. 
Ref.~\cite{Franciolini:2022tfm} fits the relation between the PBH and horizon mass using the power-law template~\eqref{eq:masspbh22} in the range of $\delta$ which most contributes to the abundance, i.e. $(\delta - \delta_c)\in [10^{-5},2\times 10^{-2}]$, and finds that deviations from the functional form used in Eq.~\eqref{eq:masspbh22} would only induce a small correction which we can neglect. 
The resulting values of ${\cal K}(M_H)$ and ${\gamma}(M_H)$ used here are shown in the right plot of Fig.~\ref{fig:MF_critical_collapse_1}.
For reference, we also show the critical collapse during the radiation dominated epoch of the early Universe, indicated with a black dashed line.
A general trend is observed: for $M_H\lesssim3 M_\odot$, there is a tendency to generate heavier PBHs, while the opposite is found when $M_H    \gtrsim 3 M_\odot$. This can be seen in the left panel of Fig.~\ref{fig:MF_critical_collapse_1}, where orange (light blue) lines fall above (below) the dashed black line indicating the result for a radiation perfect fluid. The fitted values of ${\cal K}(M_H)$ and ${\gamma}(M_H)$ are shown in the right panel of Fig.~\ref{fig:MF_critical_collapse_1}.

\begin{figure*}[t!]
	\centering
	\includegraphics[width=0.49\textwidth]{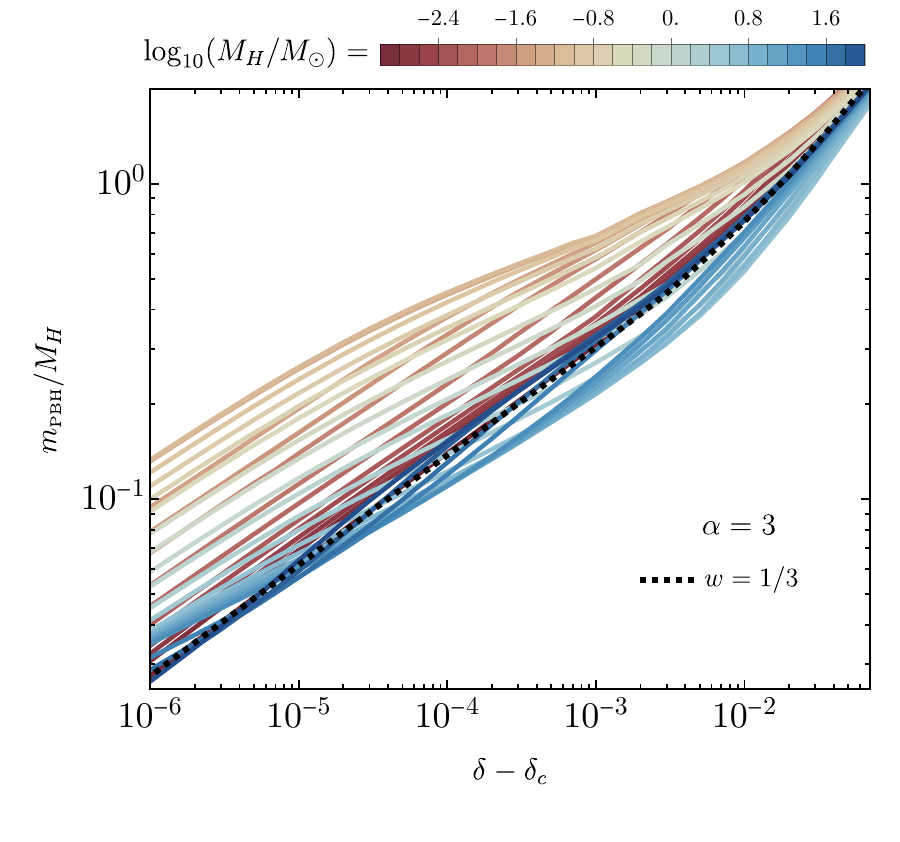}
	\includegraphics[width=0.49\textwidth]{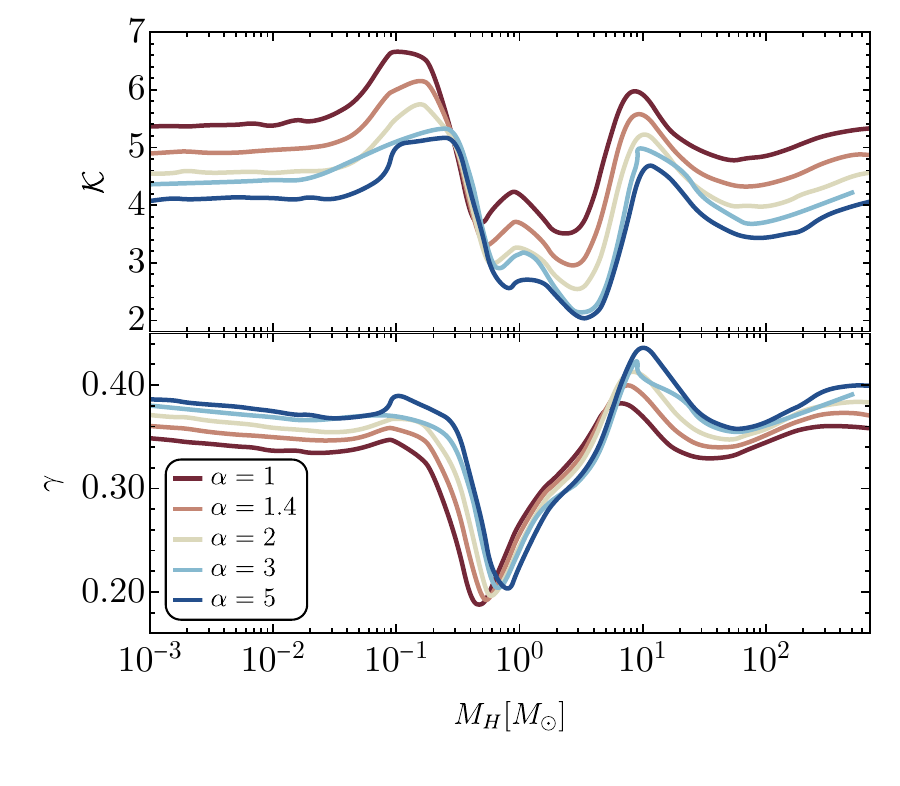}
\caption{
Figure from \cite{Franciolini:2022tfm}. {\bf Left panel:} PBH mass $m_\PBH$ plotted as a function of $\delta-\delta_c$ computed at the cosmological horizon crossing (see Ref.~\cite{Musco:2023dak} for more details). The behavior for a radiation dominated medium is plotted with a black dashed line. {\bf Right panel:} The values of the power law coefficients in Eq.~\eqref{eq:masspbh22} found by fitting the results of numerical simulations shown in the left panel.}
\label{fig:MF_critical_collapse_1}
\end{figure*}

In order to compute the full mass distribution, including both the thermal effects as well as non-linearities \eqref{NL_rel_d}, it is convenient to invert the relation between horizon and PBH mass through the critical collapse relation~\eqref{eq:masspbh22}, focusing only on the type-I branch \cite{Musco:2020jjb},  as 
\begin{equation}\label{change_var}
     \delta_l = 
2 \Phi 
\left(1 - 
\sqrt{\Lambda}
     \right),
% \end{equation}
% where
% \begin{equation}
\qquad \text{where} \qquad
    \Lambda = 
    1-\frac{\delta_c}{\Phi}
     -\frac{1}{\Phi} \left(
     \frac{m_\PBH}{{\cal K}M_H}
     \right)^{1/\gamma }.
\end{equation}
Finally, one can arrive at a compact expression for the computation of the mass distribution which is
% \begin{widetext}
\begin{align} \label{mfder}
\psi (m_\PBH) &
\equiv \frac{1}{\Omega_\PBH} 
\frac{\mathrm{d}\Omega_\PBH}
{\mathrm{d} m_\PBH}
=
\frac{8}{3 \pi\, \Omega_\PBH m_\PBH}
\int
\frac{\mathrm{d} M_{H}}{M_H}
\left( \frac{M_{\rm eq}}{M_{H}} \right)^{1/2}
\lp \frac{\sigma_1}{a H \sigma_0 } \rp  ^3
\nonumber \\
&
\times 
\frac{\Phi ^3 {\cal K}}{\gamma \sigma_0 ^4}
\left(\frac{m_\PBH}{{\cal K} M_H}\right)^{\frac{1+\gamma}{\gamma}}
\frac{\lp 1- \sqrt{\Lambda}\rp ^3}{\Lambda^{1/2}}
\exp\llp
- \frac{2 \Phi^2 }{\sigma_0^2} 
\lp 1-\sqrt{\Lambda}
\rp^2
\rrp ,
\end{align}
% \end{widetext}
and the integration range of $M_H$ is subject to the condition  $\Lambda>0$ (because we require $\delta>\delta_c$). 
An extension of this formula that includes both thermal effects and non-Gaussianities of $\zeta$, as discussed in Sec.~\ref{sec:NG_nonpert}, is presented in Ref.~\cite{Ferrante:2022mui}.
The quantities ${\cal K}(M_H)$, $\gamma(M_H)$,
 $\Phi(M_H)$, $\delta_c(M_H)$, and $\sigma_i(M_H)$ are left within the integration over the horizon mass scale, as they all explicitly depend on $M_H$ when thermal effects are included. 
 The variances can be computed integrating the curvature power spectrum as 
 \begin{align}
&\sigma^2_i (r_m)
% &
= \frac{4}{9}\Phi^2
\int\limits_0^\infty \frac{\mathrm{d}k}{k}(k r_m)^4 
\tilde{W}^2(k, r_m)
T^2 (k, r_m) 
k^{2i}
\mathcal{P}_\zeta(k),
\label{eqn:variance}
\end{align}
where $i=0,1$ and the window and transfer functions are defined in Eqs.~\eqref{eqn:window} and \eqref{eq:T}, respectively. 
If a nearly scale invariant spectrum of curvature perturbations is assumed, the smoothing scale $r_m$ is related with power spectral modes as
    $r_m k  \equiv \kappa = 4.49$, as discussed in Sec.~\ref{sec:th PBH form}.
This relation is strictly valid for a shape parameter  $\alpha = 3$ \cite{Musco:2020jjb}.
Notice that in the low mass limit, that is when we consider masses much below the smallest $M_H$,
one can find that the mass distribution~\eqref{mfder} scales as
\begin{equation} \label{tail}
    \psi(m_\PBH) \propto
    \lp 
    {m_\PBH}
    \rp^{1/\gamma},
\end{equation}
where $1/\gamma\simeq {2.8}$ if one assumes the energy density of the Universe behaving as a relativistic fluid with $w=1/3$, which gives $\gamma \approx 0.36$~\cite{Niemeyer:1997mt}.

\begin{figure}[!t]
	\centering
	\includegraphics[width=0.6\textwidth]{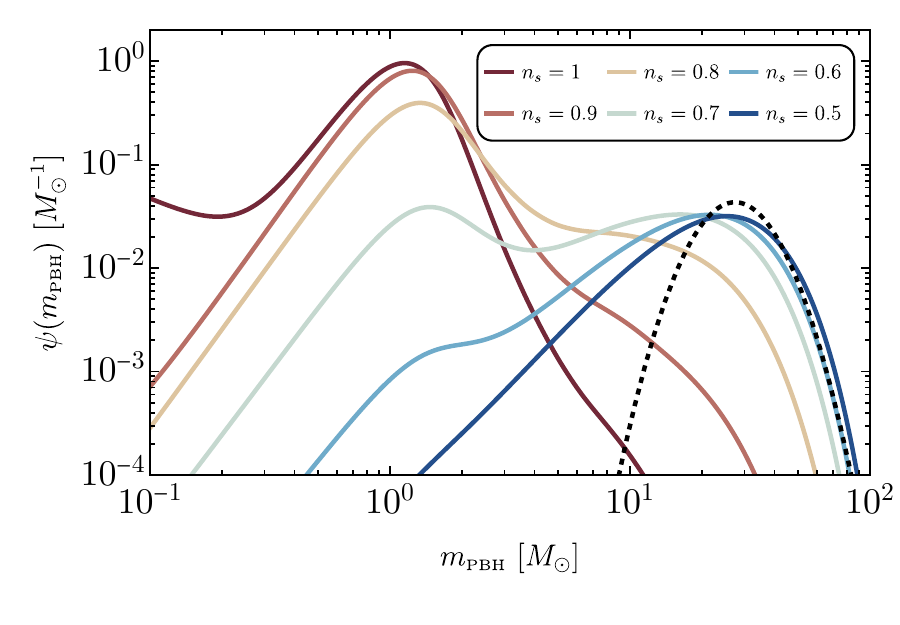}
	\caption{
	Plot taken from \cite{Franciolini:2022tfm}.
	Mass function obtained with a few choices of the curvature power spectrum. 
	This plot assumes $f_\PBH = 10^{-3}$, 
	the minimum horizon mass to be 
	$ \lesssim 10^{-2.5} M_\odot$, the largest mass  $M_H^\text{max} = 10^{2.8} M_\odot$ and a variable tilt $n_s$.
The black dashed line reports the lognormal mass distribution found as the best fit in the analysis of Ref.~\cite{Franciolini:2021tla}.
	}
\label{fig:MF_full}
\end{figure}

%{\bf ILIA: We need to revise this last paragraph including the results from \cite{Musco:2023dak} compared with \cite{Escriva:2022bwe}} 
{We conclude this section with a note of caution. 
While it is true that the QCD era produces a natural boost of formation around the solar mass, the physics of collapse alone is not predictive. This is because all the model dependence is retained in the spectrum and in the statistics of curvature perturbations.  Without strong assumptions on the tilt and running of the spectrum, for instance, no prediction on the height of the QCD peak at the solar mass compared to other scales can be obtained. 
This is  reflected in Fig.~\ref{fig:MF_full}, 
where different mass distributions are derived, assuming a power-law curvature power spectrum with varying tilt $n_s$ (not necessarily related to the one at large CMB scales).  The different ways one can introduce the curvature or density profiles can explain some of the differences observed between \cite{Musco:2023dak} and \cite{Escriva:2022bwe}, as well as the still low degree of accuracy to which we know the evolution of the equation-of-state from QCD lattice simulations.  Overall, the PBH formation remains an intrinsically non-linear and fully relativistic process, which makes the physics quite complex, as the recent developments have shown.  The improvement of our understanding of PBH formation has been very important and will probably be pushed further ahead in the next years, with significant implications for the PBH mass distribution expected in the various proposed models.  }

%\subsubsection{Comparison between results in the literature}
%\textcolor{red}{Discussion between the different results available in the literature and new fig.~\ref{fig:fofmPBH_newsims}.}
%SC:  we should discuss whether to include this section and decide what to put here.
  
\color{black}

\subsection{Evolution of the mass function through accretion}

The primordial PBH mass function would be preserved up to the present epoch unless phenomena affecting the PBH masses modify it during the cosmological history. In particular, it was shown that PBH mass accretion may be relevant for the PBH evolution \cite{DeLuca:2020bjf}. 
In this Subsection and in the following, we will describe the impact on the mass and spin of PBHs by a phase of accretion.

\subsubsection{Accretion onto isolated PBHs}
PBHs can accrete mass if particles from the surrounding environment fall into the gravitational potential well generated by the compact object and eventually get eaten by the PBH.
It can be shown that the strongest impact on the PBH mass is coming from baryonic accretion. In case PBHs are not accounting for the entirety of the DM, there exists an additional DM fluid in the Universe. Such a secondary DM component however would only act as a catalyst of baryonic accretion while it marginally contributes to the PBH mass growth \cite{Ricotti:2007au}.

During the cosmological history,  one can model the mass variation using the Bondi-Hoyle accretion rate~\cite{Bondi:1952ni,1939PCPS...35..405H, 1940PCPS...36..325H, 1940PCPS...36..424H,Ricotti:2007jk,Ricotti:2007au}
\be
\dot{m}_{\rm PBH} = 4 \pi \lambda m_{\rm p} n_{\rm gas} v_{\rm eff}^{-3} m_{\rm PBH}^2,
\label{mdotBH}
\ee
where the PBH effective velocity $v_{\rm eff} = \sqrt{v_{\rm rel}^2 + c_{\rm s}^2}$ is given in terms of the
relative velocity $v_{\rm rel}$ with respect to the surrounding gas of hydrogen with mass $m_{\rm p}$, sound speed  $c_{\rm s}$ and number density $n_{\rm gas}$. We denote with the Bondi radius $r_{\rm B}$ the region of space under the influence of the PBH potential well which is also determining the characteristic scale of the gas cloud accreting onto the PBHs.
The parameter $\lambda$, whose explicit expression is summarized for example in Appendix~B of 
Ref.~\cite{DeLuca:2020bjf}, takes into account the effects of gas 
viscosity, Hubble expansion and the coupling of the CMB radiation to the gas through Compton scattering~\cite{Ricotti:2007jk}. 
Notice that the aforementioned treatment is based on newtonian mechanics, and 
general-relativistic effects may lead to a significant increase in the mass accretion rate, depending on the specific environmental settings, see for example Ref.~\cite{Cruz-Osorio:2020dja,Cruz-Osorio:2021qbr}.

The presence of observational constraints for PBHs with masses larger than ${\cal O}(M_\odot)$ suggest that
they can  comprise only a fraction of the DM in the Universe, and therefore an additional DM halo has to be considered when modelling gas accretion~\cite{Mack:2006gz, Adamek:2019gns}. The presence of a DM halo catalyses the accretion of gas onto the PBHs, with a halo characteristic size and  mass which grows with time as~\cite{Ricotti:2007au}
\be
r_{\rm halo} = 0.019 \, {\rm pc} \lp \frac{M_{\rm halo}}{M_\odot} \rp^{1/3} \lp \frac{1+z}{1000} \rp^{-1}, \qquad M_{\rm halo} = 3 m_{\rm PBH} \lp \frac{1+z}{1000} \rp^{-1}
\ee
 and enhances the accretion rate~\cite{Ricotti:2007jk}. In the limit in which the characteristic halo radius is smaller than the PBH Bondi radius, accretion occurs onto a PBH with effective point mass $M_{\rm halo}$, while for comparable sizes or in the opposite limit, the proper contribution is captured by corrections to the accretion parameter $\lambda$, see Appendix~B of  Ref.~\cite{DeLuca:2020bjf} and references therein for additional details.

For PBHs with masses larger than $ {\cal O}(10) M_\odot$, accretion can reach super-Eddington values before the reionization epoch and play an important role in both the mass and spin evolution of PBHs, see Refs.~\cite{Ricotti:2007jk,DeLuca:2020bjf}. 
The increase of the PBH masses due to accretion also affects their mass distribution function in a non-linear fashion \cite{Garcia-Bellido:2018leu,DeLuca:2019buf}. Indeed, for an initial mass function $\psi (m_{\rm PBH}^{\rm i},z_{\rm i})$ at formation redshift $z_{\rm i}$, 
its evolution is governed by the identity \cite{DeLuca:2020bjf,DeLuca:2020fpg}
\begin{equation}
	\psi(m_{\rm PBH}(m_{\rm PBH}^{\rm i},z),z) \d m_{\rm PBH} = \psi(m_{\rm PBH}^{\rm i},z_{\rm i}) \d m_{\rm PBH}^{\rm i},
\end{equation}
where $m_{\rm PBH}(m_{\rm PBH}^{\rm i},z)$ is the final mass at redshift $z$ for a PBH with mass $m_{\rm PBH}^{\rm i}$ at redshift $z_{\rm i}$. 
The main effect of accretion on the mass distribution is to make the latter broader at high masses, producing a 
high-mass tail that can be orders of magnitude above its corresponding value at formation~\cite{DeLuca:2020fpg}.

Accretion may be also particularly relevant for the formation of SMBHs with masses $m_{\rm PBH} \gtrsim 10^9 M_\odot$, from PBHs lighter than $10^4 M_\odot$ formed at high redshift~\cite{Serpico:2020ehh}. In particular, assuming Gaussian initial conditions and PBH abundance $f_{\rm PBH} \lesssim 10^{-9}$ in order to be compatible with the strong observational constraints by CMB spectral distortions~\cite{Kohri:2014lza, Carr:2018rid,Nakama:2017xvq,Unal:2020mts}, epochs of Eddington accretion onto PBHs with masses $\sim 200 M_\odot$ may be responsible for the formation of SMBHs at redshift $z \gtrsim 6$~\cite{Serpico:2020ehh}, whose merging will possibly be detected by LISA.

At the onset of structure formation and reionization epoch, the increase of the PBH characteristic velocities and the gas speed of sound reduce the accretion efficiency~\cite{Ali-Haimoud:2017rtz, Hasinger:2020ptw, Hutsi:2019hlw}. One can adopt an agnostic view and consider a cut-off redshift  $z_{\rm cut-off}$ below which accretion is negligible~\cite{DeLuca:2020bjf,DeLuca:2020qqa}. Its value is relatively unconstrained due to the large uncertainties in modelling accretion at relatively small redshift, such as X-ray pre-heating~\cite{Oh:2003pm}, details of the structure formation and feedback effects.
In particular, the emission of X-rays from accreting PBHs may heat the plasma locally (local type) or produce Str\"omgren spheres of ionised H-II which may impact the gas sound speed and thus the PBH accretion rate (global type)~\cite{Ricotti:2007au, Ali-Haimoud:2016mbv}, while the generation of outflows may reduce the accretion efficiency depending on the outflow velocity and direction (mechanical type)~\cite{Bosch-Ramon:2020pcz}.

\subsubsection{Accretion onto PBH binaries}
When accretion occurs onto binary PBHs, one has to take into account both global accretion processes (i.e., of the binary as a whole) and  local accretion processes (i.e., onto the individual components of the binary). In particular, 
as the typical binary separation is smaller than the binary Bondi radius, the infall of gas is driven by the total mass $M_{\rm tot}$ of the binary system. In this configuration, both PBHs experience accretion from the gas with an enhanced density, and their individual accretion rates are modulated by their masses and orbital velocities~\cite{DeLuca:2020qqa}
\begin{align}
	&\dot m_1 = \dot M_{\rm tot}  \frac{1}{\sqrt{2 (1+q)}}, 
	\qquad
	\dot m_2 = \dot M_{\rm tot}  \frac{\sqrt{q} }{\sqrt{2 (1+q)}}
	 \label{M1M2dotFIN}
\end{align}
where the accretion rate $\dot M_{\rm tot}$ is computed using Eq.~\eqref{mdotBH} adopting the binary system parameters.
 For binary systems, one always finds that the relative accretion rate for the secondary component is larger than the one for the primary component, resulting in a growth of the mass ratio of the binary towards unity according to the equation~\cite{DeLuca:2020qqa}
\be
\dot q = q \lp \frac{\dot m_2}{m_2} -  \frac{\dot m_1}{m_1} \rp > 0.
\ee

\subsection{Spin Distribution}\label{sec: spin dist}
In this Subsection, we discuss the theoretical expectation for the PBH initial spin and how it evolves when accretion is effective.

The collapse of density perturbations generating a PBH is expected to be a rare event due to the fact that their cosmological abundance cannot exceed that of the DM, and applying the peak theory formalism one  finds that the high (and rare) peaks in the density contrast giving rise to PBHs are primarily spherical~\cite{Bardeen:1985tr}. At first order in perturbation theory, however, small asymmetries introduce torques induced by the surrounding matter perturbations, which are ultimately responsible for the generation of a small angular momentum before collapse.
The small time of collapse makes however  the action of these torques limited in time. At formation, the mean  PBH dimensionless Kerr parameter $\chi \equiv J/m_{\rm PBH}^2$ is estimated to be~\cite{DeLuca:2019buf} 
\be
\chi_{\rm i} \sim 10^{-2} \sqrt{1-\gamma_{\rm PS}^2},
\ee
in terms of the power spectrum shape parameter  $\gamma_{\rm PS}$, {defined as the ratio between the cross and auto-correlations of the density contrast and its spatial derivatives} (close to unity for very narrow power spectra). 
The initial PBH spin is therefore expected to be below the percent level (see also Refs.~\cite{Mirbabayi:2019uph, Harada:2020pzb})\footnote{The formation of PBHs in non-standard scenarios, like during an early matter-dominated epoch  following inflation~\cite{Harada:2017fjm} or from the collapse of Q-balls~\cite{Cotner:2017tir}, may lead to higher values of the initial  spin.}.
The distribution of the dimensionless Kerr parameter at formation takes the form  as found in \cite{DeLuca:2019buf}
\begin{equation}
P(\chi_{\rm i})\d \chi_{\rm i} = 
u
\exp \llp -2.37 - 4.12 \ {\ln 
\lp u \chi_{\rm i} \rp } - 
 1.53 \ {\ln^2 \lp u\chi_{\rm i} \rp} 
 - 0.13 \ {\ln^3 \lp u\chi_{\rm i} \rp} 
 \rrp \d \chi_{\rm i},
\end{equation}
where for simplicity we defined 
\begin{equation}
u \equiv 	\lp \frac{5}{2^{7/2}} \frac{\gamma_{\rm PS}^6 \nu}{\sigma_\delta \sqrt{1-\gamma_{\rm PS}^2} } \rp,
\end{equation}
in terms of the rescaled critical peak amplitude $\nu \equiv \delta_{\rm c}/ \sigma_\delta$ and the density contrast variance $\sigma_\delta$.

The initial PBH spin will be retained until the time of merger unless an efficient phase of accretion underwent during the PBH cosmological evolution.
Indeed, since the accretion rate and the geometry of the accretion flow are intertwined, for accretion rates slightly above the super-Eddington limit, the angular momentum carried by the infalling accreting gas may lead to the formation of a geometrically thin accretion disk along the equatorial plane of  the PBH~\cite{Shakura:1972te,Novikov:1973kta, Ricotti:2007au}\footnote{For accretion rates slightly smaller than the Eddington limit, accretion is non-spherical and an advection-dominated accretion flow~(ADAF) may form~\cite{Narayan:1994is}. In the opposite case, the accretion luminosity might be so strong that the disk ``puffs up'' and becomes thicker, resulting in less efficient angular momentum transfer.}.
Apart from efficient accretion, the formation of a disk happens if the typical gas velocity is larger than the Keplerian velocity near the PBH, which translates into a condition on the minimum PBH mass \cite{DeLuca:2020bjf}
\be
m_{\rm PBH} \gtrsim 6\cdot  10^2 M_\odot \,D^{1.17} \xi^{4.33}(z) \frac{\lp 1+z/1000\rp^{3.35}}{\llp 1 + 0.031 
	\lp1+z/1000\rp^{-1.72} \rrp^{0.68}},
\ee
in terms of the constant factor $D \sim \mathcal{O} (1 - 10)$, which takes into account relativistic effects, and the parameter $\xi (z) = {\rm Max}[1, \langle v_{\rm eff} \rangle/c_{\rm s}]$, which describes the effect of the PBH motion with respect to the gas.

In such a configuration, mass accretion is accompanied by an increase of the PBH spin perpendicularly to the disk plane, whose growth rate can be described following  a geodesic model for  circular disk motion as~\cite{Bardeen:1972fi, DeLuca:2020bjf}
 (see also Refs.~\cite{Bardeen:1972fi,Thorne:1974ve,Brito:2014wla,Volonteri:2004cf})
\be
\dot \chi = \lp {\cal F} (\chi) - 2 \chi \rp \frac{\dot{m}_{\rm PBH}}{m_{\rm PBH}},
\ee
in terms of the combination ${\cal F} (\chi) \equiv L(m_{\rm PBH},J)/m_{\rm PBH} E(m_{\rm PBH},J)$, which is a function of the energy and angular momentum per unit mass~\cite{Bardeen:1972fi}
\be
E(m_{\rm PBH},J) = \sqrt{1- 2 \frac{m_{\rm PBH}}{3 r_{\rm ISCO}}}
\qquad 
\text{and}
\qquad
L(m_{\rm PBH},J) = \frac{2 m_{\rm PBH}}{3 \sqrt{3} } \lp 1+ 2 \sqrt{ 3 \frac{r_{\rm ISCO} }{m_{\rm PBH}}-2}\rp\,,
\ee
depending on the ISCO radius
\be
r_{\rm ISCO}(m_{\rm PBH},J) = m_{\rm PBH} \llp 3 + Z_2 - \sqrt{\lp 3-Z_1\rp \lp 3+Z_1+2 Z_2\rp } \rrp,
\ee
with $Z_1= 1+ \lp 1- \chi^2 \rp ^{1/3} \llp \lp 1+\chi\rp^{1/3}+\lp 1-\chi\rp^{1/3} \rrp$ and $Z_2= \sqrt{3 \chi^2 + 
	Z_1^2}$.
The spin evolution continues within the relevant accretion timescale until it reaches the extremal limit  $\chi_{\rm max}=0.998$ dictated by radiation effects~\cite{Thorne:1974ve}. 
An analytical fit of the spin as a function of mass induced by accretion onto PBH binaries can be found in Ref.~\cite{Franciolini:2021xbq}. 
This fit is based on the accretion model presented in Ref.~\cite{DeLuca:2020bjf,DeLuca:2020qqa} which is valid in the mass range currently observed by LVKC experiment. Detailed modelling of accretion on PBH with masses above around ${\cal O}(10^2) M_\odot$, particularly relevant for PBH mergers observable by LISA, is still lacking in the literature, due to the difficulty in modelling accretion with the inclusion of feedback effects at such large masses. 

If instead accretion occurs onto a PBH binary, the fact that  total mass of the binary drives accretion leads to 
a much more efficient angular momentum transfer on each PBH~\cite{DeLuca:2020bjf, DeLuca:2020qqa}. In this case, the accretion onto the binary components is never spherical and disks can form easier around the PBHs. Moreover, since the lighter component of the binary experiences a stronger accretion rate, this results into a higher spin with respect to the one of the primary and heavier component~\cite{DeLuca:2020qqa}.

An important parameter measurable through GW observations is the effective spin 
\begin{equation}
    \chi_{\rm  eff} \equiv 
    \frac{\chi_1 \cos{\alpha_1} + q \chi_2 \cos{\alpha_2}}
    {1+q},
\end{equation} 
which is a function of the mass ratio $q$, of both BH spin magnitudes $\chi_j$ ($j=1,2$), and of their orientation with respect to the orbital angular momentum, parametrized by the tilt angles $\alpha_j$. In Fig.~\ref{mass_spin}, we show the expected distribution of $\chi_{\rm eff}$  by averaging over the spin angles, as a function of PBH masses in binaries for various choices of $z_{\rm cut-off}$. 

\begin{figure*}[t!]
\centering
\includegraphics[width=0.49\textwidth]{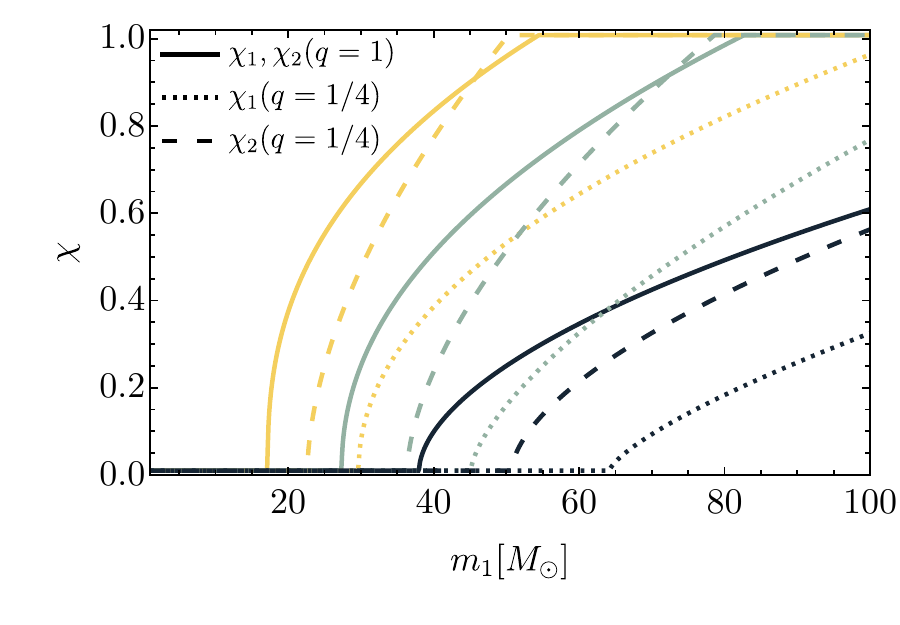}
\includegraphics[width=0.49\textwidth]{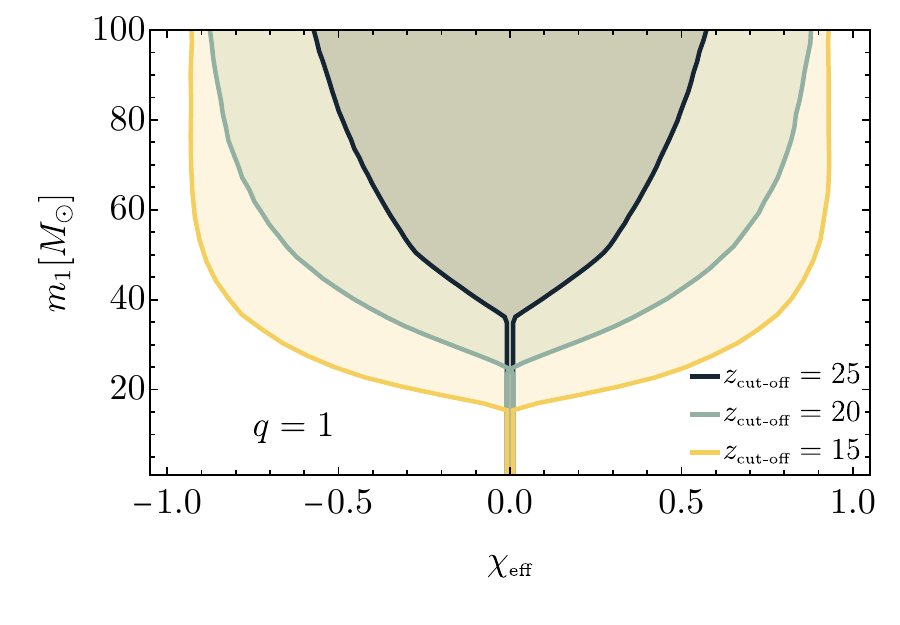}
\caption{ 
Figures taken from Ref.~\cite{Franciolini:2021xbq}.
\textbf{Left:}
Predicted primary ($\chi_1$) and secondary ($\chi_2$) spins as a function of primary mass $m_1$ and mass ratio $q$ for various values of $z_{\rm cut-off}$ (indicated by colors specified in the right panel).
\textbf{Right:}
Predicted distribution of $\chi_{\rm eff}$ as a function of PBH mass $m_1$ (assuming equal mass binaries) for three choices of $z_{\rm cut-off}$.
}\label{mass_spin}
\end{figure*}

An alternative scenario for inducing spin on PBH occurs in the context of dense PBH clusters, where close hyperbolic encounters~\cite{Garcia-Bellido:2017qal,Garcia-Bellido:2017knh} may spin up both black holes~\cite{Nelson:2019czq,Jaraba:2021ces} due to the fundamental frame-dragging effect of general relativity, analogous to magnetic induction in electromagnetism, in this case due to a current of matter as one black hole scatters off another. In the case of highly asymmetric binaries with large mass ratios, the induced spin can reach up to $\chi \simeq 0.8$ for the most massive one~\cite{Jaraba:2021ces}. Nevertheless, the distribution of spins of PBH in dense clusters is still characterized by a Gaussian around zero-spin with a width of order $\sigma\simeq0.2$~\cite{Garcia-Bellido:2020pwq}.

\subsection{Comments on the clustering from Poisson initial conditions}

In the absence of non-Gaussian initial conditions, PBHs are Poisson distributed, as shown in Ref.~\cite{Desjacques:2018wuu}. In such a case, the PBH clustering evolution has been studied with a cosmological N-body simulations in Ref.~\cite{Inman:2019wvr} up to redshift of ${\cal O}({10^2})$ and confirmed analytically in Ref.~\cite{DeLuca:2020jug} assuming a monochromatic PBH mass distribution. In this Subsection we briefly summarise the main conclusions.

To characterize the PBH two-point correlation function as a function of the comoving separation $x=|\vx|$, we introduce the overdensity of discrete PBH centers at position $\vx_i$ with respect to the total background DM energy density,
\begin{eqnarray}
\frac{\delta\rho_\PBH({\vec x},z)}{f_\PBH\overline\rho_{\text{\rm DM}}}=\frac{1}{\npbh}\sum_i \delta_{\rm D}(\vx-\vx_i(z))-1,
\end{eqnarray}
where $\delta_{\text{\rm D}}(\vx)$ is the three-dimensional Dirac distribution, {and 
\be
\npbh\simeq 3.2 \,\fpbh\,\left(\frac{20\,M_\odot/h}{ \mpbh}\right)(h/{\rm kpc})^{3}
\ee
is the average number density of PBHs per comoving volume.} Here, $i$ runs over the  positions of PBHs.
The two-point correlation function of this discrete point process takes the general form
\begin{eqnarray}
\Big< \frac{\delta\rho_\PBH({\vec x},z)}{\overline\rho_{\text{\rm DM}}}\frac{\delta\rho_\PBH(0,z)}{\overline\rho_{\text{\rm DM}}} \Big>=
\frac{f^2_\PBH}{\npbh}\delta_{\text{\rm D}}(\vx)+ \xi(x,z).
\label{eq:PBH2pt}
\end{eqnarray}
This expression emphasizes that $\xi(x,z)$ is the so-called reduced PBH correlation function and, thus, is distinct from the additive Poisson noise proportional to the Dirac delta.
As mentioned before, initially the reduced correlation function is negligible. 
The corresponding PBH  power spectrum  
\begin{eqnarray}
\Delta^2(k,z)= \frac{k^3}{2 \pi^2}\int \d ^3 x\,  e^{i {\vec k} \cdot {\vec x}}\, \Big< \frac{\delta\rho_\PBH({\vec x},z)}{\overline\rho_{\text{\rm DM}}}\frac{\delta\rho_\PBH(0,z)}{\overline\rho_{\text{\rm DM}}} \Big>,
\end{eqnarray}
is conveniently defined relative to the total cold dark matter average density.

After a linear growth, the PBH power spectrum enters in the so called "Quasi-Linear" regime during which
\begin{eqnarray}
  \Delta_\text{\rm QL}^2 (k) &\simeq&
  0.04\,\fpbh^{3/4}
\left(\frac{20\,M_\odot/h}{\mpbh}\right)^{-3/4} \left[1+ 26\fpbh \lp \frac{100}{1+z} \rp\right]^{3/2} \left(\frac{k}{h/{\rm kpc}}\right)^{9/4}.
\label{L-QL}
\end{eqnarray}
When the power spectrum becomes larger than around $200$, it enters in the "Non-Linear" regime during which
\be
\Delta_{\text{\rm NL}}^2 (k)
\simeq 0.2\,\fpbh^{3/5}\left( \frac{\mpbh}{20 M_\odot/h}\right)^{3/5}\left[1+ 26 \fpbh \left( \frac{100}{1+z} \right) \right]^{6/5} \left(\frac{k}{h/{\rm kpc}}\right)^{9/5}.
\ee

\begin{figure}
\centering
\includegraphics[width=0.49\textwidth]{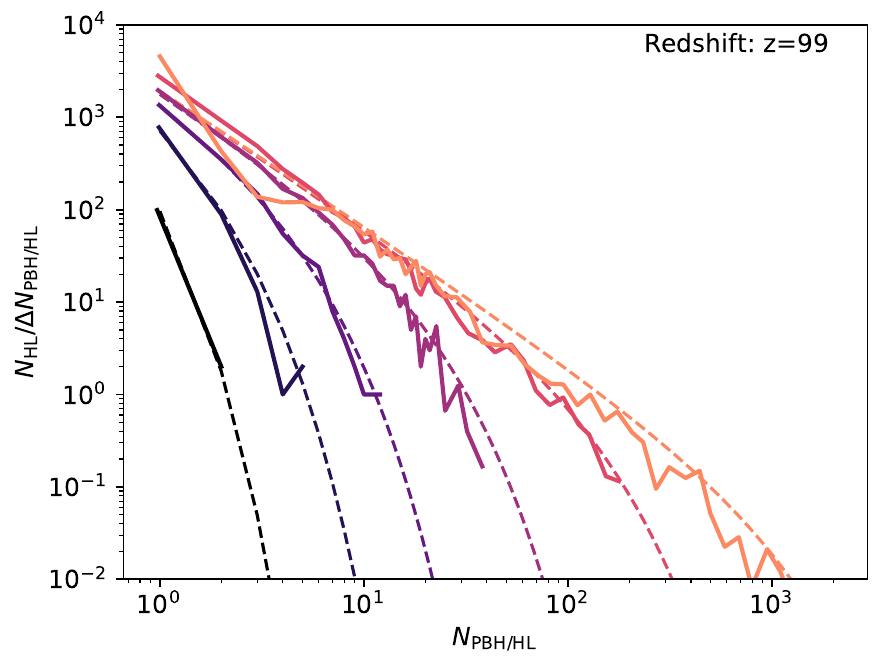}
\caption{
Figure taken from Ref.~\cite{Inman:2019wvr}.
The abundance of halos $N_{\rm HL}$ containing a given number of PBHs,  $N_{\rm PBH/HL}$, i.e. halo mass function. 
Solid lines report the results of the $N$-body simulations while dashed lines are
indicates the theoretical prediction assuming Poisson statistics.
}
        \label{fig:hl_number_fn}
\end{figure}

PBHs form halos whose mass distribution may be described by the Press-Schechter theory \cite{Press:1973iz} with an initial Poisson power spectrum. 
The resulting number density of PBH halos with mass between $M$ and $(M+\d M)$ reads
\be
\label{halo function}
\frac{\d n(M,z)}{d M} =\frac{\overline{\rho}_{\text{\rm PBH}}}{\sqrt{\pi}}\left(\frac{M}{M_*(z)}\right)^{1/2}\frac{e^{-M/M_*(z)}}{M^2},
\ee
where $\overline{\rho}_{\rm PBH}$ is the average PBH energy density and \cite{Hutsi:2019hlw}
\be
\label{dd}
M_*(z)=N_*(z)\cdot  \mpbh\simeq \fpbh^2\left(\frac{2600}{1+z}\right)^2\mpbh
\ee
is the typical mass of halos collapsing at redshift $z$.
This prediction agrees with the results of cosmolocial N-body simulations, as shown in Fig.~\ref{fig:hl_number_fn}.

If we restrict ourselves to small scales, both members of a PBH pair are almost certainly drawn from the same PBH halo. In this limit, if the PBH density  profile is $\rhopbh(x)\sim x^{-\epsilon_\text{\rm PBH}}$, then the two-point correlation function must behaves like $\sim x^{-2\epsilon_\text{\rm PBH}+3}$ \cite{1974A&A....32..197P,1977ApJ...217..331M} as it is proportional to the square density profile. Imposing $(-2\epsilon_\text{\rm PBH}+3)=-9/5$, we infer that the PBH density profile should satisfy
\be
\rhopbh(x)\sim x^{-12/5},
\label{profile}
\ee
which is confirmed by numerical simulations presented in Ref.~\cite{Raidal:2018bbj}.
Notice that since clustering is hierarchical, each halo has a certain survival time and, therefore, a given probability to be absorbed by a bigger halo formed at a later redshift. 
	As shown in Ref.~\cite{DeLuca:2020jug} the evaporation time (due to dynamical relaxation) of PBH halos is typically larger than their survival time, which implies that PBH halos are stable against evaporation.
We will discuss later on the impact of clustering on the merger rate of PBH binaries. 
% The considerations presented in this subsection indicate that: {\it i)} for a small fraction of PBH contribution to the DM, PBH clustering does not affect the standard calculation of the merger rates;  
% {\it ii)} the evaporation phenomenon is not likely to change the clustering properties of PBHs;
% {\it iii)}the early Universe merger rate is decreased in the presence of clustering for large $f_\PBH$, yet still falling above the LIGO/Virgo detection band;
% {\it iv)} the late Universe merger rate is increased and can fall within the detection band.

\subsection{Summary}

In this Section, we did not only review the standard formalism to compute the PBH mass function from the curvature or density power spectrum, but also the most recent developments on the computation of the critical overdensity threshold and of the relation between curvature and density fluctuations, which depend on the perturbation statistics (Gaussian, non-Gaussian), the primordial power spectrum and the evolution of the equation-of-state through different phases like the QCD transition.  The impact of these effects is important on the PBH abundance and mass function.
Even if it is possible to rescale the power spectrum amplitude in order to obtain the required PBH abundance when one only focuses on a single observable, this is not the case anymore when one tries to correlate multiple observations (e.g. PBH abundance and induced GWs, as discussed in the following sections).  In particular, the effects of the QCD transition may have greatly imprinted on the mass function for stellar-mass PBHs with direct consequences for GW observations that will be discussed in the next Sections. We also went beyond the common assumption that the PBH mass and spin distributions do not change much with time, by considering effects of accretion on isolated PBHs and PBH binaries.  Finally we emphasized the importance of the inevitable clustering of PBHs due to their inevitable Poisson fluctuations.

\section{PBH merging and encounter rates} \label{sec:rates}
\label{Sec:merging_rate}

In this Section, we calculate the expected PBH merging rates and their evolution with redshift for two different binary formation channels:  before matter-radiation equality and due to tidal capture in clusters.  For this purpose, we use a general mass distribution $f(m_{\rm PBH})$ that must be specified by the underlying theoretical model and assumptions about PBH formation, as explained in the previous sections.  For the two channels, we also discuss some effects linked to PBH clustering induced by the inevitable Poisson fluctuations in their spatial distribution.  This can cause a merging rate suppression for early binaries due to their tidal disruption by early-forming clusters and a merger rate boost for late-binaries due to an enhanced clustering compared to common halo mass functions in $\Lambda$-CDM cosmology.  We also review the expected rates of disrupted binaries.  Finally, we study the rate of hyperbolic encounters in PBH clusters that can induce GW bursts, possibly observable in Earth-based and space-based detectors.  

\subsection{Early binaries} 

\subsubsection{General rate formula}

PBH binaries may have formed in the early Universe before matter-radiation equality.  Because of their random spatial distribution at formation,  two PBHs can form sufficiently close to each other for their dynamics to decouple from the Universe expansion.  They form a binary instead of directly merging, because of the gravitational influence of one or several PBHs nearby.  Possibly, the binary is sufficiently stable and it takes of the order of the age of the Universe for the two black holes to merge.  

Under the hypothesis that there is no local non-Gaussianity cross-correlating large and small scales, 
PBHs do not form in clusters, but follow a Poissonian distribution~\cite{Desjacques:2018wuu, Ali-Haimoud:2018dau, Ballesteros:2018swv,MoradinezhadDizgah:2019wjf,Inman:2019wvr}. The merging rates (per unit logarithmic mass of the two binary components) of such early binaries (denoted EB) at a time $t$ were first estimated in~\cite{Ioka:1998nz} and are given by~\cite{
Sasaki:2016jop,Ali-Haimoud:2017rtz,
Kocsis:2017yty,Raidal:2018bbj,DeLuca:2020qqa,Hutsi:2020sol}
\bea
        R_{\rm EB} &=& %\frac{{\rm d}\tau}{{\rm d} \ln m_1 d \ln m_2} \nonumber \\ &\approx &
        \frac{1.6 \times 10^6}{\rm Gpc^3 yr} \times f_{\rm sup}(m_1,m_2,f_{\rm PBH}) f_{\rm PBH}^{53/37} f(m_1) f(m_2) \nonumber \\   & \times & \left(\frac{t}{t_0}\right)^{-34/37} \left(\frac{m_1 + m_2}{M_\odot}\right)^{-32/37} \left[\frac{m_1 m_2}{(m_1+m_2)^2}\right]^{-34/37} ~,  \label{eq:cosmomerg}
\eea
where $m_1$ and $m_2$ are the two black hole masses and $t_0$ is the age of the Universe.  Compared to the predictions of~\cite{Sasaki:2016jop}, one has to include a suppression factor $f_{\rm sup}$ to take into account several physical effects that have been initially neglected but were revealed by N-body simulations of~\cite{Raidal:2018bbj}. 

Some analytical prescriptions have been proposed, e.g. in~\cite{Raidal:2018bbj,Hutsi:2020sol}, to calculate the suppression factor $f_{\rm sup}(f_{\rm PBH}, m_1,m_2)$.  It can be written as the product of $S_1(f_{\rm PBH},m_1,m_2)$ and $S_2(f_{\rm PBH})$ that take into account the rate suppression due to nearby PBHs or matter fluctuations and PBH clusters seeded by Poisson fluctuations, respectively.  These analytical prescriptions have been compared with N-body simulations, but only in the case of monochromatic and log-normal PBH mass distributions.  Therefore, one must be cautious when applying them to broad mass functions, even if they exhibit a peak, like the one from the QCD transition.

\subsubsection{Suppression from close PBHs and matter inhomogeneities}

The first suppression factor we consider ($S_1$) takes into account the binary disruption by either matter fluctuations, with a (rescaled) variance $\sigma_{\rm M}^2 $, usually around $0.005$, or by the number $\bar N$ of PBHs within a sphere centered on the binary of radius corresponding to the maximum comoving distance at which a nearby PBH would fall onto the binary to disrupt it, before matter-radiation equality.   An analytical prescription for $S_1$ has been proposed in~\cite{Raidal:2018bbj,Hutsi:2020sol}, and is given by
\be
S_1 \approx 1.42 \left[ \frac{(\langle m_{\rm PBH}^2 \rangle/\langle m_{\rm PBH} \rangle^2)}{\bar N + C} + \frac{\sigma_{\rm M}^2}{f_{\rm PBH}^2}\right]^{-21/74} {\rm e}^{-\bar N},
\label{eq:S1}
\ee
with 
\be
\bar N = \frac{m_1+m_2}{\langle m_{\rm PBH} \rangle} \frac{f_{\rm PBH}}{f_{\rm PBH}+\sigma_M}~.  \label{eq:Nbar}
\ee
Note that in Eqs.~(\ref{eq:S1}) and~(\ref{eq:Nbar}), the mean PBH mass and their variance are calculated from the mass function through
\bea
\langle m_{\rm PBH} \rangle & = &  \frac{\int m_{\rm PBH} {\rm d} n_{\rm PBH}}{n_{\rm PBH}} =  \left[\int \frac{f(m_{\rm PBH})}{m_{\rm PBH}} {\rm d} \ln m_{\rm PBH} \right]^{-1} \\
\langle m_{\rm PBH}^2 \rangle & =& \frac{\int m_{\rm PBH}^2 {\rm d} n_{\rm PBH}}{n_{\rm PBH}}
 =  \frac{\int m_{\rm PBH} f(m_{\rm PBH}) {\rm d} \ln m_{\rm PBH}}{\int \frac{f(m_{\rm PBH}) }{m_{\rm PBH}}{\rm d} \ln m_{\rm PBH}}
\eea
where $n_{\rm PBH}$ denotes the total PBH number density.   The function $C$ encodes the transition between small and large $\bar N$ limits.  A good approximation is given by~\cite{Hutsi:2020sol}
\begin{align}
C &\simeq \frac{f_{\rm PBH}^2 \langle m_{\rm PBH}^2 \rangle}{\sigma_{\rm M}^2 \langle m_{\rm PBH} \rangle^2}  
% \nonumber \\
% &
\times \left\{ \left[ \frac{\Gamma(29/37)}{\sqrt{\pi} } U\left( \frac{21}{74},\frac{1}{2} , 
\frac{5 f_{\rm PBH}^2}
{6 \sigma_{\rm M}^2}
\right) 
\right]^{-74/21}  -1 \right\}^{-1},
\end{align}
where $\Gamma$ is the Euler function and $U$ is the confluent hypergeometric function.  

The value of $S_1$ as a function of $f_{\rm PBH}$ has been represented in Fig.~\ref{fig:fsup} for different cases. Assuming $\bar N = 2$, as expected for a close-to-monochromatic distribution, the value of $S_1$ ranges from $10^{-2}$ when $f_{\rm PBH} \simeq 10^{-3}$ to $0.1$ when $f_{\rm PBH} \simeq 1$.   In particular, one can also obtain a maximum value $S_1^{\rm max}$ that is independent of the mass distribution when $\bar N \rightarrow 0$.  This maximal value is around unity for $f_{\rm PBH} \gtrsim 0.1$ and goes down to $10^{-1}$ for lower values.  In this limit, one recovers the results from~\cite{Chen:2018czv,Ali-Haimoud:2017rtz} that only take into account the effect of matter inhomogeneities.   

Finally, we would like to comment on the calculation of $S_1$ using different definitions of the PBH mass function, which in particular modify the way to compute $\langle m \rangle$ and $ \langle m^2 \rangle$ depending on whether the mass distribution is defined with respect to the PBH number density  $n$  or the PBH density $\rho$, or if the normalisation changes.  The possible definitions, the way they are related, and the corresponding calculation rules are summarized in Table~\ref{tab:S1rules} from~\cite{Escriva:2022bwe}.    It is important to do these calcualtions carefully because they can significantly impact the resulting rates, and because some inconsistancies were found in the litterature.   As pointed out in~\cite{Escriva:2022bwe}, there was a conversion factor $m / \langle m \rangle $ not considered in~\cite{Franciolini:2022tfm} which led to an inconsistency with~\cite{Hutsi:2020sol} which uses exactly the same definition and normalization of the PBH mass distribution, but a different merger rate formula.  There was also a typo in the Eq. A4 of~\cite{Clesse:2020ghq} that has an incorrect mass dependence even if the correct formula was used for the calculations and paper figures.  
\begin{table}[t]
    \centering
    \begin{tabular}{|c|c|c|c|}
       \hline
        Escriv\`a et al~\cite{Escriva:2022bwe} & Raidal et al.~\cite{Raidal:2018bbj} &  &  \\
        Clesse et al.$^{*}$~\cite{Clesse:2020ghq}  & Franciolini et al.$^{*}$~\cite{Franciolini:2022tfm} & Hutsi et al.~\cite{Hutsi:2020sol} & Kocsis et al.~\cite{Kocsis:2017yty} \\
        Bagui et al.~\cite{Bagui:2021dqi} & Hall et al.~\cite{Hall:2020daa} &  & \\
        \hline
        $f\equiv \dfrac{1}{\rho_{\rm PBH}} \dfrac{{\rm d} \rho_{\rm PBH} }{{\rm d}\ln m}   $  & $ \psi_{1} \equiv \dfrac{1}{\rho_{\rm PBH}} \dfrac{{\rm d} \rho_{\rm PBH} }{{\rm d} m}   $  & $ \psi_{2} \equiv \dfrac{1}{n_{\rm PBH}} \dfrac{{\rm d} n_{\rm PBH} }{{\rm d} \ln m}   $ & $ \psi_{3}\equiv \dfrac{1}{n_{\rm PBH}} \dfrac{{\rm d} n_{\rm PBH} }{{\rm d}  m}   $ \\
        \hline
        $f =  m \psi_1  $ & $\psi_1 = f/m $ & $ \psi_2 = f \langle m \rangle /m$   & $ \psi_3= \langle m \rangle f / m^2  $  \\ 
        $ = m \psi_2 / \langle m \rangle $ & $= \psi_2 / \langle m \rangle $  & $ = \langle m \rangle \psi_1$ & $  = \langle m \rangle \psi_1 /m $ \\
        $ = m^2 \psi_3 / \langle m \rangle   $  & $ =  m \psi_3 / \langle m \rangle   $  & $= m \psi_3 $  & $ = \psi_2/m $ \\
       \hline
       $\int f {\rm d} \ln m = 1$ & $\int \psi_1 {\rm d} m = 1$ & $\int \psi_2  {\rm d} \ln m = 1$ & $ \int \psi_3 {\rm d} m = 1$  \\
        \hline
       $\langle m \rangle = \left( \int \dfrac{f}{m} {\rm d} \ln m \right)^{-1} $  & $\left( \int \dfrac{\psi_1}{m} {\rm d} m \right)^{-1}$  & $ \int m \psi_2 {\rm d} \ln m $ & $ \int m \psi_3 {\rm d} m $ \\
       \hline
       $ \langle m^2 \rangle = \langle m \rangle \int  m f {\rm d} \ln m $ & $ \langle m \rangle \int m \psi_1 {\rm d } m $   &  $ \int m^2 \psi_2 {\rm d} \ln m  $ &  $ \int m^2 \psi_3 {\rm d}{m } $ \\
       \hline
    \end{tabular}
    \caption{Different definitions of the normalized PBH mass distribution proposed in various references with their conversion, their normalisation rule and the corresponding $\langle m \rangle$ and $\langle m^2 \rangle$.  The asterisk denotes the references in which an inconsistency has been found (see details in the text).   In this review, we considered both $f$ and $\psi_1 \equiv \psi$ with our notations.  Table adapted from~\cite{Escriva:2022bwe}.  }
    \label{tab:S1rules}
\end{table}

Following these rules for an extended mass function, one gets that the abundance of light PBHs can strongly impact $\langle m \rangle$ and the resulting values of $\bar N$ that typically increases.  In turn, $S_1$ and the resulting value of $f_{\rm sup}$ can be strongly suppressed due to the exponential dependence in $\bar N$, and their exact value is dependent on the chosen lower mass cut for the PBH mass distribution.   Physically, this suppression comes from the numerous light black holes, reducing the value of the average black hole mass, at proximity of the binary that are considered as being able to disrupt it when one uses the presented conservative approach.   Such an issue was pointed out in~\cite{Gow:2019pok,Clesse:2020ghq,Escriva:2022bwe} and suggests that the rates of early binaries should probably be revised in the case of a very broad mass distribution.
In absence of a clear solution, the approach of~\cite{Clesse:2020ghq,Escriva:2022bwe} was to consider the limit obtained for a monochromatic mass function, motivated by the QCD-induced peak, and assume that $\bar N \approx 2$.   In~\cite{Franciolini:2022tfm}, which used another averaging procedure, there was a strong dependence on the high-mass PBH distribution and a high-mass cut-off was introduced that may naturally arise from the transition in the primordial power spectrum.    The different approaches however lead to different merger rates and different possible interpretations of the LIGO/Virgo observations, as well as different limits on the possible PBH abundance.

\subsubsection{Suppression from early Poisson-induced clustering} \label{sec:poisson_clust_supp}

The second factor $S_2(f_{\rm PBH})$ accounts for the effect of binary disruption in PBH clusters that rapidly form after PBH formation if their abundance is large enough \cite{Inman:2019wvr}, due to their initial Poisson fluctuations.  The interaction of binaries with other PBHs in early-Universe clusters typically induces a modification of their semi-major axis and eccentricity, which leads to longer merger timescales and a suppression of the merger rate \cite{Trashorras:2020mwn, Jedamzik:2020omx,Jedamzik:2020ypm}. 
In order to have a conservative estimate of the merger rate,  Ref.~\cite{Vaskonen:2019jpv} (and \cite{DeLuca:2020jug} includes an analytical  modelling of clustering evolution) computed the fraction of binaries which never enter in dense enough environments.  This fraction $S_2$ today can be approximated by~\cite{Hutsi:2020sol}
\be
S_2 \approx \min \left(1,9.6 \times 10^{-3} f_{\rm PBH}^{-0.65} {\rm e}^{0.03 \ln^2 f_{\rm PBH}} \right).
\label{eq:S2}
\ee
The value of $S_2$ as a function of $f_{\rm PBH} $ has been represented and compared to $S_1$ and the resulting total suppression factor $f_{\rm sup}$ in Fig.~\ref{fig:fsup}.  When it is considered at a time $t$, a good approximation is to replace $f_{\rm PBH}$ by $f_{\rm PBH}[t(z)/t_0]^{0.44}$, valid for $z \leq 100$~\cite{Hutsi:2020sol}.

%SC:  to be checked if we agree or not to remove this sentence
%The combination of abundance-dependent factors in Eqs.~\eqref{eq:cosmomerg}, \eqref{eq:S1} and 
%\eqref{eq:S2}, one finds that the early binary merger rate scales as 
%\begin{align}
%R_{\rm EB}
%\propto
%	\begin{cases}
%		f_\PBH^{2/3} &,\quad f_\PBH\gtrsim 10^{-3},
%		\\
%		f_\PBH^{2} &,\quad f_\PBH \lesssim 10^{-3}.
%	\end{cases}
%\end{align}

\subsubsection{Merging rates of disrupted binaries}

Numerical studies find that binaries are likely to be disrupted\footnote{In this context by ``disrupted" it is meant binaries whose semi major axis and eccentricity are modified but remain bound after the interactions with surrounding perturbers.} after formation if PBHs constitute most of the DM. This can happen in two ways: 1) the initial configuration contains a third PBH close to the initial PBH pair that will very likely collide with the forming binary; 2) binaries are absorbed by dense PBH clusters that form early in the history of the Universe. A conservative calculation of the possible merging rates of PBH binaries that have been disrupted in early-forming halos has been proposed in~\cite{Vaskonen:2019jpv}.  Moreover, let us note that estimates indicate that most PBH binaries are not perturbed in typical DM halos larger than 10 PBHs, and that nearly all initial binaries might be disrupted within the age of the Universe, since the fraction of disrupted binaries is relatively high at $z = 1100$ when $f_{\rm PBH} \gtrsim 0.1$. 

For a monochromatic mass function, the merger rate of perturbed binaries at time $t$ is given by \cite{Vaskonen:2019jpv}
\be
\label{disruptedMR}
R_{\rm p}(t) \propto f_{\rm PBH}^{\frac{144\gamma}{259} + \frac{47}{37}} t^{\frac{\gamma}{7} - 1}m_{\rm PBH}^{\frac{5\gamma-32}{37}},
\ee
where $\gamma \in [1,2]$, based on the angular momentum distribution for perturbed binaries in the early Universe.
The main result of this analysis is that the rate of perturbed binaries can exceed the rate of the unperturbed ones when $f_{\rm PBH} \gtrsim 0.1$.  However, when $f_{\rm PBH} \lesssim 0.02$,  binaries are not expected to be disrupted. It is still unclear how this formula and results are generalized to any broad PBH mass distribution.  

%In a cluster of $N$ PBHs, if one assumes the Press-Schechter theory for the collapse of halos at redshift $z_c$ and the Virial theorem to get velocity dispersions, one obtains that the typical interaction time is
%\be
%t_{\rm p} \approx 2 {\rm Myr} \times N^{1/3} f_{\rm PBH}^{-32/111} \left( \frac{m_{\rm PBH}}{M_\odot} \right)^{-10/111} \left(\frac{1+z_c}{1000}\right)^{-5/2}.
%\ee
%However, as already pointed out in~\cite{Vaskonen:2019jpv}, this estimate can only be applied if  small halos, formed at high redshifts, can survive, i.e. they do not expand much due to their dynamical heating induced by regular PBH encounters, and they are not absorbed into larger halos. Both effects reduce the frequency of binary-PBH encounters and, as a consequence, the merger rate of disrupted binaries.  This is the case for dark matter halos and stellar-mass PBHs.   A conservative approach is to calculate the rate of disruption in a time smaller than the typical relaxation time of the PBH cluster.  Finally, one has to include binaries that are not in clusters but that are disrupted by nearby PBHs.  
%Altogether, one gets a conservative estimate of the merging rate of disrupted binaries of order (for equal-mass mergers)
%\be
%R^{\rm dist} \simeq 3 \times 10^3 \times f_{\rm PBH} 
%\ee
%It is however unclear how this formula is generalized to any broad PBH mass distribution.  

\subsubsection{Link with observations of compact binary coalescences}

In Fig.~\ref{fig:fsup}, we represented the values of $S_1$, $S_2$ and the resulting suppression factor $f_{\rm sup}$ as a function of $f_{\rm PBH}$ for the choice of the broad mass function of Ref.~\cite{Clesse:2020ghq}, at the mean masses of the exceptional GW events GW190425, GW190814 and GW190521.\footnote{We note that the parameter estimation of GW events characterised by a small signal-to-noise ratio at the LVKC detectors may still be sensitive to the priors assumed in the analysis (see e.g. Ref.~\cite{Vitale:2017cfs}).
Therefore, the interpretation of some events may change if PBH motivated priors are assumed \cite{Bhagwat:2020bzh}, in particular when focusing on the binary mass ratio and effective spin.}  We also calculated $S_1^{\rm max}$ in the limit $\bar N \ll \min(C,1) $ that is independent of the two binary component masses.  

\begin{figure*}[t]
    \centering
    \includegraphics[width=16cm]{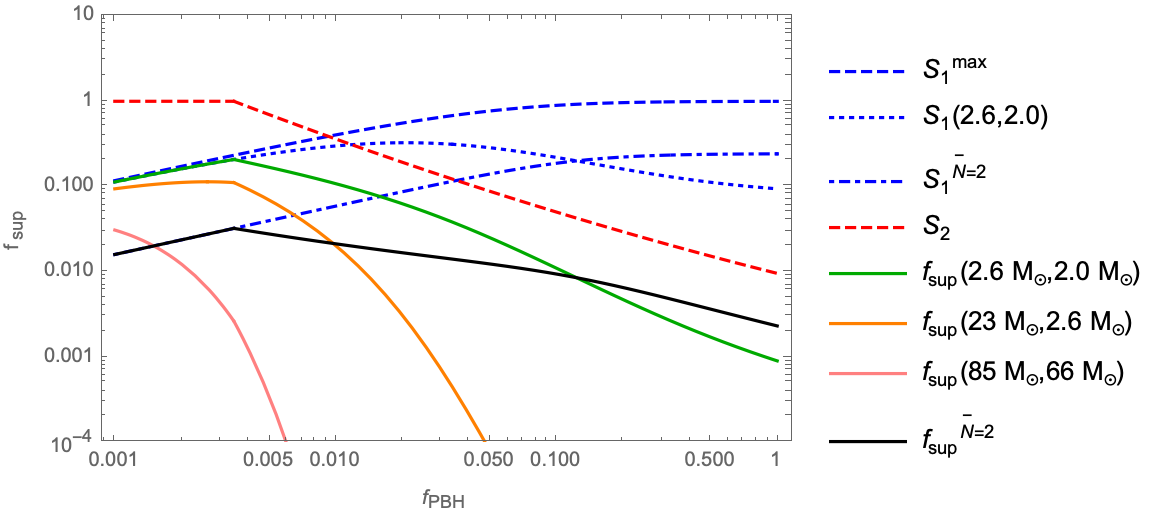}
    \caption{Suppression factor  $f_{\rm sup}$ and its possible contributions $S_1$ and $S_2$ in different cases, in terms of the PBH abundance (from~\cite{Clesse:2020ghq}).  When PBH masses are specified, it assumes a PBH mass function from a nearly-scale invariant primordial power spectrum of curvature fluctuations and including QCD-induced features.}
    \label{fig:fsup}
\end{figure*}

For an arbitrary broad mass distribution, one critical difference with respect to a monochromatic or a relatively sharp log-normal mass function comes from the large number density of tiny black holes, implying that $\langle m_{\rm PBH} \rangle \ll M_{\odot}$ and $\langle m_{\rm PBH}^2 \rangle / \langle m_{\rm PBH} \rangle^2 \ll 1$.  This implies that generically $\bar N \gg 1$.  If the analytical prescription is strictly applied, this leads to a huge exponential suppression of the EB merging.  However, this rate suppression is very likely overestimated because PBHs that are much lighter than the binary components are probably not able to bring enough energy to the system in order to disrupt it.   

Instead, it is probably more realistic to consider only the disruption by nearby PBHs whose mass is similar to the mean of the binary component masses.  By doing so, one gets $\langle m_{\rm PBH} \rangle \sim (m_1+m_2)/2$ and $\langle m_{\rm PBH}^2 \rangle / \langle m_{\rm PBH} \rangle^2 \sim 1$.  In such a case, one gets $\bar N \approx 2$ (just as in the monochromatic case) and the suppression factor obtained when $f_{\rm PBH} \simeq 1$ becomes independent of the mass, slightly below $S_2^{\rm max}$, depending on the exact value of $\bar N$.  Different values of $\bar N$ (but still of the same order) typically lead $f_{\rm sup}$ to be of order $S_2^{\rm max} \times \mathcal O(0.1-1)$.  In particular, for $\bar N = 2$ and $f_{\rm PBH}= 1$, one gets $f_{\rm sup} \simeq 0.002$.  We have chosen this value to compute the EB merging rate for two typical mass functions impacted by the thermal history.  They are represented on Fig.~\ref{fig:rates}.  One can see that these merging rates can in principle accommodate some of the LIGO/Virgo GW events and are significantly enhanced when the binary includes a PBH from the QCD peak.

\subsubsection{Limitations}

Even if the merging rates given by Eq.~\ref{eq:cosmomerg} are consistent with N-body simulations for monochromatic and log-normal mass functions in the stellar-mass range, there are still a series of uncertainties that could limit the applicability of this formalism.  Hereafter we list some of these possible limitations. 

\begin{enumerate} 
\item So far, no N-body simulations have been performed on PBH models that include thermal features, whereas it is proven that they have an important impact for stellar-mass PBHs.  In general, broad-mass functions could see also deviations from those estimations, in particular for binaries with very low mass ratios.
\item Eq.~\ref{eq:cosmomerg} including the effect described in Eq.~\ref{disruptedMR} should be improved by more appropriate calculation of the distribution of the binary parameters as interactions have occurred, encoded in the parameter $\gamma$~\cite{Vaskonen:2019jpv}.  Analytical prescriptions for non-equal mass binaries should also be further developed and checked.    
\item The fraction of PBH binaries that are disrupted as a function of the mass of an infalling PBH is uncertain.  This question is particularly relevant for binaries with strong differences in mass components, and for broad mass functions with lots of small PBHs that would dominate the calculation of $\bar N$, without being massive enough to disrupt a massive binary.  The same question arises in the case of sub-dominant intermediate-mass binaries that would hardly be disrupted by a dominant population of stellar-mass or asteroid-mass PBHs.
\item The effect of heavy PBHs in the tail of a broad mass distribution would be to seed PBH clusters, an effect that would be superimposed to the Poisson clustering, but that has been neglected so far in merging rate computations.
%  \item The effect of Poisson clustering may become largely dominant after matter-radiation equality and therefore additionally suppress the rates of EBs.
\item Slightly different results and another possible dependence in $f_{\rm PBH}$ have been obtained in~\cite{Kocsis:2017yty} using analytical methods.  
\item The suppression due to PBH clusters depends on the exact clustering history, on the cluster relaxation time, on the evaporation as well as on internal dynamics (see e.g. \cite{DeLuca:2020jug}).
\item PBH clustering at formation may either boost or suppress the merger rates of early binaries, but this effect is very model dependent and still debated (see e.g.~\cite{Raidal:2017mfl,Young:2019gfc,DeLuca:2021hde} for different models and conclusions).
\item It has recently been claimed in~\cite{Boehm:2020jwd} that subtle general relativistic effects may highly suppress this PBH binary formation channel, but this result has been disputed in~\cite{DeLuca:2020jug, Hutsi:2021nvs, Hutsi:2021vha}.  This problem is related to the question of which metric is physically relevant to describe the PBH environment (Takhurta metric, generalized Mc-Vittie, other).
\end{enumerate}

In general, one should not forget that early binaries can be impacted by their environment during the whole cosmic history, and this environment has a complex evolution, influenced by the clustering after matter-radiation equality, accretion, dynamical heating, etc.  Therefore one should be cautious and strong claims relying on these merging rates are probably still premature.  Nevertheless, Eq.~\ref{eq:cosmomerg} likely gives good estimations for some models, at least in some regimes.  To that end, they are relevant for estimating not only the merging rates but also the resulting SGWB background, based on our current (but rapidly evolving) knowledge.

\begin{figure}
    \centering
    Model 1:  Carr 2019, $n_{\rm s}=0.97$\\
	\includegraphics[width = 0.48\textwidth]{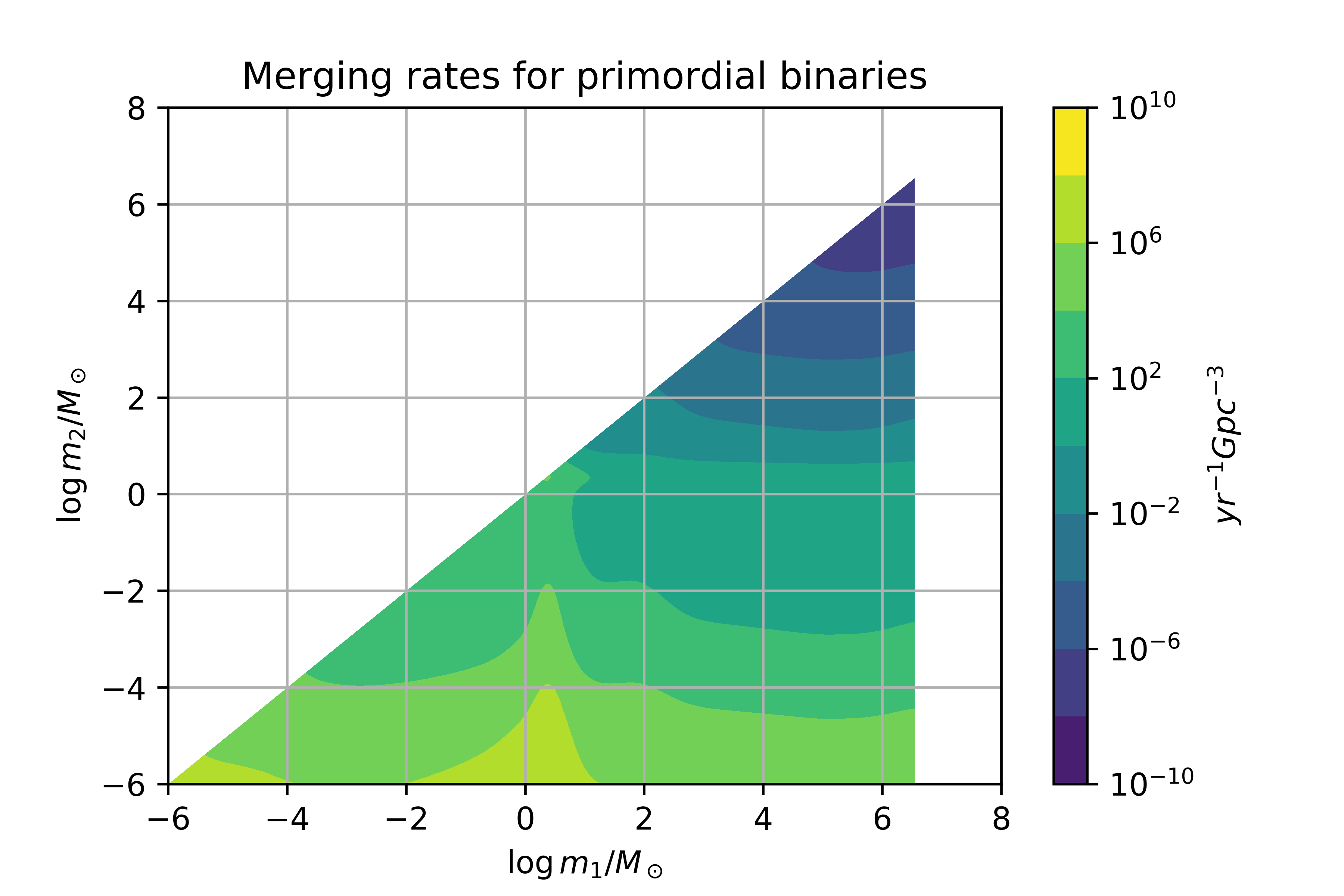} \includegraphics[width = 0.48\textwidth]{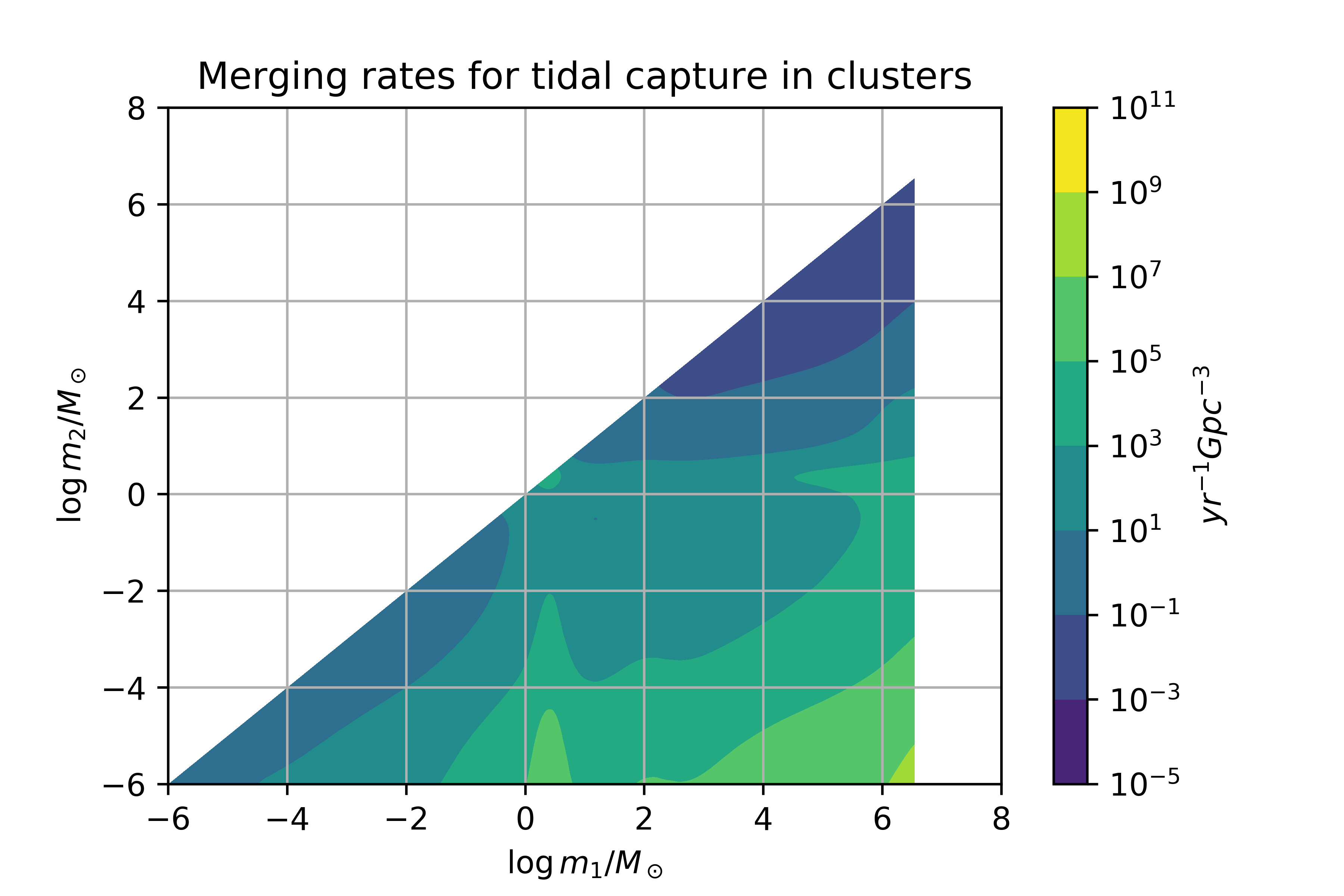}\\
	Model 2: De Luca 2020, $n_{\rm s}=1$\\
		\includegraphics[width = 0.48\textwidth]{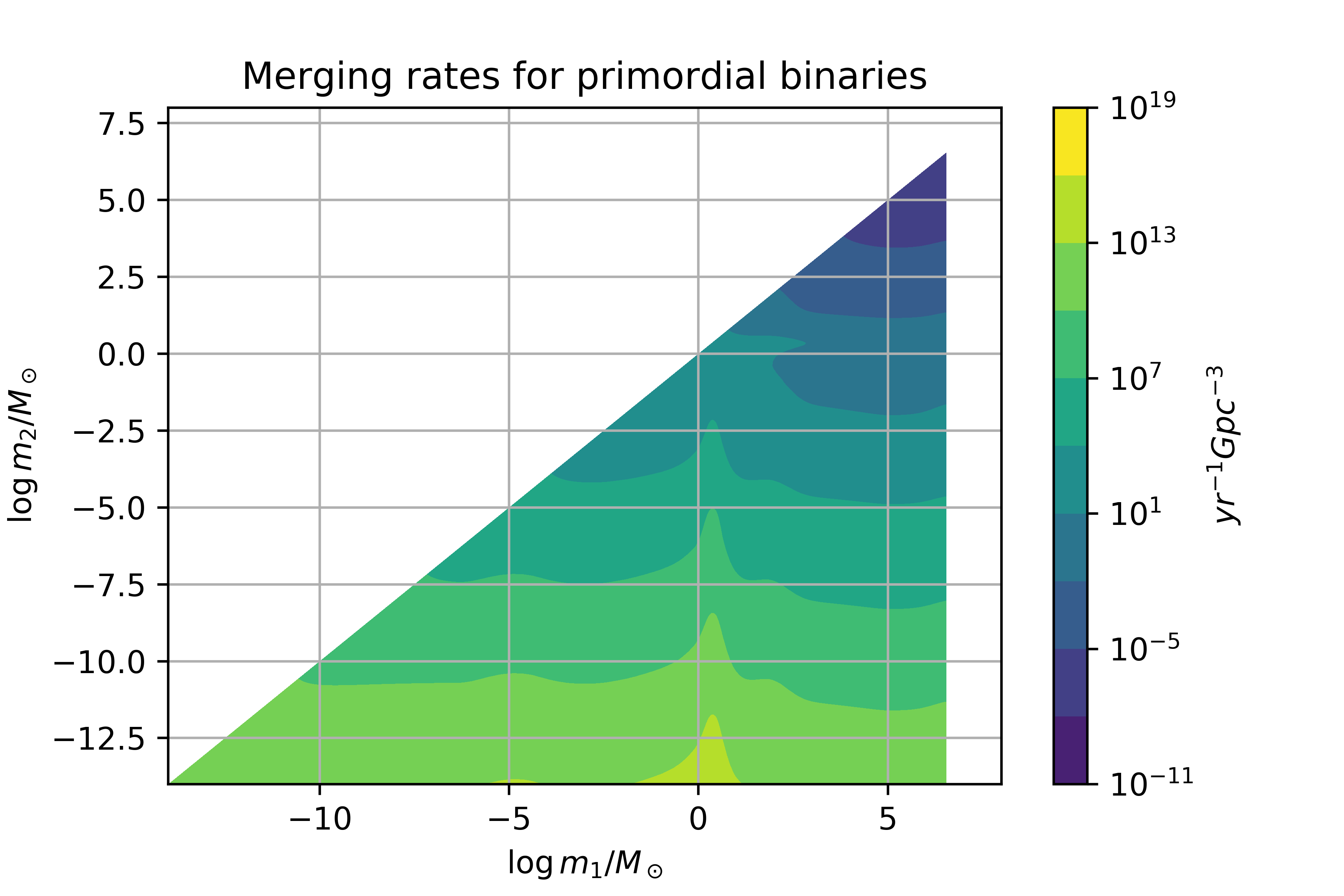}	\includegraphics[width = 0.48\textwidth]{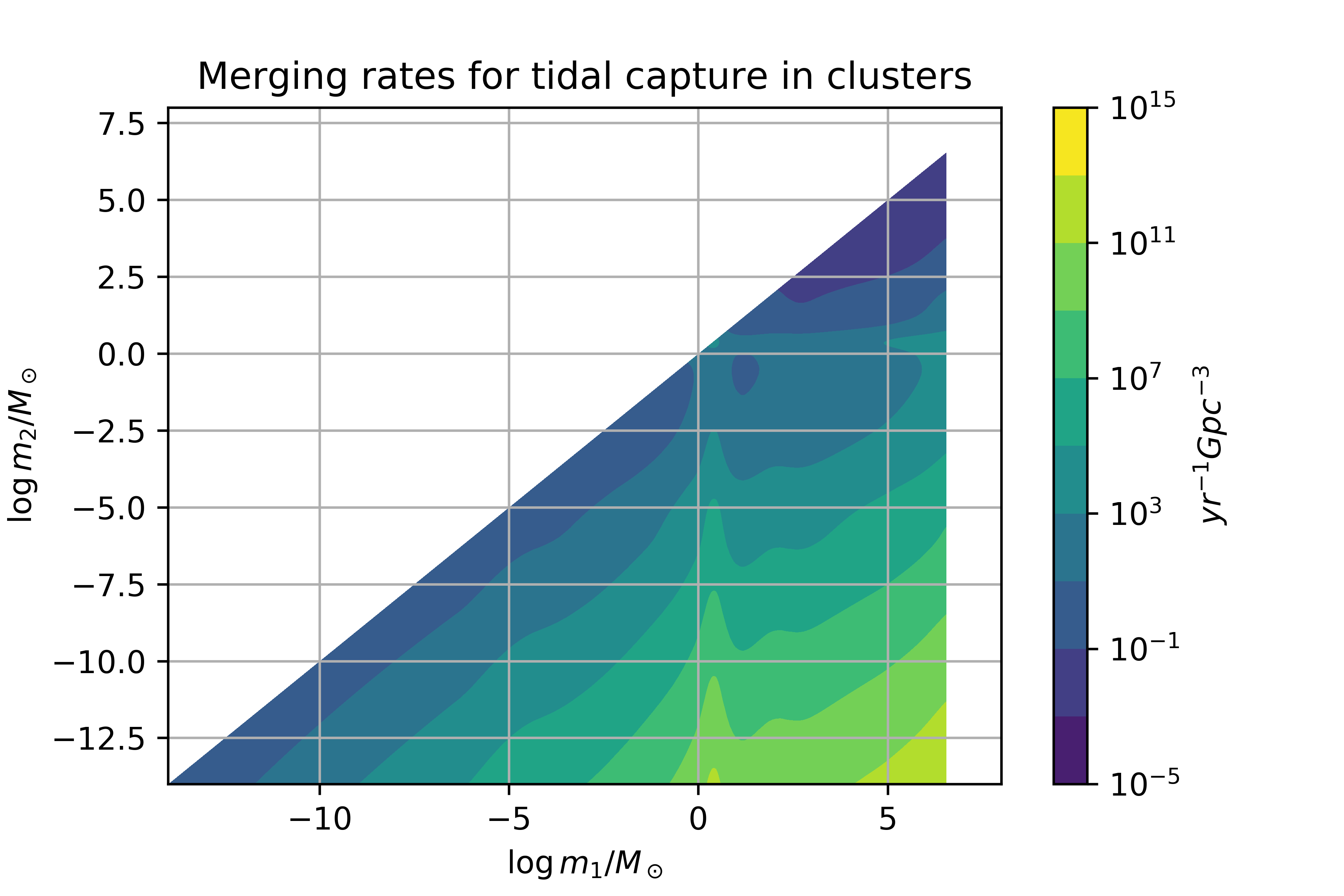}
	\caption{Expected merging rates of PBH of masses $m_1$ and $m_2$, for the two mass models represented on Fig.~\ref{fig:fPBH2models} (top panels: Model 1, bottom panels: Model 2), for the two considered binary formation channel:  primordial binaries (see Eq.~\ref{eq:cosmomerg}) on the left panels, and tidal capture in halos (see Eq.~\ref{eq:ratescatpure2}) on the right panels. Figure produced for~\cite{LISACosmologyWorkingGroup:2022jok}.}
    \label{fig:rates}
\end{figure}

\subsection{Late Binaries} \label{sec:latebinaries}
 
\subsubsection{General formula}
The second binary formation channel is through dynamical capture in dense environments. We will give a particular emphasis to the case in which PBHs comprise the totality of the dark matter in the universe and investigate the dynamical formation in dense
PBH halos.  As any other DM candidate, PBHs are expected to form halos during the cosmic history.  The clustering properties typically determine the overall merging rate. For a generic PBH mass function $f(m_{\rm PBH})$, an effective formula has been proposed~\cite{Carr:2019kxo},
   \bea 
    R_{\rm LB} &\approx& R_{\rm clust} f_{\rm PBH}^2 f(m_1) f(m_2)  \frac{(m_1 + m_2)^{10/7}}{(m_1 m_2)^{5/7}} \rm{yr^{-1}Gpc^{-3}},
     \label{eq:ratescatpure2}
    \eea
where $R_{\rm clust}$ is a scaling factor that depends on the PBH clustering properties and velocity distribution.  This formula comes from the two-body capture in a cluster and assumes that the time it takes for the binary to merge is much shorter than cosmological times.  

Hereafter, we present the estimation for $R_{\rm clust} $ in the case of Poisson clustering.  We then discuss the implications in the context of the present LIGO/Virgo observations.  Finally, we present the possible limitations of these approaches and how different models could lead to different merging rates.

% \subsubsection{Standard halo-mass function}

% The case of a standard halo-mass function was considered in~\cite{Bird:2016dcv} for a monochromatic mass function around $m_{\rm PBH} = 30 M_\odot $, but it can be extended to more general mass functions using Eq.~\ref{eq:ratescatpure2}.   In this scenario, one gets a realistic range for $R_{\rm clust}$ between $1$ and $2$ for a Press-Schechter and a Tinker halo mass function, respectively. A lower limit on the PBH halos of $400 M_\odot$ is introduced due to halo evaporation.  

%\subsubsection{Poisson clustering}

% In the previous scenario, one neglects the additional clustering due to Poisson fluctuations, an hypothesis that is only valid for $f_{\rm PBH} \ll 1$.  In the opposite case, 
Poisson clustering introduces a natural clustering scale, leading to $R_{\rm clust}$ in the range between $10^2$ and $10^3$ for solar-mass PBHs and $f_{\rm PBH} \simeq 1$~\cite{DeLuca:2020jug}. Other clustering scenarios (e.g. clusters at formation) could induce even larger values of $R_{\rm clust}$, fixed by the model as well as the evolution of these clusters (dynamical heating, mergers, etc.) through the cosmic history.

In general, one can consider a simplified picture where the PBH halo density is described by the local density contrast $\delta_{\rm loc}$. The number density of PBHs can be given as $n(m_{\rm PBH})\equiv\delta_{\rm loc}\overline{\rho}_{\rm DM}/m_{\rm PBH}$, with the mean dark matter cosmological energy density $\overline{\rho}_{\rm DM}=\Omega_{\rm DM}\rho_{\rm c}$, where $\Omega_{\rm DM}\simeq 0.25$ is the density parameter for dark matter. In this case, $R_{\rm clust}$ was estimated as~\cite{Clesse:2016ajp,Garcia-Bellido:2021jlq}
\bea 
R_{\rm clust} \approx 3.6  h^4 \left(\frac{\Omega_{\rm DM}}{0.25}\right)^2 \left(\frac{\delta_{\rm loc}}{10^8}\right) \left(\frac{v_0}{10~{\rm km/s}}\right)^{-11/7},
\eea
where $v_0$ is the velocity of PBHs and $h$ is the Hubble parameter today in units of $H_0=100~h~{\rm km/s/Mpc}$.   If one assumes $v \sim O(1-10) {\rm km/s} $ and local densities $\delta_{\rm loc} \sim 10^8$ typical to the ones of ultra-faint-dwarf galaxies, as one may also expect for Poisson clusters of size around $\sim 10-100 \rm pc$, one typically get  $R_{\rm clust} \sim 10^2 - 10^3$.   This provides an effective formula but in realistic scenarios, several effects could impact the exact value of $R_{\rm clust}$ such as the dynamical heating, merger and disruption history of PBH clusters, how extended is the PBH mass function leading to different efficiencies of the Poisson clustering effect, the radial distribution of PBHs inside a cluster and their possible mass segregation for extended mass functions.

\subsubsection{Merging rates of binaries induced by three-body interactions}

PBH binary formation is also expected to take place in the Poisson induced PBH small scale structure at high redshift. After the matter-radiation equality, the seeding effect of Poisson fluctuations in the PBH distribution give rise to the formation of small scale clusters (see Ref.~\cite{Inman:2019wvr} for a N-body simulation and Ref.~\cite{DeLuca:2020jug} for an analytical treatment). In this clusters, three-body interactions may efficiently produce binaries which can then merge in a time-scale comparable to the age of the Universe and be visible with present day GW experiments \cite{Franciolini:2022ewd} (at odds with mergers induced by GW capture which merge promptly). In this sense, this formation mechanism is similar to the one taking place in the early Universe out of binary directly decoupling from the Hubble flow.

The rate density for three bodies to interact and form PBH binaries can be expressed as \cite{Rodriguez:2021qhl}
\begin{align}
\gamma_\text{\rm 3b}(\eta\ge\eta_\text{\rm min}) & =
{3^{9/2} \pi^{13/2} \over 2^{25/2}}
	\eta_\text{\rm min}^{-{11\over2}}(1+2\eta_\text{\rm min})\left(1+3\eta_\text{\rm min}\right)
	\times{n_\PBH^3(Gm_\text{\rm PBH})^5\over\sigma_v^9}
	\nonumber \\
	&\simeq3.8\times10^{-2}{\rm Gyr}^{-1}{\rm pc}^{-3}
	\left({n_\PBH\over{\rm pc}^{-3}}\right)^{3}\left({m_{\rm PBH}\over30M_\odot}\right)^5\left({\sigma_{v}\over{\rm km/s}}\right)^{-9},
	\label{eq:gamma3b}
\end{align}
where $n_\PBH$ is the number density of PBHs.
We also  defined the hardness ratio 
$\eta = {Gm_\text{\rm PBH}}/{a\sigma_v^2}$ 
as the binding energy of a binary with size $a$ normalized to the average cluster kinetic energy, where $\sigma_v$ is the PBH velocity dispersion within the clusters. Also, in order to require the efficiency of binary formation to be close to unity and PBH binaries resulting from this channel to be in the region of parameter space where binaries are hard, one should set $\eta_\text{\rm min}=5$ as used in the literature (see e.g. Refs.~\cite{Morscher:2014doa,Rodriguez:2021qhl}).

One can compute the merger rate density of binaries produced by 3-body interactions by integrating the binary formation rate over the PBH cluster distribution throughout the evolution of Universe and by multiplying by the fraction of binaries merging within the remaining time window (See Ref.~\cite{Franciolini:2022iaa} for more details).
Depending on the assumed cluster core size and eccentricity distribution $e \equiv \sqrt{1-j^2}$ for the 3-body induced binaries \cite{1937AZh....14..207A,1975MNRAS.173..729H,2019Natur.576..406S,Raidal:2018bbj,Vaskonen:2019jpv} (i.e.
$f(j) \simeq j^\gamma$ with $\gamma = 1$ for thermal  or $\gamma = 0$ for super-thermal), this channel was found to give at least a  contribution comparable to the one from dynamical capture discussed in the following Section \cite{Franciolini:2022iaa}. 

One can also estimate how this contribution scales with the PBH abundance. Following Ref.~\cite{Franciolini:2022iaa}, this was estimated to be
$R^\text{3b}   \propto
f_\text{\rm PBH}^{2+(1+\gamma)/2},
$
showing this channel may only be relevant, compared to the EB contribution, in case of large values of the PBH abundance $f_\text{\rm PBH}$.

\subsubsection{Link with observations of compact binary coalescences}

If PBHs significantly contribute to the dark matter, late binaries merging rates are enhanced around $30-100 M_\odot$, where most compact binary mergers have been observed. Assuming Poisson clustering and $R_{\rm clust} = 450$, the corresponding merging rates for the PBH mass function of Fig.~\ref{fig:fPBH2models} that includes thermal features have been represented on Fig.~\ref{fig:rates}.  Typically one can obtain the LIGO/Virgo rates in this range with values of $f_{\rm PBH}^2 \times f(m_{\rm PBH})^2 \sim 10^{-3}$.   At the solar-mass scale, the rates for early and late binaries compete at a similar level and it is possible to get the LIGO/Virgo rates inferred for the latest events observed. These are canonically interpreted as neutron star binary mergers, even though the absence of the observation of tidal effects may allow for a primordial interpretation, with the exception of GW170817 for which the electromagnetic counterpart has also been observed. 
% if $f_{\rm PBH} \times f(m_{\rm PBH})^2 \sim 1$.  This can be obtained with $f(m_{\rm PBH} \sim M_\odot) = 1$ and $f_{\rm PBH} = 1$.  

\subsubsection{Limitations}
   
For realistic, extended mass functions, a series of effects can either boost or suppress the merging rate of late binaries and make it a rather complex and model-dependent process, subject to large uncertainties (see e.g.~\cite{Chisholm:2005vm,Chisholm:2011kn,Ali-Haimoud:2018dau,DeLuca:2020jug,Inman:2019wvr,Clesse:2020ghq,Ali-Haimoud:2018dau,Clesse:2016vqa,MoradinezhadDizgah:2019wjf,2014MPLA...2940005B,Young:2019gfc,DeLuca:2021hcf,Padilla:2020xlo,Suyama:2019cst,Ballesteros:2018swv} for studies on PBH clustering).  Some of these limitations or effects are listed below:

\begin{enumerate}
    \item The Poisson clustering is modified by the mass function and could be dominated by other mass scales than the dominant PBH masses. Therefore, one in principle needs to consider the whole mass spectrum in order to derive the importance of Poisson clustering.  \item The mass function also impacts the process of dynamical heating and dilution of sub-halos inside larger halos, which is essential to derive the natural clustering scale and the $R_{\rm clust}$ parameter.
    \item Poisson clustering should induce a broad range of cluster masses, evolving and interacting with each others in possibly complex ways through dynamical heating, dilution, collisions, tidal disruptions, etc, resulting in possibly complex time dependence and various possible PBH local environments. 
    \item The existence of intermediate-mass and supermassive PBHs leads to a seed effect that should also impact the clustering properties. 
    % \item In addition to Poisson clustering, the clustering induced by the power spectrum on small scales, enhanced compared to cosmological scales, may also play a significant role. 
    \item Clustering at formation, obtained e.g. for non-Gaussian models~\cite{Atal:2020igj}, also have an uncertain impact.  
    \item For extended mass function, the role of mass segregation in a cluster should be considered.  In average, more massive PBHs falling at the center should have a higher local density than lighter PBHs, leading to an additional mass dependence in the merger rates that was not considered so far in most analysis.
  
\end{enumerate}

\subsection{Hyperbolic encounters}
\label{sec:CHE}
In the scenario of {\em clustered} PBH of Ref.~\cite{Clesse:2015wea}, it is expected that a large fraction of BH encounters will not end up producing bound systems, which would then inspiral, but rather produce a single scattering event, via an hyperbolic encounter. This could happen, e.g. if the relative velocity or relative distance of the two PBHs is high enough that capture is not possible. The emission of GWs in close encounters of compact bodies has been studied in Refs.~\cite{Kocsis:2006hq,OLeary:2008myb,Capozziello:2008ra,DeVittori:2012da,Garcia-Bellido:2017knh,Garcia-Bellido:2017qal,Grobner:2020fnb,Mukherjee:2020hnm}. These events generate {\em bursts} of gravitational waves, which can be sufficiently bright to be detected at distances up to several Gpc. For clustered PBHs, the waveform and characteristic parameters of the GW emission in hyperbolic encounters are different to those of the inspiralling binaries, and both provide complementary information that can be used to determine the evolved mass distribution of PBHs as a function of redshift, as well as their spatial distribution.

Hyperbolic encounters are single scattering events where the majority of the energy is released near the point of closest approach, and have a characteristic peak frequency which is a function of only three variables: the impact parameter $b$, the eccentricity $e$ and the total mass of the system $M_{\rm tot} = m_1 + m_2$. Furthermore, the duration of such events is of the order of a few milliseconds to several hours, depending on those parameters. The case of inspiralling and merging PBH has been studied extensively, see e.g. Refs.~\cite{Clesse:2016vqa,Clesse:2016ajp}, and estimated the production of a few tens of events/year/Gpc$^3$ in the range of $m_{\rm PBH} \sim {\cal O}(10-100)\,\Msun$. In Refs.~\cite{Garcia-Bellido:2017qal,Garcia-Bellido:2017knh} it was shown that, within the parameter space of the clustered PBH scenario~\cite{Clesse:2015wea,Clesse:2016vqa}, we can expect a similar but somewhat lower rate of GW burst events in the millisecond range.

\begin{figure}[t]
\centering
%\hspace*{-4mm}
\includegraphics[width = 0.6\textwidth]{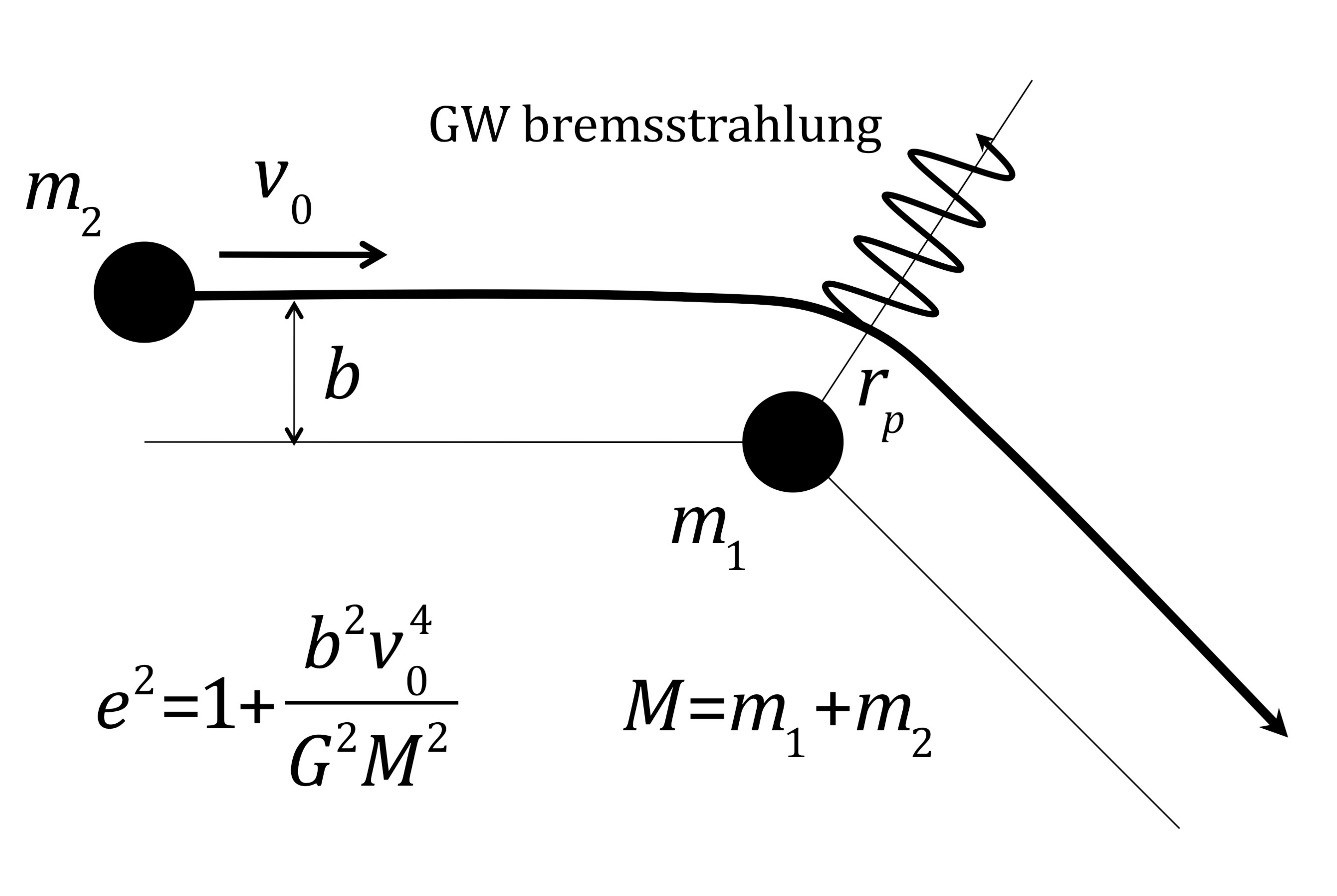}
\caption{The scattering of one BH of mass $m_2$ on another of mass $m_1$ induces the emission of gravitational waves which is maximal at the point of closest approach, $r_{\rm p}$.}
\vspace*{-1mm}
\label{fig:hyperbolic}
\end{figure}

%{\bf JGB. I am still working here...}

The waveforms of the GW emission in hyperbolic encounters are very different from
those of the inspiralling binaries, since the majority of the energy is released near the
point of closest approach, generating a burst of GWs with a characteristic ``tear-drop" shape of the emission in the time-frequency
domain~\cite{Garcia-Bellido:2017knh,Garcia-Bellido:2017qal}.
The burst has a characteristic peak frequency 
\be
f_{\rm peak} = 0.32 \, {\rm mHz} \times \frac{\beta (e + 1)}{ (e -1) } \frac{{\rm AU}}{b } ,
\ee
which corresponds to the maximum GW emission and depends only on the impact parameter $b$, the total mass of the system $M_{\rm tot}$ and the eccentricity of the hyperbolic orbit $ e = \sqrt{1+b^2 v_0^4/G^2M_{\rm tot}^2}$, where $v_0$ is the asymptotic relative velocity of the encounter and $\beta \equiv v_0 / c$. 
The maximum strain amplitude and power of the GW burst is given by 
\bea
h^{\rm max}_{\rm c} &=& 3.24\times10^{-23}\frac{R_{\rm S} ({\rm km})}{d_{\rm L}({\rm Gpc})} \frac{q \beta^2 g_{\rm max}}{(1 + q)^2}, \\
P_{\rm max}(e) &=& 5.96\times 10^{26} L_\odot \frac{q^2 \beta^{10}}{(1 + q)^4} \frac{(e + 1)}{(e -1)^5},
\eea
where 
$L_\odot$ is the solar luminosity, 
$R_{\rm S}$ is the Schwarzschild radius associated to the total mass, $q \equiv m_1/m_2$ is the binary mass ratio, $m_1 = q m_2 \geq m_2$
and $g_{\rm max} = 2 \sqrt{18(e + 1) + 5e^2/(e -1)}$.

GW signals in the LISA range could be generated by close hyperbolic encounters (CHE) of an IMBH and a SMBH as expected in the centers of galaxies, as well as from encounters of two SMBHs that could occur during galaxy collisions at low redshift.
%\textcolor{blue}{(SC: is this really possible and frequent?)}.
In the first case, an IMBH of mass $m_2 = 10^3 M_\odot$ and a SMBH of mass $m_1 = 10^6 M_\odot$, with an impact parameter $b = 1$AU and velocity $v_0 = 0.05 \, c$ 
%\textcolor{blue}{(SC: isn't it too high for a PBH cluster where $v\sim {\rm km/s}$?)} 
gives an eccentricity parameter of $e = 1.031$  with a duration of the event of 440 s. 
%The maximum
%power emitted would be $P_{max} = 1.66\times10^{49}$ erg/s and 
The maximum
stress amplitude would be $h^{\rm max}_{\rm c} = 1.02\times10^{-19}$ at a distance $d_{\rm L} = 1$, with the peak at frequency $f_{\rm peak} = 1.05$ mHz, well within the sensitivity band of LISA.
In a hyperbolic encounter between two SMBHs of equal masses $m_1 = m_2 = 10^6 M_\odot$ with impact parameter $b = 10$ AU and relative velocity $v_0 = 0.015 c$,  the eccentricity is low, $e = 1.01$, and the stress amplitude is huge, $h^{\rm max}_{\rm c} = 2.22\times10^{-17}$ at $f_{\rm peak} =1.51\times10^{-4} {\rm Hz}$, which falls again in the LISA sensitivity range. 

The event rate can be estimated by the cross-section of a CHE event, which is given by $\sigma = \pi b^2 = \pi (GM_{\rm tot}/v_0^2)^2(e^2-1)$. In the picture of the local density contrast $\delta_{\rm loc}$, the total event rate for a generic PBH mass function is given by~\cite{Garcia-Bellido:2021jlq}
\begin{equation}
\frac{{\rm d}\tau^{\rm CHE}}{{\rm d} m_1\,{\rm d} m_2}  
\approx 6.1 \times 10^{-8}~{\rm yr}^{-1}{\rm Gpc}^{-3}\,h^4
\left(\frac{\Omega_{\rm DM}}{0.25}\right)^2
\left(\frac{\delta_{\rm loc}}{10^8}\right)
\frac{f(m_1)}{m_1}\frac{f(m_2)}{m_2}\frac{M^2}{m_1\,m_2}
\frac{e^2-1}{(v_0/c)^3}\,.
\label{Eq:CHErate}
\end{equation}

These CHE events are very common in dense clusters formed soon after recombination, while the PBH scatter off each other and puff-up the cluster. Some of these events loose so much energy in their GW emission that they end up in bounded systems. Many also are responsible for the disruption of previously formed binaries. All in all, there is a significant production of gravitational waves from CHE that can contribute to a SGWB early on. In the next Sections we will explore such a stochastic background from CHE events across the history of the universe, and will see that it can be significant for a wide PBH mass function, even in the LISA range of frequencies.

\subsection{How to distinguish PBH from astrophysical binaries}

In this Section we elaborate on which observable may be used to assess or rule out the primordial origin of an individual merger event. 
We follow the comprehensive description given in Ref.~\cite{Franciolini:2021xbq}.
A systematic strategy to use these discriminators is summarized in the flowchart of Fig.~\ref{fig:flowchart}, based on the predictions of the standard PBH scenario where binaries are assembled in the early Universe. 
Notice that most of predictions would still apply to a much larger class of models, see Ref.~\cite{Franciolini:2021xbq} for more details. 

\begin{figure*}[t!]
\centering
\includegraphics[width=.95\textwidth]{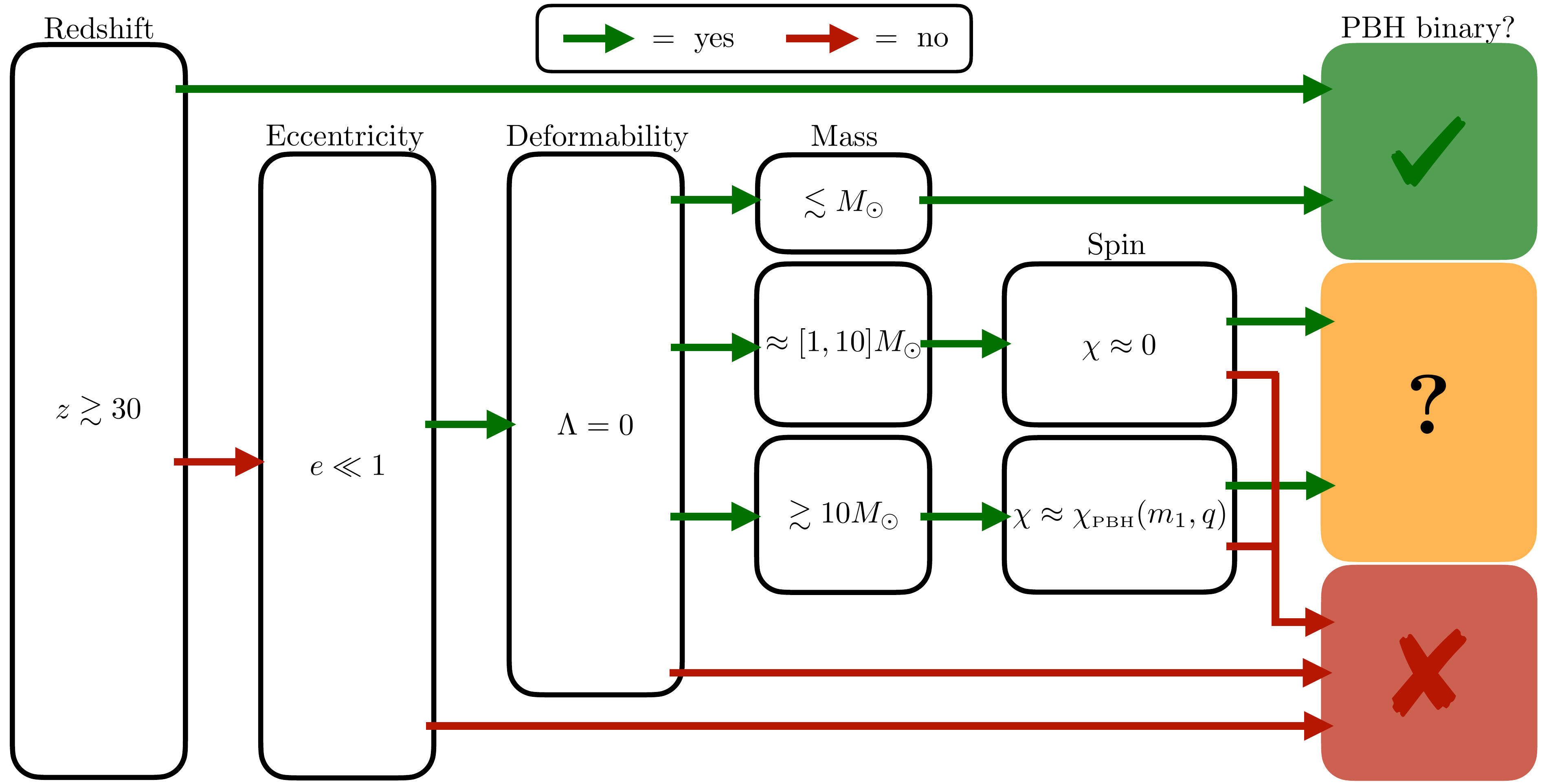}
\caption{Figure taken from Ref.~\cite{Franciolini:2021xbq}.
Schematic flowchart representing how to
systematically rule out or potentially assess the primordial origin of a binary merger. 
These criteria are based on measurements of the redshift $z$, eccentricity $e$, tidal deformability $\Lambda$, component masses $m$, and dimensionless spin $\chi$. 
Each arrow indicates if the condition in the box is met (green) or violated (red), while the marks indicate: 
\cmark) likely to be a PBH binary; 
\xmark) cannot to be a PBH binary;
\textcolor{Black}{\textbf{?}}) may be a PBH  binary. 
} 
\label{fig:flowchart}
\end{figure*}

One can identify two ``smoking-gun" signatures of the primordial scenario, which are high redshift mergers and sub-solar masses. 
A general prediction of the PBH model is a merger rate density which grows monotonically with redshift~\citep{Ali-Haimoud:2017rtz,Raidal:2018bbj,DeLuca:2020qqa}. Focusing on early binaries, one expects $
R^{\rm EB} (z) \propto(t(z)/{t (z=0)})^{-34/37}$,
extending up to redshifts $z\gtrsim{\cal O}(10^3)$. 
On the contrary, astrophysical-origin mergers are generically not expected to occur at $z\gtrsim 30$, even though the redshift corresponding to the epoch of first star formation is still poorly known. To give a conservative estimate, 
theoretical calculations and cosmological simulations suggest 
star formation does not precede $z\sim 40$~\cite{Schneider:1999us,Schneider:2001bu,Schneider:2003em,Bromm:2005ep,Tornatore:2007ds,Trenti:2009cj,deSouza:2011ea,Koushiappas:2017kqm,Mocz:2019uyd,Liu:2020ufc}. Additionally, the time delay between Pop~III star formation and BBH mergers was estimated to be around $\mathcal{O}(10)~\rm Myr$~\cite{Kinugawa:2014zha,Kinugawa:2015nla,Hartwig:2016nde,Belczynski:2016ieo,Inayoshi:2017mrs,Liu:2020lmi,Liu:2020ufc,Kinugawa:2020ego,Tanikawa:2020cca,Singh:2021zah}, implying merger redshifts $z\gtrsim30$ to represent smoking guns for primordial binaries~\cite{Koushiappas:2017kqm,DeLuca:2021wjr,Ng:2021sqn}.
Observations of such distant events may be characterised by large measurement uncertainties on the inferred luminosity distance due to the low SNR at 3G detectors \cite{Ng:2021sqn,Franciolini:2021xbq,Martinelli:2022elq,Ng:2022vbz}, including LISA.
However, it was recently shown that focusing on constraining the merger rate evolution at redshift larger than ${\cal O}(10)$
may allow to constrain PBH populations up to abundances as low as $f_{\rm PBH} \approx 10^{-5}$ \cite{Ng:2022agi} in the solar mass range, even accounting for the contamination of Pop III binaries.

Sub-solar BHs are not expected in standard stellar evolution \footnote{See, however, Ref.~\cite{Shandera:2018xkn} for models in which subsolar BHs are born out of dark sector interactions or capture of asteroidal mass PBHs \cite{Oncins:2022djq}.}.
Other compact objects like white dwarfs, brown dwarfs, or exotic compact objects~\cite{Cardoso:2019rvt} (e.g. boson stars~\cite{Guo:2019sns}) may be misinterpreted as sub-solar PBHs. Distinguishing PBHs from other compact objects requires taking into account tidal disruption and tidal deformability measurements. Less compact objects like brown and white dwarfs are expected to be tidally disrupted well before the contact frequency. This leads to a different GW signature, see Ref.~\cite{Franciolini:2021xbq} and Refs.~therein for more details on the observability of these effects. 

Another key prediction of the primordial model involves the eccentricity $e$ of PBH binaries.
While formed with large eccentricity at high redshift, PBH binaries then have enough time to circularize before the GW signal can enter the observation band of current and future detectors. Therefore, observing a non-zero eccentricity would rule out the interpretation as a primordial binary formed in the early Universe, while it may still be compatible with a PBH binary formed in the late-time universe \cite{Cholis:2016kqi,Wang:2021qsu}. Finally, the spin distribution of PBHs inherits characteristic mass-spin correlations induced by accretion effects, as discussed in Sec.~\ref{sec: spin dist}.
Using this criterion for determining the possible primordial nature of individual GW events would require reducing uncertainties on the accretion model. However, searching for signatures of this correlation can also be performed on a population level. This was recently done in Ref.~\cite{Franciolini:2022iaa} focusing on the GWTC-3 catalog, finding a similar correlation (indistinguishable at this stage from the one present in dynamically formed astrophysical binaries) may be present in the data.

In Sec.~\ref{sec:GWandLSS}, dedicated to correlating GWs with LSS, we will mention further possibilities of distinguishing astrophysical and primordial merger populations through measurements of the event bias. 

\subsection{Summary}

The calculation of PBH merger rates or encounter rates has been a very active line of research during the last few years, with some important twists.  Two main binary formation channels have been envisaged:  early binaries that formed quickly in the radiation era when two PBHs formed sufficiently close to each other, and late binaries that can form inside PBH clusters, when two PBHs pass sufficiently close to each other to form a bound system.  Today, it is still unclear which binary formation channel provides the most important merger rates on the mass scales probed by LVK, especially since these rates depend on the exact shape of the PBH mass function, on the dark matter fraction made of PBHs, on the clustering properties, on accretion, possible hierarchical mergers, on binary disruptions, etc.  It is also debated if the limits on those rates coming from the latest GW observations favor or disfavor PBHs as the dark matter. 

%SC:  to be checked if we agree or not to remove this sentence
%Nonetheless, assuming the standard formation scenario, current LVK observations force the PBH abundance to be below $f_\text{\tiny PBH} \lesssim 10^{-3}$ in the stellar mass range. For such small PBH abundance, the early universe binary formation channel is the dominant contribution to the overall merger rate \cite{Ali-Haimoud:2017rtz}. It remain to be proven whether different scenarios are able evade this bound. 

We tried to list all these limitations and we now attempt to summarize the current status:  at first, the merger rates of early binaries were found to importantly surpass the ones of late binaries in clusters, at least in the standard scenario of cosmological structure formation.  However, on the one hand, N-body simulations have revealed several mechanisms that importantly suppress those rates if PBHs constitute a significant fraction of the dark matter, due to binary disruption by other PBHs, by matter fluctuations, and above all by the PBH clusters seeded by the inevitable Poisson fluctuations in the initial PBH distribution.  On the other hand,  the same effects should be responsible for a boost of the merger rates of late binaries in these PBH clusters, at a level comparable to that of early binaries on solar-mass scales, if PBHs significantly contribute to the dark matter.   An important difference between early and late binary merger rates is the mass dependence, roughly going like $1/m_{\rm PBH}^{32/37}$ for early binaries and $1/m_{\rm PBH}^{11/21}$, by taking into account the halo mass distribution, for late binaries~\cite{Bird:2016dcv}.  Another difference is the redshift evolution of those rates.  In both cases, however, it seems very disfavored that all the dark matter is made of $30 M_\odot$ black holes.
The scenario where the PBH mass function peaks at the solar mass scale, where the QCD transition should have favored PBH formation, may represent an alternative possibility, 
provided the overall abundance is below the current observational constraints in that mass range, see in particular Refs.~\cite{Petac:2022rio,Gorton:2022fyb,Juan:2022mir,
DeLuca:2022uvz,Franciolini:2022tfm}.

\section{Stochastic backgrounds} \label{sec:SGWBs}

\subsection{From second order curvature fluctuations}
\label{sec:secondordergw}
If PBHs are generated by the collapse of large density perturbations, they are unavoidably associated to the emission of induced GWs at second order by the same scalar perturbations due to the intrinsic nonlinear nature of gravity \cite{Tomita:1975kj,Matarrese:1993zf,Acquaviva:2002ud,Mollerach:2003nq,Ananda:2006af,Baumann:2007zm}~\footnote{We stress 
that we do not refer here to the GWs produced only in the overdense regions that contract at horizon crossing due to the gravitational force \cite{DeLuca:2019llr}, but from everywhere in the Universe, due to the general increase of the power of the density perturbations at small scales. See also here~\cite{Lin:2020goi,Chen:2021nio,Kawai:2021edk,Lin:2021vwc,Papanikolaou:2021uhe,Ivanov:2021chn,Zhang:2021rqs,Yi:2022anu,Cheong:2022gfc,Feng:2023veu,Zhang:2022xmm,Arya:2023pod,Tzerefos:2023mpe} for studies within modified gravity setups.}.
The phenomenological implications have been investigated in various contexts also associated to PBHs \cite{Saito:2008jc,Bugaev:2009zh,Saito:2009jt,Garcia-Bellido:2017aan,Ando:2017veq,Bartolo:2018qqn,Bartolo:2018rku,Bartolo:2018evs,Clesse:2018ogk,Unal:2018yaa,Wang:2019kaf,Domenech:2019quo,Domenech:2020kqm,Pi:2020otn,Ragavendra:2020sop,Fumagalli:2020nvq,Yuan:2019udt,Fumagalli:2021mpc,Domenech:2021and} (see recent reviews \cite{Yuan:2021qgz,Domenech:2021ztg}).
Such a source of GWs is present at all scales, but due to the necessary enhancement of the scalar power spectrum responsible for the generation of PBHs around the characteristic scale $k_\star$, this stochastic background can become detectable by GW experiments like LISA.

There are several current and future experiments searching for a SGWB in various frequency ranges.
 In the ultra-low frequency range (around nHz), the observations at Pulsar Timing Array (PTA) experiments 
 like {NANOGrav~\cite{NG15-SGWB,NG15-pulsars}, EPTA (in combination with InPTA)\,\cite{EPTA2-SGWB,EPTA2-pulsars,EPTA2-SMBHB-NP}, PPTA\,\cite{PPTA3-SGWB,PPTA3-pulsars,PPTA3-SMBHB} and CPTA\,\cite{CPTA-SGWB}, give rise to the most stringent constraints on the GWs abundance. 
 }
 Future experiments like SKA \cite{2009IEEEP..97.1482D} (see also \cite{Moore:2014lga}) will greatly improve the sensitivity.
 
 In the LVK frequency range, an additional constraint has been set by the non-observation of a stochastic background after O1-O2 runs \cite{LIGOScientific:2019vic}.
 These searches can be translated into a constraint on the amplitude of the comoving curvature perturbation at the corresponding scales \cite{Inomata:2018epa,Unal:2020mts}. Those bounds are also affecting the maximum allowed PBH fraction of DM  with the hypothesis that they originate from the collapse of density perturbations. Detailed studies with the LVK data affecting the mass range $\llp 10^{-20},10^{-18} \rrp M_\odot $ are reported in Ref.~\cite{Kapadia:2020pir}. Also, in \cite{Cai:2019elf}, the dependence of the result on non-Gaussianities is also investigated, finding that local non-Gaussianity can for example alleviate the bounds (see Section \ref{Impact of non-gaussianity} for details). This is possible for two main reasons : i) Non-Gaussianity can allow PBHs to be produced more efficiently for the same power spectra, as the tail of the non-Gaussian probability distribution function of the perturbations has more area/probability for the perturbations greater than the threshold. ii) Due to several contractions, there is a higher symmetry factor. Hence, non-Gaussian perturbations can produce a large fraction of PBHs with a smaller amplitude power spectrum, and therefore with a smaller amount of sourced GWs. 
 
 SKA \cite{Moore:2014lga,Zhao:2013bba} is the next generation PTA experiment that will probe primordial perturbations very sensitively \cite{Byrnes:2018txb,Inomata:2018epa,Kalaja:2019uju,2020arXiv200803289G}. It has been recently shown in Ref \cite{Unal:2020mts} that PTA-SKA combined with CMB distortions will robustly test the PBHs from inflationary fluctuations, namely they will detect the stochastic GW background or distortion signatures and possibly make a extraordinary discovery or constrain the PBHs heavier than a solar mass for 13 orders of magnitude ($1-10^{13}M_\odot$) to a completely negligible amount  $\frac{\rho_{\rm PBH}}{\rho_{\rm DM}} < 10^{-10}$. 
  
 The origin of these multimessenger signals is not coming from PBHs themselves but from the GW background induced at second order by cosmological curvature perturbations (including ones that do not lead to PBHs).  As a consequence, the conclusions will be robust to changes in 
 the astrophysical evolution of PBHs (accretion and merger history) and clustering effects\cite{Unal:2020mts}. %i) statistical properties of inflationary perturbations (Gaussian or  Non-Gaussian), ii) astrophysical evolution (accretion and merger history) and iii) clustering effects \cite{Unal:2020mts}. 
 This will be also a conclusive test for the intriguing proposal that the seeds of the SMBHs are formed by PBHs \cite{Duechting:2004dk,2014MPLA...2940005B, 2015PhRvD..92b3524C,2016PhRvD..94j3522N,Garcia-Bellido:2016dkw}.

The LISA experiment will be able to provide insights in the intermediate frequencies and the corresponding masses.  Since the emission mostly occurs when the corresponding scales cross the horizon, one can relate the GW frequency to the PBHs mass $m_{\rm PBH}$ as (see for example \cite{Saito:2009jt,Garcia-Bellido:2017aan})
\begin{equation} \label{eq:f_to_mPBH}
	f \simeq 6  \, \text{mHz} \sqrt{\gamma} \lp \frac{m_{\rm PBH}}{ 10^{-12}M_\odot}\rp ^{-1/2}.
\end{equation}
Notice that the peak frequencies fall within the LISA sensitivity band for PBH masses around $m_{\rm PBH} \sim {\cal O} \lp 10^{-15} - 10^{-8} \rp M_\odot$ \cite{Saito:2008jc,Garcia-Bellido:2017aan,Bartolo:2018rku,Cai:2018dig,Unal:2018yaa,Bhaumik:2020dor}.

In the following, we review the procedure to compute the induced SGWB spectrum from primordial scalar perturbations.
First of all, let us define the scalar and tensor perturbations in the Newtonian gauge as
\begin{equation}
	\d s^2 = a^2 \left \{ - \lp1+ 2 \Psi  \rp \d \eta ^2 + \llp \lp1 - 2 \Psi   \rp \delta _{ij} +\frac{1}{2} h_{ij} \rrp  \d x^i \d x^j \right \} , 
\end{equation}
where we neglected the anisotropic stress.
The GWs emission is captured by the equation of motion for the GWs \footnote{We do not consider the free-streaming effect of neutrinos on the GW amplitude \cite{Weinberg:2003ur}. } as
\begin{equation}
	h_{ij}''+2\mathcal H h_{ij}'-\nabla^2 h_{ij}=-16 \mathcal T_{ij}{}^{ \ell m}
	\llp 
\Psi\partial_\ell \partial_m\Psi+2\partial_\ell\Psi\partial_m\Psi-\partial_\ell\left(\frac{\Psi'}{\mathcal H}+\Psi\right)\partial_m\left(\frac{\Psi'}{\mathcal H}+\Psi\right)
\rrp,
	\label{eq: eom GW1}
\end{equation}
where derivatives are taken with respect to conformal time  $\eta$,  $\mathcal H \equiv a'/a$ is the conformal Hubble parameter, and the source is evaluated assuming a radiation-dominated (RD) epoch \footnote{
The projector  $\mathcal T_{ij}{}^{\ell m}$ to the transverse and traceless component  is defined in Fourier space 
in terms of the polarisation tensors  $e_{ij}^{\lambda}(\vk)$ the chiral basis $({\rm L},{\rm R})$ as
$
	\tilde{\mathcal T}_{ij}{}^{\ell m}(\vk)=e_{ij}^\text{\tiny L}(\vk)\otimes e^{\text{\tiny L} \ell m}(\vk)
	+ e_{ij}^\text{\tiny R}(\vk)\otimes e^{\text{\tiny R}\ell m}(\vk)
$. 
The normalisation adopted is 
$e_{ij,\lambda}( \vec{k}) e_{ij,\lambda'}^*( \vec{k}) = \delta_{\lambda\lambda'}$.
}.
The solution of the first order equation of motion in a radiation-dominated Universe relates the Bardeen's potential $\Psi$ to the gauge invariant comoving curvature perturbation through \cite{Lyth:1998xn}
\begin{equation}
	\Psi(\eta,\vk)=\frac 23 T( k \eta ) \zeta(\vk) ,
\qquad \text{where} \qquad 
	T(z)= \frac{9}{z^2}\left[ \frac{\sin (z/\sqrt 3)}{z/\sqrt 3} -\cos(z/\sqrt 3) \right].
	\label{eq: transfer}
\end{equation}
One can decompose the tensor field in Fourier space as
\begin{equation}
h_{ij} \left( \eta ,\, \vec{x} \right) = \int \frac{{\rm d}^3 k}{\left( 2 \pi \right)^3} \sum_{\lambda = R,L} h_\lambda ( \eta ,\, \vec{k} )\,   e_{ij,\lambda}  ({\hat k} ) \, {\rm e}^{i \vec{k} \cdot \vec{x} }, 
\end{equation} 
where $h_\lambda$ are the two helicity modes. Introducing the dimensionless variables $x=p/k$ and $y=|\vec{k}-\vec{p}|/k$, the GWs emitted take the form, see for example \cite{Espinosa:2018eve}, 
\begin{equation} 
h_\lambda ( \eta ,\, \vec{k} ) = \frac{4}{9 k^3 \eta} \int \frac{{\rm d}^3 p}{\left( 2 \pi \right)^3}  
\, {\rm e}_\lambda^* ( \vec{k} ,\,\vec{p} ) \zeta ( \vec{p} )  \zeta ( \vec{k} - \vec{p} ) %\nonumber\\ 
%&& \quad\quad \times 
\left[ {\cal I}_{\rm c} ( x, y )  \cos \left( k \eta \right) 
+  {\cal I}_{\rm s} (x, y) 
\sin \left( k \eta \right) \right],
\label{h-sourced}
\end{equation} 
where ${\rm e}_\lambda ( \vec{k} ,\,\vec{p} ) \equiv {\rm e}_{ij,\lambda} ({\hat k} ) \vec{p}_i \vec{p}_j$, and two functions ${\cal I}_{{\rm c},{\rm s}}$ are \cite{Espinosa:2018eve,Kohri:2018awv}
\begin{align}
\label{eq: Ic, Is tau0=0}
{\cal I}_{\rm c}(x,y) &= -36\pi\frac{(s^2+d^2-2)^2}{(s^2-d^2)^3}\theta(s-1)\ ,\\
{\cal I}_{\rm s}(x,y) &= -36\frac{(s^2+d^2-2)}{(s^2-d^2)^2}\left[\frac{(s^2+d^2-2)}{(s^2-d^2)}\log\frac{(1-d^2)}{|s^2-1|}+2\right], 
\end{align}
with 
\begin{equation}
d \equiv \frac{1}{\sqrt{3}}|x-y|, \qquad  s \equiv \frac{1}{\sqrt{3}}(x+y) ,\qquad  (d,s) \in [0,1/\sqrt{3}]\times[1/\sqrt{3},+\infty).
\label{eq: xy to ds}
\end{equation}
The definition of the energy density associated to GWs is \cite{Misner:1974qy,Maggiore:1999vm,Flanagan:2005yc} 
\begin{equation}
\rho_{\rm GW} = \frac{M_{\rm Pl}^2}{4} \left\langle \dot{h}_{ab} \left( t ,\, \vec{x} \right)  \dot{h}_{ab} \left( t ,\, \vec{x} \right) \right\rangle_T,
\label{rho}
\end{equation} 
where the angular brackets denote a time average performed on a timescale $T$ much greater than the GW phase oscillations ($T k_i \gg 1$) but much smaller than the cosmological time ($T H \ll 1$).
%, and $M_{\rm p} = 1/\sqrt{8\pi G}$ is the reduced Planck mass. 

Assuming that the scalar perturbations $\zeta$ are Gaussian (see the next Sections for a possible relaxation of this assumption), one finds 
\begin{align} 
&\left\langle \rho_{\rm GW} \left( \eta ,\, \vec{x} \right) \right\rangle
\equiv \rho_{\rm c} ( \eta) \, \int {\rm d} \ln k \; \Omega_{\rm GW} \left( \eta ,\, k \right) \nonumber\\
&=  
  \frac{2 \pi^4 M_{\rm p}^2}{81  \eta^2 a^2}  \, \int \frac{{\rm d}^3 k_1 {\rm d}^3 p_1}{\left( 2 \pi \right)^{6} } 
\frac{1}{k_1^4}\, 
\frac{\left[ p_1^2 -  ( \vec{k}_1 \cdot \vec{p}_1)^2/k_1^2 \right]^2}{p_1^3 \, \left\vert \vec{k_1} - \vec{p}_1 \right\vert^3} \, 
 {\cal P}_\zeta ( p_1) 
{\cal P}_\zeta( |\vec{k_1} - \vec{p}_1|)   \llp{\cal I}_{\rm c}^2( \vec{k}_1 ,\, \vec{p}_1) + {\cal I}_{\rm s}^2( \vec{k}_1 ,\, \vec{p}_1) \rrp,
\label{rho1-par}
\end{align} 
as a function of the curvature perturbation power spectrum $ {\cal P}_\zeta$. In the first line, we have defined the fractional GW energy density for log interval $\Omega_{\rm GW}$, in terms of the critical energy density of a spatially flat Universe, $\rho_{\rm c} = 3 H_{0}^2 M_{\rm p}^2$.

The current GW abundance can then be obtained as
\begin{equation}
	\Omega_{\rm GW}(\eta_0,k) =  \frac{a_{\rm f}^4\rho_{\rm GW}(\eta_{\rm f},k)}{\rho_{\rm r}(\eta_0)}\Omega_{{\rm r},0} =
\frac{g_*(\eta_{\rm f})}{g_*(\eta_0)} \left(\frac{g_{*{\rm S}}(\eta_0)}{g_{*{\rm S}}(\eta_f)}\right)^{4/3}
\Omega_{{\rm r},0} 	\Omega_{\rm GW}(\eta_{\rm f},k),
	\label{eq: Omega GW}
\end{equation}
in terms of the present radiation energy density fraction $\Omega_{{\rm r},0}$ if the neutrinos were massless and the effective degrees of freedom for entropy density $g_{* \rm{S}}$.
For the frequencies of interest, using 
$
	f \simeq 8\,  {\rm mHz}  \lp{g_*}/{10} \rp ^{1/4} \lp T/{\rm 10^6 GeV} \rp $,
	one can show that the emission of GWs takes place at $\eta_{\rm f}$ (with corresponding scale factor $a_{\rm f}$) well before the time at which top quarks start annihilating, above which we can assume a radiation-dominated Universe with constant effective degrees of freedom.

As one can see, the GW's abundance depends on the curvature perturbation power spectrum.
One possibility is represented by a monochromatic power spectrum  with support at the momentum scale $k_\star$
\begin{equation}
{\cal P}_\zeta (k) = A_{\rm s} \, k_\star \delta \left( k - k_\star \right),
\label{Pz-delta}
\end{equation}
enhanced with respect to the power spectrum on large CMB scales, for which the current GWs abundance can be computed analytically as (see also Refs.~\cite{Saito:2009jt,Bugaev:2009zh})
\begin{equation}
\Omega_{\rm GW}(\eta_0,k) =  \frac{\Omega_{{\rm r},0} A_{\rm s}^2}{15552} \frac{g_*(\eta_{\rm f})}{g_*(\eta_0)} \left(\frac{g_{*S}(\eta_0)}{g_{*S}(\eta_{\rm f})}\right)^{4/3}  \left( \frac{4 k_\star}{k} -  \frac{k}{k_\star} \right)^2 \theta\left( 2 k_\star  - k \right) \left[ {\cal I}_{\rm c}^2 \left( \frac{k_\star}{k} ,\,  \frac{k_\star}{k}  \right) + {\cal I}_{\rm s}^2 \left(  \frac{k_\star}{k} ,\,  \frac{k_\star}{k}  \right) \right],
\label{OGW-delta}
\end{equation}
where $\theta$ is the Heaviside step function.

A more realistic spectrum is provided by a log-normal shape with width $\sigma$
\begin{equation}
	\label{Pz-gauss}
{\cal P}_\zeta (k)=A_\zeta\,{\rm exp}\left(-\frac{\ln^2(k/k_\star)}{2\sigma^2}\right),
\end{equation}
whose abundance is shown by the black line of  Fig.~\ref{fig:inducedSGWBspectrum}, compared to the monochromatic shape in red.
For illustrative purposes, the power spectra peak scale has been chosen to be the one at which LISA has its maximum sensitivity, i.e. $f_\star = f_{\rm LISA} \sim 3.4 \, {\rm mHz}$ where  $f_\star \equiv k_\star / 2 \pi $, corresponding to PBH with masses around $10^{-12} M_\odot$,  while the parameters were chosen in order to have PBHs which account for the totality of the DM. One can see that the curves fall well within the sensitivity curves of LISA. It is thus clear that if PBHs of such masses form the totality (or a fraction) of dark matter, LISA will be able to measure the GWs sourced during the PBH formation time for a wide variety of PBH masses.

The spectral shapes present differences at high frequencies, because of the absence of the sharp cut-off at $2f_\star$ characteristic of the Dirac delta case, and at lower frequencies, since physical spectra would typically give a white-noise ($\propto f^3$) behaviour \cite{Espinosa:2018eve,Cai:2019cdl}, while the slow fall-off of the monochromatic case is an unphysical effect of assuming such a power spectrum. Furthermore, in the latter a resonant effect at $f \sim 2 f_\star / \sqrt{3}$ produces the spike, see for example Ref.~\cite{Ananda:2006af}.
\begin{figure}[t!]
	\centering
	\includegraphics[width=0.6\columnwidth]{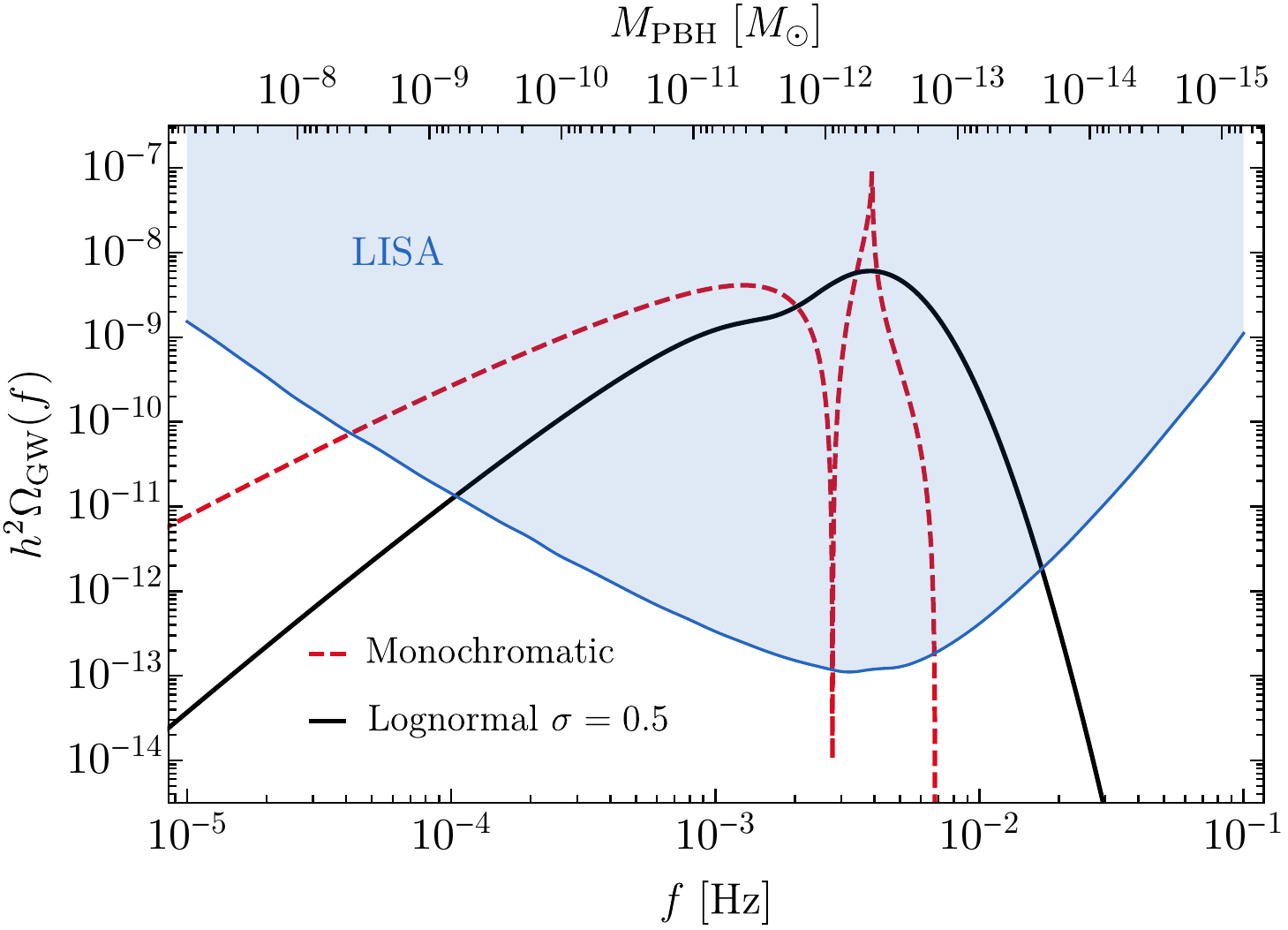}
	\caption{
		Induced GWs spectra for both examples considered in Eqs.~\eqref{Pz-delta} and \eqref{Pz-gauss}, for the parameters choice $A_{\rm s} = 0.033$, $A_\zeta = 0.044$ and $\sigma = 0.5$.
		For comparison, we also show  the estimated sensitivity for LISA \cite{Audley:2017drz}, {following the proposed design (4y, 2.5 Gm of length, 6 links)}. The PBH mass corresponding to the characteristic frequency is depicted on the top horizontal axis, according to Eq.~\ref{eq:f_to_mPBH}.
	}
	\label{fig:inducedSGWBspectrum}
\end{figure}

 As one can appreciate from Eq.~\eqref{h-sourced}, the non-linear coupling to the curvature perturbation naturally leads to an intrinsically non-Gaussian GW signal imprinted in phase correlations. However, it has been shown in Refs. \cite{Bartolo:2018evs,Bartolo:2018rku} (see also \cite{Bartolo:2018qqn,Margalit:2020sxp}), that the coherence is washed out by the propagation of the waves in the perturbed Universe mainly due to time delay effects originated from the presence of large scale variations of gravitational potential. This is simply a consequence of the central limit theorem applied to a number $N \sim \lp k_\star \eta_0 \rp^2 \ggg 1$ of independent lines of sight \cite{Bartolo:2018evs}. Possible small deformations smearing the GW spectrum can also arise from similar effects \cite{Domcke:2020xmn}.

\subsection{Gauge invariance of the SGWB spectrum}
The fact that tensor modes are generated at second-order in perturbation theory 
raises the issue of the possible gauge dependence of the result commonly computed in the Newtonian gauge. Indeed, second order tensor modes, contrary to the first order ones, are not gauge invariant. This issue has been recently highlighted in the literature \cite{Hwang:2017oxa,Domenech:2017ems,Gong:2019mui,Tomikawa:2019tvi,Lu:2020diy}. 

A physical observable is however not dependent on gauge choices by definition and it can be identified by understanding how the measurement in GW experiments is performed. 
The description of the detector's response to the GW signal can be best performed in the so-called transverse-traceless (TT) frame \cite{maggiore2008gravitational,DeLuca:2019ufz}, particularly in space based experiments like LISA. The TT frame is defined as the one where the coordinates are fixed with the position of the mirrors and the effect of the passing GWs is captured by the delay of the arrival times between the arms of the experiment. The peculiarity of the LISA experiment is due to the relation between the characteristic frequency observable and the arm length $L$, which is $f_{\rm LISA} L \sim {\mathcal O} (1)$. The TT choice is optimal because it avoids keeping track of the otherwise large corrections ${\mathcal O} \lp f L \rp$ appearing if expanding around a locally inertial reference frame. Indeed, the projected sensitivity curves for the interferometer LISA  are provided in such a frame. 
%If the emission takes place in a matter dominated universe, for example,
% A thorough identification of the correct degrees of freedom propagating the observable GWs needs to be performed along this line.

One should remark that, during the period in which the source of the tensor mode is active and therefore the tensor field is coupled to the scalar perturbations at second-order in perturbation theory, it retains a gauge dependence and cannot be identified with the freely-propagating GW, which can be treated linearly and is therefore gauge invariant. 

In the case of the production of tensor degrees of freedom from scalar perturbation during a radiation-dominated phase of the Universe, the source is active only when the scalar perturbations re-enter the horizon. Once produced, the GWs effectively decouple from the second order source and behave like linear perturbations of the metric in the late-time limit well within the cosmological horizon. Therefore, the initial gauge dependence of the result is lost \cite{DeLuca:2019ufz,Inomata:2019yww,Yuan:2019fwv} and does not affect the spectrum shown in Fig.~\ref{fig:inducedSGWBspectrum}.

\subsection{Impact of Primordial non-Gaussianity on Scalar Induced SGWB}
\label{Impact of non-gaussianity}

As discussed above,  the existence of amplified primordial fluctuations at small scales can lead to PBHs and also accompanying scalar induced (secondary) SGWB. Depending on the amplification/enhancement scale, this background can be detected at different frequencies with various GW experiments including LISA, PTA-SKA, DECIGO and BBO. Since the physical scales for PBHs of mass less than million solar mass correspond to very small sub-galactic distances (${\rm perturbation \, wavelength} < {\rm kpc}$), the nonlinearity of scalar fluctuations in the current Universe prevents us from extracting primordial information about such small scales. However, the freely propagating SGWB sourced by primordial curvature fluctuations could be an excellent probe for these scales. The spectral shape of the scalar induced (or secondary) SGWB can be influenced by the statistical properties of the scalar density perturbations. Therefore, the detection and spectral analysis of the induced/secondary GW background can also shed light on the properties of the inflationary era.

\begin{figure}[t!]
	\centering
	\includegraphics[width=0.3\columnwidth]{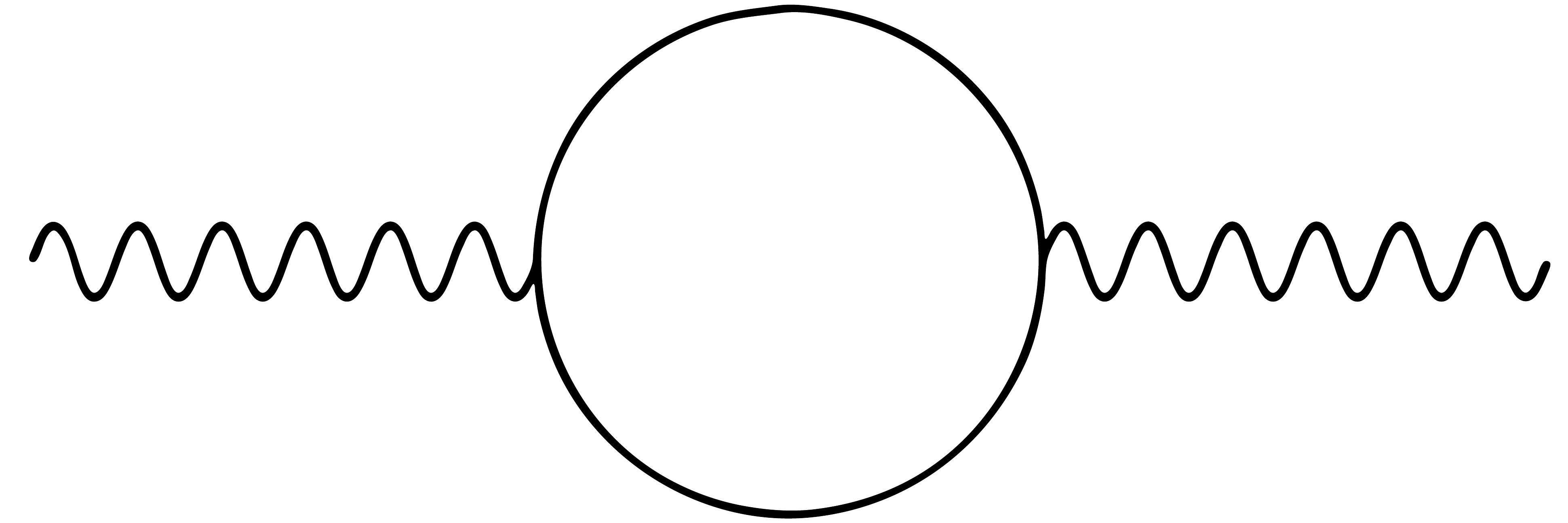}
	\includegraphics[width=0.3\columnwidth]{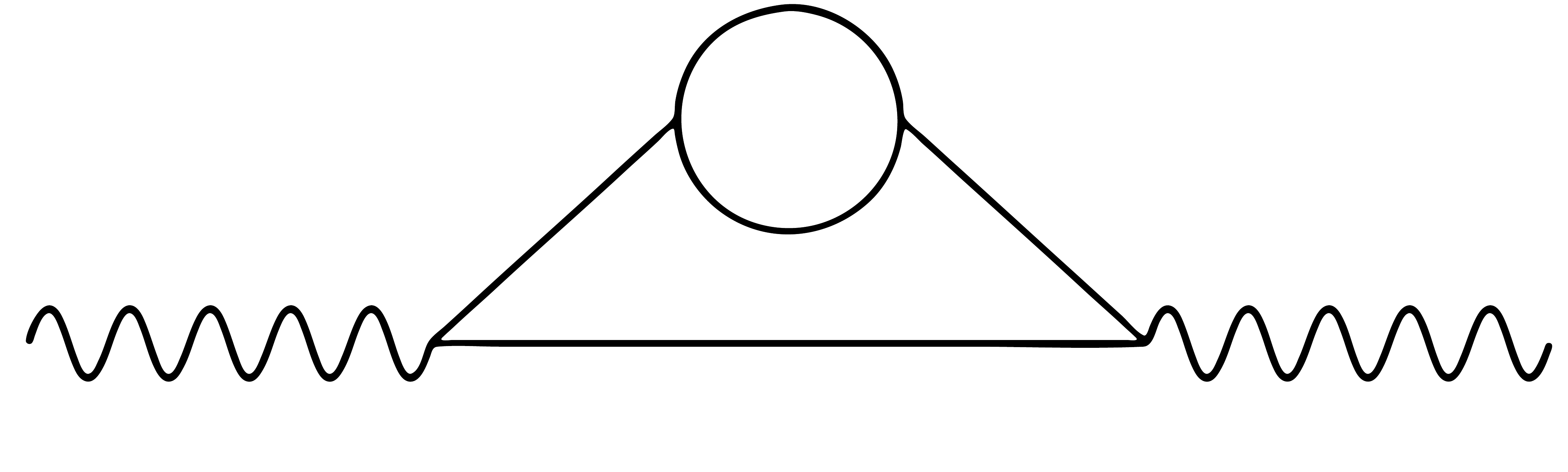}
	\includegraphics[width=0.3\columnwidth]{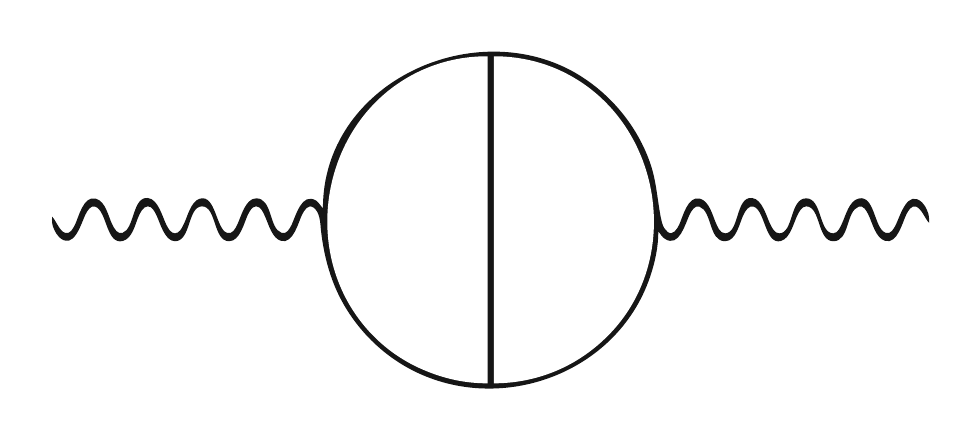}
	\caption{Nonvanishing diagrams for $ {\cal O} (f_{\rm NL}^{0})$ (left) and $ {\cal O} (f_{\rm NL}^{2})$ : center and right, called Hybrid and Walnut diagrams due to their topology and shape \cite{Unal:2018yaa}. 
	}
	\label{fig:SGWBtopologiesfnl2}
\end{figure}

\begin{figure}[t!]
	\centering
	\includegraphics[width=0.297\columnwidth]{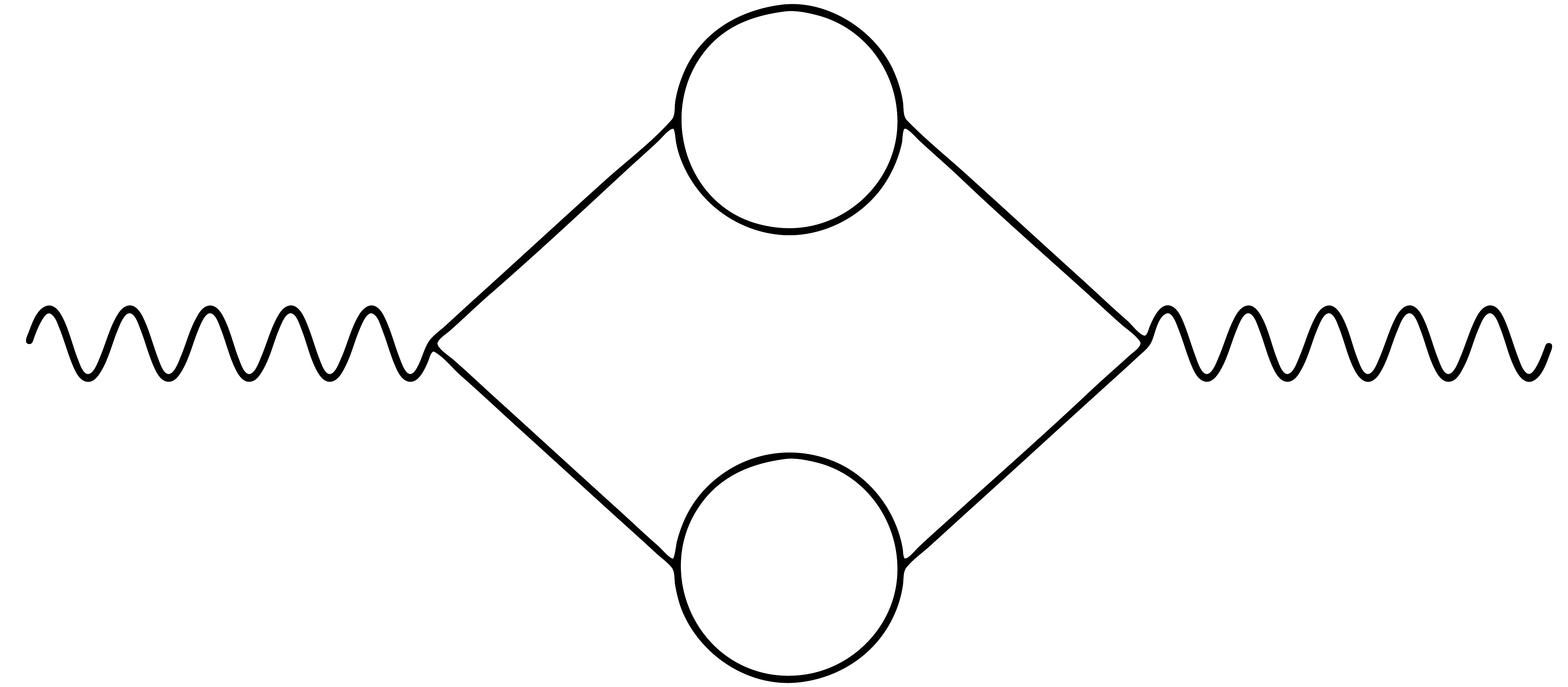}
	\includegraphics[width=0.297\columnwidth]{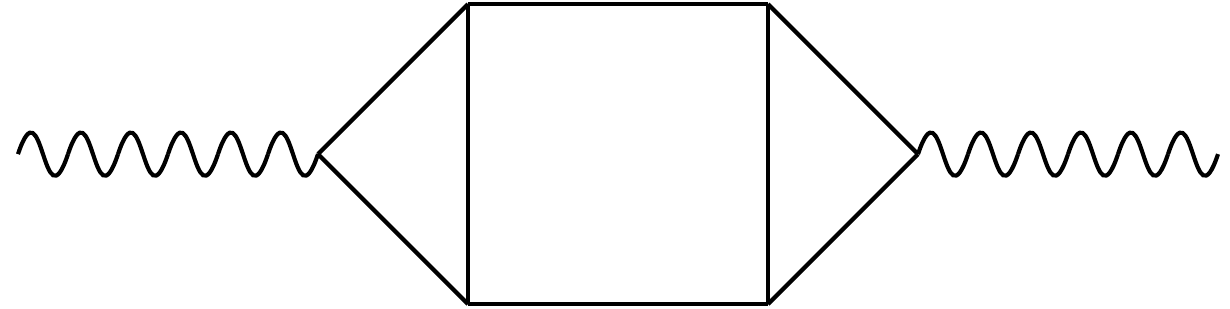}
	\includegraphics[width=0.297\columnwidth]{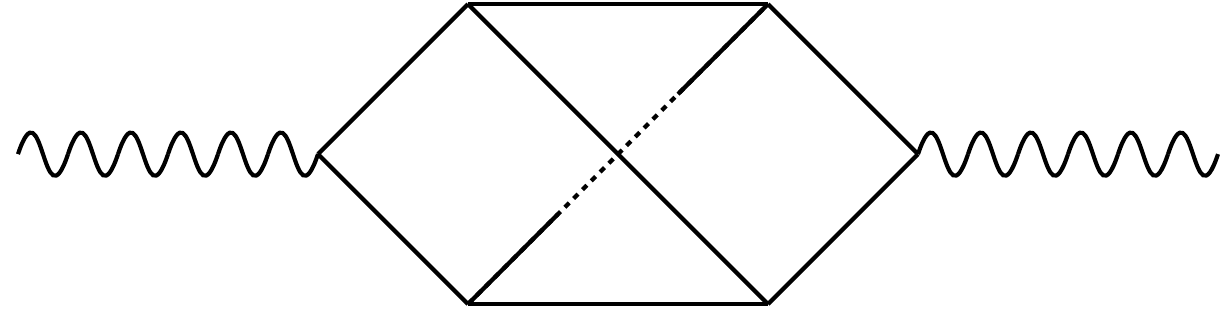}
	\caption{Diagrams of ${\cal O} (f_{\rm NL}^{4})$, called (from left to right) Reducible, Planar and non-Planar due to their topological properties  \cite{Garcia-Bellido:2017aan,Unal:2018yaa}.
	}
	\label{fig:SGWBtopologiesfnl4}
\end{figure}

Since the SGWB is sourced directly by the primordial curvature fluctuations in the horizon re-entry, it is natural to expect to observe some statistical properties of these scalar fluctuations in the SGWB spectrum. The SGWB includes a 4-pt function of the primordial curvature perturbations ($P_h \propto \langle h^2 \rangle \propto \langle \zeta^4 \rangle$ see also Eqs. \eqref{h-sourced} and \eqref{rho1-par}), hence for a Gaussian case, it can be reduced to only powers of a 2-pt function (see left panel of Fig. \ref{fig:SGWBtopologiesfnl2}). In the presence of non-Gaussianity, a 4-pt correlator can not be written in terms of power spectra exactly. 

Curvature perturbations with non-Gaussianity that can be expressed in terms of quadratic order Gaussian curvature perturbations contribute to GWs with 5 more diagrams given in Figs. \ref{fig:SGWBtopologiesfnl2}  and \ref{fig:SGWBtopologiesfnl4}. Note that the interaction could be of local form, or might include time and position derivatives.  Hence, the topologies of these five diagrams are independent of the interaction type and universal for curvature perturbations with the form 
\begin{equation}
    %\zeta = \zeta_{\rm G} + {\pazocal O} \cdot \zeta^2_{\rm G} .
    \zeta = \zeta_{\rm G} +   O \cdot \zeta^2_{\rm G} .
\label{eq:generalized-local-ansatz}    
\end{equation}
Here $O$ is an operator acting on a quadratic order Gaussian field.  It can be constant or depend on space and time. For two powers of interaction, we have 2 diagrams in Fig. \ref{fig:SGWBtopologiesfnl2}
\begin{itemize}
    \item Hybrid, since the power spectrum of one leg is tree level and the other one is the interaction;
    \item Walnut, due to the walnut shape. Note that the walnut topology diagram can be obtained by two different contractions.
\end{itemize}

\noindent For 4 powers of interaction, there are 3 diagrams given in Fig. \ref{fig:SGWBtopologiesfnl4} 
\begin{itemize}
    \item Reducible, since this diagram can be reduced to a 1 loop diagram;
    \item Planar and  Non-planar, respectively, as their topology can or cannot be drawn on a plane. 
\end{itemize}

The properties of the spectrum have been studied in a number of references. For instance, Ref.~\cite{Nakama:2016gzw} estimated the order of magnitude of the SGWB in the presence of non-Gaussianity. In Ref.~\cite{Garcia-Bellido:2017aan}, the SGWB spectrum was computed in the presence of large non-Gaussianities, and in such a case the main contribution to the SGWB comes from the 3 diagrams given in Fig.~\ref{fig:SGWBtopologiesfnl4}. The SGWB spectrum in the presence of mild and small non-Gaussianity has been computed in~\cite{Unal:2018yaa} with the diagrams in Figs.~\ref{fig:SGWBtopologiesfnl2} and~\ref{fig:SGWBtopologiesfnl4}. Mild and small non-Gaussianities have also been studied in Ref.~\cite{Atal:2021jyo,Cai:2018dig}. 

It is important to note that Walnut diagram topology has two distinct contributions from two distinct contractions. This missing point in previous studies has been identified  and computed carefully in ~\cite{Adshead:2021hnm}. With these results, the whole diagrams have been completed up to quartic order in interaction ($f_{\rm NL}^4$) and quadratic order in scalar density perturbations. The scalar induced SGWB with scalar non-Gaussianities has been worked out in Ref.~\cite{Yuan:2020iwf} up to third order in density perturbations, results are unchanged around the peak but due to convolution of 3 modes, decay has not been sharp after the wavenumber 2$k_{peak}$, where $k_{peak}$ is the peak scale of the scalar perturbations.

\begin{figure}[t!]
	\centering
	\includegraphics[width=0.49\columnwidth]{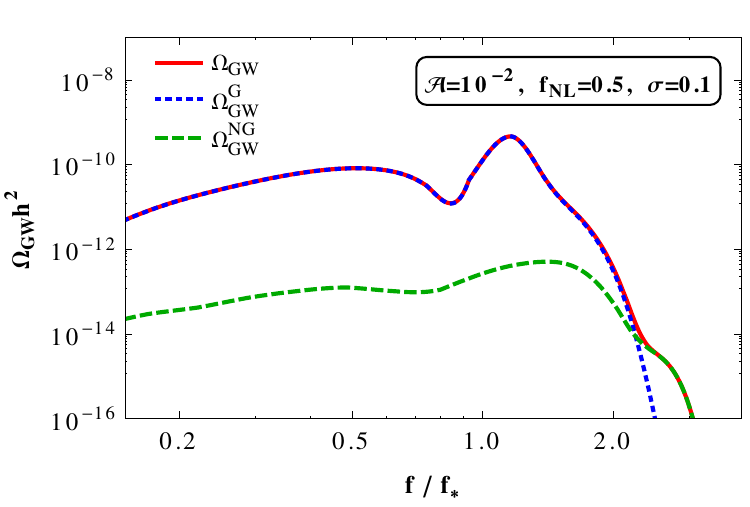}
	\includegraphics[width=0.49\columnwidth]{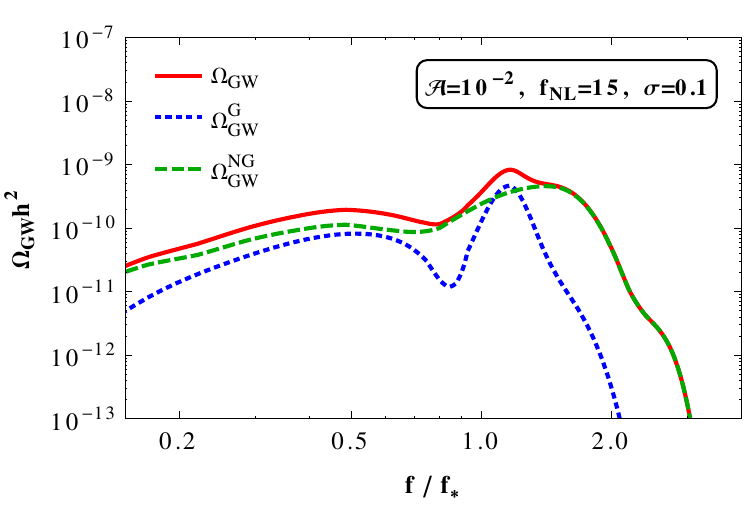}
	\caption{Second peak produced by primordial non-Gaussian component of curvature perturbations. Figure taken from \cite{Unal:2018yaa}.
	}
	\label{fig:SGWBasNGdetector} 
\end{figure}

The following results about the imprints of non-Gaussianity on the secondary/induced GW spectrum have been found: 
\begin{itemize}
    \item Non-Gaussianity can enhance the production efficiency of the SGWB.  Namely, for the same amplitude of Gaussian or non-Gaussian scalar fluctuations, the resulting SGWB has a larger amplitude for non-Gaussian than for Gaussian fluctuations.
    \item Because non-Gaussianity is a result of interactions, this leads to a convolution in the momentum space and to a broader distribution of scalar fluctuations, with a peak at frequencies about ${\mathcal O}(1)$ higher, especially if the curvature fluctuations have a narrow spectrum.
    \item In the limit of large non-Gaussianities, the diagrams in Fig.~\ref{fig:SGWBtopologiesfnl4} are important and dominate the signal.  For small and mild non-Gaussianities, the signal is dominated by the diagrams in Fig.~\ref{fig:SGWBtopologiesfnl2}.
\end{itemize}

Some of the results above can be visualized in Fig.~\ref{fig:SGWBasNGdetector}, where $\Omega_{\rm GW}$ indicates the total GW spectrum, $\Omega^{\rm G}_{\rm GW}$ indicates GWs from Gaussian scalar perturbations, and $\Omega^{\rm NG}_{\rm GW}$ GWs from non-Gaussian scalar perturbations \footnote{Recall that induced GWs are sourced by scalar sources at quadratic order hence they are inherently non-Gaussian. By G and NG symbols we indicate the statistical properties of scalar perturbations sourcing GWs.}. Due to the interactions of the scalar perturbations, the high frequency regime of the SGWB is dominated by the NG scalar perturbations. This fact can lead to a second peak whose amplitude is set by the strength of the interactions, ie. the non-Gaussianity parameter $f_{\rm NL}$. If the NG contribution is considerable, the SGWB leads to two peaks, as shown in the right panel of Fig.~\ref{fig:SGWBasNGdetector}. The second peak, which results from interactions/non-Gaussian perturbations, is at a frequency about $\mathcal O$(1) times higher with respect to the first peak from Gaussian perturbations. It can be seen on the left panel of the same Figure that even if the NG contribution is tiny, it still dominates the large frequency regime of the GW spectrum and produces an abrupt second bump at smaller amplitudes which can be detected with sensitive next generation GW experiments, such as LISA, Cosmic Explorer, DECIGO and PTA-SKA, and allow probing non-Gaussianity as sensitive as $f_{\rm NL} \sim  0.5$~\cite{Unal:2018yaa}. This level of non-Gaussianity could be probed even better than using next generation CMB and LSS experiments.\\

In \cite{Garcia-Saenz:2022tzu}, the impact of primordial non-Gaussianity on the scalar induced SGWB has been studied beyond the generalized local ansatz \eqref{eq:generalized-local-ansatz}, paying attention to whether it can 
self-consistently be important in concrete early-universe scenarios. As pointed out above, the gravitational waves spectrum can be split into two contributions: 
the one set by the power spectrum, and the one determined by the connected four point function of the curvature perturbation, namely the primordial trispectrum.
First, it is shown that this trispectrum-induced gravitational wave spectrum can always be written as a sum of three ``channels'' contributions, independently of the precise shape of the trispectrum. Second, the question addressed in \cite{Garcia-Saenz:2022tzu} is: can the trispectrum be observationally relevant for the scalar induced SGWB in conventional set-ups where non-Gaussianity provides subleading corrections to the Gaussian scalar signal?, i.e.~in inflationary scenarios where interactions generating non-Gaussianity maintain perturbative control. Formulating a precise quantitative criterion ensuring perturbative control in
strongly scale-dependent theories is difficult and model-dependent. Hence, the focus of \cite{Garcia-Saenz:2022tzu} is on scale-invariant theories. In this context, it is shown that neither regular trispectrum shapes peaking in the so-called equilateral configurations, nor local trispectrum shapes diverging in soft momentum limits, can contribute significantly. Indeed, those contributions are always bound to be smaller than an order-one
(or smaller) number multiplying the relative one-loop correction to the scalar power spectrum,
necessarily much smaller than unity in order for the theory to be under perturbative control. This result is shown to be also valid in a toy-model for the phenomenologically more relevant situation of a scale-dependent scalar spectrum, calling for more in-depth investigations of this question.

\subsection{Induced SGWB Anisotropies}
It is interesting to note that the induced SGWB associated with PBH production \footnote{The generation of a SGWB in the early Universe may also happen due to other mechanisms, possibly with characteristic anisotropic signatures, see for example Refs.~\cite{Garcia-Bellido:2016dkw,Cook:2011hg, Guzzetti:2016mkm,Bartolo:2016ami,Ricciardone:2017kre,Caprini:2018mtu,Geller:2018mwu,Dimastrogiovanni:2019bfl}.}  will present some angular anisotropies, which can be quantified by 
computing the two-point correlation function of the density field in different directions.
The LISA experiment will be able to detect anisotropies with an angular resolution of $\ell \lesssim 15$ \cite{Contaldi:2020rht} (see also \cite{Baker:2019ync}), which will require the spatial points to be separated by a non-negligible fraction of the present horizon.
Since the characteristic scales of the perturbations generating the SGWB are much smaller than those distances, 
$k_* \, \vert \vec{x} - \vec{y} \vert \gg 1$, and that the emission takes place near horizon crossing with the assumption of Gaussian scalar perturbations, the Equivalence Principle dictates that the  anisotropies decay as $(k_* \vert \vec{x} - \vec{y} \vert )^{-2}$. Therefore, the anisotropy coarse grained with the resolution of the experiment will be undetectable \cite{Bartolo:2019zvb}.
Another possibility, as far as anisotropies imprinted at formation are concerned, is given by the presence of non-Gaussianity which, by providing a correlation between small and large scales, may allow large scale modulation of power and lead to a large-scale anisotropy \cite{Bartolo:2019zvb}. Finally, the propagation of GWs across disconnected regions of the Universe leads to large-scale anisotropies at detection \cite{Alba:2015cms,Contaldi:2016koz,Cusin:2017fwz,Cusin:2018avf,Jenkins:2018nty,Bartolo:2019oiq, Bartolo:2019yeu,Renzini:2019vmt, Bertacca:2019fnt}.

Assuming a local, scale-invariant, shape of non-Gaussianity $\zeta = \zeta_{\rm g} + \frac{3}{5} f_{\rm NL} \, \zeta_{\rm g}^2$ and taking into account propagation effects, one can compute the two-point correlation function of the direction dependent GW energy density contrast { $\delta_{\rm GW} (\eta, \vec x, k) = \Omega_{\rm GW} (\eta, \vec x, k)/\Omega_{\rm GW} (\eta, k) - 1$, defined as the relative difference between the direction dependent GW energy density and its monopole contribution,} in a spherical harmonics decomposition as ({following the notation used in Ref.~\cite{Bartolo:2019zvb}})
 \begin{equation}
 \label{GWdensitycontrast}
\left\langle \delta_{{\rm GW},\ell m}  \delta_{{\rm GW},\ell' m'}^* \right\rangle = \delta_{\ell \ell'} \delta_{m m'} \, {\hat C}_\ell \left( k \right),
\end{equation} 
where~\cite{Bartolo:2019zvb}
\begin{equation}
\sqrt{\frac{\ell \left( \ell+1 \right)}{2 \pi} \, {\hat C}_{\ell} \left( k \right)}  \simeq \frac{3}{5} 
\left\vert 1 + {\tilde f}_{{\rm NL}} \left( k \right) \right\vert \, \left\vert  4-\frac{\partial \ln {\Omega}_{{\rm GW}} (\eta ,\, k ) }{\partial \ln k} \right\vert \,  {\cal P}_{\zeta_{\rm L}}^{1/2},
\end{equation}
{and $k$ denotes the momentum on small-scales.}
Here ${\cal P}_{\zeta_{\rm L}}$ denotes the power spectrum at large scales, while the momentum dependent non-Gaussian parameter is defined as
\begin{equation}
 {\tilde f}_{\rm NL} \left( k \right) \equiv  \frac{8 \, f_{\rm NL}}{4-\frac{\partial \ln {\Omega}_{{\rm GW}} (\eta ,\, k ) }{\partial \ln k}}.
\label{GammaI-time}
\end{equation}
The presence of  non-Gaussianity in the curvature perturbation, constrained by the Planck collaboration to be $- 11.1 \leq f_{\rm NL} \leq 9.3$ at $95\% \, {\rm C.L.}$ \cite{Akrami:2019izv}, would also induce a significant variation on large scales of the PBH abundance through the modulation of the power on small scales induced by the long modes. 
If PBHs compose a non-negligible fraction of the DM, the presence of isocurvature modes in the DM density fluid associated to the non-Gaussianity \cite{Young:2015kda} is strongly constrained by the CMB observations \cite{Akrami:2018odb}.

In Fig.~\ref{fig:anisotropiesPBH}, we show the expected GW anisotropy for the choice of a monochromatic and log-normal small-scale power spectrum of the curvature perturbation peaked at the LISA maximum sensitivity frequency. The dot-dashed lines identify the corresponding GW abundance at the present epoch evaluated at the peak frequency, while the colored region identifies the range of parameters allowed by the Planck constraints, and we choose the non-linear parameter $f_{\rm NL} \gtrsim -1/3$ to avoid the failure of the perturbative approach in the computation of the PBH abundance happening at larger negative values, see for example \cite{Young:2013oia,Yoo:2019pma}.
One finds that if a large fraction of the dark matter is composed of PBH, one expects a highly isotropic and Gaussian SGWB, up to propagation effects. On the contrary, the detection of a large amount of anisotropy in the signal associated to the PBH formation scenario would imply that only a small fraction of the DM can be  accounted by PBHs \cite{Bartolo:2019zvb}.
\begin{figure}[t!]
	\centering
	\includegraphics[width=0.45\columnwidth]{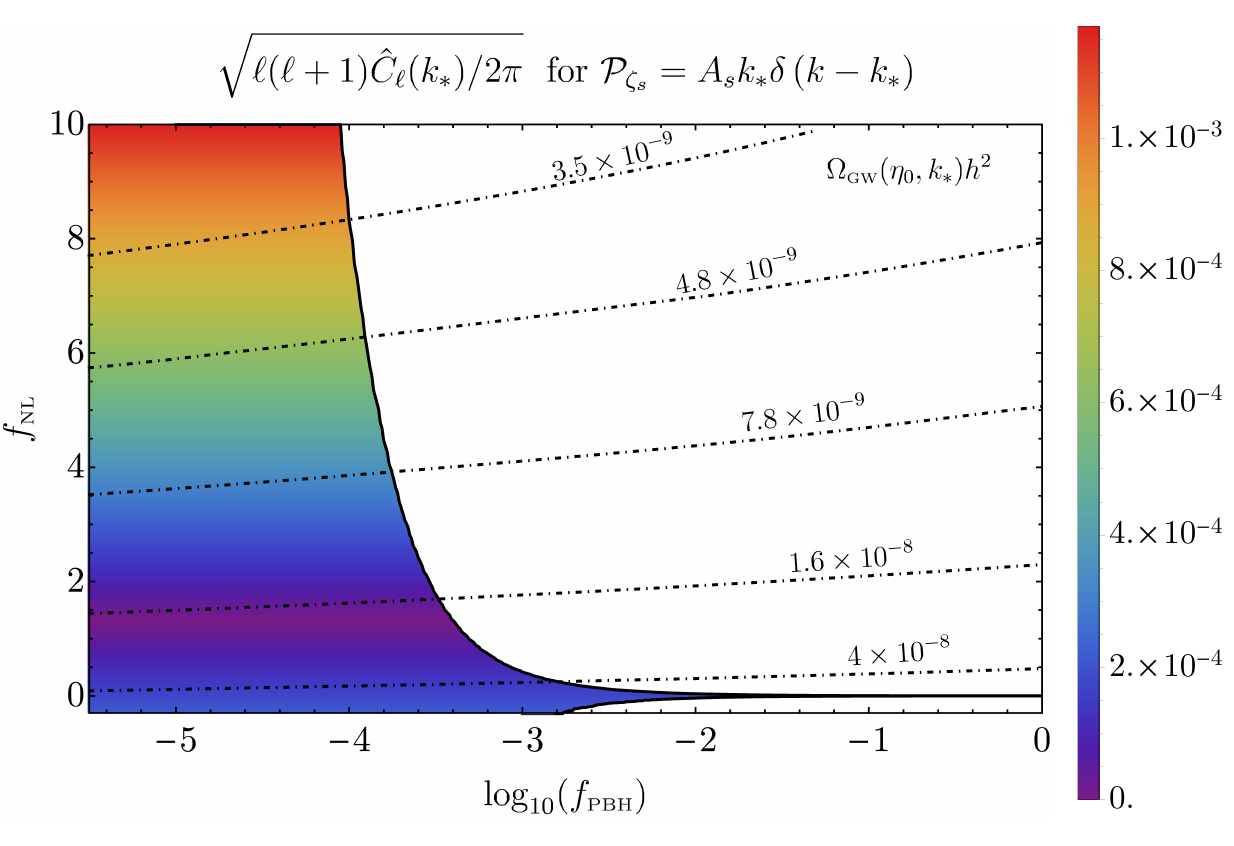}
	\hspace{.2 cm} 
	\includegraphics[width=0.45\columnwidth]{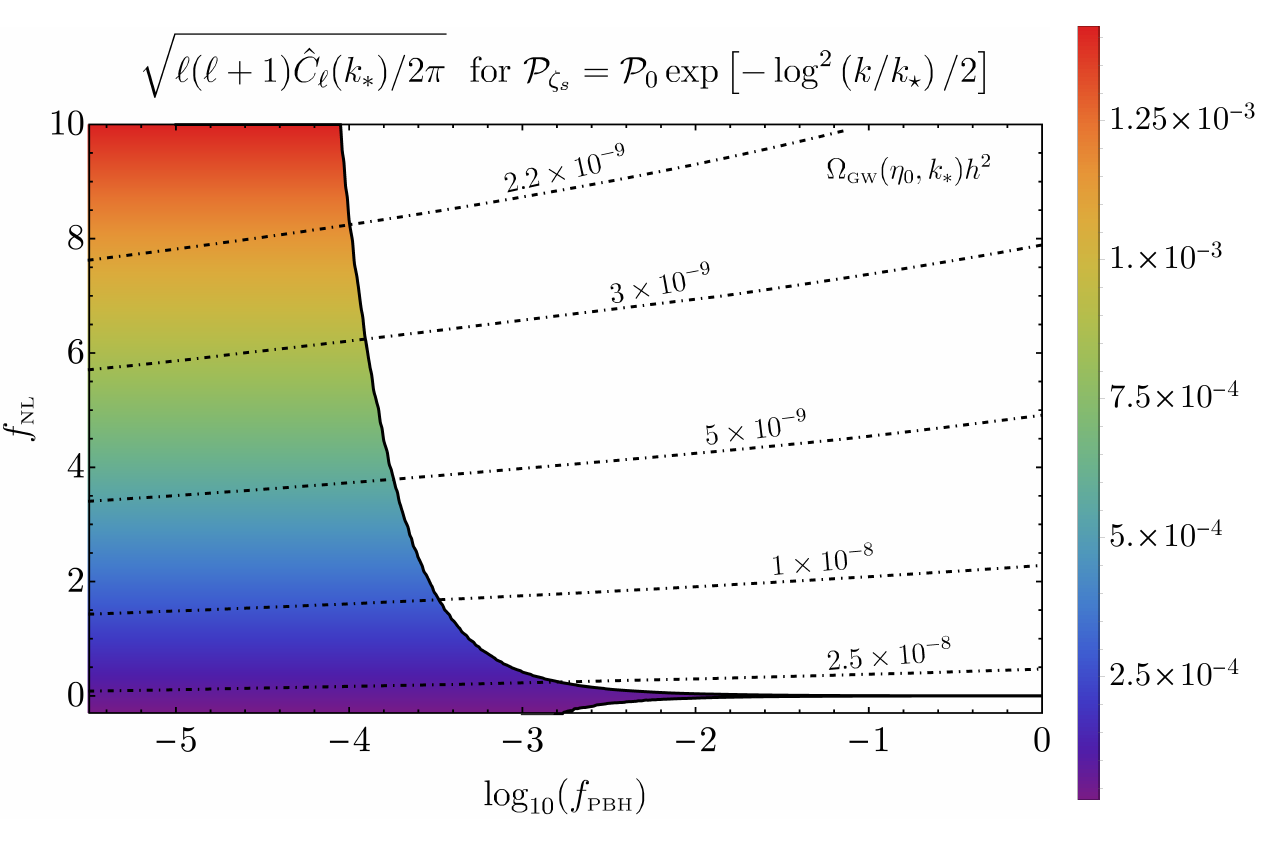}
	\caption{Contour plot showing the amount of GWs anisotropy in the parameter space of $f_{\rm PBH}$ and $f_{\rm NL}$ allowed by the Planck constraints for the choice of a monochromatic and log-normal small-scale power spectrum, respectively, with peak frequency around the maximum sensitivity of LISA. The dot-dashed lines identify the corresponding present GWs abundance. Figure taken from Ref.~\cite{Bartolo:2019zvb}.
	}
	\label{fig:anisotropiesPBH}
\end{figure}

\subsection{From PBH Poisson fluctuations}
\label{secPBHPoissonflctns}
At small scales, the gravitational interaction between individual PBHs (scattering, merging, etc.) results in the production of GWs. In this Section, we are interested in the large-scale counterpart of this signal, i.e. at distances much larger than the mean separation length between PBHs, for which PBHs can be described by a pressureless fluid. This fluid is endowed with density perturbations, which can be treated within the framework of cosmological perturbation theory. In particular, these PBH density perturbations induce the production of GWs through second-order effects studied both in the context of general relativity~\cite{Papanikolaou:2020qtd, Domenech:2020ssp} as well in the context of modified gravity~\cite{Papanikolaou:2021uhe,Papanikolaou:2022hkg}.

To be explicit, we assume that PBHs form in a radiation-dominated, homogeneous Universe (at large scales, primordial curvature perturbations provide a contribution to inhomogeneities but they are negligible compared to the ones generated by PBHs, at least in the range of scales we are interested in).  One can therefore view PBHs as a dust fluid, formed from the transition of a fraction of the radiation energy density into PBH dust matter.  Assuming that PBHs are randomly distributed at formation time (i.e. they have Poisson statistics),  the energy density associated to them is inhomogeneous while the total energy density of the background is homogeneous.  Consequently,  the energy density perturbation of the PBH matter field can be described by an isocurvature Poisson fluctuation.  If the initial abundance of PBHs is large enough,  PBHs can potentially dominate the Universe energy density content and in that case the isocurvature PBH energy density perturbation in the radiation era is converted into an adiabatic curvature perturbation deep in the PBH dominated era. These early PBH dominated eras can be naturally driven by ultralight PBHs, which evaporate before BBN~\cite{GarciaBellido:1996qt, Hidalgo:2011fj, Martin:2019nuw, Zagorac:2019ekv} and addressing a plethora of cosmological problems, most importantly that of Hubble tension~\cite{Hooper:2019gtx,Papanikolaou:2023oxq}.

At second-order in cosmological perturbation theory, the gravitational potential of these Poisson distributed PBHs induce a SGWB which may be detected by LISA~\cite{Papanikolaou:2020qtd, Domenech:2020ssp}.  More precisely, as found in~\cite{Papanikolaou:2020qtd}, this gravitational potential gives rise to the following power spectrum for $\Phi$:
\begin{equation}
\label{eq:PowerSpectrum:Phi:PBHdom}
\mathcal{P}_\Phi(k) = \frac{2}{3\pi} \left( \frac{k}{k_{\rm{UV}}} \right)^3 \left(5+\frac{4}{9}\frac{k^2}{k_{\rm{d}}^2} \right)^{-2}\, ,
\end{equation}
where $\Phi$ is the Bardeen gravitational potential,  $k_{\rm{d}}$ is the comoving scale exiting the Hubble radius at PBH domination time and $k_{\rm{UV}}\equiv a/\bar{r}$, where $\bar{r}$ corresponds to the mean PBH separation distance. Note that $k>k_{\mathrm{UV}}$ corresponds to distances within the mean separation distance, where the granularity of the PBH matter field and the associated non-linear effects become important, and where the gas of PBHs cannot be described by a fluid anymore. This is why the above expression should be restricted to $k<k_{\rm{UV}}$ with $k_{\rm{UV}}$ acting as a UV cutoff (hence the notation). Since $\bar{r}\propto a$, $k_{\rm{UV}}$ is a fixed comoving scale. From Eq. \eqref{eq:PowerSpectrum:Phi:PBHdom}, one can see that $\mathcal{P}_\Phi$ is made of two branches: when $k\ll k_{\rm{d}}$, $\mathcal{P}_\Phi \propto k^3$, while $\mathcal{P}_\Phi \propto 1/k$ when $k\gg k_\mathrm{d}$. It reaches a maximum when $k \sim k_\mathrm{d}$, where $\mathcal{P}_\Phi$ is of order $(k_{\rm{d}}/k_{\rm{UV}})^3$. 

The energy density of the associated induced GWs can be computed according to the prescription described in Sec.~\ref{sec:secondordergw}, where one should use the linear transfer function during a matter (PBH) dominated era. One can show that it peaks at the comoving scale that exits the Hubble radius at the onset of the PBH domination era~\cite{Papanikolaou:2020qtd}. This scale corresponds to the frequency displayed in Fig.~\ref{fig:Omega_GW_newtonian_gauge_RD_to_MD} as a function of the fraction of the Universe made of PBHs at the time they form, and their mass (assuming an initial monochromatic distribution of PBHs). One can see that a substantial fraction of parameter space lies within the LISA detection band. This would provide a unique access to ultralight PBHs, which otherwise cannot be detected through the PBH merging channel or the second-order induced GW channel presented in Sec.~\ref{sec:secondordergw}.

\begin{figure}[t!]
\begin{center}
  \includegraphics[width=0.696\textwidth, clip=true] {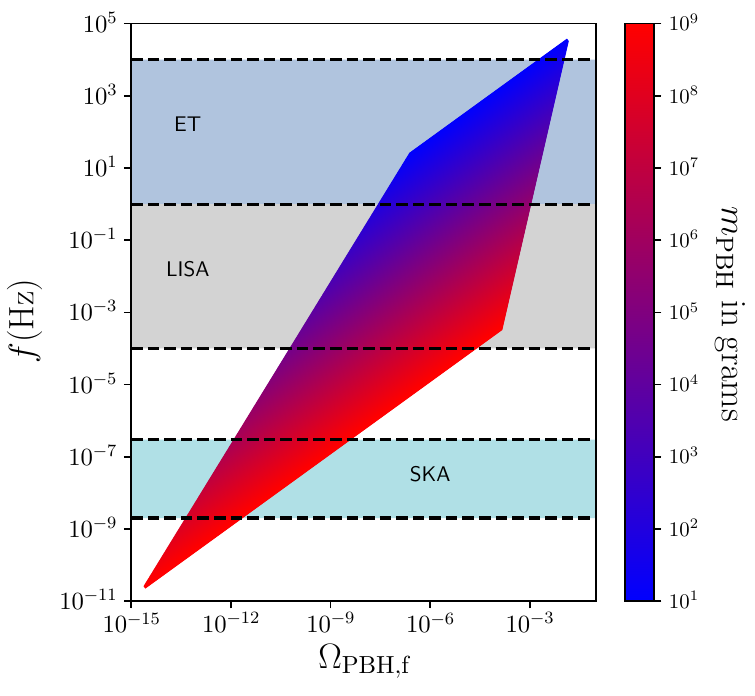}
\caption{Frequency at which the GWs induced by a dominating gas of PBHs peak, as a function of  
their energy density fraction at the time they form, $\Omega_{\mathrm{PBH,f}}$ (horizontal axis), and their mass $m_{\mathrm{PBH}}$ (colour coding).
The region of parameter space that is displayed corresponds to values of $m_{\mathrm{PBH}}$ and $\Omega_{\mathrm{PBH,f}}$, such that black holes form after inflation, dominate the Universe content for a transient period and Hawking evaporate before BBN. We also impose that the induced GWs do not lead to a backreaction problem before they evaporate, see~\cite{Papanikolaou:2020qtd} for more details.  For comparison, the frequency detection bands of ET, LISA and SKA are shown. Figure credited to~\cite{Papanikolaou:2020qtd}.
} 
\end{center}
\label{fig:Omega_GW_newtonian_gauge_RD_to_MD}
\end{figure}

Let us point out that the calculation of the precise amplitude of the signal is more complex as it implies to follow its transition from the PBH-dominated era to the subsequent radiation-dominated (RD) era. This is why~\cite{Papanikolaou:2020qtd} was followed up by~\cite{Inomata:2020lmk, Domenech:2020ssp} where, by considering a monochromatic PBH mass function and a sudden transition from the PBH-dominated era to the radiation era after evaporation,  it was found that the GW signal discussed here is not suppressed due to the $\mathrm{PBH}\rightarrow\mathrm{RD}$  transition and it can be detectable from LISA and other GW experiments.  In fact, the fate of the signal through the transition depends on whether the transition is sharp or slow.  In particular, on the one hand, when the transition is gradual,  the gravitational potential oscillates with a decreasing amplitude~\cite{Inomata:2019zqy}, suppressing in this way the signal.  On the other hand,  if the transition is sudden, the gravitational potential oscillates very fast with large amplitudes up to the end of the transition and therefore the GW signal is expected to be enhanced in this case~\cite{Inomata:2019ivs}. 

It is also important to mention that another calculation was performed in~\cite{Kozaczuk:2021wcl}, where the contribution from scales for which the density contrast becomes non linear during the PBH-dominated era was removed (note that the gravitational potential always remains linear).  Assuming again a monochromatic PBH mass function, a reduction was found in the GW signal which removed it from the reach of LISA, while remaining accessible by other GW experiments like BBO.  However, to fully access the signal, one should take into account more realistic extended mass functions which will make PBHs evaporate at different times, leading in this way to a longer period of PBH-domination, a fact which can potentially enhance the detectability of the signal discussed here.  However, if the transition is gradual due to different PBH evaporation times, this will give an extra suppression as found in~\cite{Inomata:2019zqy}.  One therefore is met with two competitive effects which should be considered together in the case of a calculation with realistic extended mass distributions. Interestingly, as it was shown recently in~\cite{Papanikolaou:2022chm} if one accounts for a cosmologically motivated power-law primordial curvature power spectrum for the generation of an extended PBH mass function, they can find that despite the gradualness of the transition the GW signal can be detectable by LISA and can serve as a novel probe to constrain cosmological parameters of the early Universe.

\subsection{From PBH mergers}
\label{sec:SBWBmergers}

\subsubsection{Formulation}
Overlapped GWs from PBH mergers form a SGWB. The spectral shape of the SGWB, its constraints by GW observations, and its implication to PBH physics have been discussed in the literature \cite{Mandic:2016lcn,Clesse:2016ajp,Wang:2016ana,Raidal:2017mfl,Chen:2018rzo,Wang:2019kaf,DeLuca:2020qqa,Bavera:2021wmw,Bagui:2021dqi}. For a binary system with a circular orbit, the single source energy spectrum $\frac{{\rm d} E_{\rm GW}}{{\rm d} \ln f_r}$ for the inspiral phase is given by
\begin{equation}
  \frac{{\rm d} E_{\rm GW}}{{\rm d} \ln f_{\rm r}}=\frac{\pi^{2/3}}{3}{\cal M}_{\rm c}^{5/3}(Gf_{\rm r})^{2/3},
\end{equation}
where 
%${\cal M}_{\rm c}=\frac{(m_1 m_2)^{3/5}}{(m_1+m_2)^{1/5}}$ is the chirp mass 
%and 
$f_{\rm r}=(1+z)f$ is the GW frequency in the source frame with $f$ being the frequency at the observer. One may use a more accurate fitting formula, which includes the contributions of the inspiral, merger, and ringdown parts of the BBH waveform  (see \cite{Ajith:2009bn} for the details).
The amplitude of the SGWB is given by summing up the energy spectrum of each binary system and by taking into account the merger rate distribution. At the end, the GW spectral abundance is written as 
\begin{equation}
  \Omega_{\rm GW}(f)=\frac{1}{\rho_{\rm c}}\int_0^{z_{\rm max}}{\rm d} z ~ \frac{N(z)}{(1+z)} \frac{{\rm d} E_{\rm GW}}{{\rm d} \ln f_{\rm r}},
  \label{eq:OGW}
\end{equation}
where $\rho_{\rm c}$ is the critical density of the Universe, $\rho_{\rm c}=\frac{3H_0^2}{8\pi G}$.
Here $N(z)$ is  the number density of GW events at redshift $z$ and can be related to the merger rate in a comoving volume from the previous section $R_{\rm EB/LB}$ 
%$\tau_{\rm merg}$ 
\begin{equation}
  N(z) = \frac{ R_{\rm EB/LB}  }{(1+z)H(z)}.
\end{equation}
In the case of a broad mass distribution, we have to integrate Eq.~\eqref{eq:OGW} over the BH masses, $m_1$ and $m_2$, and we get \cite{Clesse:2016ajp}
\begin{eqnarray}\label{omegaGW}
  \Omega_{\rm GW}(f)
  &=& \frac{8\pi^{5/3}G^{5/3}}{9H_0^2}f^{2/3}
  \int_0^{z_{\rm max}}{\rm d} z ~ 
  \frac{1}{H(z)(1+z)^{4/3}} \nonumber\\
  &\times&
  \int {\rm d} \ln m_1
  \int {\rm d} \ln m_2 ~
  %\frac{{\rm d} \tau_{\rm merg}}{{\rm d} \ln m_1 {\rm d} \ln m_2} 
   R_{\rm EB/LB}
  {\cal M}_{\rm c}^{5/3}(m_1,m_2).
\end{eqnarray} 
The maximum frequency at the observer is determined by the innermost stable circular orbit which is given by 
\be
f_{\rm ISCO}\approx 4.4 {\rm kHz} ~ \frac{M_\odot}{m_1+m_2} \frac{1}{1+z}.
\ee 
This can be translated to the maximum redshift as
\begin{equation}
  z_{\rm max}=\frac{f}{f_{\rm ISCO}}-1.
\end{equation}
As described in Sec.~\ref{Sec:merging_rate}, there are two binary formation channels, dubbed early and late binaries. Below, we discuss the SGWB for each case.

\subsubsection{Early binaries}

The SGWB from early binaries is obtained by inserting the corresponding merging rate distribution of Eq.~\ref{eq:cosmomerg} into the above expression, Eq.~\ref{omegaGW}. Since the merging rate depends on the redshift $z$, $R_{\rm{EB}}(z,m_1, m_2) = R_{\rm{EB}}(m_1, m_2) \times (t(z)/t_{0})^{-34/37-0.29}$ for $f_{\rm PBH} = 1$, with $t(z)$ being the elapsed time between the Big-Bang and a certain redshift $z$, this redshift-dependent factor can be included in the integral over $z$~\cite{Bagui:2021dqi}, 
\begin{equation} \label{eq:redshiftEB}
\int^{z_{\rm{max}}}_{0} \frac{(t(z)/t_{0})^{-34/37 - 0.29}}{(1+z)^{4/3 }H(z)} \thinspace {\rm{d}}z = 5.92 \thinspace H^{-1}_{0},
\end{equation}
where $z_{\rm max} \simeq 100$, as discussed in Sec.~\ref{sec:poisson_clust_supp}.
The integral on the masses can be performed by direct numerical integration or by using a Monte-Carlo method as in \cite{Clesse:2016ajp}, e.g. for a synthetic population of $10^{6}$ BHs. Expressing the merging rate in $\rm{yr^{-1}Gpc^{-3}}$, the chirp mass in solar masses $M_\odot$ and the frequency in Hz, one obtains
\begin{equation} \label{eq:finalomegaEB}
\Omega_{\rm GW}(f)h^2 \simeq \thinspace  6.46 \times 10^{-14} \int {\rm d} \ln m_1  \int {\rm d} \ln m_2 \left(\frac{R_{\rm{EB}}(m_1, m_2)}{{\rm{yr}}^{-1} {\rm{Gpc}}^{-3}} \right) \left(\frac{f}{{\rm{Hz}}}\right)^{2/3}\left(\frac{\cal M_{\rm{c}}}{M_\odot}\right)^{5/3},
\end{equation}
where the constant $h$ takes the value $h = 0.67$. The resulting SGWB coming from early PBH binaries following a log-normal mass function (with mean mass $\mu = 2.5 M_\odot$ and width $\sigma = 1 $) as well as a mass distribution with thermal features are displayed in Fig.~\ref{fig:EB_ns}, as a function of the frequency. In the log-normal case, one notices that the spectrum reaches its highest point at 100 Hz and is suppressed at lower frequencies, whereas in the broad mass distribution case including QCD-induced features, it peaks at $10^{-6}, 10^{-3}$ and around 100 Hz. 
In the latter case, for ground-based detectors, one observes that the SGWB is above the projections for the LVK and ET detector designs, but just below the current limits imposed by the third observing run of LVK. The next observing runs of LVK may therefore be decisive for the detection of this SGWB. In addition, the detected signal by NANOGrav located within the interval $10^{-8} - 10^{-7}$ Hz with $\Omega_{\rm GW}h^{2} \sim 10^{-9}$ (gray window, {see sec.~\ref{sec:PTAs_NG} for more details}) could be interpreted as a background of GWs in the model for late binaries in clusters (see next Subsection). However, in the early binaries model, the amplitude of the SGWB is always lower than the signal detected by NANOGrav, and therefore this model can not explain the observed signal. The detectability with LISA will be discussed in Sec.~\ref{sec:LISA}.
The dependence of the SGWB on the PBH masses for the case of a broad mass function from the thermal history of the Universe is shown in Fig.~\ref{fig:Omegah2_m1m2}.

\begin{figure*}[t!]
\includegraphics[width=0.7\textwidth]{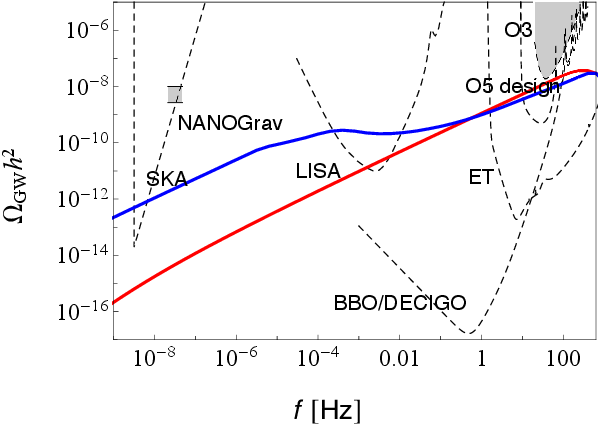}
\centering
\caption{The SGWB spectrum $\Omega_{\rm GW}h^{2}$ for early PBH binaries with a log-normal mass function (in red, with central mass $\mu = 2.5 M_\odot$ and width $\sigma = 1 $) and a broad mass distribution with scalar spectral index $n_{\rm s}$ = 0.970 (in blue), and $f_{\rm PBH}= 1$. The numerical spectrum also shows the sensitivities of the ground-based interferometers: {the LVK O3 Run, the final LVK} and the Einstein Telescope (ET). The sensitivity of future space-based interferometers is also shown (LISA, BBO/DECIGO). The Pulsar Timing Array (PTA) considered here is the Square Kilometer Array (SKA) \cite{SchmitzK}. 
The NANOGrav 12.5 signal is represented by the gray square {(see sec.~\ref{sec:PTAs_NG} for more details)}.}
\label{fig:EB_ns}
\end{figure*}

\begin{figure*}[t!]
\includegraphics[width=0.7
\textwidth]{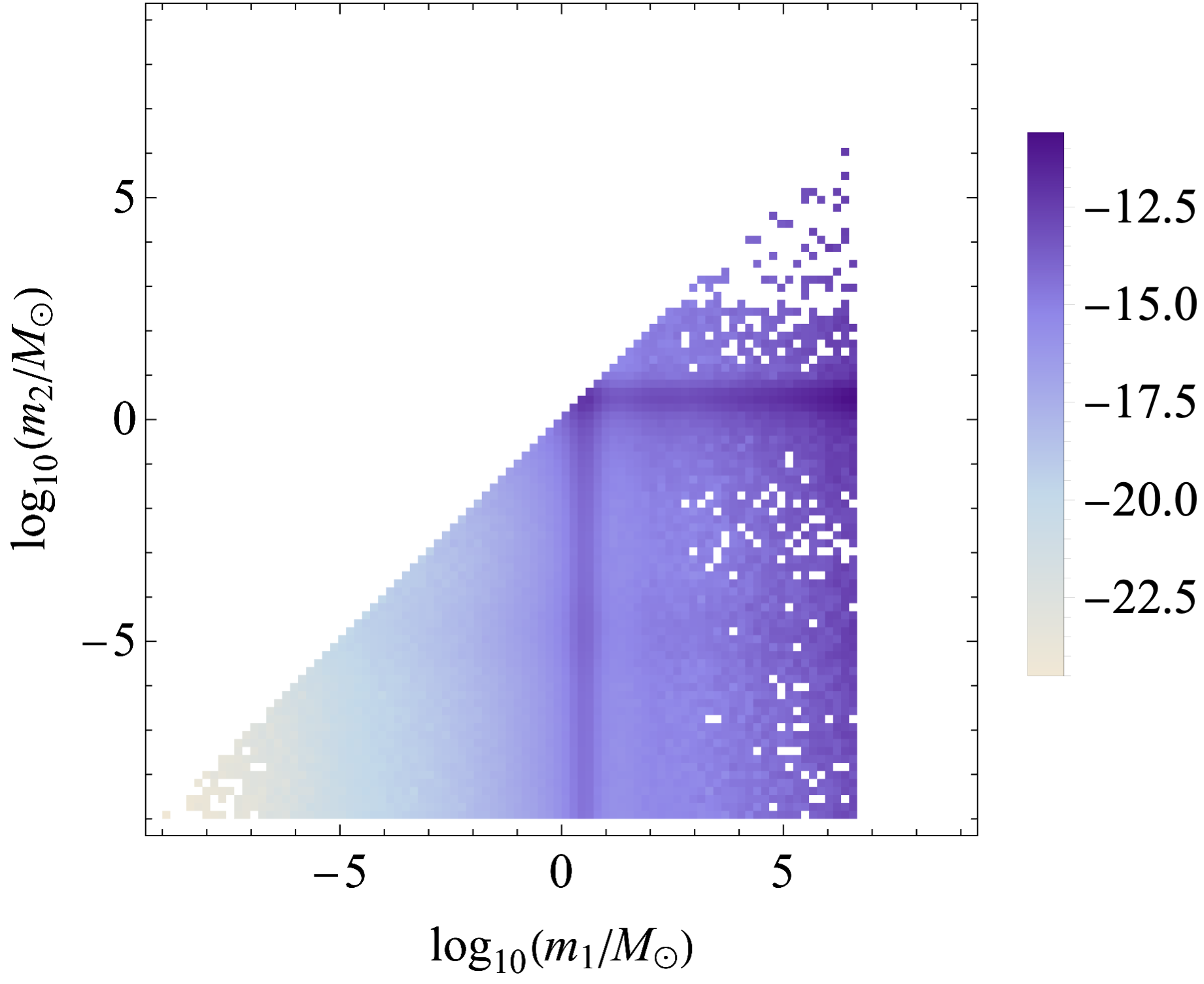}
\centering
\caption{Contribution to different PBH masses to the SGWB from early PBHs binaries (with $f_{\rm PBH} = 1$). The color bar indicates the values of the quantity $\log_{10}(\Omega_{\rm GW}h^{2})$, which is represented as a function of the logarithm of the masses $m_1$ and $m_2$ of the PBHs following a mass distribution from the thermal history of the Universe (with $n_{\rm s}$ = 0.97) for LISA frequencies ($10^{-3}$ Hz). Figure from~\cite{Bagui:2021dqi}.}
\label{fig:Omegah2_m1m2}
\end{figure*}

\subsubsection{Late binaries in clusters}
In the case of late binary formation channel discussed in Sec.~\ref{sec:latebinaries}, the merger rate is given by Eq.~\ref{eq:ratescatpure2}. Substituting it into Eq.~\eqref{omegaGW}, the energy spectrum for the low-frequency inspiral regime can be estimated as
\begin{equation}
\label{eq:OGWBBH}
\OGW(f) h^2
\approx
2.41\times 10^{-14}
R_{\rm clust}
\left(\frac{f}{{\rm Hz}}\right)^{2/3} 
\times\int {\rm d} m_1\,{\rm d} m_2 
\frac{f(m_1)\,f(m_2)\,
(m_1+m_2)^{23/21}}
{(m_1\,m_2)^{5/7}}
\,,
\end{equation}
with the masses in solar masses and we use $h=0.67$.
In Fig.~\ref{fig:Omegah2_late}, we show the SGWB spectrum generated by late PBH binaries in clusters (we have chosen the virial velocity in those clusters to be $v_{\rm{vir}} = 5$ km/s, compatible with velocity dispersions in faint dwarf galaxies) for the log-normal mass function (with the central mass $\mu=2.5\Msun$ and the variance $\sigma=1$) and the thermal history mass distribution (with $n_{\rm s}=0.97$ and no running). Here, the normalization is taken by requiring that the integral of the merger rate at $z=0$ over the mass range [5, 100]$\Msun$ gives a total rate of $38 \, {\rm yr}^{-1} {\rm Gpc}^{-3}$. Note that the SGWB spectrum at lower frequency is highly uncertain, as it can be suppressed by many factors such as the tilt and running of the primordial curvature perturbation spectrum, and by the eventual presence of a cutoff in the merger rate due to the fact that massive PBHs become isolated \cite{Braglia:2021wwa}.  We can see that, in the case of the log-normal mass function, the spectrum peaks around the ISCO frequency of PBH binaries (for $m_{\rm PBH}=2.5\Msun$, we get $f_{\rm ISCO}\sim 900$ Hz), while in the case of the thermal history mass function, we see that the SGWB is produced over a wide range of frequencies because of the wide mass distribution. In Fig.~\ref{fig:Omegah2_m1m2_clust}, we show which PBH masses contribute to the GW amplitude $\Omega_{\rm GW}$ at the LISA frequency $10^{-3}$Hz. We can see that $2 \Msun$ mass PBHs paring with more massive ones are giving large contribution to the SGWB.

\begin{figure*}[t!]
\includegraphics[width=0.7\textwidth]{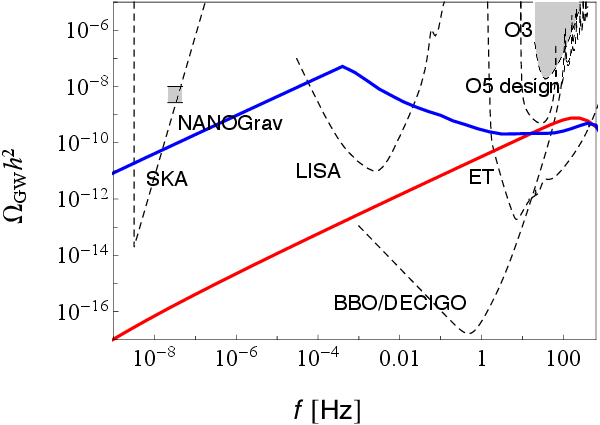}
\centering
\caption{The SGWB spectrum $\Omega_{\rm GW}h^{2}$ for late PBH binaries in clusters with $f_{\rm PBH} = 1$ for a log-normal mass function (in red, with the central mass {$\mu=2.5\Msun$} and the width $\sigma=1$) and a broad mass distribution (in blue, with $n_{\rm s}=0.97$ and no running). Note that here we use \cite{Ajith:2009bn} for the single source GW energy spectrum instead of Eq.~\eqref{omegaGW}.
}
\label{fig:Omegah2_late}
\end{figure*}

\begin{figure*}[t!]
\includegraphics[width=0.7\textwidth]{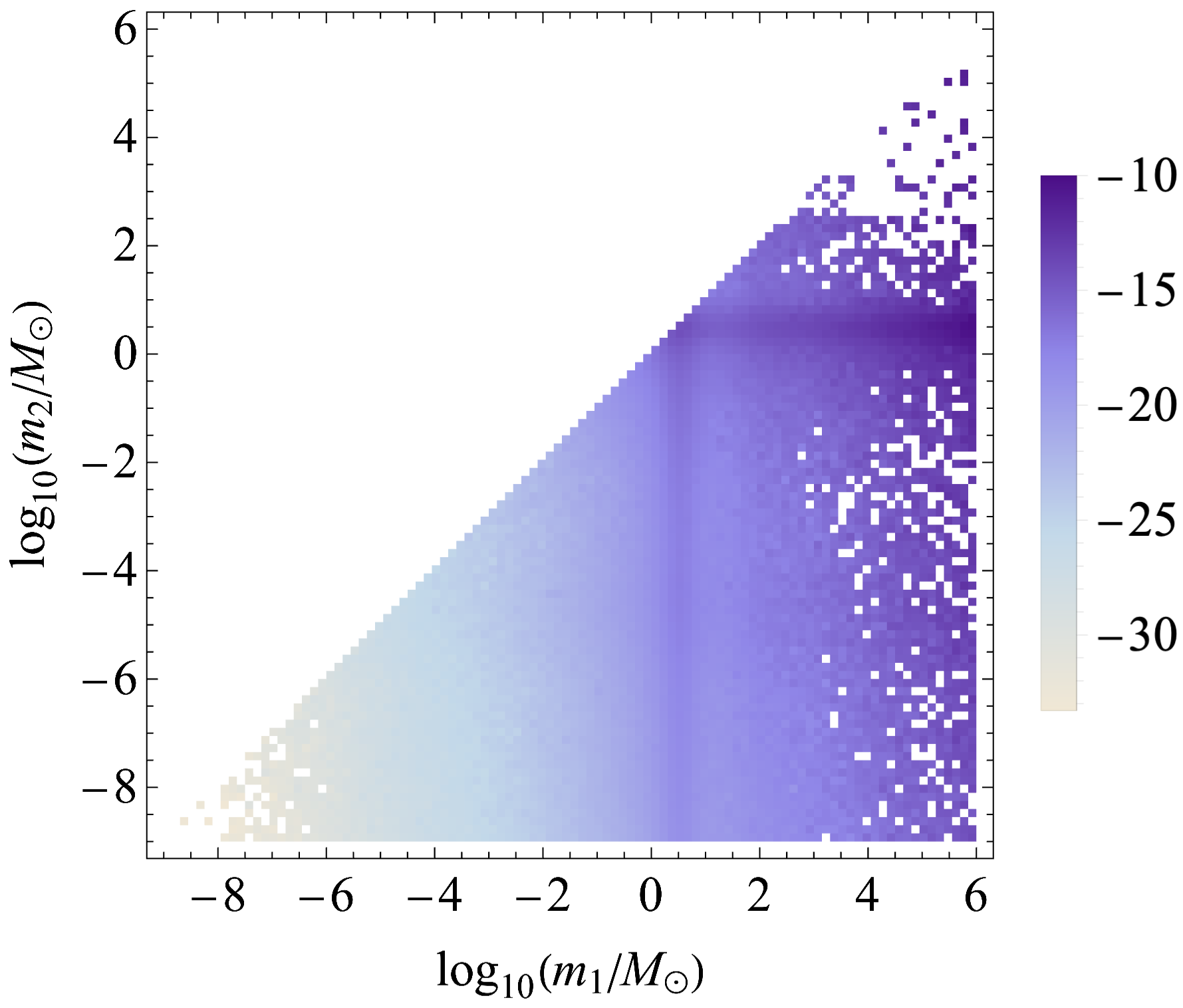}
\centering
\caption{The mass contribution to the SGWB from late PBHs binaries in clusters (with $f_{\rm PBH} = 1$). The color bar indicates the values of the quantity $\log_{10}(\Omega_{\rm GW}h^{2})$, which is represented as a function of the logarithm of the masses $m_1$ and $m_2$ of the PBHs following a mass distribution from the thermal history of the Universe (with $n_{\rm s}$ = 0.97 and $v_{\rm vir}$ = 5 km/s) for LISA frequencies ($10^{-3}$ Hz).}
\label{fig:Omegah2_m1m2_clust}
\end{figure*}

\subsection{SGWB duty cycle}
The SGWB of binary origin may have a characteristic property which could be used to infer the origin of BHs. If the interval between BBH events is larger than the typical duration of the signal, the waveforms do not overlap, and the SGWB becomes strongly non-Gaussian (sometimes referred to as intermittent or popcorn signal).  

One of the quantities to characterize such a non-Gaussian characteristics of the SGWB is the so-called astrophysical duty cycle. The duty cycle gives the average number of events present in the frequency bin and can be defined using the event rate ${\rm d} R/{\rm d} z$ and the duration of the signal staying in the frequency bin ${\rm d} \bar\tau/{\rm d} f$ as  
\begin{equation}
\label{eq:duty_cycle}
\frac{{\rm d} D}{{\rm d} f}=\int {\rm d} z \frac{{\rm d} R}{{\rm d} z} \frac{{\rm d} \bar\tau}{{\rm d} f}.
\end{equation}
Here, ${\rm d} \bar\tau/{\rm d} f$ is determined by the chirp mass ${\cal M}_c^z=(1+z){\cal M}_c$ as
\begin{equation}
 \frac{{\rm d} \bar\tau}{{\rm d} f}  = \frac{5}{96\pi^{8/3}}(G{\cal M}_c^z)^{-5/3}f^{-11/3},
\end{equation}
where ${\cal M}_c^z=(1+z){\cal M}_c$. When the duty cycle is larger than unity, the GW events overlap, and the background is in the so-called {\em continuous} regime. On the other hand, if the interval between events is comparable or larger than the typical duration of the signal, the duty cycle becomes smaller than unity, and the statistical properties are strongly non-Gaussian. The value of the duty cycle can differ depending on the mass function and the redshift distribution. Thus, the non-Gaussian property could be used to distinguish primordial and astrophysical scenarios~\cite{Mukherjee:2019oma,Braglia:2022icu} together with the spectral shape~\cite{Mukherjee:2021ags}.

\subsection{From close encounters}

Overlapped GWs from PBH close encounters can also form a SGWB~\cite{Garcia-Bellido:2021jlq}. The formulation is the same as the PBH mergers, Eq.~\eqref{omegaGW}. In the case of close hyperbolic encounters (CHE), the energy emitted per logarithmic frequency bin is given by~\cite{DeVittori:2012da,Garcia-Bellido:2017qal}
\begin{equation}
\frac{{\rm d}E^{\rm CHE}_{\rm GW}}{{\rm d}\ln f_{\rm r}} = \nu\frac{{\rm d}E_{\rm GW}}{{\rm d}\nu} =
\frac{4\pi}{45}\,\frac{G^3m_1^2m_2^2}{a^2c^5\nu_0}\,\nu^5F_{\rm e}(\nu)\,,\\[1mm]
\end{equation}
where we have defined $\nu \equiv 2\pi\nu_0\,f_{\rm r}$ and $\nu_0^2 \equiv a^3/GM$, and the semi-major axis $a$ is related to the initial velocity as $a=G M/v_0^2$. The function $F_{\rm e}(\nu)$ describes the dependence on eccentricity $e$ and it is given by~\cite{Garcia-Bellido:2017knh} 
\begin{align}
& \nu^5F_{\rm e}(\nu) \simeq \frac{12F(\nu)}{\pi\,y\,(1+y^2)^2}\,e^{-2\nu\,\xi(y)} \,,  \nonumber \\
& F(\nu) = \nu^2\left(1-y^2-3\,\nu\,y^3+4\,y^4
+ 9\,\nu\,y^5+6\,\nu^2 y^6\right) \,,  \nonumber \\
& \xi(y) = y - {\rm tan}^{-1}y \,,  \nonumber\\
& y = \sqrt{e^2-1} \,.
\label{Eq:CHEpower}
\end{align}
Substituting Eqs.~\eqref{Eq:CHErate} and \eqref{Eq:CHEpower} into Eq.~\eqref{omegaGW}, we can estimate the peak
frequency 
\begin{equation}
\label{eq:fmax}
f_{\rm peak} \simeq 4.3\,{\rm Hz}\,\left(\frac{y}{0.01}\right)^{-3}
\left(\frac{M}{2000\Msun}\right)^{1/2}
\left(\frac{a}{1~{\rm AU}}\right)^{-3/2}\,,
\end{equation}
and the peak amplitude
\begin{eqnarray}
\label{eq:OGW_CHE_max}
\OGW(f_{\rm peak}) &\approx&
0.9\times 10^{-13}\,
h 
\left(\frac{\Omega_{\rm M}}{0.3}\right)^{-1/2}
\left(\frac{\Omega_{\rm DM}}{0.25}\right)^2
\left(\frac{\delta_{\rm loc}}{10^8}\right)
\left(\frac{a}{0.1{\rm AU}}\right)^{-2} \nonumber \\
&&\times \left(\frac{y}{0.01}\right)^{-5}\, \frac{m_1}{100\Msun}\,\frac{m_2}{100\Msun}\,\frac{m_1+m_2}{200\Msun}\,.
\end{eqnarray}
In Fig.~\ref{fig:Omegah2_CHE}, we show the SGWB spectrum originating from close hyperbolic encounters, compared with the one from binary PBHs. We find that, for the modest choice of the parameter values, it is difficult to be reached by LISA. This difficulty can easily be understood if we rewrite Eq.~\eqref{eq:OGW_CHE_max} in terms of the peak frequency,
\begin{align}
\OGW^{\rm CHE}(f_{\rm peak}) & \approx 3.1\times 10^{-13}\,
h \nonumber \\
& \times\left(\frac{\Omega_{\rm M}}{0.3}\right)^{-1/2}
\left(\frac{\Omega_{\rm DM}}{0.25}\right)^2
\left(\frac{\delta_{\rm loc}}{10^8}\right)
\left(\frac{f_{\rm peak}}{50~{\rm Hz}}\right)^{4/3} \nonumber \\
& \times \left(\frac{y}{0.01}\right)^{-1}\, \frac{m_1}{100\Msun}\,\frac{m_2}{100\Msun}\,\left(\frac{m_1+m_2}{200\Msun}\right)^{1/3}.
\end{align}
We can see that the maximum amplitude $\OGW^{\rm CHE}(f_{\rm peak})$ grows with $f_{\rm peak}^{4/3}$. Thus, the peak amplitude of the background decays significantly at low frequencies. However, we have observed that the peak amplitude tends to get enhanced when we make the log-normal distribution of the mass function wider. Thus, by considering a different PBH mass function, we may be able to find a case where the SGWB from CHEs gives an interesting contribution at the LISA frequency.

It is worth mentioning that it could be tested by third-generation ground-based GW detectors such as Einstein Telescope and Cosmic Explorer. Although the CHE curves are below the BBH curves, loud BBH events can be detected individually and subtracted from the data, which may allow us to probe the other SGWB component below the green curve.

\begin{figure*}[t!]
\includegraphics[width=0.7\textwidth]{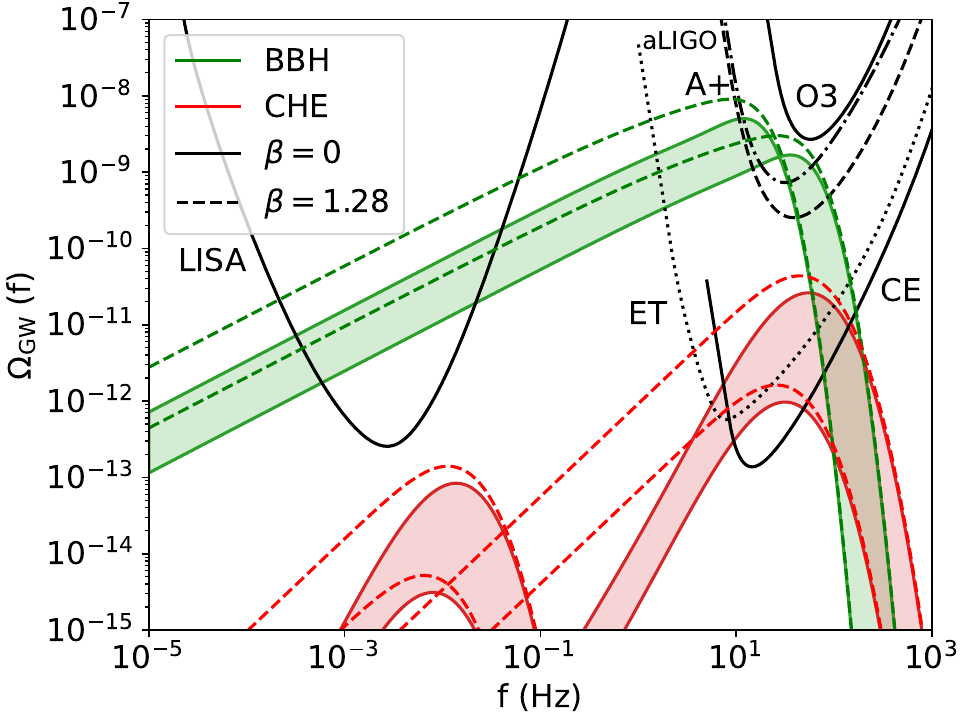}
\centering
\caption{Comparison of the SGWB spectrum originating from BBHs and CHEs, both for $\beta=0$ (solid) and $\beta=1.28$ (dashed), where $\beta$ is a parameter characterizing the redshift dependence of the merger rate as $\tau^{\rm BBH}\propto(1+z)^\beta$. For the BBH curves, we take $m_1=m_2=100-300~\Msun$ and $v_0=30$km/s. The CHE curves correspond to the same range of masses with $a_0=5$AU, $y_0=2\times 10^{-3}$ for frequencies around 10 Hz, and $a_0=5\cdot 10^7$AU, $y_0=10^{-5}$ in the mHz range. For all cases, we assume log-normal distributions for $a_0$, $y_0$ and the PBH mass, of respective widths $\sigma_a,\sigma_y = 0.1$, $\sigma_m = 0.5$, 
as well as $f_{\rm PBH}=1$.  {The bands come from the possible range of parameters $a_0$ and $y_0$.}  For a smaller fraction of PBHs, the GW spectral amplitude simply scales as $\Omega_{\rm GW}\propto f^2_{\rm PBH}$. Figure and estimated LISA sensitivity from~\cite{Garcia-Bellido:2021jlq}.}
\label{fig:Omegah2_CHE}
\end{figure*}

\subsection{Summary}

In this section we have explored the different sources of SGWBs related to PBHs:
\begin{itemize}
    \item The SGWB induced at second order by the large curvature fluctuations leading to PBH formation.  Interestingly, it peaks at PTA frequencies for stellar-mass PBHs, which makes it a very complementary probe to test a possible primordial origin of GW events seen in ground-based detectors.  {This SGWB could be connected to the recent pulsar timing array observations, see sec.~\ref{sec:PTAs_NG} for more details}.   Future GW detectors like LISA, Cosmic Explorer and Einstein Telescope will probe small PBHs, in particular in the asteroid-mass range that remains so far unconstrained (see the section on the limits on the abundance of PBHs).
    \item The SGWB from the Poisson fluctuations in the initial PBH distribution, even if they were so small that they have totally evaporated.  This mechanism could allow to probe the possible PBH formation at very high energy scales, up to the reheating phase and the GUT energy.
    \item The stochastic background from early binaries, in particular if PBHs have an extended mass function.  In such a case the SGWB is boosted by the merging of binaries with extreme mass ratios.  
    \item The stochastic background from late binaries in clusters, which for stellar-mass PBHs is smaller than the SGWB from early binaries at frequencies corresponding to ground-based GW detectors.  Because the merger rates are not suppressed with the PBH mass, it would however be the dominant background at low frequencies, i.e. in the LISA and PTA frequency range.
    \item The stochastic background from close hyperbolic PBH encounters in clusters, probably undetectable with LISA but in the range of future ground-based GW detectors.
\end{itemize}
The SGWB is therefore expected to be an interesting complementary signal that will help to distinguish between PBH formation models, mass functions, clustering histories and binary origins.

\section{Gravitational-waves and Large Scale Structure correlations}\label{sec:GWandLSS}

\subsection{Introduction}
Primordial black holes, both in the case where they comprise a large part of the dark matter or if they exist only in small numbers, are part of the Large-Scale Structure (LSS) of the Universe, and their observational relation with the other matter-energy components can be generally different with respect to black holes of stellar origin.
By the time LISA will be online, there will be data from several large-scale structure experiments, both preceding and overlapping in time with LISA. These include observations from the ground and from space, such as DESI~\cite{DESI:2016fyo,DESI:2016igz}, the Vera Rubin Observatory~\cite{Blum:2022dxi}, Euclid~\cite{Amendola:2016saw}, SPHEREx~\cite{Dore:2014cca}, Nancy Roman~\cite{Wenzl:2021rrq}, ATLAS~\cite{ATLAS-cosmology} and the SKAO~\cite{Weltman:2018zrl}, all together providing galaxy catalogs over a very wide range of scales both in width and deepness, at unprecedented accuracy.
It will therefore be paramount to take advantage of such data for PBH science.
In this Section we discuss a few possible ways to use gravitational wave observations in combination with the LSS to constrain the existence and abundance of PBHs.
LSS observations can be cross-correlated with different types of GW maps, both from the resolved mergers of compact objects and the stochastic backgrounds.

Cross-correlations between GWs from resolved compact object mergers and galaxies have been first investigated in~\cite{Laguna:2009re}. Later, it was suggested that the cross-correlation of mergers with EM counterparts could be used to constrain dark energy and modified gravity models~\cite{Camera:2013xfa}.
After the first detections of gravitational waves,~\cite{Oguri:2016dgk} investigated the possibility to constrain the distance-redshift relation using such correlations.

It was then shown that the cross-correlation of resolved mergers with LSS catalogs can be used to test primordial black hole scenarios~\cite{Raccanelli:2016cud, Raccanelli:2016fmc, Scelfo:2018sny, Scelfo:2021fqe} and different astrophysical models~\cite{Scelfo:2020jyw}. More recently, an additional technique for constraining PBH abundances, formation channels and redshift distributions using only GW data has been proposed and studied~\cite{Libanore:2021jqv}. 
%\textcolor{red}{Rename the references and add them to the main.bib file}

The GWxLSS probe became very popular and several different approaches have been suggested and investigated (see e.g.,~\cite{Raccanelli:2016fmc, Scelfo:2018sny,Canas-Herrera:2019npr,Calore:2020bpd}), looking at both correlations with resolved sources and the SGWB~\cite{Cusin:2018rsq,Jenkins:2018uac,Jenkins:2018kxc, Bertacca:2019fnt,Yang:2020usq,Mukherjee:2019oma,Alonso:2020mva}.
Here we present the main ideas and formalism developed so far; specific predictions for LISA and investigations involving similar analyses for extreme mass-ratio inspirals (EMRIs) and intermediate mass black holes (IMBHs) are currently being developed.

\subsection{GW-galaxy cross-correlations for resolved events}
The cross-correlation of LSS catalogs with maps of gravitational waves coming from the merger of compact objects will provide information on several astrophysical and cosmological parameters and models, and has been investigated in a variety of works from exploring ways to indirectly detect GWs to proper data correlations.

The idea of cross-correlations with the LSS in order to constrain the existence of PBHs and their abundances was first explored in~\cite{Raccanelli:2016cud} and then further investigated in e.g.,~\cite{Raccanelli:2016fmc, Scelfo:2018sny, Scelfo:2020jyw}.
The possibility of using the GWxLSS correlation to test the existence of PBHs comes from the fact that primordial and astrophysical binary black holes trace the LSS in different ways. This reflects into the fact that the bias of the hosts of the binary system will be different when the BHs have different progenitors and formation channels.

In~\cite{Bird:2016dcv}, it was shown that primordial black holes preferentially merge in lower biased objects and thus have a lower cross-correlation with luminous galaxies.
This happens because binary PBHs would preferentially merge in halos with low velocity dispersion, which are low-mass halos and hence with very little or absent star formation; these objects  have a bias $b<1$. On the other hand, mergers of compact objects that are the endpoint of stellar evolution naturally happen for the vast majority within star forming-rich halos, which have larger galaxy bias values, $b>1$. In the linear biasing scheme, $b$ quantifies the relation between the underlying matter distribution and observed sources, $\delta_{\rm O}=b\delta_{\rm m}$.
This reasoning is valid for PBH binary systems that form by capture 
%and we call this the 
{(\textit{late binary} formation scenario)}, as these systems form in the local Universe.

However, there is also another way to have mergers of PBH binaries, which we {have referred to as } \textit{early binary} mergers. 
In this case, within the standard formation scenario (i.e.,~the collapse of primordial curvature perturbations at horizon re-entry), PBHs
should form in the peaks of the initial matter distribution, and therefore they will trace the dark matter field.
As a consequence, such mergers should be unbiased tracers of the underlying density field.

%Details on early PBH binaries, their abundance, distribution and merger rate are still under study. In particular, it has been shown that primordial non-Gaussianity can affect the early merger rate, and three-body interactions can also disrupt early binaries (see e.g.,~\cite{Ali_Ha_moud_2018, Ballesteros_2018, DeLuca:2021hcf, DeLuca:2020jug}).

The overall catalog of PBH mergers will consequently be composed of a mixture of early and late binaries, with the total bias being the weighted average of the two.

Therefore, by measuring the amplitude of the angular cross-correlation of galaxy maps with catalogues of compact-object mergers, which directly depends on the bias of the mergers' hosts, one can statistically probe the abundance of PBH mergers.
This information can also be used to discriminate between different astrophysical models (see e.g.,~\cite{Scelfo:2020jyw}).

In order to measure the mergers' hosts bias, one can use the 3D angular power spectrum:
\begin{equation}
\label{eq:cls}
    C_\ell^{\rm GW\times LSS} (z_1,z_2) = \frac{2}{\pi} \int {\rm d}k \, k^2 P(k) \, \Psi_\ell^{\rm GW}(k) \Psi_\ell^{\rm LSS}(k) \, ,
\end{equation}
where $P(k)$ is the matter power spectrum and the kernels $\Psi_\ell^X(z)$ include the observational window functions and the relevant physical effects:
\begin{align}
%  \Psi_\ell^{X}(k)  &= \int \mathcal{R}(z) T_{\rm obs} \frac{4 \pi \chi^2(z)}{(1+z)H(z)} b(z) W(z) \Delta_\ell^X(k,z) \, dz \\
  \Psi_\ell^{X}(k) &= \int N^X(z) b^X(z) D(z) W(z) \Delta_\ell^X(k,z) \, {\rm d}z \, ;
\end{align}
here $X$ will be either GW or LSS, and $W$ are observational window functions related to the experiment specifications (i.e.,~they include information on the redshift distribution and survey geometry and sensitivity).
The $\Delta_\ell$ terms are the (gauge-independent) observed overdensities and include effects from intrinsic clustering, peculiar velocity and Doppler contributions, lensing and gravitational potentials containing in this way information information on the underlying cosmological and astrophysical models. They are usually expressed as:
\begin{equation}
\Delta_\ell(k) = \Delta^\mathrm{den}_\ell(k,z) + \Delta_\ell^\mathrm{rsd}(k,z) + \Delta_\ell^\mathrm{dop}(k,z) +  \Delta^{\mathrm{len}}_\ell(k,z) +  \Delta^{\mathrm{gr}}_\ell(k,z) \, ,
\label{eq:relative contr}
\end{equation}
where one has that:
\begin{equation}
\begin{aligned}
\Delta_\ell^\mathrm{den}(k,z) &= b_X \delta(k,\tau_z) j_\ell,	\\
%\Delta_\ell^\mathrm{vel}(k,z) &= \Delta_\ell^\mathrm{rsd}(k,z) + \Delta_\ell^\mathrm{dop}(k,z),	\\
\Delta_\ell^\mathrm{rsd}(k,z) &=  \frac{k}{\mathcal{H}}j''_\ell  V(k,\tau_z),	\\
\Delta_\ell^\mathrm{dop}(k,z) &= \left[(f^\mathrm{evo}_X-3)\frac{\mathcal{H}}{k}j_\ell + \left(\frac{\mathcal{H}'}{\mathcal{H}^2}+\frac{2-5s_X}{r(z)\mathcal{H}}+5s_X-f^\mathrm{evo}_X\right)j'_\ell \right]  V(k,\tau_z),	\\
\Delta_\ell^\mathrm{len}(k,z) &= \ell(\ell+1) \frac{2-5s_X}{2} \int_0^{r(z)} {\rm d}r \frac{r(z)-r}{r(z) r} \left[\Phi(k,\tau_z)+\Psi(k,\tau_z)\right] j_\ell(kr),	\\
\Delta_\ell^\mathrm{gr}(k,z)  &= \left[\left(\frac{\mathcal{H}'}{\mathcal{H}^2}+\frac{2-5s_X}{r(z)\mathcal{H}}+5s_X-f^\mathrm{evo}_X+1\right)\Psi(k,\tau_z) + \left(-2+5s_X\right) \Phi(k,\tau_z) + \mathcal{H}^{-1}\Phi'(k,\tau_z)\right] j_\ell \\
&+ \int_0^{r(z)} {\rm d}r \frac{2-5s_X}{r(z)} \left[\Phi(k,\tau)+\Psi(k,\tau)\right]j_\ell(kr) \\
&+ \int_0^{r(z)} {\rm d}r \left(\frac{\mathcal{H}'}{\mathcal{H}^2}+\frac{2-5s_X}{r(z)\mathcal{H}}+5s_X-f^\mathrm{evo}_X\right)_{r(z)} \left[\Phi'(k,\tau)+\Psi'(k,\tau)\right] j_\ell(kr).
\end{aligned}
\end{equation}
For details on these terms, their derivation and physical meaning, see e.g.,~\cite{Bonvin:2011bg, Challinor:2011bk, Raccanelli:2015vla, Scelfo:2018sny}.

\subsection{GW-galaxy cross-correlations for the stochastic background}
In addition to the signal from mergers detected by interferometers, there is another signal coming from the superposition of many unresolved sources. These will form a stochastic background of gravitational waves, as discussed in Section~\ref{sec:SGWBs}. Obviously, primordial black holes, if they exist, would contribute to this signal, and their associated GW backgrounds have been discussed in Section~\ref{sec:SGWBs}. This will be particularly relevant for LISA, as the SGWB is expected to be a product of LISA's observations.

In the same fashion as for the catalogs of resolved events described above, we can cross-correlate the stochastic background with the LSS; such correlation will contain information on the redshift distribution, mass function and clustering behavior of the sources. Moreover, the observed signal will be the combination of the stochastic GW background from astrophysical objects, denoted AGWB, the one from primordial black holes, and the one generated from inflation; disentangling the signal and distinguishing between the different sources will be of paramount importance to obtain information about black hole and early Universe physics, and one of the most promising ways to do so is thanks to the SGWB--LSS correlation.

Given that the GW energy density depends not only on astrophysical properties but also on cosmological perturbations, it will correlate with other cosmological probes; moreover, GWs will experience projection effects that will need to be accounted for in order to observe the signal in the appropriate frame~\cite{Bertacca:2019fnt, Bellomo:2021mer}.
Some preliminary works on this have been done in the last few years, including forecasts for the cross-correlation signals between GW observatories and future galaxy surveys, as e.g.,~Euclid and SKA
(see, e.g.,~\cite{Contaldi:2016koz, Cusin:2017fwz, Cusin:2017mjm, Cusin:2018rsq,Jenkins:2018kxc,Jenkins:2018uac, Cusin:2019jpv, Cusin:2019jhg, Jenkins:2019uzp, Jenkins:2019nks, Bertacca:2019fnt, Pitrou:2019rjz, Mukherjee:2019oma, Alonso:2020mva}).

In a similar way to the case of resolved mergers, the most natural observable is the angular cross power spectrum that correlates the energy density of the AGWB with galaxy number counts: 
\begin{equation}
C^{\rm AGWB\,\times\,LSS}_\ell = \frac{2}{\pi} \int {\rm d}k \, k^2 P(k) \Psi^{\rm AGWB}_\ell(k) \Psi^{\rm LSS}_\ell(k),
\end{equation}
where most quantities are the same of Eq.~\eqref{eq:cls}, and
%\AR{check formalism with resolved section} xxx
the dependence on astrophysical models and parameters can be included in the expression for the AGWB energy density. {This is usually written as the total GW energy density (denoted by the superscript TOT)} per logarithmic frequency $f_{\rm o}$ and solid angle~$\Omega_{\rm o}$ along the line-of-sight~${\bf \hat{n}}$ of a SGWB, defined in e.g.,~\cite{Cusin:2017fwz, Bertacca:2019fnt} as
\begin{equation}
\Omega^{\rm TOT}_{\rm GW} \left(f_{\rm o}, {\bf \hat{n}}\right)= \frac{f_{\rm o}}{\rho_{{\rm c}}} \frac{ {\rm d} \rho^{\rm TOT}_{\rm GW}}{ {\rm d} f_{\rm o} {\rm d} \Omega_{\rm o} } \; ;
\end{equation}
this contains both a background, which is homogeneous and isotropic, and a direction-dependent contribution; the total relative fluctuation can thus be defined as {(similarly to Eq.~\eqref{GWdensitycontrast})}:
\begin{equation}
    \Delta^\mathrm{TOT}_\mathrm{AGWB}(f_{\rm o}, \hat{\mathbf{n}})= \frac{\Omega^\mathrm{TOT}_\mathrm{AGWB}-\bar{\Omega}^\mathrm{TOT}_\mathrm{AGWB}}{\bar{\Omega}^\mathrm{TOT}_\mathrm{AGWB}} \; .
\end{equation}
Following~\cite{Bertacca:2019fnt}, we report the expression for the contributions to the AGWB in the Poisson gauge:
\begin{equation}
\begin{aligned}
&\Delta\Omega_{\rm AGWB}(f_{\rm o}, \hat{\mathbf{n}}, \mathbf{\theta}) =   \sum_{[i]}  \int {\rm d}z \, W(z) \, \mathcal{F}^{[i]}(f_{\rm o} , z, \mathbf{\theta})
\Bigg\{ b^{[i]} D  \\& + \left(b^{[i]}_\mathrm{evo} - 2 - \frac{\mathcal{H}'}{\mathcal{H}^2}\right) \hat{\mathbf{n}}\cdot\mathbf{V} - \frac{1}{\mathcal{H}}\partial_\parallel(\hat{\mathbf{n}}\cdot\mathbf{V}) - (b^{[i]}_\mathrm{evo}-3) \mathcal{H} V  \\
&\quad + \left(3 - b^{[i]}_\mathrm{evo} + \frac{\mathcal{H}'}{\mathcal{H}^2}\right)\Psi + \frac{1}{\mathcal{H}}\Phi' + \left(2 - b^{[i]}_\mathrm{evo} + \frac{\mathcal{H}'}{\mathcal{H}^2}\right) \int_0^{\chi(z)} {\rm d}\tilde{\chi} \left(\Phi'+\Psi'\right)  \\
&\quad + \left(b^{[i]}_\mathrm{evo} - 2 - \frac{\mathcal{H}'}{\mathcal{H}^2}\right) \left( \Psi_{\rm o} - \mathcal{H}_0 \int_{0}^{\tau_0} {\rm d}\tau \left.\frac{\Psi(\tau)}{1+z(\tau)}\right|_{o} -  \left(\hat{\mathbf{n}}\cdot\mathbf{V} \right)_{o} \right) \Bigg\},
\end{aligned}
\label{eq:fluctuation_poissongauge}
\end{equation}
where the \textit{density}, \textit{velocity}, \textit{gravity} and \textit{observer} terms, in the first, second, third and fourth line, respectively contain all the cosmological information. On the other hand, the function $\mathcal{F}^{[i]}(f_o, z, \mathbf{\theta})$ contains all the astrophysical dependencies, i.e.,~the mass and spin distribution of the binary, the emitted GW energy spectrum, the clustering properties of GW events and the details of the GW detectors; $b^{[i]}(z,\mathbf{\theta})$ and~$b^{[i]}_\mathrm{evo}(z,\mathbf{\theta}):=-d\log\left[(1+z)\mathcal{F}^{[i]}\right]/d\log(1+z)$ are the bias and the evolution bias of the $i$-th type of GW source, which specify the clustering properties of GW sources and characterise the formation of the sources.

Studies of the cross-correlation between the AGWB with galaxy number counts are presented in~\cite{Cusin:2018rsq, Mukherjee:2019oma, Canas-Herrera:2019npr, Alonso:2020mva, Cusin:2019jpv,Yang:2020usq, Bellomo:2022}. This allows a tomographic reconstruction of the redshift distribution of sources, and can be also useful for the shot noise characterization (in the case of ground-based detectors)~\cite{Alonso:2020mva}.
The cross-correlation shows that the combination of galaxy surveys with the AGWB can be a powerful probe for GW physics and can be a robust observational probe for multi-messenger cosmology.

\subsection{GW$\times$LSS forecasts}
Constraints on the presence of PBHs from measurements of the GWxLSS have been forecasted for the case of resolved GW signals, correlated with both galaxy surveys~\cite{Raccanelli:2016cud, Raccanelli:2016fmc, Scelfo:2018sny, Bosi:2022} and intensity mapping experiments~\cite{Scelfo:2021fqe}. Constraints coming from measurements of the stochastic background signal are currently being worked out. In this Section we introduce the formalism for obtaining such forecasts and report some of the results present in literature.

The correlation of two tracers ($X,Y$) can be estimated from tomographic maps of their number counts and evaluating their N-point statistics; in the case we consider here, this is the angular power spectrum~$C_\ell^{XY}(z_i, z_j)$ of the tracer $X$ in redshift bin $z_i$ and the tracer $Y$ in $z_j$. Assuming that the noise comes only from the standard shot noise contribution, the total signal extracted from the maps is:
\begin{equation*}
\left\langle a_{\ell m}^{X}\left(z_{i}\right) a_{\ell^{\prime} m^{\prime}}^{Y^{*}}\left(z_{j}\right)\right\rangle=\delta_{\ell \ell^{\prime}} \delta_{m m^{\prime}} \tilde{C}_{\ell}^{X Y}\left(z_{i}, z_{j}\right)=\delta_{\ell \ell^{\prime}} \delta_{m m^{\prime}}\left[C_{\ell}^{X Y}\left(z_{i}, z_{j}\right)+\delta_{X Y} \delta_{i j} \mathcal{N}_{\ell}^{X}\left(z_{i}\right)\right] \, ,
\end{equation*}
where $\mathcal{N}_{\ell}$ is the noise angular power spectrum.

The observed signal will receive contributions from several physical effects, coming from the cosmological perturbations described above, from density, velocity, lensing and gravity effects. It is important to include all of them to avoid mis-estimating parameters, best fit values and errors when doing parameter inference, as shown in~\cite{Bellomo:2021mer, Bernal:2020pwq}.

Forecasts obtained so far have been based on the~\texttt{Multi\_CLASS}~\cite{Bellomo:2020pnw} and~\texttt{GW\_CLASS} codes~\cite{Bellomo:2021mer}.
\texttt{Multi\_CLASS} is the first public Boltzmann code that allows to compute the angular power spectrum for multiple galaxy (and other tracers) populations. The code lets the
user specify specific properties of different tracers, such as its number density redshift distribution and bias, magnification bias and evolution bias parameters, while~\texttt{GW\_CLASS} calculates the stochastic signal from astrophysical mergers and accounts for all the astrophysical and projection effects contributions.

The first suggestion that this observable can be used to constrain PBHs was presented in~\cite{Raccanelli:2016cud}, where after introducing the concept, it is shown that the cross-correlation between galaxies mapped by future radio surveys and GW interferometers could distinguish between primordial and stellar origin of merging black holes. Such possibility depends on the merger rate of primordial black holes and the angular resolution of the GW interferometer.
In that work, to obtain the results, a minimum-variance estimator was introduced for the {\it effective correlation amplitude}, 
$A_{\rm c}\equiv r \times b_{\rm GW}$, where $r$ is the cross-correlation coefficient of Eq.~\eqref{eq:cls}.
This cross-correlation coefficient parametrises the extent to which two biased tracers of the matter field 
are correlated~\cite{Tegmark:1998wm}.
%\textcolor{red}{Rename reference and add it to the main.bib file.}
 
The minimum-variance estimator for the effective correlation amplitude is given by (see e.g.~\cite{Jeong:2012fossils}):
\begin{equation}
\label{eq:Ac}
\widehat{A_{\rm c}} = \frac{\Sigma_\ell {\tilde{C}_\ell} F_\ell / {\rm Var}[\tilde{C}_\ell] 
}{\Sigma_\ell F_\ell^2 / {\rm Var}[\tilde{C}_\ell]} \, ,
\end{equation}
where $\tilde{C}_\ell$ is the measured power spectrum and $F_\ell \equiv{d \tilde{C}_\ell}/{d \widehat{A_{\rm c}}}\propto b_{\rm g}$. 
The variance of this estimator is then:
\begin{equation}
\sigma^2_{\widehat{A_{\rm c}}} = \left[\sum_\ell \frac{F_\ell^2}{{\rm Var}[\tilde{C}_\ell]}\right]^{-1} \, ,
\end{equation}
which can be used to forecast the measurement error when neglecting that of other parameters.

More generally, the measurement error for specific parameters in a given experiment can 
be estimated using Fisher analysis. For this case, we write the Fisher matrix as:
\begin{equation}
\label{eq:Fisher}
F_{\alpha\beta} = \sum_{\ell} \frac{\partial C_\ell}{\partial 
\vartheta_\alpha}
\frac{\partial C_\ell}{\partial
\vartheta_\beta} {\sigma_{C_\ell}^{-2}} \, , 
\end{equation}
where $\vartheta_{\alpha} = \{ A_{\rm c}, b_{\rm g} \}$;
the derivatives of the power spectra $C_\ell$ are evaluated at fiducial values 
$\bar \vartheta_{\alpha}$ corresponding to the scenario at hand, 
and $\sigma_{C_\ell}$ are errors in the power spectra.

The results in~\cite{Raccanelli:2016cud} were obtained by computing the $2\times2$ Fisher matrix for the 
parameters $\{A_{\rm c}, b_{\rm g}\}$, using a prior on the galaxy bias corresponding 
to the precision reached by fitting the amplitude of the galaxy auto-correlation 
function $C_\ell^{\rm gg}$, using a value of $\ell_{\rm max}=200$, which yields a 
$\sim 10\%$ precision in the measurement of the bias $b_{\rm g}$.
For more details on this analysis, see~\cite{Raccanelli:2016cud}.

Subsequently, in~\cite{Raccanelli:2016fmc}, such analysis was updated to include improved merger rate estimates from the LVK collaboration, for the correlation with different possible surveys with the SKA. In Figure~\ref{fig:LSS1} we show forecasts for different experiments, considering different versions of LVK and ET correlated with HI and continuum SKA surveys.
The bars show the constraints on $A_{\rm c}$ for different values of a parameter $R$ that encapsulates uncertainties on the merger rate, which is degenerate with observation times from the gravitational wave interferometers.

\begin{figure}[t!]
  \centering
    \includegraphics[width=0.81\textwidth]{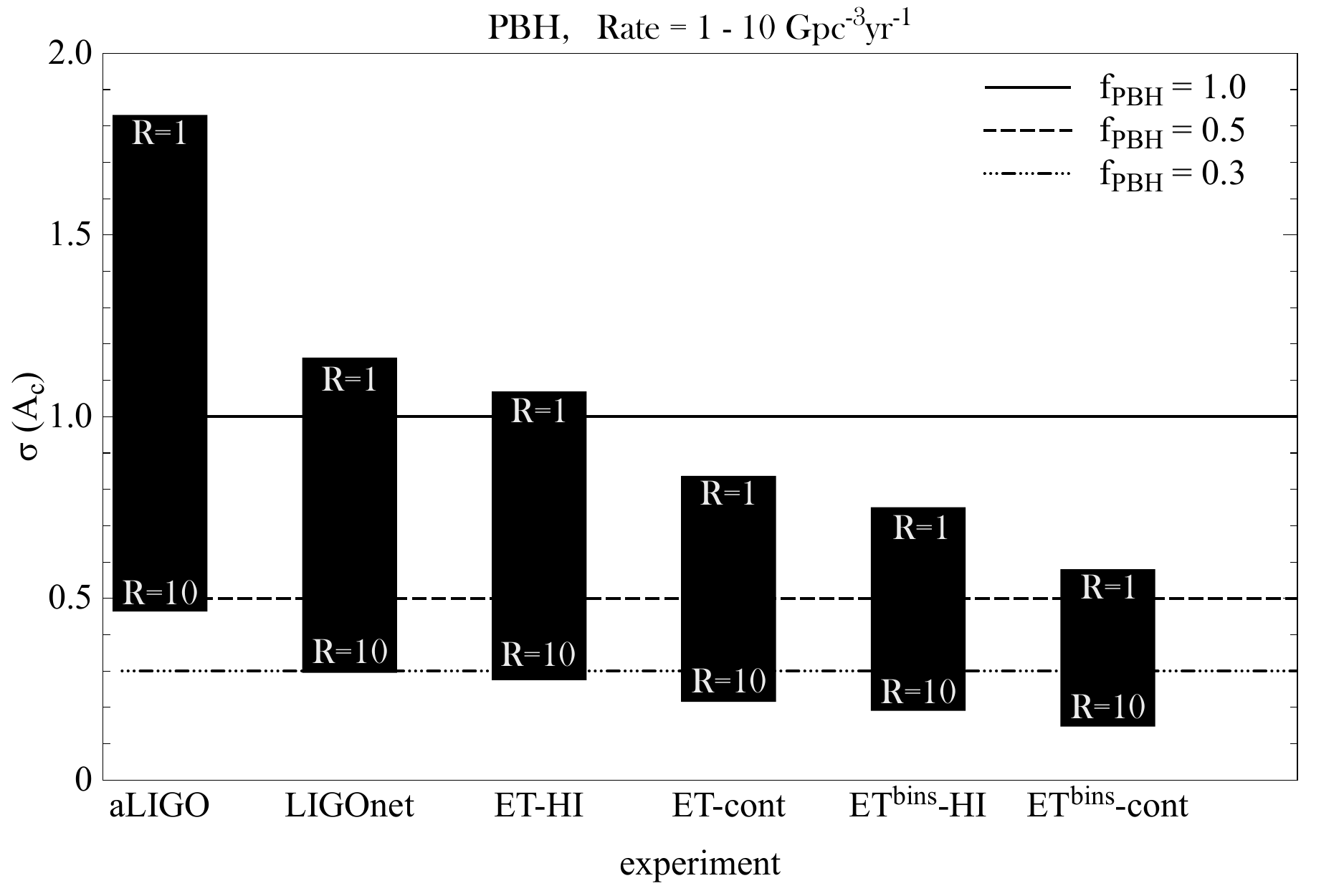}
    \caption{Forecast errors on the cross-correlation amplitude, $A_{\rm c}$, for different experiment combinations, varying merger rates and years of observations. 
Each column corresponds to a GW detector experiment, for merger rates from 1 to 10 Gpc$^{-3}$yr$^{-1}$.
Horizontal lines show the expected difference in the cross-correlation between (late binary) PBH and stellar binaries, for different values of $f_{\rm PBH}$.}
\label{fig:LSS1}
\end{figure}

Following this, in~\cite{Scelfo:2018sny} these forecasts are updated by including all projection effects, and inserted in an analysis of a null hypothesis test by comparing two models, one in which the BHs origin is stellar, the other in which it is primordial. One model is assumed as a fiducial model. Then, by computing the signal-to-noise ratio (SNR), it is checked whether the alternative model can be differentiated from the fiducial one. The null hypothesis is that the model is indistinguishable from the fiducial, which happens for low values of the signal-to-noise ratio.

For instance, the distance of an alternative model from the fiducial can be quantified using a $\Delta \chi^2$ statistics. The $\Delta \chi^2$ is given by the logarithm of a likelihood, and the SNR can be written as:
\begin{equation*}
\left(\frac{S}{N}\right)=f_{\mathrm{sky}} \sum_{2}^{\ell_{\max }}(2 \ell+1)\left(\mathbf{C}_{\ell}^{\mathrm{Alternative}}-\mathbf{C}_{\ell}^{\mathrm{Fiducial}}\right)^{T} \operatorname{Cov}_{\ell}^{-1}\left(\mathbf{C}_{\ell}^{\mathrm{Alternative}}-\mathbf{C}_{\ell}^{\mathrm{Fiducial}}\right) \, ,
\end{equation*}
where $\mathbf{C}_{\ell}^{T}=\left(C_{\ell}^{\mathrm{gg}}\left(z_{1}, z_{1}\right), \cdots, C_{\ell}^{\mathrm{gGW}}\left(z_{1}, z_{1}\right), \cdots, C_{\ell}^{\mathrm{GWGW}}\left(z_{1}, z_{1}\right), \cdots\right)$, $f_{\mathrm{sky}}$ is the sky fraction covered by both GW and galaxy surveys and $\mathrm{Cov}_\ell$ is the covariance matrix, computed from the angular power spectra of the fiducial model.

In Fig.~\ref{fig2}, we show the results where the fiducial model is taken to that of stellar BH mergers (See~\cite{Scelfo:2018sny}).
%\textcolor{red}{Rename reference and add it to the main.bib file.}. 
The bars span values between 0.1 and 10 for the factor $r$, (similarly to the case of Fig.~\ref{fig:LSS1}, this parametrises the uncertainties on the BBH merger rates, which is degenerate with the observation time of the experiment).

\begin{figure}[t!]
  \centering
    \includegraphics[width=0.81\textwidth]{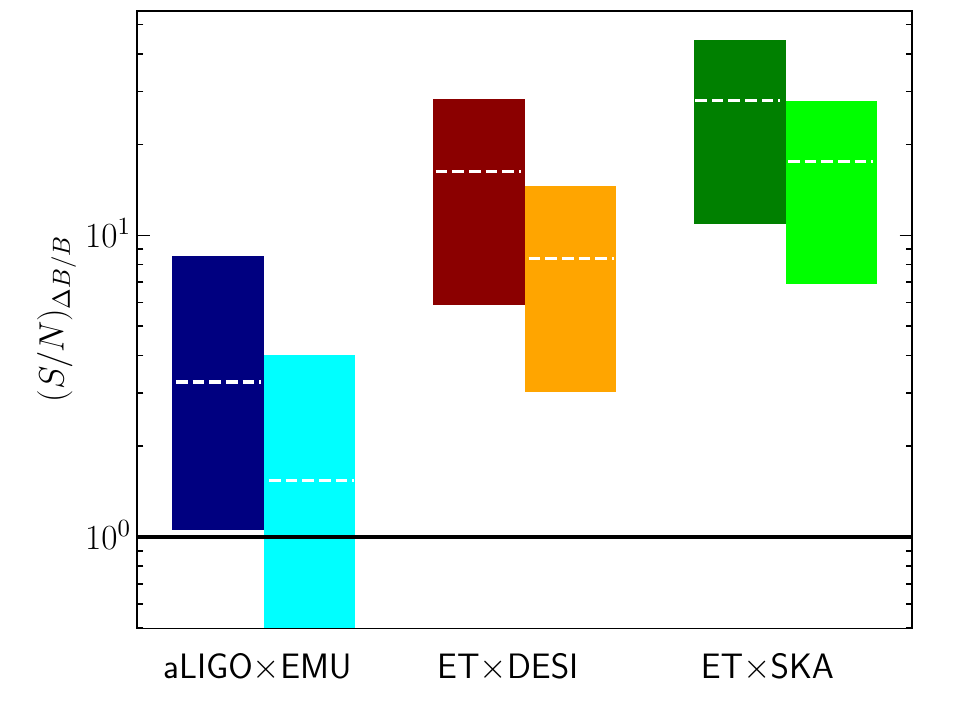}
    \caption{Expected signal-to-noise ratio $(S/N)_{\Delta B/B}$ from the Fisher matrix analysis, for cross-correlations between GW observations (from either LVK or Einstein Telescope) and high-redshift LSS surveys (EMU, DESI and SKA), assuming stellar black holes for the fiducial scenario.  In each case, the left bar corresponds to late PBH binaries and the right bar to early PBH binaries.  Figure from~\cite{Scelfo:2018sny}. }
    \label{fig2}
\end{figure}
After updates on the BBH merger rate and the specifications for the DESI survey~\cite{DESI:2016fyo}, forecasts for the SNR for detecting a component of PBHs are shown in~\cite{Bosi:2022} and a summary plot can be found in Fig.~\ref{fig3}.
Here the color code is for the SNR, while on the x-axis is the fraction of PBH mergers from the total of observed BBH mergers observed by the Einstein Telescope, and on the y-axis the parameter $r$, as defined above.

\begin{figure}[t!]
  \centering    \includegraphics[width=0.47\textwidth]{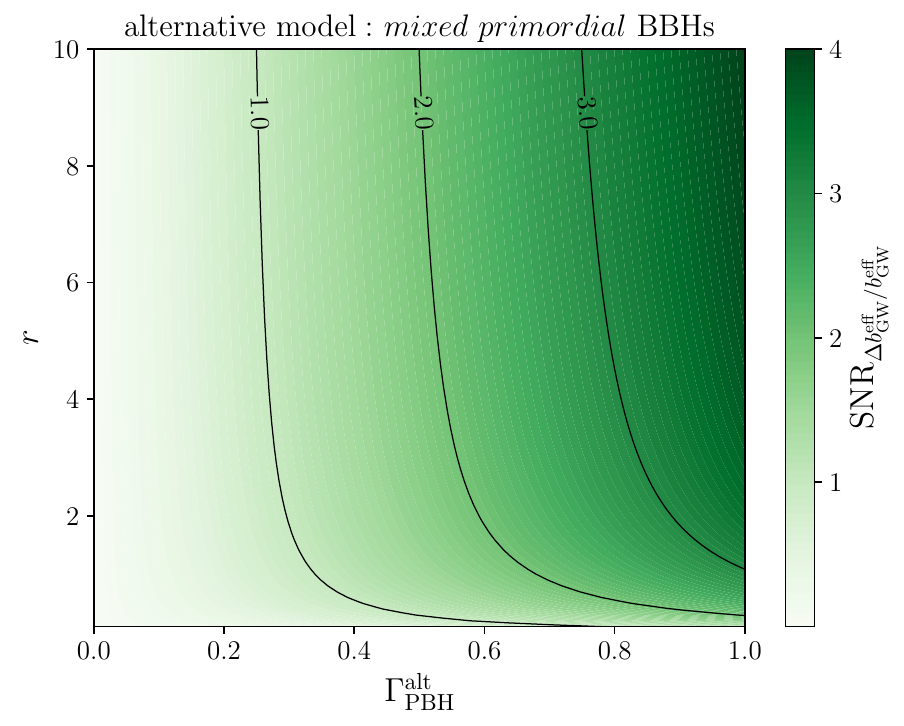}    \includegraphics[width=0.47\textwidth]{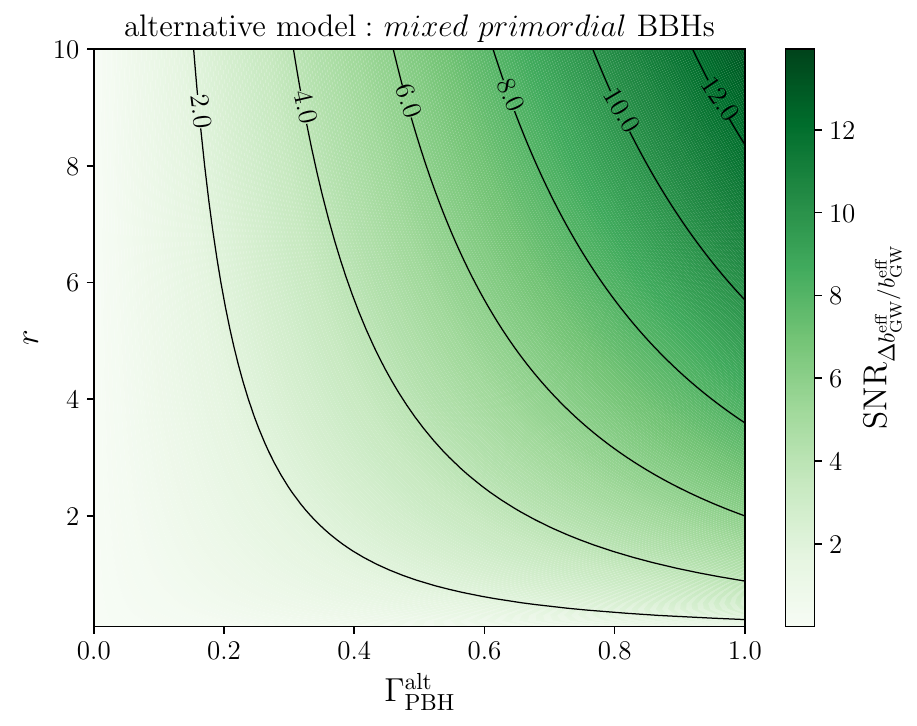}
    \caption{SNR for detecting a fraction $\Gamma$ of PBH mergers out of the total BBH mergers observed by the ET, as a function of the parameter $r$, cross-correlated with DESI. The left panel shows the result using just DESI x ET, while the right panel includes Planck priors. Results from~\cite{Bosi:2022})}
    \label{fig3}
\end{figure}

The cross-correlation forecasts shown so far are for the correlation of GWs with resolved galaxies; naturally, a similar analysis can be done for observations of intensity mapping.
A first investigation for the 21cm Intensity Mapping (IM)  observed with the SKAO is presented in~\cite{Scelfo:2021fqe}, focusing on the IM  of the neutral hydrogen (HI) from the proposed 21cm IM survey with the SKAO and on resolved GW events from the merger of BBHs as detected by the Einstein Telescope (ET) \cite{Sathyaprakash:2012jk}.
%\textcolor{red}{Rename reference and add it to the main.bib file.}.
A cross-correlation signal is expected, as both HI and GWs trace the cosmic density field, and once again they do so in different ways depending on astrophysical and cosmological models.
In particular, for constraining the PBH abundance, results were obtained for the predicted IM from the SKAO correlated with GWs from ET.

\begin{figure}[t!]
  \centering
    \includegraphics[width=0.91\textwidth]{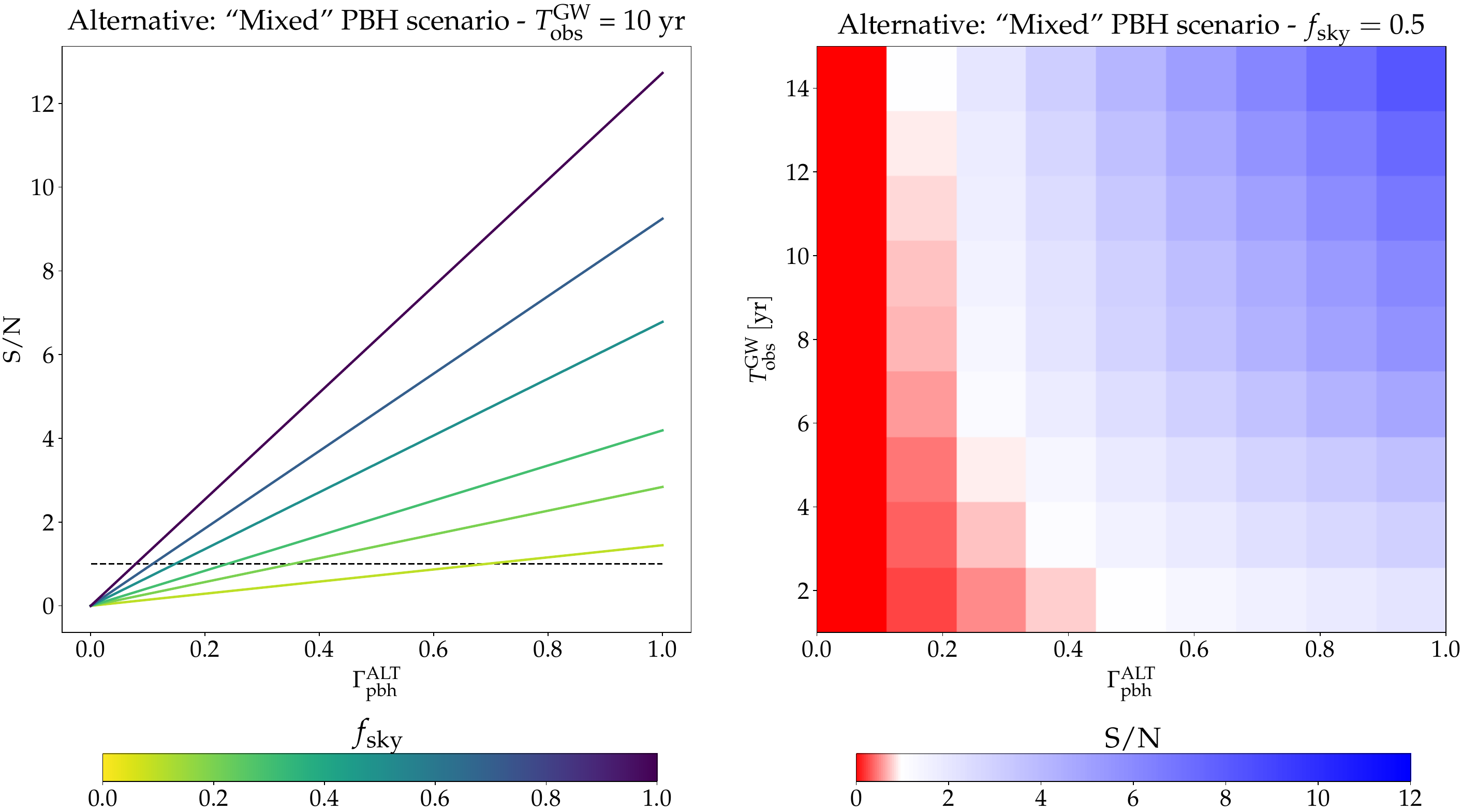}
    \caption{SNR for different values of the GW observation time $T_{\mathrm{obs}}^{\rm GW}$ (from 1 yr to 15 yr), $f_{\mathrm{sky}}$ (from 0.1 to 1.0) and mixed PBH scenarios $\Gamma_{\mathrm{pbh}}^{\mathrm{ALT}}$ (from 0.0 to 1.0), assuming the astrophysical model as fiducial. The fiducial model assumed is astrophysical black hole binaries. Results on the right-side plots are normalized to white at SNR=1.}
\end{figure}

\subsection{Summary}

Gravitational-wave observations of resolved black hole mergers and of the stochastic background can be cross-correlated with the LSS to constrain the existence and abundance of PBHs.  This technique may provide additional information to test and constrain the different PBH scenarios, the PBH abundance within a broad range of masses, as well as binary formation channels.   This will be of particular interest for the next generation of LSS surveys, such as DESI, Euclid and SKA, and gravitational-wave detectors like LISA, Einstein Telescope or Cosmic Explorer.

\section{Current limits}
\label{sec:limits}

The most stringent limits on $f_{\rm PBH}$ as a function of the PBH mass for a monochromatic mass function are represented in Fig.~\ref{fig:PBHlimits}.    Zooms on evaporation-based limits (between $10^{-19} M_\odot$ and $10^{-15} M_\odot$), on microlensing limits (between $10^{-11} M_\odot$ and $1 M_\odot$) and on the multiple limits above $1 M_\odot$ are shown in  Fig.~\ref{fig:PBHlimitszoom}.
These astrophysical and cosmological probes of PBHs can be divided in six categories, each related to the physical process at the origin of a PBH signature in observations:
\begin{enumerate}
\item Observations impacted by \textbf{black hole evaporation}
\item \textbf{Microlensing} observations
\item Observational signatures of the black hole \textbf{dynamics}
\item Observations impacted by the black hole \textbf{accretion}
\item Indirect constraints from \textbf{density fluctuations}
\item Signatures of PBHs in \textbf{Gravitational-wave} observations 
\end{enumerate}

In this Section, we review these current limits on the PBH abundance and comment on possible caveats and model dependences.  
We also discuss the possible observational evidences for the existence for PBHs, in relation with these limits.   

\begin{figure}[t] 
    \centering
    \includegraphics[width=\textwidth]{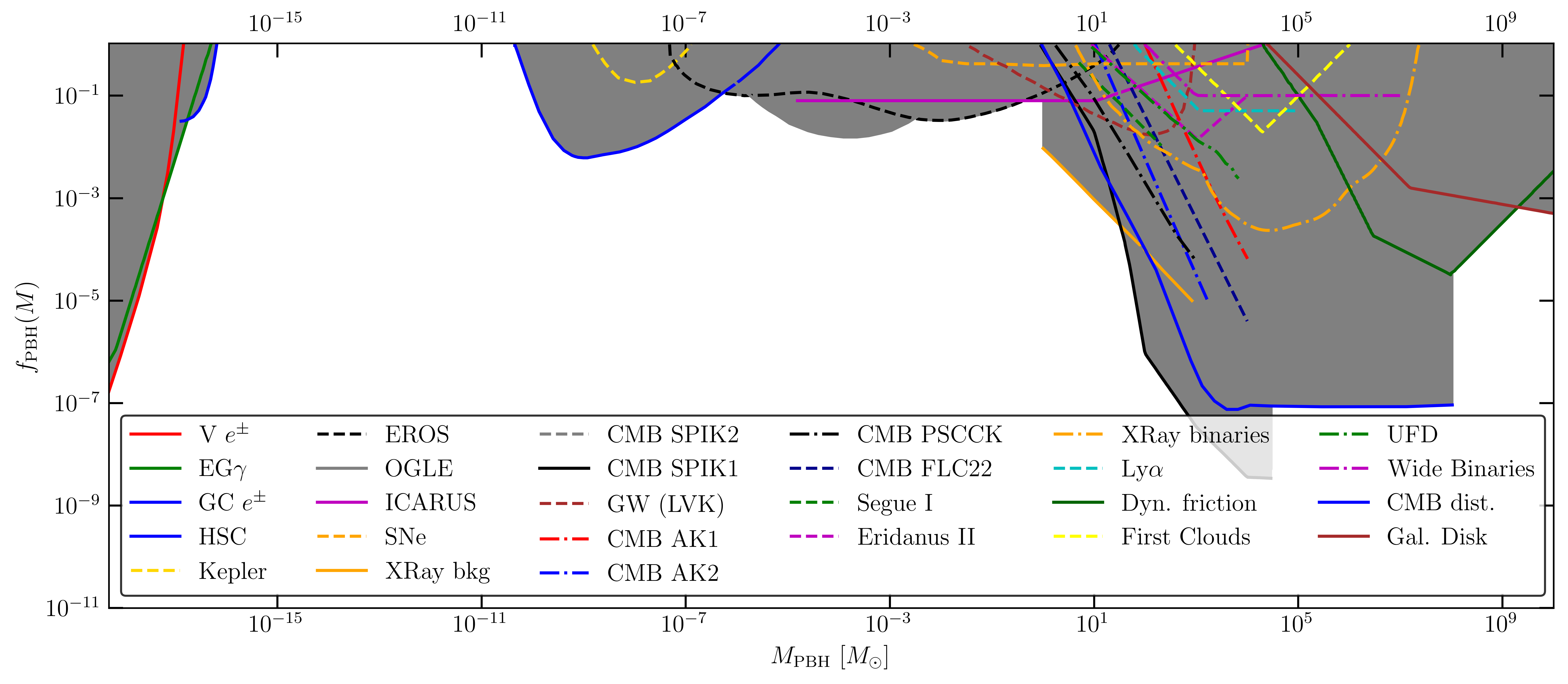}
    \caption{We show the most stringent claimed constraints in the mass range of phenomenological interest. They come from the Hawking evaporation producing
extra-galactic gamma-ray (EG $\gamma$) \cite{Arbey:2019vqx}, $e^\pm$ observations by Voyager 1 (V $e^\pm$) \cite{Boudaud:2018hqb}, positron annihilations in the Galactic Center (GC $e^+$) \cite{DeRocco:2019fjq} and gamma-ray observations by INTEGRAL (INT) \cite{Laha:2020ivk} (for other constraints in the ultra-light mass range see also \cite{Carr:2009jm, Ballesteros:2019exr,Laha:2019ssq, Poulter:2019ooo,Dasgupta:2019cae,Laha:2020vhg}). We plot microlensing searches by Subaru HSC \cite{Niikura:2017zjd, Smyth:2019whb}, MACHO/EROS (E) \cite{Alcock:2000kd, Allsman:2000kg}, Ogle (O) \cite{Niikura:2019kqi} and Icarus (I) \cite{Oguri:2017ock}. 
%The different lines correspond to different assumptions on the fraction
%of PBHs in small clusters observable through lensing (from $0\%$ to $75\%$). 
%We only shade the most conservative version.
Other constraints come from CMB distortions. In black dashed, we show the ones assuming disk accretion (Planck D in Ref.~\cite{Serpico:2020ehh} and Ref.~\cite{Poulin:2017bwe}, from left to right)
while in black solid the ones assuming spherical accretion 
(Planck S in Ref.~\cite{Serpico:2020ehh} and
both photo- and collisional ionization in Ref.~\cite{Ali-Haimoud:2016mbv}, from left to right).
Only Ref.~\cite{Serpico:2020ehh} includes the effect of the secondary dark matter halo in catalysing accretion.
Additionally, constraints coming from X-rays (Xr) \cite{Manshanden:2018tze} and X-Ray binaries (XrB)\cite{Inoue:2017csr} are shown. 
Dynamical limits coming from the disruption of wide binaries (WB) \cite{2009MNRAS.396L..11Q}, survival of star clusters in Eridanus II (EII)
\cite{Brandt:2016aco} and Segue I (SI) \cite{Koushiappas:2017chw,Stegmann:2019wyz} are also shown.
LVC stands for the constraint coming from LVK measurements  \cite{Ali-Haimoud:2017rtz,Raidal:2018bbj,Vaskonen:2019jpv,Wong:2020yig}. Constraints from Lyman-$\alpha$ forest observations (L$\alpha$) come from Ref.~\cite{Murgia:2019duy}.
We neglect the role of accretion which has been shown to affect constraints on masses larger than ${\cal O}(10) M_\odot$ \cite{DeLuca:2020fpg,DeLuca:2020qqa} in a redshift dependent manner.
See Ref.~\cite{Carr:2020gox} for a comprehensive review on constraints on the PBH abundance. Notice that there are no stringent bounds in the asteroid mass range \cite{Katz:2018zrn, Montero-Camacho:2019jte}
where LISA may constrain PBHs through the search of a second order SGWB.
%\SC{(suggestion:  extend the figure to larger masses, and make a second one zooming in the stellar-IMBH mass range)}
}
 \label{fig:PBHlimits}
\end{figure}

\begin{figure}[t]
    \centering
    \includegraphics[width=0.6 \textwidth]{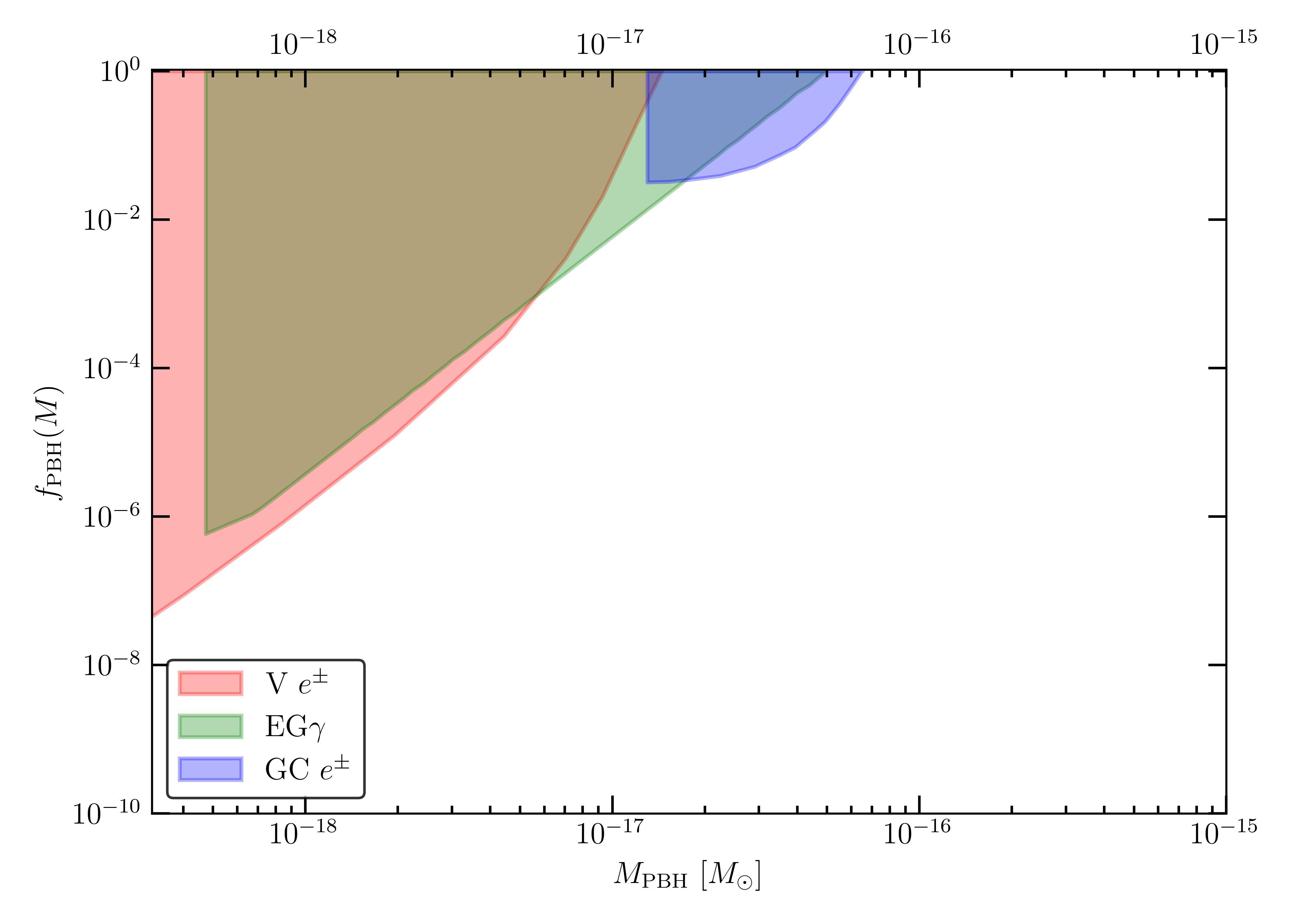} \\
    \includegraphics[width=0.6 \textwidth]{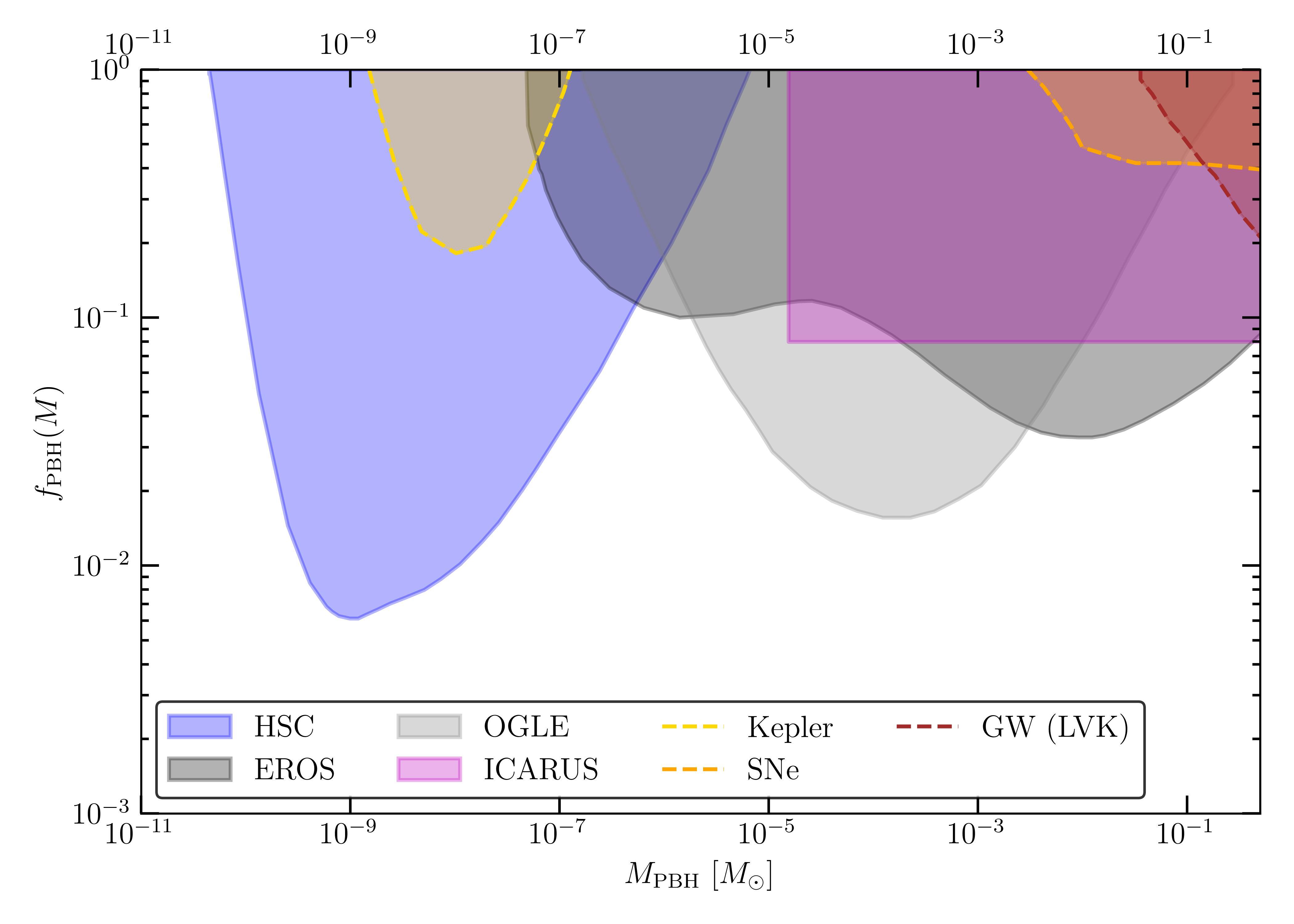} \\
    \includegraphics[width=0.93 \textwidth]{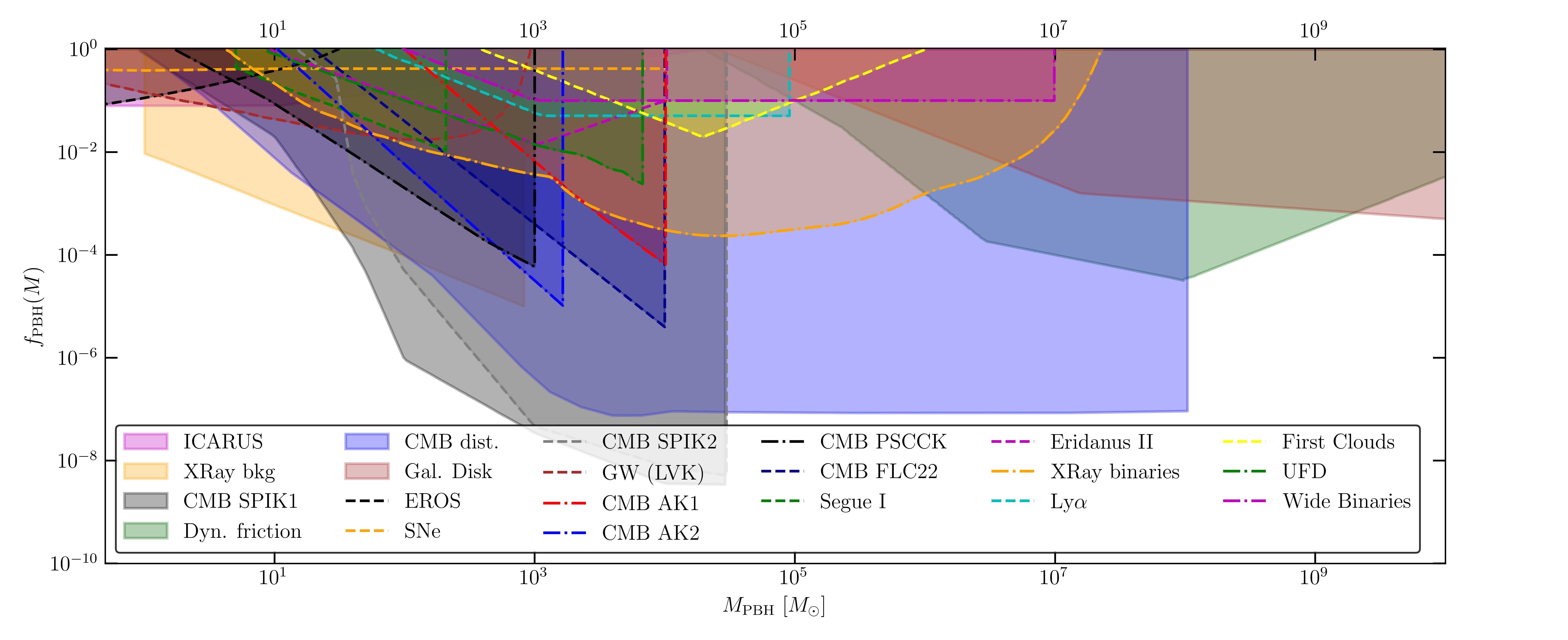}
    \caption{Zooms over some claimed limits on the PBH abundance $f_{\rm PBH}$ for a monochromatic distribution of mass $m_{\rm PBH}$, in the asteroid-mass range where constraints are dominated by various probes of PBH Hawking evaporation (top panel), in the planetary-mass and low stellar-mass range up to $m_{\rm PBH }\sim 10 M_\odot$ coming from microlensing surveys (middle panel), and in the range from stellar-mass up to the supermassive PBHs, from a combination of accretion, dynamical, GW and indirect constraints (bottom panel).  The legend indicates the origin of each represented limit.  It is worth noticing that all those limits are subject to important uncertainties and can be highly model dependent, moving up and down with different model and astrophysical assumptions.  The possible limitations and sources of uncertainties are discussed in the text. } 
\label{fig:PBHlimitszoom}
\end{figure}

\subsection{PBH evaporation}

We review here the PBH constraints relying on their evaporation.  {Let us note that the PBH mass is expressed in grams through this Section, and not in solar mass as for all the other probes, for a better consistency with the literature  and because these constraints concern PBHs much lighter than a solar mass. }  

\subsubsection{Big-bang nucleosythesis (BBN)}  

PBHs with masses $10^9-10^{13}\mathrm{g}$ can affect the abundance of light elements produced during 
the Big-Bang Nucleosynthesis through Hawking radiation and the related emitted 
particles, allowing to set constraints on the abundance of low-mass PBHs~\cite{Carr:2020gox}. In particular, for PBHs with masses $m_\mathrm{PBH} \approx 10^9-10^{10}\mathrm{g}$, Hawking radiated mesons and antinucleons induce extra interconversion processes of protons to neutrons, increasing the neutron-to-proton ratio at the time of freeze-out of the weak interaction~\cite{1978SvA....22..138V}, and consequently leading to an increase in the final $^4\mathrm{He}$ abundance~\cite{1978PThPh..59.1012M}. Regarding PBHs with masses $m_\mathrm{PBH}\approx 10^{10}-10^{12}\mathrm{g}$,  long-lived high energy hadrons produced out of PBH evaporation, such as pions, kaons and nucleons, remain long enough in the ambient medium to trigger dissociation processes of light elements produced during BBN~\cite{Kohri:1999ex}, reducing the abundance of $^4\mathrm{He}$ and increasing the one of $\mathrm{D}$,$^3\mathrm{He}$,$^6\mathrm{Li}$ and $^7\mathrm{Li}$. Finally, for PBHs with $m_\mathrm{PBH}\approx 10^{12}-10^{13}\mathrm{g}$, photons produced from the particle cascade process further dissociate $^4\mathrm{He}$, increasing the abundance of light synthesized elements~\cite{Keith:2020jww,Acharya:2020jbv}. However, it is important to stress that the BBN constraints carry out some uncertainties regarding the baryon-to-photon ratio, the reaction and the decay rates of the elements produced during the BBN processes~\cite{Kawasaki:2004qu,Kawasaki:2004yh}.

\subsubsection{Gamma-rays and neutrinos from Hawking radiation }

The lifetime of a PBH due to its evaporation in %the SM of particle physics 
standard model particles can be approximated as \cite{Carr:2009jm}
\begin{equation}
    \tau\approx400\left(m_{\rm PBH}\over 10^{10}\, {\rm g}\right)^3 {\rm s}.
\end{equation}
For a PBH with mass $5\times10^{14}$ g, 
%that formed shortly after the Big Bang, 
its lifetime is close to the age of Universe. Thus, such low-mass PBHs  can be probed by looking for bursts of gamma rays. Such searches have been performed by the Water Cherenkov Observatories, Imaging Atmospheric Cherenkov Telescopes (IACTs) and gamma-ray space observatories. So far, no gamma-ray signal that could be due to PBH evaporation has been detected. The current limits are summarized in Table~\ref{tab:pbh_gamma}. These are expressed in terms of the PBH burst rate densities $\dot B_{\rm PBH}$, describing the number of bursts from PBH evaporation in the local Universe in a unit of volume per  unit of time.

The observations of Galactic gamma-ray emission by INTEGRAL~\cite{Laha:2020ivk} also put a bound on the mass above which PBHs could constitute the entirety of DM, $m_\PBH > 1.2\times10^{17}$g, under some relatively conservative assumptions. 
New analysis of the diffuse soft gamma-ray emission towards the inner Galaxy measured by INTEGRAL/SPI over 16 years \cite{Berteaud:2022tws}, implementing a  spatial template fit of SPI data, an improved instrumental background model, and including the spatial distribution of MeV photons emitted by the evaporating PBHs into the fitting procedure, strengthens this lower bound to $ m_\PBH \gtrsim 4\times10^{17}$g.

The prospects for detection of gamma-ray signals from PBH evaporation for the future Southern Wide Field of View Gamma-ray Observatory (SWGO) have been estimated in \cite{Lopez-Coto:2021lxh}: it will be sensitive to local PBH burst rate densities $\dot B_{\rm PBH}$ of the order of $\sim$ 50 pc$^{-3}$yr$^{-1}$, improving by more than one order of magnitude the limits from current experiments.

In addition to gamma-rays, high energy extra-galactic neutrinos can also be used to put constraints on the local PBH burst rate density $\dot B_{\rm PBH}$ for both non-rotating ($a^{\star}=0$) and rotating ($a^{\star}=0.99$) black holes. The 90\% C.L. limits from 10-years $\nu_{\mu}$ IceCube data~\cite{2021JCAP...12..051C} are presented in Table~\ref{tab:pbh_gamma}. 

\begin{table}
\begin{tabular}{|c|c|c|c|c|}
\hline
Experiment&$\dot B{\rm PBH}$ Upper Limit [pc$^{-3}$yr$^{-1}$]&Burst duration [s]&$T_{\rm obs}$&Reference\\
\hline
Milagro&$3.1\times10^5$ (99\% CL)&0.001&1673 days&\cite{Abdo:2014apa}\\
Milagro&$1.2\times10^4$ (99\% CL)&0.01&1673 days&\cite{Abdo:2014apa}\\
Milagro&$5.4\times10^4$ (99\% CL)&0.1&1673 days&\cite{Abdo:2014apa}\\
Milagro&$3.6\times10^4$ (99\% CL)&1&1673 days&\cite{Abdo:2014apa}\\
Milagro&$3.8\times10^4$ (99\% CL)&10&1673 days&\cite{Abdo:2014apa}\\
Milagro&$6.9\times10^4$ (99\% CL)&100&1673 days&\cite{Abdo:2014apa}\\
VERITAS&$2.22\times10^4$ (99\% CL)&30&747 hours&\cite{Archambault:2017asc}\\
H.E.S.S.&$4.9\times10^4$ (95\% CL)&1&2600 hours&\cite{Glicenstein:2013vha}\\
H.E.S.S.&$1.4\times10^4$ (95\% CL)&30&2600 hours&\cite{Glicenstein:2013vha}\\
H.E.S.S.&$5.6\times10^4$ (99\% CL)&10&2860 hours&\cite{Tavernier:2019exh}\\
H.E.S.S.&$2.0\times10^3$ (95\% CL)&120&4816 hours&\cite{HESS:2023zzd}\\
Fermi-LAT&$7.2^{+8.1}_{-2.4}\times10^3$ (99\% CL)&$1.26\times10^8$&$1.26\times10^8$ s&\cite{Fermi-LAT:2018pfs}\\
HAWC&$3.3^{+0.3}_{-0.1}\times10^3$ (99\% CL)&0.2&959 days&\cite{Albert:2019qxd}\\
HAWC&$3.5^{+0.4}_{-0.2}\times10^3$ (99\% CL)&1&959 days&\cite{Albert:2019qxd}\\
HAWC&$3.4^{+0.4}_{-0.1}\times10^3$ (99\% CL)&10&959 days&\cite{Albert:2019qxd}\\
IceCube&$1.7\times10^7$(90\% CL, $a^{\star}$=0)&10&10 years&\cite{2021JCAP...12..051C}\\
IceCube&$8.2\times10^6$ (90\% CL, $a^{\star}$=0)&$10^2$&10 years&\cite{2021JCAP...12..051C}\\
IceCube&$4.1\times10^6$ (90\% CL, $a^{\star}$=0)&$10^3$&10 years&\cite{2021JCAP...12..051C}\\
IceCube&$1.5\times10^7$ (90\% CL, $a^{\star}$=0.99)&10&10 years&\cite{2021JCAP...12..051C}\\
IceCube&$6.2\times10^6$ (90\% CL, $a^{\star}$=0.99)&$10^2$&10 years&\cite{2021JCAP...12..051C}\\
IceCube&$3.3\times10^6$ (90\% CL, $a^{\star}$=0.99)&$10^3$&10 years&\cite{2021JCAP...12..051C}\\
\hline
\end{tabular}
\caption{Current limits on the local PBH burst rate density from different experiments.}
\label{tab:pbh_gamma}
\end{table}

\subsubsection{Positron annihilation in the Galactic Center}

PBH limits can be obtained with observations of the 511 keV line from the Galactic Center with INTEGRAL.  Positrons produced by evaporating PBHs  subsequently propagate throughout the Galaxy.  As they annihilate, they  contribute to the Galactic 511 keV line.  Both monochromatic and log-normal mass distributions have been considered, e.g. giving the corresponding limit of $f_{\rm PBH}<1$ for $m_{\rm PBH} \lesssim 2 \times 10^{17}$ g for a log-normal mass distribution with $\sigma=0.5$ and $m_{\rm PBH}< 2 \times10^{17}$ g with $\sigma=0.1$ \cite{DeRocco:2019fjq}.
%\SC{why not same ref as in the figure?}
Estimates for different background assumptions \cite{Ballesteros:2019exr} show that for the low ($\times0.1$)
%\SC{where does this come from}
power law background model, $f_{\rm PBH}$ can reach 1 up to $m_{\rm PBH}\sim 5\times10^{17}$g. Estimates for different mass distributions and spins \cite{Dasgupta:2019cae} give the mass limit $m_{\rm PBH} \sim3-5\times10^{18}$ g for the log-normal mass distribution with $\sigma=1$, $a^*=0.5-1$ and $f_{\rm PBH}= 1$.
It has been recently shown \cite{Keith:2022sow} that future MeV-scale gamma-ray observatories e-ASTROGAM or AMEGO would be able not only to detect the Hawking radiation from the Inner Galaxy, but also to precisely measure the abundance and mass distribution of PBHs responsible for the Galactic 511 keV line.

\subsubsection{$e^\pm$ observations by Voyager 1}

The measurements of interstellar low-energy (sub-GeV) $e^\pm$ flux are used to constrain the PBH fraction of DM in the Galaxy \cite{Boudaud:2018hqb}. These electrons and positrons would be due to PBH evaporation.   The obtained limits are based on local measurements, independent of the cosmological model.  This flux must be shielded for Earth-bounded detectors by the magnetic field of the Sun.  However, limits can be obtained from Voyager-1 measurements since it is now beyond the heliopause. 
 
 The Hawking radiation flux of $e^\pm$ is computed within the fully general diffusion-convection-reacceleration  model  of  propagation with the parameters adjusted to the AMS-02 data. For the log-normal mass distribution of PBHs with a width $\sigma$ larger than 1 and a central value $\mu\lesssim10^{17}$ g, a PBH fraction larger than 1\% of DM is excluded by the Voyager 1 data.

\subsubsection{Ultra-faint dwarf galaxies}

The ultra-faint dwarf galaxies (UFDGs), such as Leo T, also provide some limits on light, evaporating PBHs, through the heating of the interstellar medium gas~\cite{Kim:2020ngi}.  The emitted energetic particles can heat and ionize the gas, in particular the fast electrons can deposit a substantial fraction of their energy and heat the interstellar gas through the Coulomb interaction.  The ultra-faint dwarf galaxy Leo T is of particular interest, not only because it is highly dark matter dominated but also because it has a large amount of neutral hydrogen and a relatively low metal abundance.  In~\cite{Kim:2020ngi}, limits are obtained for non-spinning PBHs, with masses between { $2 \times 10^{15} $ g and $ 10^{17} $~g} as low as $f_{\rm PBH} \lesssim 10^{-3}$, i.e. more stringent than the limits from galactic and extragalactic emission.  The possible spin of PBHs would affect the Hawking radiation, generally increasing the emission of particles, in particular ones with large spins. In comparison, stronger limits are obtained for spinning PBHs~\cite{Laha:2020vhg} than for non-spinning ones.  In both cases, $f_{\rm PBH}=1$ is claimed to be excluded for $m_{\rm PBH}\sim (2-2.5)\times10^{16}$g. 

\subsubsection{CMB limits on Dark radiation}

Evaporating PBHs in the mass range $[3\times 10^{13}, 5\times 10^{16}]$ g should have injected an exotic amount of electromagnetic energy in the Universe, which has an impact on the CMB anisotropies.  This effect has been computed in~\cite{Stocker:2018avm}, using the \texttt{ExoCLASS} code that includes electromagnetic cascade calculations, initially developed in the context of WIMP annihilation.   For a monochromatic mass function, the obtained bounds exclude PBHs as the main source of dark matter between $3 \times 10^{13}$ and $2.5 \times 10^{16}$g, and are up to several orders of magnitude more stringent than the limits from the galactic gamma-ray background between $3 \times 10^{13}$ and $3 \times 10^{14}$ g.  A similar analysis was conducted in~\cite{Acharya:2020jbv}, leading to limits at slightly lower masses, around $1.1 \times 10^{13}$ g.  See also~\cite{Lucca:2019rxf} for the connection with CMB spectral distortions induced by the energy injection from evaporating PBHs.  The non-standard energy injection problem has also been reformulated to account for an extended PBH mass distribution in~\cite{Poulter:2019ooo}.  In comparison with a monochromatic mass function, the marginalised and log-normal ones have tighter exclusion bounds for heavier PBH masses, while at low PBH masses these two distributions exhibit weaker bounds than the delta distribution. In the future, experiments dedicated to the detection of the CMB spectral distortions or of the 21cm signal from the dark ages could improve these limits. Finally, it has been pointed out in~\cite{Nesseris:2019fwr} that the time-dependence of the matter density, as the PBH density decreases as they lose mass via Hawking radiation, can be formulated as an equivalent early dark energy model with varying equation of state $w(z)$.  For instance, a population of ultra-light PBHs, decaying around neutrino decoupling, leads to a dark matter-radiation coupling altering the expansion history of the Universe, and alleviating the $H_0$ tension.

The rate of mass loss from particle emission by the PBHs through Hawking radiation is then computed, and found to be negligible compared to the total PBH mass. The electrons, positrons and photons are the particles emitted from PBHs that interact most strongly with the primordial plasma, so actually the matter of concern is the effect the emitted particles have on the early Universe. After obtaining the formula for the total energy injected into the primordial plasma by evaporating PBHs, the effects on CMB anisotropies are then examined, where only the ionisation and excitation of hydrogen atoms are considered (helium recombines earlier and has a smaller effect in the CMB).

\subsubsection{Neutrinos}

For PBH masses between $10^{15}$ and $10^{17}$ g, black hole evaporation via Hawking radiation leads to sizeable fluxes of MeV neutrinos that could be detected with the next generation of detectors like Hyper-Kamiokande~\cite{Hyper-Kamiokande:2018ofw}, JUNO~\cite{JUNO:2015zny}, DUNE~\cite{DUNE:2016hlj} and THEIA~\cite{Theia:2019non}.  Current experiments like SuperKamiokande already provide some limits on the PBH abundance~\cite{Dasgupta:2019cae,Bernal:2022swt}, but they are not yet competitive with gamma-ray limits.  
%\SC{Add them in the figure.} 
Those future experiments will have the ability to set complementary limits in the asteroid-mass range.  The expected sensitivity of these experiments (Hyper-Kamiokande, JUNO, DUNE and THEIA) with various designs has been calculated in~\cite{Wang:2020uvi,DeRomeri:2021xgy,Bernal:2022swt} for both monochromatic and log-normal PBH mass functions.   Abundances as low as $10^{-6}$ could be probed for PBH with zero spin and even lower if PBHs have significant spin.  DUNE and THEIA would also complement each other, because they are respectively sensitive to neutrino and antineutrino fluxes.   
Finally, neutrinos resulting from PBHs could also be observed in direct detection experiments, through coherent
elastic neutrino-nucleus scattering, see~\cite{Calabrese:2021zfq}.

\subsubsection{21-cm signal}

The 21cm line is an electromagnetic radiation spectral line created by an atomic transition between the two hyperfine levels of the $1s$ ground state of neutral hydrogen.  This spectral line is of great interest and importance in cosmology as it seems to be the only known way to probe the dark ages of the Universe from the epoch of recombination to reionization. Observations of 21cm are known to be very difficult to make, primarily because it is a faint signal and is usually plagued by interferences from various sources. However, a global 21cm signal 
was claimed to be detected by the Experiment to Detect the Global Epoch of Reionization Signature (EDGES), which shows an absorption feature with an amplitude of $T_{21} \sim 500\, {\rm mK}$ centered at redshift $z \sim 17$, %and is% 
much stronger (by about a factor of 2) than expected in the $\Lambda$CDM model~\cite{Bowman:2018yin}. Various explanations, such as interactions between the dark matter particles and baryons, have been put forward to explain this anomalous signal. 
 
In general, during that epoch, any additional sources, e.g. the dark matter annihilation or decay, will heat up the intergalactic medium (IGM) and increase the kinetic temperature that eventually determines the strength of the 21cm signal. Therefore, observations of the EDGES experiment can, in principle, impose constraints on the properties of a dark matter scenario. 
It has been discussed that, in the context of 21cm cosmology, PBHs can lead to multiple distinct observable effects: (i) the heating of
the IGM due to their Hawking evaporation; (ii) the modification of the primordial power spectrum (and hence the halo mass function) induced by Poisson noise due to PBHs being point sources; (iii) a uniform heating and ionization of the IGM due to X-rays produced during accretion and (iv) a local modification of the temperature and density of the ambient medium surrounding isolated PBHs.
Due to their Hawking evaporation, PBHs can affect the evolution of the IGM, and therefore their mass fraction can be constrained by the global 21cm signal. Using this, the initial mass fraction of PBHs is constrained to be $\beta_{\rm PBH} \sim 2 \times 10^{-30}$ for very light PBHs in the mass range $6\times 10^{13}\, {\rm g} \lesssim m_{\rm PBH} \lesssim 3 \times 10^{14} \,{\rm g}$~\cite{Yang:2020egn}.  
%\SC{Add them in the figure.  Why $\beta$ and not $f_{\rm PBH}?$}
%Since the dominant 21cm signal is expected to arise from star-forming halos, it has been shown that the Poisson noise in the power spectrum is not relevant and thus can be ignored. 
%For accretion onto isolated PBHs, the accretion effects only become noticeable for PBHs with mass $M > {\cal O}(10) M_\odot$. 
%Evaporating PBHs would inject energy into the intergalactic medium and heat the gas at high redshifts, during the dark ages and the cosmic dawn.  As a result, this would modify the absorption or emission line of the global 21-cm signal.  The possible absorption feature in the 21-cm signal observed by EDGES can therefore provide a bound on the abundance of PBHs. 
 Considering PBH evaporation, in~\cite{Mittal:2021egv} the following limits have been found : $f_{\rm PBH}=10^{-6.84}(m_{\rm PBH}/10^{15}\,$g)$^{3.75}$ in absence of X-ray heating and $f_{\rm PBH} \leq 10^{-9.73}(m_{\rm PBH}/10^{15}\,$g)$^{3.96}$ in presence of X-ray heating. The mass and spin are fundamental properties of a black hole, and they can substantially affect the BH evaporation rate.  In~\cite{Natwariya:2021xki}, the lower limits on masses of PBHs allowed to entirely constitute the DM have been found to be 1.5$\times10^{17}$ g, 1.9$\times10^{17}$ g, 3.9$\times10^{17}$ g and 6.7$\times10^{17}$ g for PBH spins 0, 0.5, 0.9 and 0.9999, respectively. 

\subsubsection{Limits from DM particle production}

If evaporating PBHs are at present times a source of the boosted light DM particles, then the constraints on PBH mass and abundance can be obtained from the XENON1T data~\cite{Calabrese:2021src}. Considering DM masses smaller than 1 MeV, for low DM-nucleon cross-sections, the constraints on $f_{\rm PBH}$ are almost inversely proportional to the cross-section. However, for larger cross-sections the results might depend on the model of the DM propagation through Earth and atmosphere. The constraints are valid for the mass range $0.5\lesssim m_{\rm PBH}/(10^{15} g)\lesssim 8.0$: the kinetic energy of particles emitted by more massive PBHs is too low, the less massive PBHs have evaporated nearly completely. For the DM particles interacting with electrons similar analyses are presented in \cite{Li:2022jxo,Calabrese:2022rfa}.
%\SC{We should expand this subsection.}

\subsubsection{Discussion and limitations}

It is important to notice that all the limits on the PBH abundance in the asteroid mass range rely on the common assumption that PBHs evaporate through Hawking radiation.  If the process of black hole evaporation is well established theoretically, its existence is not proven observationally.  Furthermore, it is still unclear whether quantum gravity effects would affect or not this process (see for example \cite{Anchordoqui:2022txe,Anchordoqui:2022tgp}, where it is shown that the existence of one mesoscopic extra-dimension leads to a longer life-time of the PBHs and one can have an all-dark-matter interpretation in terms of PBHs in the mass range $10^{14}\lesssim m_{\rm PBH}/g\lesssim 10^{21}$).   The efficiency of the evaporation through the different channels can also depend on the underlying particle physics theory.  It is therefore important to keep in mind these possible limitations.  Nevertheless, it is also worth noticing that such very light PBHs, if they exist, are the ideal target to probe and test the Hawking radiation hypothesis for the first time. 

The existence and importance of the Hawking radiation are not the only source of uncertainty on the abundance limits from PBH evaporation.   The different source of uncertainties have been recently reviewed and critically analyzed by J. Auffinger in~\cite{Auffinger:2022dic}, especially coming from the instrument characteristics, the prediction of the (extra)galactic photon flux, the statistical method of signal-to-data comparison and the computation of the Hawking radiation rate.  The constraints on PBHs are found to vary by several orders of magnitude, depending on the hypothesis.  Even with an "ideal" experiment, PBHs can only be probed through Hawking radiation at masses below $10^{20}$ g, which is an intrinsic limitation of the phenomena.   

In order to obtain those results, study the model dependence of the limits related to Hawking radiation and include all these uncertainties in the data analysis, the \texttt{BlackHawk}~\cite{Arbey:2019mbc} and \texttt{Isatis}~\cite{Auffinger:2022dic} codes have been developed, allowing to compute the evaporation primary and secondary spectra of stable or long-lived particles, for any black hole distribution.  We envisage to interface the PBH numerical tool developed in the context of the present review paper with these key tools. 

\subsection{Microlensing searches}

One of the most stringent constraints on PBH as constituents of the dark matter in the Universe comes from the microlensing amplification of light from a distant star as a Massive Compact Halo Object (MACHO) moves across its line of sight. Microlensing events from lenses of less than one solar mass towards the Large Magellanic Cloud (LMC) where reported in the 1990s by several groups, but in the subsequent decades only upper bounds were published, until SUBARU and later OGLE reported events towards the galactic bulge and nearby Andromeda. The constraints on the halo abundance of those objects depend very much on the mass, velocity and spatial distribution in the halo. Most of these constraints are reported for a standard Navarro-Frenk-White (NFW) halo with virial velocities and a spatially uniform distribution of monochromatic (same mass) MACHOs, which put stringent constraints in the range of masses from $10^{-9}$ to $30\,M_\odot$. Above 10 $M_\odot$, it is essentially impossible to derive bounds on MACHOs due to the fact that the amplification light curve lasts more than a decade, while no microlensing survey has monitored continuously the light from stars for so long. Below $10^{-9}\,M_\odot$, the wavelength of visible light is comparable to the Schwarzschild radius of the compact object, and we enter the regime of wave optics, where diffraction makes the constraints essentially disappear. One could search for PBHs with $m_{\rm PBH} < 10^{-9}\,M_\odot$ with light of shorter wavelength, like X-rays or gamma-rays, but sources that emit in those bands are typically not as abundant and stable as stars.

In the following Subsections, we give a brief summary of the different microlensing constraints coming from different surveys.

\subsubsection{Microlensing from stars in the Magellanic clouds}

Microlensing observations of stars in the Large and Small Magellanic Clouds probe the fraction of the Galactic halo in MACHOs in a certain mass range \cite{Paczynski:1985jf}. The optical depth of the halo towards LMC and SMC is related to %the fraction $f(M)$%
{$f_{\rm PBH}$} 
by $\tau^{\rm SMC}_{\rm L} = 1.4 \tau^{\rm LMC}_{\rm L} = 6.6 \times 10^{-7} f_{\rm PBH}$ 
%f(M)$ 
for the standard halo model \cite{Alcock:2000kd}. The MACHO project detected lenses with mass {around} $0.5 M_\odot$ but concluded that their halo contribution could be at most $10\%$ \cite{Hamadache:2006fw}, while the EROS project excluded $6\times 10^{-8} M_\odot < m_{\rm PBH} < 15 M_\odot$ from dominating the halo \cite{Allsman:2000kg,EROS-2:2006ryy}. Since then further limits in the range $0.1 M_\odot < m_{\rm PBH} < 20 M_\odot$ have come from the OGLE experiment \cite{Wyrzykowski:2010mh,Wyrzykowski:2011tr}. 

\subsubsection{Microlensing from stars in the galactic bulge}

{Recently Niikura et al. \cite{Niikura:2019kqi} have used data from a 5-year OGLE survey of the Galactic bulge to place much stronger limits in the range $10^{-6}M_\odot < m_{\rm PBH} < 10^{-4}M_\odot$, although they also claim some positive detections.}
{The galactic bulge is a dense region of stars towards the galactic center where microlensing events are likely to occur. In fact, the OGLE collaboration has detected numerous microlensing events towards the bulge~\cite{Wyrzykowski:2015ppa}, searching specifically for dark objects that amplify the light of stars behind them. Moreover, these stars are sufficiently close to Earth that one can use paralax measurements by GAIA to break degeneracies between mass and distance which plague magnification microlensing events. This use of the so-called astrometric microlensing has allowed Ref.~\cite{Wyrzykowski:2019jyg} to conclusively detect black holes in the mass gap, in the range $2 - 5~M_\odot$. Niikura et al.~\cite{Niikura:2019kqi} also report microlensing events towards the galactic bulge in a lower mass range of Earth-mass primordial black holes. These are tantalizing hints of a multimodal population of primordial black holes in the halo of our galaxy, where non of the mass ranges are sufficient to comply 100\% of all the dark matter halo.}

\subsubsection{Microlensing from stars in Andromeda}

{The Andromeda galaxy (M31) is too far away to resolve individual stars with a typical dedicated (small, 2 meter) telescope, as most microlensing surveys do. Therefore they use a different technique called pixel lensing, which corresponds to single pixels brightening up over the length of the survey with a characteristic Paczinsky amplification curve due to a passing compact object. This technique was used by the POINT-AGAPE collaboration~\cite{POINT-AGAPE:2005swi}, who reported six microlensing events, which they argued could be interpreted as self-lensing of stars and this allowed them to conclude that 20\% of the Milky Way halo in the direction of M31 could be in the form of $0.5 - 1.0~M_\odot$ primordial black holes.}

{More recently, it was reported by~\cite{Niikura:2017zjd} the observation, with the Hyper Supreme Cam (HSC) on Subaru Telescope, of a microlensing event towards Andromeda by a compact body with mass in the range $10^{-10} - 10^{-5}~M_\odot$. Give the detailed statistics of the observation, they claimed this could provide evidence for a non-zero component, at the level of few percent, for the dark matter halo to be in the form of primordial black holes.}

\subsubsection{Microlensing from quasars}

Hawkins has claimed for many years that quasar microlensing data suggest the existence of PBH dark matter \cite{Hawkins:1993yud}.  He originally argued for Jupiter-mass PBHs but later increased the mass estimate to $0.4 M_\odot$ \cite{Hawkins:2006xj}. More recently, the detection of 24 microlensed quasars \cite{Niikura:2019kqi} suggests that up to  25\% of galactic halos could be in PBHs with mass between $0.05$ and $0.45 M_\odot$. These events could also be explained by intervening stars, but in several cases the stellar region of the lensing galaxy is not aligned with the quasar, which suggests a population of subsolar halo objects with $f_{\rm PBH} \gtrsim 0.01$. Indeed, Hawkins has argued that the most plausible microlensers are PBHs, either in galactic halos or distributed along the lines of sight to the quasars \cite{Hawkins:2020zie}. 

\subsubsection{Microlensing of supernovae}

One can also constrain the abundance of PBHs with stellar masses using supernovae surveys.  A certain number of supernova light curves are expected to be lensed by PBHs, exhibiting a magnification of order $\mu \sim 0.1$ if PBHs constitute all the dark matter~\cite{Zumalacarregui:2017qqd}.  Nevertheless, it has been debated whether the current JLA and Union data already exclude $f_{\rm PBH} = 1$, especially in the case of broad PBH mass distributions~\cite{Garcia-Bellido:2017imq}. The main discrepancy has to do with the assumed statistics and the treatment of outliers and other systematics, like the probability of amplification due to discrete sources along the line of sight~\cite{Bosca:2022viy}. Interestingly, the SNe limits are of the same order from $10 M_\odot$ up to $10^4 M_\odot$.

\subsubsection{Femtolensing}

Additional limits on the PBH abundance have been established based on the femtolensing of gamma-ray bursts.   Compared to microlensing, the terminology of femtolensing is used to refer to the very small angular distance between lensed images and applies to objects of very small masses.  The effect is probed by searching for interferometry patterns in the energy spectrum of lensed objects~\cite{1992ApJ...386L...5G}. One must nevertheless take into account that the wavelength of the electromagnetic radiation can become comparable to the Schwarzschild radius of the PBH, which implies a lower bound on the mass range probed with this technique.   In~\cite{Barnacka:2012bm}, a search for femtolensing events has been performed on the FERMI observations of gamma-ray bursts for which the redshifts were known.  The absence of femtolensing events has been used to set constraints on the dark matter fraction in PBHs of mass between $5 \times 10^{17}$ g and $10^{20}$ g, with a limit down to $f_{\rm PBH} \lesssim 0.04$.

\subsubsection{Discussion and limitations }

The robustness of microlensing limits is of very high importance and still controversial because this is the only observational technique to probe the abundance of PBHs smaller than a solar mass, down to $10^{-10} M_\odot$.  Another debated question is whether microlensing surveys of quasars or towards the galactic center only provide a limit on PBHs due to the lack of observations or also hint at their existence due to a series of observations.

The contribution of PBHs to the dark matter is also an interesting question.  Deriving limits from microlensing surveys of stars in the Magellanic clouds depends on numerous astrophysical assumptions (velocity dispersion, galactic dark matter profile, etc...)  and varying limits can be obtained with different hypothesis~\cite{Green:2016xgy,Green:2017qoa,Hawkins:2015uja,Hawkins:2020zie,Calcino:2018mwh}.  Wide mass distributions can also significantly alter the limits on their abundance, but without totally removing them.  
The clustering of PBHs was the main argument invoked to totally suppress the microlensing limits~\cite{Garcia-Bellido:2017xvr,Carr:2019kxo}, even though it has been shown that this is not the case assuming initial Poisson distributed PBHs~\cite{Petac:2022rio, Gorton:2022fyb}. 
Furthermore, in the case of a large initial spatial clustering distribution beyond Poisson, it has been also recently shown that the combination of constraints from microlensing and Lyman-$\alpha$ forest completely rules out the parameter space for stellar-mass PBHs to be a dominant constituent of the dark matter~\cite{DeLuca:2022uvz}.

\subsection{Dynamical limits}

\subsubsection{PBH capture by neutron stars or white dwarfs }

The asteroid-mass range, between $10^{-16} M_\odot$ and $10^{-10} M_\odot$ ($10^{17} - 10^{23}$g), is the only one for which there is a general consensus that PBHs can contribute up to the entirety of the dark matter.  Nevertheless, if this mass window remains open nowadays, this has not been always the case.  Some constraints had been proposed, coming from observations of neutron stars and white dwarfs (WD) in globular clusters~\cite{Capela:2012jz,Capela:2013yf,Pani:2014rca,Defillon:2014wla} that can capture asteroid-mass PBHs falling at the center of the star and rapidly swallowing its material until it becomes a black hole.  Such constraints therefore arise from the combination of a dynamical and an accretion process. But those limits have been relieved due to large uncertainties. It was initially claimed that $f_{\rm PBH}$ cannot exceed between $10^{-2}$ and $0.2$, depending on the DM density in globular clusters.  However, this approach is limited by our knowledge of the DM constitution of globular clusters.  Using updated prescriptions, the limits from~\cite{Capela:2013yf,Pani:2014rca} have been then relieved in~\cite{Montero-Camacho:2019jte}, re-opening the asteroid-mass range to full PBH-DM scenarios.

In \cite{Graham:2015apa}, it was proposed that the transit of PBHs through a white dwarf causes localized heating around the BH trajectory due to dynamical friction, eventually triggering a runaway thermonuclear fusion and causing the WD to explode as a supernova, even if the WD mass is below the Chandrasekhar limit. The constraints on PBHs were derived by using two classes of observations. One class comes from direct observations of WDs with known masses, whose existence places bounds on PBHs that are abundant enough that they would have transited the WD with high probability. The second class arises from the measured rate of type Ia supernovae, that constrain PBHs with a low abundance and thus have a low probability of transit through a WD. The shape of the observed distribution of WD masses (with masses up to $1.25 M_\odot$) excludes PBHs with masses $\sim 10^{19}-10^{20}$g from being a dominant constituent of the local DM density, whereas the type Ia supernova rate disfavors BHs with masses between $10^{20}-10^{22}$g. However, the latter is not a robust exclusion due to uncertainties in the methods used to determine the WD population with masses larger than $\sim 0.85 M_\odot$ (necessary condition for a WD to explode due to a BH), and this mass could possibly change if the current simulation data used to derive it is refined. Measurements of WD binaries in GW observatories could be used in the future to further strengthen the BH bounds. 

{More recently, limits on PBH in the interesting asteroid-mass range between $10^{18}$ g and $10^{21}$ g based on the capture of PBHs by Sun-like stars in ultra-faint dwarf galaxies leading to its destruction, have been calculated in~\cite{Esser:2022owk}.  If one requires that no more than a fraction of stars $\xi = 0.5$ is destroyed, one obtains a limit on the PBH abundance of about $f_{\rm PBH} \lesssim 0.3$ in the above-mentioned mass range.  But the constraints strongly depend on the possible value of $\xi $ that is still fairly unknown.  In~\cite{Smirnov:2022zip}, it was suggested that the recent discovery of a population of faint supernovae (Ca-rich gap transients) may come from the explosion of white dwarfs, possibly due to PBH captures.  Their unusual spatial distribution, predominantly at large distances from their presumed host galaxies, supports this hypothesis and the authors have shown that their spatial distribution matches well the distribution of dwarf spheroidal galaxies.   }
%\SC{Include exclusion line in figure?}

%and BHs with masses of $10^{20}-10^{22}$g can also contribute importantly to the observed rate of type 1a supernovae.

\subsubsection{Ultra-faint dwarf galaxies} 

If PBHs above $10 M_\odot$ constitute an important fraction of the dark matter, their regular encounters in highly dark matter dominated ultra-faint dwarf galaxies (UFDGs), such as in Segues I or in Eridianus II, should have induced their dynamical heating, allowing them to reach a half-light radii larger than 10 parsecs.  This argument also applies to star clusters in UFDGs, like for Eridanus II. Relatively stringent limits from UFDGs have been first established by Brandt in~\cite{Brandt:2016aco}, for monochromatic mass functions.  These have been extended to log-normal PBH mass functions, including the effect of a central intermediate-mass BH, by Green in~\cite{Green:2016xgy}, or by Li et al (DES collaboration) in~\cite{DES:2016vji}.  All these works basically exclude $f_{\rm PBH} = 1$ for $m_{\rm PBH} > 10 M_\odot$.  These limits are relatively solid because they rely on well-known Newtonian gravitational dynamics.  Nevertheless, one should point out that these limits can be shifted if UFDGs host a central intermediate-mass black hole.  

Another way to constrain PBHs from UFDGs is to consider the radial luminosity profile, given that stars must also be heated up by PBHs.   In this case, Segue 1 provides the most stringent limit, excluding $f_{\rm PBH} =1$ for $m_{\rm PBH} \sim 2 M_\odot$~\cite{Koushiappas:2017chw,Stegmann:2019wyz}.  However, these limits rely on astrophysical assumptions about the stellar mass distribution and they could change for wide-mass PBH models, depending on the existence and mass of a central intermediate-mass black hole, etc.  Finally,  UFDGs also provide an argument in favor of solar-mass PBHs constituting an important fraction of the dark matter, given that there is no observation of such objects with a radius less than about 20 parsecs, whereas they would still be above the magnitude limit of the current instruments.  
%\SC{Add refs.}

\subsubsection{Dark Matter profile of dwarf galaxies (core-cusp problem) }

The regular PBH-PBH interactions in a dense environment should prevent the formation of cusps in the central region of dwarf galaxies, possibly explaining the core-cusp problem.  As first pointed out in~\cite{Clesse:2017bsw}, the gravitational scattering cross-section for PBHs in the stellar-mass range are of order $(0.1-1) {\rm cm^2/g}$ and so PBHs are in essence a well motivated self-interacting dark matter model.  But contrary to particle dark matter, this does not need exotic physics but relies on the only well-known interaction that is not included in the standard model of particle physics, gravitation.  The core-cusp problem in the context of PBHs was then further explored in~\cite{Boldrini:2019isx}.  They found that the mechanism works well for PBHs of mass between $20$ and $100 M_\odot$, if they have $f_{\rm PBH} \sim 0.01$, but it would be less efficient and would allow larger abundances for PBHs of the order of the solar mass.  Therefore, observations of low-mass dwarf galaxies provide at the same time a new limit on the PBH abundance and a possible hint for their existence.  These limits imposed on the PBH abundance have been represented in Fig.~\ref{fig:PBHlimits}. 
%\SC{is the limit included?}.

\subsubsection{Wide halo binaries }

Wide stellar binaries can be destructed by the encounters with PBHs, therefore the existence of the undisrupted ones in the galactic halo also provides constraints on the PBH abundance \cite{1999ApJ...516..195C}. %\SC{Add one line to explain the mechanism.}  

Using the analysis of \cite{Quinn:2009zg}, the following constraint at 2$\sigma$ confidence level has been found \cite{Carr:2020gox}: $f_{\rm PBH}<500 M_\odot/m_{\rm PBH}$ flattens above $M\sim 1000 M_\odot$ with $f_{\rm PBH} \lesssim 0.5$. Using the results of \cite{2014ApJ...790..159M}, where the low-mass limit was reduced from 500$M_\odot$ to $21-78 M_\odot$, $f_{\rm PBH}$ flattens at $m_{\rm PBH}\sim 100 M_\odot$ with $f_{\rm PBH} \lesssim 0.1$ \cite{Carr:2020gox}. In \cite{2020ApJS..246....4T}, by using a sample of 4351 halo wide binaries from the Gaia catalog, a break at $\sim$ 0.1 pc in the separation distribution has been detected. Any break in the power law for the separation distribution of wide halo binaries can be attributed to a  disruption of the binary as a result of encountering MACHOs (e.g. PBHs). Therefore, this break implies $m_{\rm PBH} > 10 M_\odot$. 
These limits come from the expected disruption of such wide binaries when they encounter PBHs. They therefore strongly depend on the number density of PBHs, and they could vary if, for instance, a majority of PBHs live in clusters. 

\subsubsection{Disruption of stellar streams}

{Stellar streams have been observed in the Milky Way halo and are likely formed by the tidal stripping of progenitors.   These would be perturbed by encounters with dark matter subhalos, such as PBH clusters.  Some methods have been proposed to distinguish the origin of perturbers, in particular the PBH-DM hypothesis form standard particle CDM~\cite{Montanari:2020gcr}.  Stellar streams like GD-1 could probe the abundance of haloes down to $10^3 M_\odot$, providing a test of the existence of PBHs. }

\subsubsection{PBHs in the Solar System}

{The dynamics of objects in the Oort cloud suggests the possible existence of a not yet observed planet, referred as Planet Nine.  It could also be a dark compact object several times the mass of the Earth, in orbit around the Sun in the Oort cloud.  It has been suggested that Planet Nine is a PBH captured by the Sun~\cite{Scholtz:2019csj} a long time ago.  It was later proposed to search for the existence of such a PBH in the Oort cloud by probing the flares from accretion of small Oort cloud objects with LSST~\cite{Siraj:2020upy}, which could result in improved limits on PBH abundances, or by probing the associated Hawking radiation~\cite{Arbey:2020urq}. }

{Alternatively, it has been proposed to probe asteroid-mass PBHs in orbit around the Earth by searching for their gravitational signatures in gravimeter networks~\cite{Namigata:2022vry}.  But the obtained limits on the PBH abundance with the current gravimeter sensitivity are not yet relevant.  Near-Earth PBHs could also leave imprints in gravimeters or in GNSS orbit products.  One way to detect a PBH passing through the Earth would be to observe the induced seismic signatures~\cite{Luo:2012pp}.}

{All this nevertheless remains highly speculative and it is not clear, for instance, if a PBH captured by the Sun (or by the Earth) would remain on a stable orbit, or if its orbit would likely pass near the Earth.  However, it must also be pointed out that discovering a PBH inside the Solar System would have tremendous implications and could even, in principle, allow in-situ investigations of a black hole~\cite{Witten:2020ifl}.}

\subsubsection{Discussion and limitations}

Again, there are a series of limitations and uncertainties that restrict the range of applicability of the dynamical limits.  In particular, let us mention that if most PBHs are regrouped in clusters, the limits from wide halo binaries would be strongly affected and possibly suppressed. These clusters could even be the ultra-faint-dwarf galaxies, for which observations show a minimum radius around a few tens of parsecs.  Dynamical heating of UFDGs by PBHs can therefore be seen both as a limit and as a clue for the existence of such PBHs.  For UFDGs, other sources of uncertainties come from the stellar population model, the exact DM profile and the existence of a central intermediate-mass black hole that typically makes limits less stringent.   This illustrates that the border between observational limits and evidence is blurred and connected, and above all strongly model dependent.  A similar conclusion applies to the asteroid-mass region, for which uncertainties on the DM content of globular clusters and the details of the tidal capture of small PBHs by neutrons stars or white dwarfs still make this range viable for a full PBH-DM scenario.

\subsection{Accretion limits}

\subsubsection{Limits from accretion at present epoch}

Primordial black holes which experience a phase of accretion at the present epoch are subject to several constraints depending on their characteristic environment.

First, a population of PBHs with masses around $\mathcal{O}(10 - 100) M_\odot$ in the Milky Way would be responsible for radio and X-ray emission if they experience a phase of accretion from the interstellar gas.
By comparing the predicted emission with observation data in the VLA radio and Chandra X-ray catalogs, one can set a constraint on the fraction of dark matter in the form of PBHs to be of the order of $f_{\rm PBH} \lesssim (10^{-2} - 10^{-3})$ in the relevant mass range, depending on the assumed PBH mass function~\cite{Gaggero:2016dpq,Manshanden:2018tze}. This constraint is shown in Fig.~\ref{fig:PBHlimits} and denoted ``Xr".

Second, the abundance of PBHs in the stellar or intermediate mass ranges, which interact with the interstellar medium emitting significant fluxes of X-ray photons, may be constrained by using the observed number density of X-ray binaries. In particular, by comparing the emitted radiation signals with the electromagnetic data, one can constrain the PBH abundance to be smaller than $f_{\rm PBH} \lesssim (10^{-2} - 10^{-3})$ in the considered mass range~\cite{Inoue:2017csr}. This constraint is labelled ``XrB" in Fig.~\ref{fig:PBHlimits}.

Finally, a cosmology-independent constraint can be set on the PBH abundance based on the absence of gas heating in the interstellar medium due to PBH interactions. In particular, photon emission from accreting PBHs, dynamical friction, winds and jets emission from accretion disks may be constrained using data from the Leo T dwarf galaxy, setting a bound on PBHs with masses in the range $(10^{-2} - 10^6) M_\odot$ up to the percent level~\cite{Lu:2020bmd, Takhistov:2021upb}.

\subsubsection{Cosmic X-ray, infrared and radio backgrounds}

{Right after the first detection of gravitational-waves from a binary black hole merger, it has been proposed that a PBH origin would be supported by the spatial correlations between the source-subtracted infrared and X-ray backgrounds~\cite{Kashlinsky:2016sdv}, which have been measured at a $5\sigma$ level~\cite{2018RvMP...90b5006K,2005Natur.438...45K} and are unexplained by known sources.  These correlations can be induced by DM halos seeded at higher redshift than in the standard CDM scenario by the Poisson fluctuations in the PBH distribution, if one has $f_{\rm PBH} m_{\rm PBH} \sim \mathcal{O}(1)$. In turn, matter accretion onto PBHs leads to X-ray emission while star formation in these halos leads to a spatially correlated infrared radiation.  The possibility of this scenario was confirmed and explored in more details in~\cite{Hasinger:2020ptw,Cappelluti:2021usg}. } 

{Recently, it has also been investigated if the observed excess of the cosmic X-ray and radio cosmic background by Chandra and ARCADE 2, suggesting an undiscovered population of emitters, could be explained by a high-redshift population of PBHs~\cite{Ziparo:2022fnc} distributed in DM halos and in the intergalactic medium.  It was shown that the emission should dominantly come from small DM halos, with masses of order or smaller than $10^6 M_\odot$ at high redshifts $z\gtrsim 6$.  Interestingly, it was also found that even if PBHs account for the excess in the radio background, it cannot at the same time explain the X-ray cosmic background. This leads to limits on the PBH abundance, in particular in the $1-100 M_\odot$ mass range, which are stronger than the microlensing limits or the bounds from ultra-faint dwarf galaxies, with $f_{\rm PBH} \lesssim 9 \times 10^{-3} M_\odot / m_{\rm PBH} $. }

\subsubsection{21-cm signal}

There are several limits based on the EDGES detection of the absorption signal of the global redshifted 21-cm line in the range $15<z<22$. The Poisson distribution of PBHs can lead to isocurvature perturbations and consequently to early structure formation. The DM annihilation is enhanced in early formed dense halos and modifies the ionization and temperature evolution of baryons. Using the redshift dependence of the EDGES signal \cite{Tashiro:2021kcg}, the following constraint on PBH abundance has been found: $f_{\rm PBH}<10^{-3.4}$ for $f_{\rm ann}\langle \sigma v\rangle /m_{\chi}=3\times 10^{-28}$ cm$^3$/s/GeV,
where $m_{\chi}$ denotes the DM mass. 

PBHs can accrete the surrounding baryonic matter as well as DM particles from the surrounding halo. This results in the emission of high energy photons that are injected into the IGM, leading to changes in its thermal and ionization history via ionization, excitation and heating. For the PBHs accreting only baryons \cite{Hektor:2018qqw}, the following approximation for the constraint on the PBHs fraction has been found: $f_{\rm PBH}\leq C(\beta) (0.15/f_{\rm E})(\lambda/0.01)^{-(1+\beta)}(M_{\rm PBH}/10 M_\odot)^{-(1+\beta)}$, where $C(\beta)=0.00015+0.00051\beta+0.0091\beta^2$ if $\beta\leq 0.37$ and $C(\beta)=0.019\beta^{2.5}$ if $\beta>0.37$ (here $\lambda$ is the mass accretion parameter, $T_k=8\,K$ the maximal gas temperatures allowed, and $f_{\rm E} < 1$ the effective energy absorption factor, assuming a monochromatic mass function). Including also DM particles accretion, in \cite{Yang:2021idt} {for a differential 21cm brightness temperature} $\delta T_{21}\lesssim -100\,$mK  the limit $f_{\rm PBH}\lesssim 2.6\times 10^{-5}$  has been obtained for $m_{\rm PBH}=10 M_\odot$ and $f_{\rm PBH}\lesssim 2.6\times 10^{-6}$ for $m_{\rm PBH}=10^4 M_\odot$. These limits are weakened by a factor of order 3 for $\delta T_{21}\lesssim -50\,$mK.

In \cite{Mena:2019nhm}, it has been shown that future experiments like SKA and HERA could potentially improve the existing CMB bounds  by more than an order of magnitude. The prospects for constraining PBHs with the future 21-cm forest observations have been discussed in \cite{Villanueva-Domingo:2021cgh}.

In \cite{Hasinger:2020ptw} the contribution of accretion on PBHs to the observed
low-frequency cosmic radio background (CRB), and thus to the EDGES signal, has been estimated. It has been found that for radio-quiet PBHs the contribution to the CRB constitutes at 1.4 GHz a fraction of 0.1\% of the observed synchrotron radio background, mostly accumulated at $z\gtrsim20$. This increases the depth of the 21cm absorption line by only about 30\%. However, some fraction of the radio-loud PBHs (e.g. 5\% with 1000 times higher fluxes), as observed in the AGN population, will easily provide the 5\% excess high-redshift radio background flux necessary to explain the EDGES observation.
%\SC{Add a paragraph on the proposal of Hasinger et al to explain EDGES with PBHs}

\subsubsection{Cosmic Microwave Background (CMB) anisotropies}

%The aim of this section is to present the current known limits on the PBH abundance and mass coming from the cosmic microwave background (CMB) frequency spectrum, temperature and polarization anisotropies. The main idea behind these constraints is the following: in the early Universe, PBHs accrete primordial gas and then convert a fraction of the accreted mass to radiation. This results in the injection of energy into the primordial plasma, which affects its thermal and ionization histories, leading to CMB distortions in its frequency spectrum, as well as temperature and polarization power spectra. 

CMB anisotropies are also impacted by accreting PBHs around the time of the recombination and until the completion of the reionization, leading to stringent limits on the abundance of PBHs heavier than around $10-100 M_\odot$.  The main idea behind these constraints is the following: in the early Universe, PBHs accrete primordial gas and then convert a fraction of the accreted mass to radiation. This results in the injection of energy into the primordial plasma, which affects its thermal and ionization histories, leading to CMB distortions in its frequency spectrum, as well as signatures in the CMB temperature and polarization angular power spectra. 

When accretion comes into play, the first aspect to consider is its geometry, i.e. spherical accretion or accretion disk formation. The former is obtained when the characteristic angular momentum of the accreted gas at the Bondi radius (distance from the center at which the escape velocity equals the sound speed) is smaller than the angular momentum at the ISCO (Innermost Stable Circular Orbit) radius. The second aspect that must be taken into account is the local feedback. Indeed, radiation emanating from the accreting PBH may heat and ionize the accreting gas, influencing the radiative output. Moreover, as discussed in \cite{Ali-Haimoud:2016mbv}, one can assume that $1 )$ in the outermost region of the accretion flow, the ionization fraction is approximately equal to the background value. Radiative feedback is neglected, and Compton drag and cooling are included in the calculation. When the temperature of the gas reaches $\sim 10^4$ K, the gas gets collisionally ionized. This first limiting case is refered to as collisional ionization; 2) in the innermost region, the radiation from the PBH photoionizes the gas, which becomes fully ionized and adiabatically compressed. This second limiting case is refered to as photoionization. Furthermore, the accreting PBHs are moving with respect to the ambient gas with some relative velocity, which does not allow for a perfectly spherically-symmetric accretion. This relative velocity is comprised of 1) a Gaussian linear contribution on large scales $v_{\rm L}$ (derived in \cite{Ali-Haimoud:2016mbv}); 2) a small-scale contribution $v_{\rm NL}$ due to non-linear clustering of PBHs (not examined in \cite{Ali-Haimoud:2016mbv}). 

For PBHs more massive than $\sim 1 M_\odot$, strong constraints were derived by \cite{Ricotti:2007au}, while in \cite{Ali-Haimoud:2016mbv}, CMB limits on the PBH abundance were re-examined in detail, based on previous works. In particular, in order to set the most conservative constraints on the PBH abundance, it is useful to compute the minimum PBH luminosity that is physically plausible and the accretion rate, taking into account the Compton drag and Compton cooling by CMB photons. Assuming a quasi-spherical accretion flow, a steady-state flow and a Newtonian approximation, the bounds derived show that in the collisional ionization limit, CMB anisotropy measurements by \textit{Planck} exclude PBHs with masses $M \gtrsim 10^2 M_\odot$ from being the dominant DM component. In the photoionization limit, this threshold is lowered to $\sim 10 M_\odot$. These constraints are weaker than those derived in \cite{Ricotti:2007au}.

As a side note, it is shown in \cite{Ali-Haimoud:2016mbv} that local Compton heating by the radiation produced by the accreting PBH can be safely neglected, i.e. one can neglect local thermal feedback. It is also demonstrated that it is not self-consistent to assume that ionizations proceed exclusively through photoionizations or collisional ionization, i.e. the level of ionization feedback is expected to lie between the two limiting cases. Finally, it was also derived that global heating of the plasma due to accreting PBHs does not leave any (observable) imprint on CMB spectral distortions. 

The QCD phase transition, inducing a peak in the probability to form PBHs at the solar-mass scale, is invoked to be a  good-motivated scenario where PBHs could contribute to the BBH mergers detected by LVK (see for example \cite{Carr:2019kxo,Jedamzik:2020omx}). Nevertheless, the viability of this scenario has been revisited by \cite{Juan:2022mir} in view of current CMB anisotropies constraints related to the accretion onto PBHs, CMB spectral distortions, gravitational wave searches, and direct counts of supermassive black holes at high redshift. For this purpose, the mass function was computed by not only considering the QCD phase transition, but also the following particle-antiparticle annihilation processes. Their conclusion is that the scenario is not viable unless one introduces a mass evolution for the PBH mass function and a cutoff in the power-spectrum close to the QCD scale. Specifically, by defining $M_{\rm cut}$ as the mass scale above which the PBH mass function is cut by hand, the mass function was found for three different values of the cutoff mass, $M_{\rm c} = 10^{8}, 10^{4}, 10^{2} M_{\odot}$, each respectively corresponding to SMBH counting, spherical accretion, and GW production. Assuming $M_{\rm cut} = 10^{8} M_{\odot}$, this leads to $f_{\rm PBH} < 4 \times 10^{-4}$. For $M_{\rm cut} = 10^{4} M_{\odot}$, we get $f_{\rm PBH} < 3.9 \times 10^{-4}$ (disk accretion model, see below) or $f_{\rm PBH} < 7.5 \times 10^{-4}$ (spherical accretion model). Further lowering $M_{\rm cut}$ below $10^{2} M_{\odot}$ relaxes the CMB anisotropy bounds to $f_{\rm PBH} \lesssim 10^{-3}$, indicating that the PBHs from the QCD-inspired scenario have at most a tiny contribution to the detected LVK events.

%\begin{figure*}[h]
%\includegraphics[width=0.9\textwidth]{Figures/1612.05644_CMB_limits.png}
%\centering
%\caption{CMB-anisotropy constraints on the fraction of DM made of PBHs (thick black curves). The “collisional ionization” case assumes that the radiation from the PBH does not ionize the local gas, which eventually gets collisionally ionized. The “photoionization” case assumes that the local gas is ionized due to the PBH radiation. The correct result lies somewhere between these two limiting cases. For comparison, we also show the CMB bound previously derived by \cite{Ricotti:2007au} (thin dashed curve), as well as various dynamical conservative constraints: micro-lensing constraints from the EROS (purple curve) and MACHO (blue curve) collaborations, limits from Galactic wide binaries, and ultra-faint dwarf galaxies \cite{Ali-Haimoud:2016mbv}.
%}
%\label{fig:CMB_limits_Ali_Haimoud}
%\end{figure*}

In \cite{Poulin:2017bwe}, it is argued that the spherical accretion approximation is likely not valid and that an \textit{accretion disk} should form during the dark ages, between recombination and reionization. More precisely, all plausible estimates suggest that a disk forms soon after recombination, mainly due to the fact that, already before recombination, stellar-mass PBHs are in a non-linear regime (i.e. they are clustered in halos of bound objects, e.g. PBH binaries or big clumps containing thousands of PBHs). The main criterion necessary for disk formation considered in \cite{Poulin:2017bwe} is that the angular momentum of the material at accretion distance is sufficient to keep the matter in Keplerian rotation at a distance $\gg 3 r_{\rm S}$, where $r_{\rm S}$ is the Schwarzschild radius. To accumulate angular momentum, the accreted material must have significant velocity or density differences. Based on the angular momentum of the accreted gas, one can derive the condition required for a disk to form \cite{Poulin:2017bwe}. For instance, the disk formation criterion is likely satisfied if we consider an ideal, free-streaming homogeneous gas moving at a bulk motion with velocity of order $v_{\rm L}$ without dispersion, or when PBHs constitute an important fraction of the DM with $M \gtrsim M_\odot$. Another key quantity is the radiative efficiency factor $\epsilon$, which is strongly related to the geometry of the accretion flow and the accretion rate. It usually takes the value $\epsilon \sim 0.1$ for moderate or low disk accretion rate, that corresponds to the value inferred from BH observations.  In the case of spherical accretion, one gets smaller values of the order of $10^{-5}$. Finally, to compute the CMB bound, it is crucial to evaluate the total energy injection rate, by taking into consideration that not all radiation is equally effective. To compute the impact on the CMB, one has to quantify what amount of injected energy is actually deposited into the medium by using the \textit{energy deposition functions}, and the last step is to include the spectrum of the radiation emitted through BH accretion. Let's note that each accretion scenario has a distinctive energy injection history, that could in principle be distinguished in the CMB angular power spectrum~\cite{Poulin:2017bwe}.

The conservative constraints obtained in \cite{Poulin:2017bwe} (at 95\% C.L. using data from \textit{Planck} high-\textit{l} TT TE EE+lensing) are represented in Fig. \ref{fig:PBHlimits}. One notices that the presence of disks improves the CMB constraints on PBH by at least two orders of magnitude, so that PBHs with masses $M \gtrsim 2 M_\odot$ (with monochromatic mass distribution) are excluded from accounting for the totality of the DM. The constraints derived for the monochromatic case were also extended to a log-normal mass distribution, and become more restrictive for a broader distribution \cite{Poulin:2017bwe}.

%\begin{figure*}[h]
%\includegraphics[width=0.9\textwidth]{Figures/1707.04206_CMB_limits.png}
%\centering
%\caption{Constraints on accreting PBH as DM represented by the blue and light-red regions, with $v_{\rm eff}$ the effective velocity (of...?), $c_{{\rm s}, \infty}$ the sound speed far from the accreted point mass, and $v_{\rm L}$ the relative baryon-DM velocity in the linear theory. These constraints are compared to i) the CMB constraints obtained in the spherical accretion case (red full line); ii) the non observation of micro-lensing events in the Large Magellanic Cloud as derived by the EROS-2 collaboration (black dot-dashed line); iii) the non observation of disk-accreting PBH at the Galactic Center in the radio band (green long-dashed line); iv) the constraints from the disruption of the star cluster in Eridanus II (blue short-dashed line) \cite{Poulin:2017bwe}.
%}
%\label{fig:CMB_limits_Poulin_Serpico}
%\end{figure*}

In \cite{Serpico:2020ehh}, they extended previous calculation \cite{Poulin:2017bwe} to account for the accretion of  dominant, non-PBH DM particles onto PBH. The increased gravitational potential felt by the baryons enhances baryonic accretion and the PBH luminosity constrained by the CMB. The accreting DM halos around each PBH are treated via a toy model and numerical simulations, assuming that the ionization of gas in the dark ages due to accretion onto PBHs is homogeneous.

%The key input required to compute the CMB bound is the total energy injection rate, which is the product of the accretion luminosity $L_{\rm acc}$ onto a PBH of mass $M$ and the PBH number density $n_{\rm PBH}$. To find the energy deposited in the medium, one makes use of the transfer functions from \cite{Slatyer:2015kla}. The function $L_{\rm acc}$ is parameterized in terms of $\dot{M}$ (the matter accreted onto the PBH per unit time) and $\epsilon$, the efficiency factor. The latter can be computed in spherical symmetry, giving us small values of the order of $10^{-5}$. Let us remind that in the disk accretion case, a typical value is around 0.1. 

The toy model considered in \cite{Serpico:2020ehh} is used to determine the upper limit to the mass growth of the PBH via DM accretion. The most optimistic scenario for PBH growth is for a cold DM with no dispersion.
%, and the PBH being the only center of attraction in the universe. 
The time evolution of the radius of a mass-shell around a PBH is found to be $M_{\rm halo} \simeq (3000/1+z)M_{\rm PBH}$, which is expected to break down at very late times. This conjecture is then verified via cosmological N-body simulations, where in order to model the halo growth, it is required that the DM halo is composed of many DM particles when it becomes comparable to the PBH mass.  
%Typically, the PBH contribution to the halo is negligible when the DM distribution is important. 
An important result is that when the halo is compact, the accreting baryons see a BH whose effective mass is the sum of the PBH and the DM halo masses. If the halo is large, only a fraction of its mass contributes to the accretion. This is the model used in \cite{Serpico:2020ehh} to compute the impact of PBH accretion onto the CMB.

The CMB constraints of \cite{Serpico:2020ehh} are obtained by taking into account the effect of accretion within DM halos. In the simulations previously mentioned, one gets the necessary resolution for $100 < z < 1000$, encompassing the redshift range at which the deposition of electromagnetic energy has the biggest impact on the CMB ($300 < z < 600$) \cite{Slatyer:2009yq}. One important finding is that the inclusion of halos increases the impact of the PBH on the CMB angular power spectra by up to 2 orders of magnitude. A series of runs were performed assuming spherical and disk accretion, with the presence and absence of a DM halo around the PBH, and assuming a monochromatic PBH mass function (for extended ones, bounds usually become more stringent). First, supposing that PBHs are accreting at Eddington luminosity (luminosity at which accretion is balanced by radiation pressure in a spherical system), one gets the limit $f_{\rm PBH} < 2.9 \times 10^{-9}$. Secondly, in absence of DM halos the constraints at 95\% C.L. are stronger (by a factor $\sim 4$) than bounds derived in \cite{Poulin:2017bwe}. Accounting for the DM halo leads to improved bounds for $f_{\rm PBH} \lesssim 10^{-2}$ and $f_{\rm PBH} \lesssim 0.2$ for the disk and spherical accretion cases, respectively. Eventually, the bounds flatten when $M \gtrsim 10^{4} M_\odot$ as a consequence of the accretion that reaches the Eddington limit. CMB constraints exceed the ones coming from the non-observation of mergers by LVK, BBN, the non-observation of the SGWB, etc., for masses $M \gtrsim (20 - 50) M_\odot$. The CMB bounds remain dominant until $10^{3.5} M_\odot$, when they come close to the BBN ones. Specifically, it is excluded that PBH constitute the totality of the DM over the stellar mass range, and around $10 M_\odot$ no more than $f_{\rm PBH} \sim 0.1$ is allowed  \cite{Serpico:2020ehh}.

{Finally, a possible relaxation of the limits from CMB anisotropies has recently been suggested in~\cite{Facchinetti:2022kbg} and~\cite{Piga:2022ysp}.  
The work of~\cite{Facchinetti:2022kbg} relies on a more realistic accretion model based on hydrodynamical simulations and conservative assumptions for the emission efficiency.  This leads to limits that are up to 2 orders of magnitude less stringent than previously estimated, between $10$ and $10^4 M_\odot$, which reopens the possibility that PBHs might explain at the same time (at least a fraction of) the dark matter, some of the LVK binary BH mergers, and the existence of supermassive BHs.  The work of~\cite{Piga:2022ysp} takes a step towards the development of a more realistic PBH accretion by accounting the contribution of outflows, for various accretion geometries, ionization models and mass distributions in absence and in presence of mechanical feedback and non-thermal emissions due to the outflows.  In general, it seems that PBH accretion is rather complex physical process that is hard to assess, which induces large uncertainties in all the accretion-based limits.
}

\subsubsection{Discussion and limitations}

Accretion limits, in particular limits from CMB temperature  anisotropies, are claimed to set the most stringent limits on the abundance of PBHs between about $10 M_\odot$ and $10^4 M_\odot$, i.e. in the range that is particularly relevant for GW observations with ground-based and space-based detectors.  Their strengths nevertheless depend on the details of the accretion (e.g. disk or spherical accretion).  Given the possible complexity of the accretion process, it is possible that those limits significantly change if different assumptions are chosen, for instance for the velocity-dependence of the accretion process.  One should also notice that the Poisson-induced clustering of PBHs were not taken into account in the calculation of these limits.   Other sources of uncertainties come in the computation of the injected energy and the resulting changes in the evolution of the ionized fraction at high redshfit.  Limits from the X-ray and radio emission  in the galactic center have their own uncertainties, e.g. related to the dark matter distribution.  The limits that were presented were nevertheless considered as relatively conservative.  

An important implication that was recently discovered for the limits from CMB temperature anisotropies is that the existence of a particle dark matter component in addition to PBHs reinforce those limits on the abundance of PBHs, in such a way that they may become incompatible with an explanation of GW observations.

\subsection{Indirect constraints from density fluctuations}

\subsubsection{CMB distortions}

Enhanced primordial density perturbations producing PBHs will also generate CMB distortions by dumping energy into the primordial plasma in the wave-number range $1 \lesssim k$/Mpc$^{-1}$ $\lesssim 10^{4.5}$. The amount of distortion can be roughly estimated as $<\mu> \thinspace \thinspace \sim \, O(1) \, P_{\zeta} \, \Delta N$, where $ \Delta N$ is the e-fold number duration of the enhanced perturbations, which is equivalent to $\ln \left[ \frac{k_{\rm max}}{k_{\rm min}}\right]$. Assuming $\Delta N \sim O(1)$, one can probe the distortion level from the power spectrum, and current bounds are $<\mu> \thinspace \thinspace \lesssim 10^{-4}$
\cite{Kohri:2014lza,Nakama:2017xvq,Inomata:2018epa,Byrnes:2018txb,Kalaja:2019uju,Gow:2020bzo,Garcia-Bellido:2017aan}.

The PBH production is extremely sensitive to the tail of the statistical properties, i.e. probability distribution, of the primordial perturbations as discussed in Sec. \ref{sec:PBHformation}. Hence, for highly efficient PBHs producing non-Gaussian perturbations, the distortion signal can be smaller and evade the current bounds. However, future experiments will reach remarkable sensitivities to probe both Gaussian and non-Gaussian primordial signals \cite{Nakama:2017xvq,Garcia-Bellido:2017aan}.

If scalar induced stochastic GW background is combined with CMB distortions, the probed PBH seed range will be $(0.1-10^{13})M_\odot$ \cite{Unal:2020mts} (equivalently the probed wavenumber range will be $1 \lesssim k$/Mpc$^{-1}$ $\lesssim 10^7 $). These probes will also reach conclusive results about the intriguing possibility that PBHs could be the seeds of SMBHs, and with the detection of over a billion of solar mass BHs at redshift higher than 7, this has become an even more pressing question. Since CMB distortions and scalar induced GWs are global cosmic signals associated with the formation/production of PBHs, the conclusions will be robust to modifications in: i) astrophysical evolution, such as merger and accretion history; ii) statistical properties of the primordial perturbations (Gaussian or non-Gaussian); iii) clustering effects.  The combined probes are shown in Figure \ref{figrobustboundsfromskadistortion}, where non-detection of both signals will robustly constrain the PBH abundance to the negligible amount of $f_{\rm PBH}<10^{-10}$  \cite{Unal:2020mts}.

Light PBHs, in the mass range, $10^{-15}-1 M_\odot$, can experience superradiance even for tiny spin values, $10^{-3}-10^{-2}$ \cite{Unal:2023yxt}, and they can probe 17 decades of bosons (scalar, vector and spin-2) in the motivated mass range, $10^{-12}-10^5 \, {\rm eV}$. Bosons could be beyond Standard model particles such as axions and dark photons, also it could be Standard Model photon with effective mass due to plasma interactions. If spinning light PBHs are observed, this implies non-existence of superradiance for the given BHs and boson mass range, therefore this observation can imply either there are no such particles or their self/external interactions prevent them from superradiance. This results in bounds on the axion decay constant (inversely related to the strength of self-interaction) and its energy density \cite{Unal:2023yxt}. Furthermore, PBHs can also produce photons with an effective mass in the plasma, which results into an energy injection in the early universe and can lead to CMB distortion, giving rise to mass dependent bounds on PBHs \cite{Pani:2013hpa}.

\begin{figure}
    \centering
    \includegraphics[width=0.8\textwidth]{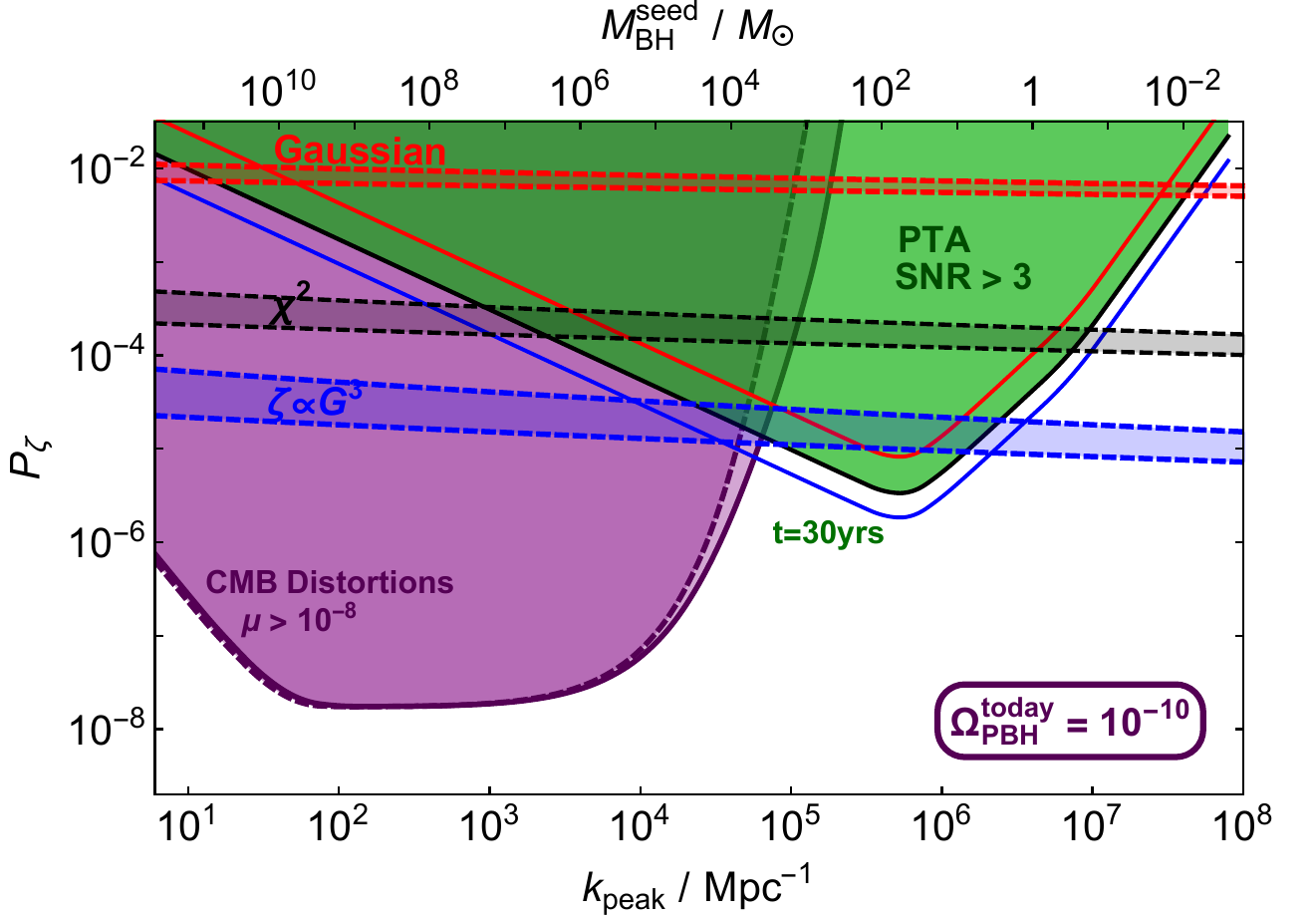}
    \caption{Probing power of future PTA (SKA) and CMB distortion (PIXIE-like) experiments on primordial fluctuations for about 7 decades in wavenumbers and 13 decades in masses. Results are shown for inflationary perturbations  that obey distinct probability distributions: Gaussian (red horizontal dashed line), chi-sqr ($\chi^2$, black horizontal dashed line), cubic-Gaussian ($G^3$, blue horizontal dashed line), detectable if $f_{\rm PBH}>10^{-10}$. Taken from \cite{Unal:2020mts}.}
    \label{figrobustboundsfromskadistortion}
\end{figure}

\subsubsection{Lyman-$\alpha$ forest}

The constraints on PBHs can be obtained from the high-resolution high-redshift Lyman-$\alpha$ forest data since the Poisson fluctuations in the PBH number density induce a small-scale power enhancement departing from the standard CDM prediction. This enhancement can be interpreted as an isocurvature perturbation with the scale-invariant spectrum and isocurvature-to-adiabatic amplitude ratio for the monochromatic distribution: $f_{\rm iso}=\sqrt{\frac{k_\star^3 m_{\rm PBH}f_{\rm PBH}}{2\pi^2 \Omega_{\rm CDM}\rho_{\rm cr}}\frac{1}{A_{\rm s}}}$ (here $k_\star=0.05$/Mpc is the pivot scale) \cite{Murgia:2019duy}. Using MIKE and HIRES/KECK samples of quasar spectra and a grid of hydrodynamic simulations exploring different values of astrophysical parameters, in \cite{Murgia:2019duy} the  marginalized upper limit $f_{\rm PBH}m_{\rm PBH}\sim60M_\odot$ at 2$\sigma$ on the PBH mass have been found for a monochromatic distribution. This has been obtained with a Gaussian prior on the reionization redshift, preventing its posterior distribution to peak on very high values that are disfavoured by the most recent estimates from the CMB and IGM observations $f_{\rm PBH}m_{\rm PBH}\sim60M_\odot$ at 2$\sigma$ with a conservative flat prior.

\subsubsection{El Gordo cluster and enhanced Halo Mass Functions}

{One of the most fascinating connections between PBHs and large scale structure (LSS) arises when one considers the effect of quantum diffusion in single field inflation with quasi-inflection points~\cite{Ezquiaga:2019ftu} as a natural way to generate large exponential tails in the density contrast PDF a fact that enhances the probability of collapse of small-scale fluctuations upon re-entry leading in this way to PBH formation. These highly non-Gaussian tails in the PDF can also be responsible at larger (galactic and cluster) scales for the enhancement of the halo mass function at high redshift~\cite{Ezquiaga:2022qpw}, possibly explaining the increased abundance of massive galaxies seen by the James Webb Space Telescope at redshifs $z\sim 9-20$~\cite{2022ApJ...940L..55F}, as well as significantly more massive clusters, like El Gordo~\cite{Asencio:2020mqh} at $z\sim 1$, than expected by the standard model $\Lambda$CDM.
It remains to be shown whether such large deviation from the $\Lambda$CDM predictions for the halo mass function are compatible with LSS constraints. 
In the near future we may have deep imaging up to redshift $z\sim20$ with both infrared and radio surveys like SKA, to test the enhanced exponential tails that could have given rise to PBHs in the early universe.}

\subsubsection{High-z galaxies from JWST}

The James Webb Space Telescope (JWST) imaging program
via the Cosmic Evolution Early Release Science (CEERS) survey recently reported 
a population of massive galaxy candidates at redshift $z\gtrsim 8$ with stellar masses of the order of $10^9 M_\odot$ \cite{2022arXiv220712338A,2022arXiv220712474F,2022arXiv220801612H,2022arXiv220802794N,Yan:2022sxd}.
A note of caution here is in order, as spectroscopic follow-up will be necessary to confirm current observation only based on photometry.
However, if confirmed, such early formation of massive galaxies reported by the JWST is hardly   reconcilable with the standard $\Lambda$CDM expectations. 
Recently, Ref.~\cite{2022arXiv220712446L}  derived the cumulative stellar mass density at $z=8$ and 10 for $M_{\star}\gtrsim 10^{10}\ M_{\odot}$
based on 14 galaxy candidates with masses in the range $\sim 10^{9} \div 10^{11}\ M_{\odot}$ at $7<z<11$ 
observed by the CEERS program. They found
\begin{align}
&\rho_*(>10^{10} M_\odot)\simeq 1.3^{+1.1}_{-0.6}\cdot 10^6 M_\odot\,{\rm Mpc}^{-3},
% \nonumber\\
% &\rho_*(>10^{10.5} M_\odot)\simeq 9^{+11}_{-6}\cdot 10^5 M_\odot\,{\rm Mpc}^{-3}.
\end{align}
at $z\simeq 10$.
These values are larger than the $\Lambda$CDM predictions, even allowing maximum Star Formation Efficiency (SFE) $\epsilon=1$, or invoking extreme value statistics \cite{Lovell:2022bhx}.

Recently, Refs.~\cite{Liu:2022bvr,Hutsi:2022fzw}
investigated the hypothesis that early structure formation was induced by the seeding effect of PBH induced Poisson isocurvature perturbations.\footnote{See also Refs.~\cite{Biagetti:2022ode,Menci:2022wia,Gong:2022qjx} for different potential explanations. } 
Their result shows that 
such early formation of massive galaxies is possible if the PBH abundance is such that
$m_\PBH f_\PBH \gtrsim  6 \times 10^6 
\ 
(2  \times 10^5)
M_\odot$ for SFE $\epsilon < 0.1\ (1)$. 
Such abundance for massive PBHs are ruled out by observations of the CMB $\mu$-distortion, 
although this conclusion relies on the assumptions that primordial density fluctuations are sufficiently Gaussian 
and that PBHs formed in the standard scenario of primordial density fluctuations not growing during the radiation-dominated era.
However, strong isocurvature perturbations 
of PBHs with $m_\PBH f_\PBH \gtrsim 170 M_\odot$ 
are ruled out by high-z Lyman-$\alpha$ 
forest data \cite{Murgia:2019duy}.

\subsubsection{Ultra-compact mini haloes}

{
It has been claimed in~\cite{Adamek:2019gns} that a small fraction of dark matter made of PBHs is not compatible with the rest of the dark matter being made of weakly interacting massive particles (WIMPs) because PBHs should seed the formation of ultra-compact mini-halos with steep density profiles.  Then, the WIMPs in the dense innermost part of these halos would annihilate, which produces a detectable gamma-ray signal.   The absence of detection allows to set limits on the PBH fraction in presence of WIMPs, of at most one billionth of the dark matter for PBH masses larger than $10^{-6} M_\odot$.  An interesting corollary is that any firm detection of PBHs implies that the remaining dark matter cannot be constituted by WIMPs.   Similar analysis, e.g. including the effect of primordial non-Gaussianity or the mass dependence of this limit in the case of light PBHs, have been performed in~\cite{Nakama:2019htb,Carr:2020mqm,Kadota:2020ahr}.    The s-wave and p-wave annihilation scenarios have been compared in~\cite{Kadota:2021jhg} where a stringent limit has been obtained for the thermal relic p-wave  with a WIMP mass of 100 GeV (see also~\cite{Chanda:2022hls}).  It has been also shown in~\cite{Tashiro:2021xnj} that WIMPs around PBHs enhance the heating and ionization in the intergalactic medium, due to WIMP annihilations, leading to comparable or even tighter bounds of the possible DM fraction made of PBHs, for instance $f_{\rm PBH} \lesssim \mathcal O (10^{-10}-10^{-8} ) $ for a WIMP mass of order $10-10^3$ GeV with a conventional annihilation cross section $\langle \sigma v \rangle = 3 \times 10^{-26} {\rm cm}^3 /{\rm s} $.   Some of those limits were recently reanalyzed in~\cite{Gines:2022qzy}.  WIMP annihilation in ultra-compact mini-halos around PBHs can also lead to a potentially detectable neutrino signal~\cite{Freese:2022ouh} and the limits can be competitive with respect to gamma-ray searches if PBHs are heavy.  Another proposed way to probe DM dresses around PBHs is by observing the gravitational strong lensing of sources like fast radio bursts in the PBH mass range from $10 M_\odot$ to $10^3 M_\odot$~\cite{Oguri:2022fir}.  }

{The specific case of SU(2)L triplet fermion “winos,” has been studied in~\cite{Hertzberg:2020kpm}.  In this case as well, after the wino kinetic decoupling, DM particles are captured by PBHs leading to dark mini-halos, constrained by the production of narrow line gamma rays from wino annihilation in the Galactic Center.    Another very specific scenario of parsec-size ultra-compact mini-halos around PBHs, in the context of the QCD axion dark matter, encountering neutron stars and leading to transient ratio emissions due to  resonant axion-photon conversion in the neutron star magnetosphere, has been explored in~\cite{Nurmi:2021xds}.}

{Finally, it has recently been argued in~\cite{StenDelos:2022jld} that the large density fluctuations required for PBH formation should also lead to ultradense DM halos (see also \cite{Ricotti:2009bs,Kohri:2014lza,Gosenca:2017ybi,Delos:2017thv,Delos:2018ueo,Nakama:2019htb,Hertzberg:2019exb,Ando:2022tpj}) that would actually host a large fraction of the DM (in much greater abundance than PBHs), enabling a variety of new observational tests. 
This conclusion is viable as long as clustering of the dark matter is possible on the relevant scales (e.g.~\cite{Bringmann:2009vf}). 
The formation of such halos can already take place before matter-radiation equality \cite{Kolb:1994fi,2010PhRvD..81j3529B,Berezinsky:2013fxa,Delos:2019tsl,StenDelos:2022jld}, which occurred at redshift $z\simeq 3400$ \cite{Planck:2018vyg}
Due to their early formation, ultradense DM halos 
could be compact enough to produce detectable microlensing signatures. }

{
In Ref.~\cite{Delos:2023fpm} it is investigated whether the EROS, OGLE, and HSC surveys can probe these halos by fully accounting for finite source size and extended lens effects.
Interestingly, current data by the EROS-2~\cite{EROS-2:2006ryy}, OGLE-IV~\cite{Niikura:2019kqi} and  Subaru-HSC~\cite{Niikura:2017zjd} surveys may already constrain the amplitudes of primordial curvature perturbations in a new region of parameter space, even though this conclusion is sensitive to details about the internal structures of these ultradense halos. 
Adopting optimistic assumptions, HSC data would constrain a power spectrum that features an enhancement at scales 
$k\sim 10^7/$Mpc, and an amplitude as low as ${\cal P}_\zeta \simeq 10^{-4}$ may be accessible.
This range of scale is particularly interesting
as it connects to primordial black hole formation in a portion of the LVK mass range and the production of scalar induced gravitational waves in the nanohertz frequency range reachable by pulsar timing arrays. Further dedicated numerical simulations are required to investigate in details the central density reached by such DM halos, to which lensing constraints are particularly sensitive. }

{In summary, the study of dark matter halos, seeded or induced by the existence of PBHs, brings support for the incompatibility of the coexistence of PBHs and WIMPs and all-or-nothing scenarios.}

\subsubsection{Discussion and limitations}

The primordial fluctuations at the origin of the formation of PBHs with masses above $60 M_\odot$ can have left detectable signatures in the form of spectral distortions of the CMB black-body spectrum, which have been constrained by FIRAS, in high-redshift Lyman-$\alpha$ forest observations, which can typically probe smaller length scales than large scale structure surveys or CMB anisotropies, and in a GW background sourced at second order in perturbation theory that was discussed in a dedicated section.  Those indirect probes are therefore independent of the details of the PBH formation and their subsequent complex evolution.  They also depend on the statistical properties of primordial fluctuations, e.g. they can be suppressed in the presence of large non-Gaussianities.  
These indirect probes are therefore very interesting to discriminate between the possible PBH formation models in a way that is independent of the complex physical processes that could have impacted the PBH properties.  This is of particular interest to constrain models with broad PBH distributions aiming at explaining LVK observations and SMBH seeds in a unified way.  The absence of any detection of CMB distortions would bring support to non-Gaussian models or to scenarios where PBHs are not formed by primordial density fluctuations.

\subsection{Gravitational waves} 

\subsubsection{LVC GW mergers}
The LVK collaboration has been searching for compact binary coalescences. By the end of the O3 run, 90 binary black hole (BBH) candidates have been found~\cite{LIGOScientific:2021djp}, providing rich information on the BH population~\cite{LIGOScientific:2021psn}. The latest catalog GWTC-3 indicates that the BBH merger rate is between 17.3 ${\rm Gpc}^{-3} {\rm yr}^{-1}$ and 45 ${\rm Gpc}^{-3} {\rm yr}^{-1}$ at a fiducial redshift $(z=0.2)$. 

Soon after the first discovery of the BBH event GW150914, it has been pointed out that the observed BBH could be of primordial origin ~\cite{Bird:2016dcv,Sasaki:2016jop,Clesse:2016vqa}.
More detection of BBH events provided information on the merger rate and the chirp mass distribution, and Ref.~\cite{Clesse:2017bsw} made the first MCMC analysis to infer the PBH mass function. Statistical tests have been performed with the GWTC-1 catalog to test the hypothesis that all or part of observed BBHs is primordial~\cite{Hall:2020daa}. The use of the effective spin parameter has also been discussed~\cite{Fernandez:2019kyb,Garcia-Bellido:2020pwq}. With more events from GWTC-2, the combination of an astrophysical and a primordial BH population have been tested,
both using phenomenological models to describe the ABH population~\cite{DeLuca:2021wjr,Hutsi:2020sol}
as well as state-of-the-art astrophysical population synthesis results \cite{Franciolini:2021tla}. 
These analyses confirm it is difficult to explain all the features observed in the GWTC-3 catalog of events by means of a single binary population, while showing shortcomings of both astrophysical and primordial scenarios, but still
allows for PBH to contribute to a 
fraction of the observed BBHs. 

Indication of particular events has been also discussed. GW190521 has attracted attention because at least the primary component mass lies in the so-called pair-instability mass gap, where there should not be direct formation of stellar black holes, and the possibility of a primordial origin has been discussed in Ref.~\cite{DeLuca:2020sae}. The events
GW190425 and GW190814 have both companions with masses $[1.6-2.5] M_\odot $ and $[2.5-2.7]M_\odot $, which is unexpected if they originated from a neutron star or a stellar black hole. Ref.~\cite{Vattis:2020iuz} discussed the possibility of PBH pairing with an astrophysical BH, and concluded that it is unlikely that such binaries form and merge within a Hubble time (see also \cite{Kritos:2020wcl,Sasaki:2021iuc}). Refs.~\cite{Jedamzik:2020omx,Jedamzik:2020ypm,Carr:2019kxo,Clesse:2020ghq} investigated the possibility that both the primary and secondary BHs are primordial considering the thermal history mass function, which indeed has a bump at a few $ M_\odot$ and around $30-50 M_\odot $ and explains the events very well, together with GW190521.

Recently, Ref.~\cite{Franciolini:2022tfm} performed the most comprehensive Bayesian population analysis of LVKC data, that includes both BH and NS merger events, while testing the hypothesis that a fraction of events may come from the primordial scenario,
where the latter is derived from first principles
with the inclusion of QCD effects on the mass distribution. 
The results of such an analysis can currently only set an upper bound on the PBH fraction, 
showing the still limited constraining power of available data, 
but interestingly peaked up some aforementioned special events with high probability of being primordial (i.e. GW190924\_021846, GW190814, GW190412, and GW190521). 
One peculiar property of the PBH channel, is that if it produced mergers in the NS mass range, due to the QCD enhancement and the critical collapse, it would necessarily produce a mass distribution which is wider the the one expected for astrophysical NS mergers \cite{Franciolini:2022tfm}. This would inevitably predicts subsolar, as well as lower mass gap event mergers. 

Due to the much improved sensitivity of future observation runs, we notice that O4 and O5 may have more chances to detect many PBH events. However, unless some of these events have smoking-gun features~\cite{Franciolini:2021xbq}, it would be hard to distinguish them from ordinary astrophysical channels.
Therefore, an interesting prediction,
reported in Table~\ref{TbPBH:O4O5}, is the number of subsolar and mass-gap events detectable in O4 and O5, assuming GW190814 is primordial. 
Such numbers can be considered conservative upper bounds, as well as complementary tests of the primordial origin of the aforementioned events. 
In particular, in O5 there could be as many as $\approx 8$ subsolar events per year (but the $90\%$ confidence interval is also compatible with zero events). Then O5 should detect one to a few dozen events per year in the lower mass-gap (and up to $\approx 50$ upper mass-gap events), which might be more difficult to interpret in astrophysical scenarios.

{
\renewcommand{\arraystretch}{1.4}
\begin{table}[!t]
% \vspace{.1cm}
\begin{tabularx}{1 \columnwidth}{|X|c|c|c|c|}
\hline
\hline
 LVKC observing run & $N_\PBH^\text{\tiny det}$ &
 $N_\PBH^\text{\tiny det}$(SS) & 
 $N_\PBH^\text{\tiny det}$(LMG) &
 $N_\PBH^\text{\tiny det}$(UMG)
\\
\hline
 O1-O3 & 
$[0.8,22.4]$ &
$[0.0,0.6]$ & 
$[0.1,2.3]$ &
$[0.0,6.1]$
\\
\hline
O4 & 
$[1.9,43.7]$ &
$[0.0, 1.3]$ & 
$[0.3, 13.0]$ & 
$[0.0, 13.1]$ 
\\
\hline
O5 & 
$[10.3,216.7]$ &
$[0.0 ,8.6]$ & 
$[0.8, 25.2]$ & 
$[0.0, 47.3]$ 
\\
\hline
\hline
\end{tabularx}
\caption{
Table adapted from Ref.~\cite{Franciolini:2022tfm}. Assuming GW190814 had primordial origin, this table reports the 90\% C.I. for the number of detected PBH events within GWTC-3, 
and predicted events (per year) with LVKC O4 and O5 sensitivity.
We also indicate forecasted detections (within the 90\% C.I.) that would fall in the subsolar ($m_1<M_\odot$, SS), 
lower mass gap 
($m_1\in[2.5,5]~{M_\odot}$ and/or $m_2\in[2.5,5]~{M_\odot}$, LMG) and
upper mass gap ($m_1>50 M_\odot$, UMG)
ranges.}
\label{TbPBH:O4O5}
\end{table}
}

Future observations at third generation experiments will be able to further constrain the abundance of PBHs in the LVKC mass range by searching for high redshift PBH mergers, where the possible astrophysical contamination from Pop III mergers fades out \cite{Ng:2022agi} (see also Refs.~\cite{Nakamura:2016hna,Koushiappas:2017kqm,DeLuca:2021hde,Pujolas:2021yaw,Ng:2021sqn,Franciolini:2021xbq,Martinelli:2022elq}).
This technique would allow to constrain the PBH abundance down to values as small as $f_\PBH \lesssim {\cal O}(10^{-5})$ in the standard scenario, see Fig.~\ref{fig:fPBHlimit_CEET} \cite{Ng:2022agi}.\footnote{Such constraint becomes much more stringent if one assumes clustered PBH initial conditions, boosting the PBH merger rate \cite{DeLuca:2021hde}.
However, further work on understanding the merger rate of initially clustered scenarios is needed.}

\begin{figure}[t!]
    \centering
    \includegraphics[width=0.7\columnwidth]{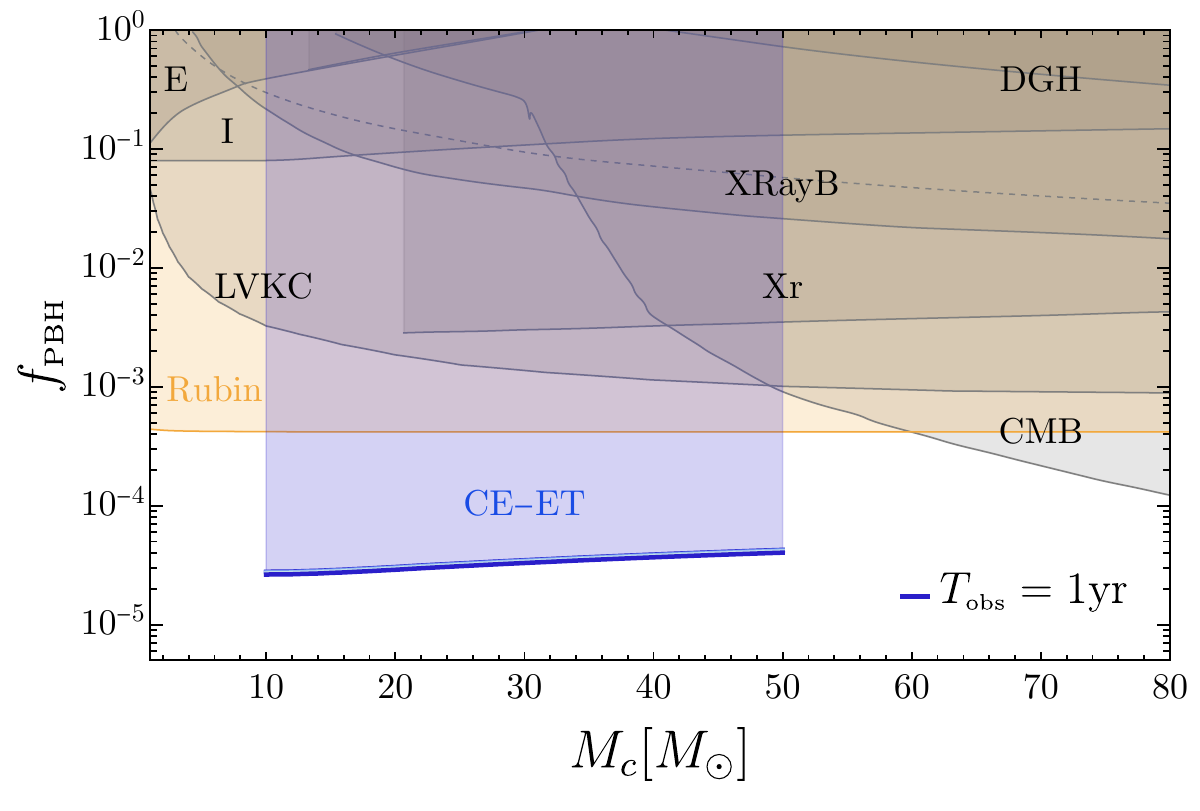}
    \caption{Figure taken from Ref.~\cite{Ng:2022agi}.
    Projected upper limit on $f_\PBH$ as a function of central PBH mass scale $M_c$ (assuming a narrow log-normal mass distribution) obtained from null high redshift merger detections with one year of observations at a CE-ET network. See the main text for a discussion of the other (non-GW constraints). In yellow, we show forecasts for the limits that will be set by microlensing searches with the Rubin observatory~\citep{LSSTDarkMatterGroup:2019mwo,2022arXiv220308967B}.
    }
\label{fig:fPBHlimit_CEET}
\end{figure}

\subsubsection{LVC sub-solar searches}
In addition to the standard compact binary coalescence search, the LVC has been updating the constraint on sub-solar mass binaries, where at least one binary component has a mass between $[0.2 - 1] \Msun$~\cite{LIGOScientific:2018glc,LIGOScientific:2019kan,LIGOScientific:2021job} (see also \cite{Nitz:2020bdb,Nitz:2021vqh,Phukon:2021cus,Nitz:2022ltl}). So far, no firm detection of such GWs has been reported, but four subsolar black hole candidates with false alarm rate smaller than $2$ yr$^{-1}$ and signal-to-noise ratio SNR $>8$ were reported in~\cite{Phukon:2021cus} after re-analysing the data from the second observing run within an extended mass range.   If such a binary system were to be found with a clear statistical significance, it could be strong support for the existence of PBHs, since neither black holes nor neutron stars are expected to form below $\sim 1\Msun$ through conventional stellar evolution. The recent analysis with the data from the first half of the LVC O3 run has provided a bound on the merger rate of subsolar binaries in the range $[220 - 24200] {\rm Gpc}^{-3} {\rm yr}^{-1}$. Assuming an isotropic distribution of equal-mass primordial black holes, the bound on the merger rate translates to the bound on the fraction of PBHs as $f_{\rm PBH} \equiv \Omega_{\rm PBH}/\Omega_{\rm DM} \lesssim 6\%$. Although the LVC collaboration papers have placed stringent constraints on the PBH abundance, they can be relaxed by two orders of magnitude if we take into account merger rate suppression due to binary disruption by early forming clusters, matter inhomogeneities and nearby PBHs \cite{Phukon:2021cus}.

\subsubsection{LVC SGWB}
As mentioned in Sec. \ref{sec:SBWBmergers}, overlapped GWs from binary PBHs form a SGWB. The most updated bound on the SGWB by the LVC O3 run is $\Omega_{\rm GW} < 5.8\times 10^{-9}$ at the 95\% credible level for a flat (frequency-independent) SGWB~\cite{KAGRA:2021kbb}. The first constraints on PBH abundance were argued in~\cite{Wang:2016ana} using the O1 data. The updated constraints by O2 and O3 data can be found in~\cite{Raidal:2017mfl,Hutsi:2020sol}. In addition, parameter estimation has been attempted by~\cite{Mukherjee:2021ags} using the O3 data by considering both astrophysical and primordial components. The analysis provided weak constraints on the PBH merger rate and time delay parameter, while it indicated the difficulty of separating astrophysical and primordial BBH contributions with the current sensitivity of the detectors (see also \cite{Bavera:2021wmw}). We may be able to obtain meaningful constraints on the PBH merger rate with the increased sensitivity of upcoming runs and future detectors.   

It is worth mentioning that the upper bound on the SGWB can provide constraints also on the ultralight PBHs in the mass range of $[10^{-20} - 10^{-19}]\Msun$ through the scalar induced GWs. Those PBHs are strongly constrained by the Hawking radiation, but with the upgraded sensitivity, the ground-based interferometers could provide constraints that are several orders of magnitude stronger than existing constraints~\cite{Kapadia:2020pir}. The first constraint was demonstrated by \cite{Kapadia:2020pnr} with the O2 data. See \cite{Romero-Rodriguez:2021aws} for the updated constraint by the O3 data.

\subsubsection{PTAs and NANOGrav}\label{sec:PTAs_NG}

As discussed in Sec.~\ref{sec:secondordergw}, enhanced density perturbations inevitably source stochastic GWs, dubbed as scalar induced GWs, or SIGW. The SIGW can be used to probe the small scale power spectrum~\cite{Nakama:2016gzw,Inomata:2018epa,Byrnes:2018txb,Cai:2019elf,Kalaja:2019uju,Unal:2020mts,Papanikolaou:2022chm}. The amplitude of this signal can be estimated as $\Omega_{\rm GW}\sim \, {\rm (symmetry \, factors)} \cdot \Omega_{\rm rad} \cdot P_\zeta^2$, where symmetry factors include possible contractions and different diagrams. Hence, it is seen that current PTA experiments can probe $P_\zeta > 10^{-2.5}$ at the nano-Hertz~(nHz) scale.

In 2020, the NANOGrav collaboration first reported evidence in their 12.5 year dataset~\cite{Arzoumanian:2020vkk} for a common spectrum of a stochastic nature, 
representing the first hint of the existence of a nHz SGWB (also later on confirmed in refs.~\cite{Goncharov:2021oub,Chen:2021rqp,Antoniadis:2022pcn}). 
However, at the time, no evidence was found for a quadrupolar spatial correlation (i.e. Hellings-Down (HD) curve~\cite{1983ApJ...265L..39H}), necessary to interpret the signal as a GW background consistent with General Relativity. 
Interestingly, the more recent PTA data released in 2023 by the NANOGrav~\cite{NG15-SGWB,NG15-pulsars}, EPTA (in combination with InPTA)\,\cite{EPTA2-SGWB,EPTA2-pulsars,EPTA2-SMBHB-NP}, PPTA\,\cite{PPTA3-SGWB,PPTA3-pulsars,PPTA3-SMBHB} and CPTA\,\cite{CPTA-SGWB} collaborations, gathered significant evidence for a HD angular correlation.

Focusing on the most stringent dataset, the latest NANOGrav $15\,{\rm yr}$, 
the reported observations can be fitted by a smooth power law with a scaling $\Omega_{\rm GW}\propto f^{(1.6, 2.3)}$ at $1\sigma$. 
It is known that supermassive black hole binaries produce a SGWB with a scaling  $\Omega_{\rm GW} \propto f^{2/3}$~\cite{Phinney:2001di} if binary evolution is GW driven, which is currently disfavoured at $2\sigma$ by the NANOGrav15 data\,\cite{NG15-SMBHB,NG15-NP}. Nevertheless, environmental and statistical effects may play a relevant role, and lead to a steeper scaling ~\cite{Sesana:2008mz,2011MNRAS.411.1467K,Kelley:2016gse,Perrodin:2017bxr,Ellis:2023owy,NG15-SMBHB,NG15-NP,Ghoshal:2023fhh}.
At the present stage, current data does not allow to rule-out (or rule-in) a cosmological origin for the observed signal, which may also be due to first-order phase transitions~\cite{NANOGrav:2021flc,Xue:2021gyq,Nakai:2020oit,DiBari:2021dri,Sakharov:2021dim,Li:2021qer,Ashoorioon:2022raz,Benetti:2021uea,Barir:2022kzo,Hindmarsh:2022awe,Gouttenoire:2023naa,Baldes:2023fsp,Ghosh:2023aum,Gouttenoire:2023bqy}, cosmic strings and domain walls~\cite{Ellis:2020ena,Datta:2020bht,Samanta:2020cdk,Buchmuller:2020lbh,Blasi:2020mfx,Ramazanov:2021eya,Babichev:2021uvl,Gorghetto:2021fsn,Buchmuller:2021mbb,Blanco-Pillado:2021ygr,Ferreira:2022zzo,An:2023idh,Qiu:2023wbs,Zeng:2023jut,King:2023cgv,Babichev:2023pbf, Kitajima:2023cek,Barman:2023fad,Blasi:2023sej,Lazarides:2023ksx,Cai:2023dls,Gouttenoire:2023ftk},
and inflationary first order tensor perturbations \cite{Vagnozzi:2023lwo} (see also for other explanations~\cite{Franciolini:2023wjm,Madge:2023cak,Figueroa:2023zhu, Garcia-Bellido:2023ser, Murai:2023gkv}).

Standard PBH formation scenarios generating a population of subsolar PBHs are related to the emission of an induced SGWB, which may explain current PTA detections~\cite{Vaskonen:2020lbd,DeLuca:2020agl,Bhaumik:2020dor,Inomata:2020xad,Kohri:2020qqd,Domenech:2020ers,Vagnozzi:2020gtf,Namba:2020kij,Sugiyama:2020roc,Zhou:2020kkf,Lin:2021vwc,Rezazadeh:2021clf,Kawasaki:2021ycf,Ahmed:2021ucx,Yi:2022ymw,Yi:2022anu,Zhao:2022kvz,Dandoy:2023jot,Zhao:2023xnh,Ferrante:2023bgz,Cai:2023uhc,Franciolini:2023pbf,Balaji:2023ehk, Liu:2023ymk, You:2023rmn,Jin:2023wri,Zhao:2023joc,Zhu:2023faa,Firouzjahi:2023lzg,Liu:2023ymk,Wang:2023ost,Basilakos:2023xof,Basilakos:2023jvp,Abe:2023yrw,Inomata:2023zup}.
In Fig.~\ref{fig:PTA_NG_abundancecomp} we show the parameter space that would be compatible with PTA signal and how this relates to PBH overproduction constraints (see also Sec.~\ref{sec:PBHformation} for a discussion on the uncertainties affecting the computation of the PBH abundance). 
\begin{figure}[t]
  \centering
  \includegraphics[width=0.8\textwidth]{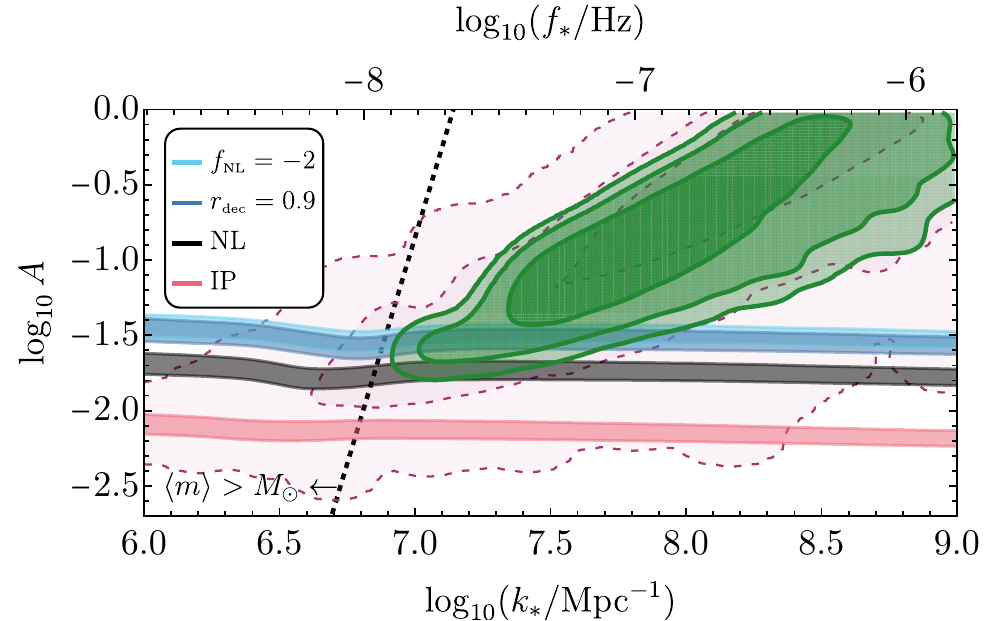}
  \caption{The green and purple posterior indicate the 1, 2, 3$\sigma$ contours obtained fitting to NANOGrav15 and EPTA dataset, respectively. This plot assumed a broken power law curvature spectrum with a fixed growth index $n =4$ and varying decay power law index after the peak located at $k_*$. $A$ indicates the power spectral peak  amplitude and we indicate on top the corresponding peak SGWB frequency.  We also show the corresponding  PBH abundance for different models: 
   Gaussian curvature perturbations including non-linear effects (black), 
   quasi-inflection-point models with $\beta = 3$ (red), 
   curvaton models with $r_{\rm dec} = 0.9$ (blue) and negative $f_\text{\tiny NL}$ (cyan) (see Tab.~\ref{tabNGmm} and Refs. therein for more details). 
   The colored bands cover values of PBH abundance in the range $f_{\rm PBH} \in (1,10^{-3})$ from top to bottom. 
   The dashed line indicates an average PBH mass $\langle m \rangle = \Msun$.
   We point the reader to Sec.~\ref{sec:PBHformation} for a discussion on the uncertainties affecting the computation of the PBH abundance. Figure taken from \cite{Franciolini:2023pbf}.}
  \label{fig:PTA_NG_abundancecomp}
\end{figure}

Due to the extreme sensitivity of PBHs to the statistics of the primordial perturbations, it is also possible to 
reduce or enhance the abundance of PBHs efficiently via non-Gaussian perturbations.
This could result in a larger/smaller amplitude of curvature perturbations and of the associated SIGWs.
On the one hand, assuming a fixed abundance of PBHs,  negative non-Gaussianities would be able to enhance the associated amplitude of the GW signal, bringing it closer to the range preferred by PTA observations.
On the other hand, positive non-Gaussianities could reduce $\Omega_\text{\tiny GW}$ allowing this scenarios to escape the sensitivity level of current PTA experiments.

 Future nHz frequency gravitational wave detectors will reach remarkable sensitivities, provided a potential astrophysical background can be tamed. Therefore, with future PTA-SKA experiments we will either detect or rule out conclusively the cosmological signals accompanying the formation of PBHs. PTA-SKA will robustly probe PBH mass formation in the range $[0.1-10^4] M_\odot$. This regime is extremely important for both astrophysical sources such as stellar mass BH binaries, intermediate BHs and SMBH seeds. On Fig.~\ref{figskadistinctstatistics}, from \cite{Unal:2020mts}, we show the PTA-SKA detection capabilities of the stochastic GW background sourced by primordial scalar perturbations which obey different statistical distributions: Gaussian, chi-sqr ($\chi^2$) and cubic-Gaussian ($G^3$) for the cases i) $f_{\rm PBH}\sim1$ (i.e PBHs constituting all the DM) and ii) $f_{\rm PBH}=10^{-10}$. 
 
One should point out here as well the possibility of future nHz PTA-SKA GW detectors to observe ultralight PBHs in the mass range $10^7 - 10^9\mathrm{g}$ given that the peak frequency associated to the scalar induced GWs sourced by Poisson PBH energy density fluctuations, related to the PBH spatial distribution, is well within the PTA-SKA frequency range~\cite{Papanikolaou:2020qtd}. This potential detectability of such small mass PBHs is very important given the fact that these ultralight PBHs evaporate before BBN so they do not leave another direct observational imprint of their existence.

\begin{figure}
    \centering
    \includegraphics[width=0.8\textwidth]{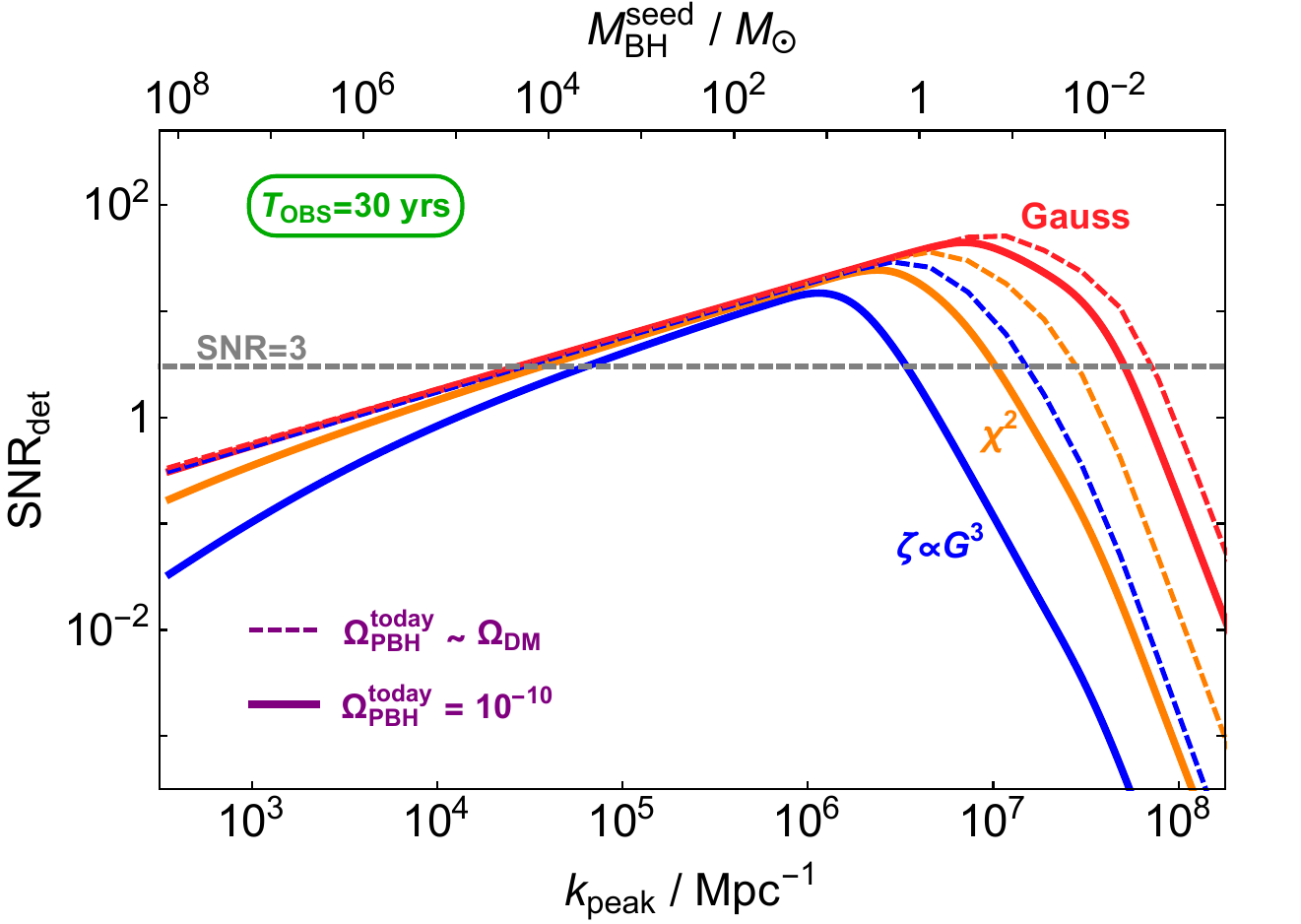}
    \caption{PTA-SKA detection capabilities of the stochastic GW background sourced by scalar perturbations with distinct primordial statistical distributions: Gaussian (red), chi-sqr ($\chi^2$, orange) and cubic-Gaussian ($G^3$, blue) for i) $f_{\rm PBH}\sim1$ (i.e PBHs constituting all the DM) and ii) $f_{\rm PBH}=10^{-10}$, from \cite{Unal:2020mts}.}
    \label{figskadistinctstatistics}
\end{figure}

\subsubsection{Continuous and high-frequency GWs from planetary-mass PBHs}

PBH binary systems with a planetary chirp mass should emit very weak continuous GWs in the frequency range of ground-based interferometers, many years before they actually merge.  It has been proposed to search for continuous waves from PBH binaries in our galaxy~\cite{Miller:2020kmv}, either towards the galactic center or in the solar system vicinity.   However, the current limits on continuous waves do not yet provide significant limits on the PBH abundance~\cite{Miller:2021knj}, even for the models with the highest merger rates. However, future instruments like Einstein Telescope should be able to probe $f_{\rm PBH}$ between $10^{-2}$ and $1$.   

When they merge, these low-mass PBH binaries emit gravitational waves in the MHz or GHz range.  These could be probed with high-frequency GW detectors like resonant microwave cavities, which could probe $f_{\rm PBH}$ down to about $10^{-3}$~\cite{Herman:2020wao,Aggarwal:2020olq}.
Additionally, one recent proposal involving resonant LC circuits may represent the best option in terms of individual merger detection prospects in the range $(1 - 100) \, \text{MHz}$. It was also shown in Ref.~\cite{Franciolini:2022htd} that a detection of the stochastic gravitational wave background produced by unresolved light PBH binaries is possible only if the theoretical sensitivity of the proposed Gaussian beam detector is achieved. Such a detector, whose feasibility is subject to various caveats, may be able to rule-out some scenarios for asteroidal mass primordial black hole dark matter (see Ref.~\cite{Franciolini:2022htd} and Refs. therein for more details).

\subsubsection{Discussion and limitations}

Gravitational wave observations already provide both limits on the PBH abundance and possible hints in favor of the PBH hypothesis.  There is a rather general agreement that the observations of black hole mergers around $30 M_\odot$ do not allow for more than about a percent of the dark matter to be made of PBHs of this mass. 

Exceptional GW events, e.g. including at least one black hole in the pair-instability mass gap (like GW190521), with one object in the low-mass gap or at least heavier than most of neutron stars (like GW190814 and GW190421) without electromagnetic counterparts, mergers with low mass ratios and small individual spins (like GW190814), with statistical evidence for a black hole spin not aligned or anti-aligned with orbital momentum, as well as the measured low effective spins, provide a series of indications that challenge most standard stellar-origin scenarios and suggest alternative formation mechanisms. Among these alternatives, the PBH hypothesis is a front-runner together with scenarios of stellar black hole binary formation in dense environments like globular clusters or nearby AGNs. All these observations are also pushing the odds in favor of PBHs when one performs model comparisons based on Bayesian statistics. But the statistical significance of these observations, as well as the large theoretical uncertainties on both PBHs and stellar black hole models, do not yet allow to draw a definitive conclusion.  

If at least some of the detected black holes are primordial, this suggests a somehow extended mass function, which is also more realistic theoretically.  This means that thermal effects induced by the QCD transition needs to be taken into account, as recently done in \cite{Franciolini:2022tfm,Escriva:2022bwe}, 
which have important effects in the stellar mass range and can modify the merger rate distribution previously derived with phenomenological lognormal models. 

Limits on GW backgrounds is another way to already probe the possible DM fraction made of PBHs, but these limits highly depend on the model parameters and are impacted by the merger rate uncertainties or by the possible non-Gaussian statistics of scalar perturbations in the case of a SGWB from second order perturbations. Nevertheless, it has been claimed that the possible signal from PTAs at nHz frequencies could come from the scalar perturbations leading to stellar-mass or planetary-mass PBHs.  

Altogether, GWs appear to be a very novel and promising way to probe and set a limit on the abundance of PBHs. Once again, one must be cautious because of the large model uncertainties, but the intriguing LVK observations provide a very strong motivation to investigate further the PBH hypothesis with improved models and Bayesian statistical methods. Among the GW signatures of PBHs, signals from the coalescence of subsolar black holes provide the most compelling way to distinguish between stellar and primordial origin.  A few candidates have been recently found but whose statistical significance is too limited to claim a detection. Upcoming data will have the ability to detect subsolar PBHs if they significantly contribute to the dark matter.

\subsection{Summary}

There are dozens of ways to probe and constrain the abundance of PBHs, which have been reviewed in this section.   If one strictly follows the claimed limits and assume a monochromatic distribution, it seems excluded that PBHs account for the entirety of the dark matter below $10^{-16} M_\odot$ due to limits based on PBH evaporation through Hawking radiation and above $10^{-10 } M_\odot$ due to microlensing limits up to the solar mass scale, and multiple other astrophysical and cosmological observations at masses above a solar mass.    This has fed the interest for the asteroid-mass region that until know remains open, even if some limits have been claimed from the capture of PBHs by stars, but still using quite specific assumptions.  

It is however worth noticing that the situation is in reality much less simplistic, and that all the limits rely on assumptions that can be debated and that very often are also model dependent.   The simplest example of the mitigation of these limits is probably the fact that all the evaporation-based limits rely on a still unproven  physical phenomena mixing quantum theory and gravity.   If for any reason our current description of the Hawking radiation is incorrect or if it is less efficient than expected, all the limits on the PBH abundance below $10^{-16} M_\odot$ disappear.    Lots of work have tried to examine the validity of the assumptions in different contexts and these were also discussed in this section.   It is also worth noticing that the strength of the microlensing limits can vary depending on the PBH clustering, on the galactic halo profiles, etc.    Another example of how limits can change depending on assumptions is provided by CMB limits, varying by many orders of magnitude depending on the details of the accretion process onto PBHs or on the dark matter content.

Besides asteroid-mass PBHs, one can also notice that in the solar-mass region, the number of probes is limited to microlensing and X-ray limits and are not extremely stringent, varying between $10^{-2}$ and $10^{-1}$.  For quite extended PBH mass functions, it could remain a viable possibility that the main peak of PBHs arises within this range as one could expect from the QCD-transition.   Scenarios combining a peak in the asteroid-mass range and a peak in the solar-mass range, possibly explaining GW observations, are also an interesting possibility.  

Finally, one should note that limits on PBH abundances cohabit with possible positive evidence~\cite{Carr:2023tpt}, e.g. from detected microlensing events, cosmic background properties supporting the high redshift existence of PBH clusters, and obviously GW observations and the intriguing properties of black hole binary mergers.   Most of these evidences point to the stellar-mass region.

Taken all these considerations into account, it seems indeed difficultly plausible that all the dark matter is made of planetary-mass or intermediate mass (above $10 M_\odot$) PBHs.  But both the asteroid-mass region and the stellar-mass region remain of very high interest, especially in the context of extended mass distributions.   But strong claims are probably premature given the remaining large level of uncertainties for all the probes.

%\begin{figure}
%    \centering 
%%   \includegraphics[width=\textwidth]{pbh_constrainsts_low.png}
%    \includegraphics[width=\textwidth]{pbh_constrainsts_high.png}
%    \caption{Constraints on the PBH dark matter fraction (from the repository %https://doi.org/10.5281/zenodo.3538999 by Bradley J. Kavanagh). {Do we %keep this?}}
%    \label{fig:pbh_limits_all}
%\end{figure}

\section{Detectability with LISA}
\label{sec:LISA}

In this section, we aim at summarizing the different ways with which the LISA mission can probe PBHs, and the possible impact that LISA will have on our understanding of the viable PBH models, in particular their mass function, merging rates, and the underlying formation scenarios.   Remarkably, we emphasize that not only LISA will be able to probe intermediate-mass PBHs but also solar-mass and subsolar mass PBHs, and even tiny PBHs formed at very high energies in the early Universe that would have entirely evaporated, due to the Poisson fluctuations and the resulting SGWB they would have left.

\subsection{SGWB from second order curvature fluctuations}

The generation of PBHs from the collapse of sizeable density perturbations is unavoidably associated with the emission of a SGWB at second order in perturbation theory, whose characteristic frequency can be related to the PBH mass~\cite{Yuan:2021qgz, Domenech:2021ztg}. It has been found that the formation of PBHs with masses around ${\cal O} \lp 10^{-15} - 10^{-8} \rp M_\odot$, where they can comprise the dark matter in the Universe, corresponds to the emission of a SGWB with frequency and abundance well within the LISA sensitivity curve~\cite{Bartolo:2018evs}.

The properties of such a SGWB can be summarised as follows:
\begin{itemize}
    \item The shape of the signal depends on the power spectrum of the curvature perturbations and is model dependent. However, the behaviour at low frequencies is predicted to be scaling like $\sim k^3$ due to causality arguments~\cite{Cai:2019cdl}\footnote{See Refs.~\cite{Domenech:2020kqm, Hook:2020phx,Witkowski:2021raz} where the impact of thermal history on the SGWB spectral shape was investigated.}.
    \item Even though GW signals have an intrinsic non-Gaussian nature at emission epoch, their coherence in phase correlations is washed out by their propagation in the perturbed Universe due to time delay effects generated by large scale variations of the gravitational potential, leading to a completely low-redshift Gaussian signal at the LISA detector~\cite{Bartolo:2018rku}.
    \item The presence of local scale-invariant primordial non-Gaussianities in the scalar curvature perturbation would be responsible in modifying the shape and abundance of the GW spectrum~\cite{Unal:2018yaa}, as well as inducing a large amount of anisotropies in the SGWB, potentially detectable by LISA~\cite{Bartolo:2022pez}, if PBHs comprise only a small fraction of the dark matter in the Universe~\cite{Bartolo:2019zvb}.
\end{itemize}
These predictions can be used to disentangle this signal from other SGWBs of different nature.

\subsection{SGWB from ultralight evaporated PBHs} 

LISA will also probe the formation of PBHs of mass between $10$ g and $10^9$ g that are too light to have survived until today, due to Hawking evaporation. In particular, a SGWB comes from the Poisson fluctuations in the spatial distribution of these PBHs at formation~\cite{Papanikolaou:2020qtd}. These Poisson fluctuations correspond actually to scalar inhomogeneities, sourcing a SGWB at second order. This will be particularly interesting in the context of modified gravity setups~\cite{Papanikolaou:2021uhe,Papanikolaou:2022hkg} that could be constrained or even ruled out using the SGWB from ultralight PBHs.   

Through this PBH-induced SGWB, LISA will indirectly probe the early Universe at energies up to the scale of grand-unified theory, including inflationary models, even if these PBHs have quickly evaporated. 

\subsection{SGWB from early binaries}

As seen in Fig.~\ref{fig:EB_ns}, the SGWB amplitude for a log-normal mass function with central mass $\mu = 2.5 M_\odot$ and width $\sigma = 1$ reaches its peak at around 100 Hz, and starts to be greatly suppressed below the LISA frequencies. For the thermal broad-mass function, one notices in Fig. \ref{fig:Omegah2_m1m2} the important contribution to the SGWB from binaries composed of intermediate and solar-mass PBH (coming from the electron peak and the QCD peak in the PBH mass function respectively, see Fig.~\ref{fig:fPBH2models}). These low-mass-ratio binaries start to boost the SGWB below 0.1 Hz by up to 3 orders of magnitude at most, compared to the log-normal case (Fig. \ref{fig:EB_ns}). In particular, one can notice a bump in the SGWB around $10^{-3}$ Hz, originating from the bump around $10^{6} M_\odot$ in the broad PBH mass function. The shape of the background could therefore give us important details about the PBH mass function and allow for precise reconstruction. The SGWB spectral index (${\rm d} \log \Omega_{\rm GW} / {\rm d} \log f$) is overall positive for the early binaries contrary to the late-time PBH binaries (see next Subsection), thus probing the SGWB spectral index at the LISA frequencies could be a way to differentiate PBH binaries in clusters from early PBH binaries.
%Furthermore, the GWB amplitude at LISA frequencies, now detectable even for the worst experimental design, originates from the merging of binaries with low mass ratios. More precisely, as can be seen in Fig. \ref{fig:Omegah2_m1m2}, one notices an important contribution from intermediate and solar-mass PBH, the former coming from the electron peak and the latter from QCD peak in the PBH mass function (see Fig. \ref{fig:my_label}). Also, the contribution from binaries with two PBHs from the QCD peak is almost as important.

\subsection{SGWB from late binaries}

Similarly to the early binaries case, the SGWB amplitude for a log-normal mass function is much lower at the LISA frequencies, but we can still expect a detection if the GW amplitude is high enough to be detected by LVK. In the case of a thermal history mass function, the SGWB amplitude is enhanced at LISA frequencies because we have more massive PBHs paring with the $2\Msun$ PBHs (see Figs.~\ref{fig:Omegah2_late} and~\ref{fig:Omegah2_m1m2_clust}). The overall amplitude is determined by the merger rate, which still has a large uncertainty. As discussed in Sec.~\ref{sec:latebinaries}, it can be enhanced by increasing the PBH density in clusters and/or lowering the PBH velocity. Further observations by LVK would provide useful implications on the merger rate to refine the prediction on the SGWB amplitude at LISA frequency.

\subsection{Intermediate-mass binary mergers}

As shown in Fig.~\ref{fig:rates} for two typical models, intermediate-mass PBHs between $10^3$ and $10^5 M_\odot$ that are contributing to the dark matter at the sub-percent level at most, due to other limits in this range, could merge with rates above $10^{-3} {\rm yr^{-1} Gpc^{-3}}$.  Interestingly, this mass range corresponds to GW frequencies where LISA has the highest range, effectively probing any merger in the observable Universe up to redshifts of 100 to 1000. This means that even if the merger rate is lower than in the stellar-range, LISA could detect more than a few of these mergers every year. Determining the merger rate and redshift distribution of intermediate-mass BH mergers is therefore a promising way to constrain the PBH mass functions and clustering properties, possibly linking them to the seeds of supermassive black holes.  

\subsection{Extreme mass ratio inspirals}
The capability of LISA of targeting low frequency regimes makes it an ideal experiment for the search of GW signals generated from the inspiral of a subsolar mass compact object that could have a primordial origin around a supermassive black hole (SMBH).
GW constraints can be set on the PBH abundance from the expected PBH-SMBH merger rate~\cite{Guo:2017njn,Kuhnel:2018mlr}. In particular, a null-detection during a 5-year operation of the experiment would constrain the PBH abundance to values smaller than $f_\text{\rm PBH} \lesssim 3 \times 10^{-4}$ for PBHs masses in the range $(10^{-2} - 1) \, M_\odot$~\cite{Guo:2017njn}. 
Furthermore, the detection of these extreme mass-ratio inspirals at LISA will be crucial in the search for subsolar-mass black holes, given the unparalleled precision in measuring the 
mass of the secondary object at subpercent level for PBHs as light as $\mathcal{O}(0.01) M_\odot$ up to luminosity distances around hundred megaparsecs~\cite{Barsanti:2021ydd}.
This would allow the claim of detection of a subsolar-mass black hole with very high statistical confidence.

\subsection{High-redshift binary mergers}

It is challenging to explain how supermassive black holes can exist in only partially reionized environments at redshifts as high as $z\gtrsim 7$~\cite{Banados:2017unc}.  If the seeds of these black holes come from the first population of stars, this typically needs super-Eddington accretion rate and large mass seeds.  Another possible mechanism is the direct collapse of gas into BHs.  Both models have their caveats.  Even if one invokes super-Eddington accretion, it is very challenging for these seeds to reach sufficiently large masses to explain observations.  PBHs are an alternative explanation to the existence of SMBHs since they can provide seeds of intermediate-mass BHs at higher redshift than for the other astrophysical mechanisms~\cite{Duechting:2004dk,Kawasaki:2012kn,Clesse:2015wea,Bernal:2017nec,DeLuca:2022bjs}. Therefore, the easiest way to distinguish PBH seeds from other candidates is to observe intermediate-mass or even supermassive black holes at $z \gtrsim 20$, prior to star formation.  

The astrophysical range of LISA will allow for the observation of IMBBH mergers at redshifts $z>20$ with a SNR larger than five, for equal-mass mergers and progenitor masses between $10^3 M_\odot$ and $10^6 M_\odot$, as shown in Fig.~\ref{fig:zhorizon}.   The possible merger rates of PBHs for a broad mass function with the imprints of the thermal history, shown in Fig.~\ref{fig:rates},  can be larger than $\mathcal O(1) {\rm yr}^{-1}$ for primordial IMBBHs that would be formed in PBH clusters at high redshift.   The existence of these clusters is relevant since they would also form in the standard Press-Schechter theory.  LISA observations will be complementary to those of Earth-based GW detectors, like CE and ET, which will probe mergers with lower masses, and to future PTA limits from SKA, which will probe eventual mergers of SMBHs at similar redshifts. 

\begin{figure}[t!]
\includegraphics[width=0.9\textwidth]{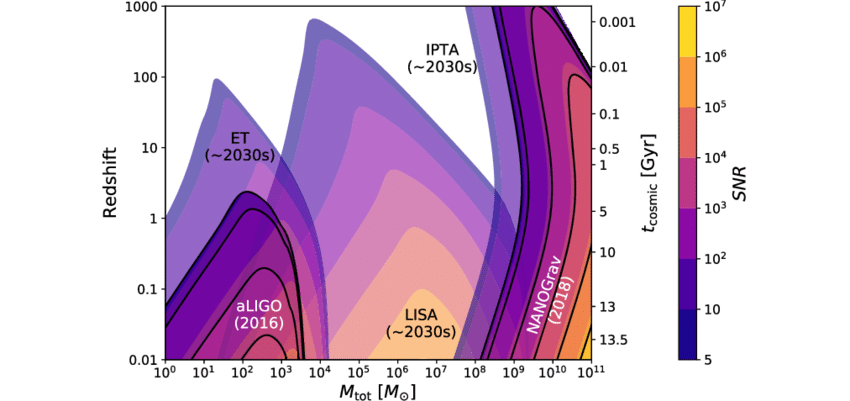}
\centering
\caption{Redshift range of LISA for equal-mass BBH coalescences as a function of the total system mass and comparison with the range of other detectors and pulsar timing arrays.  The color scale represents the expected SNR. Figure taken from Ref.~\cite{Burke-Spolaor:2018bvk}.
}
\label{fig:zhorizon}
\end{figure}

\begin{figure}[t!]
\includegraphics[width=0.8\textwidth]{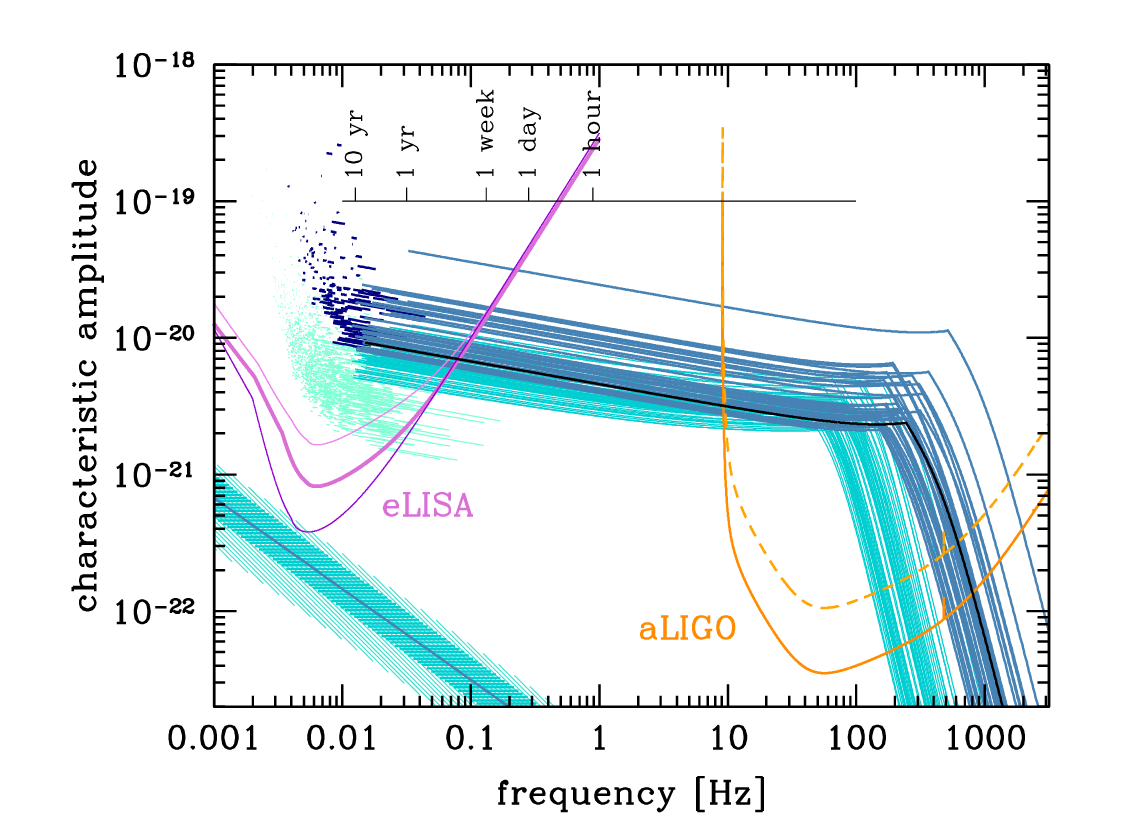}
\centering
\caption{Illustration of the concept of multi-band GW astronomy, combining (e)LISA and aLIGO (LVK).  Each blue line corresponds to a trajectory in the strain-frequency plane for black hole mergers.  The horizontal time scale corresponds to the probed time before the merger.   Figure taken from Ref.~\cite{Sesana:2016ljz}.
} \label{fig:multiband}
\end{figure}

\subsection{Combination with LVK observations }

Numerous compact binary coalescences with total mass between $10 M_\odot$ and $100 M_\odot$ and a merger frequency in the range probed by Earth-based interferometers will also be detected by LISA due to the GW emission in the inspiral phase, as illustrated in Fig.~\ref{fig:multiband}.  This provides a unique way to determine months or even years in advance when and in which sky location these mergers are going to occur.  It will be then easier to search for electromagnetic counterparts of mergers.  Parameter reconstruction will also be improved, which could also help to better determine the source location, distance and redshift.   According to~\cite{Sesana:2016ljz}, up to hundreds of binary black hole mergers could be pre-determined with an accuracy of about 10 seconds and 1 deg$^2$, as well as a chirp mass reconstruction at the $10^{-6}$ level. The latest analysis \cite{Klein:2022rbf} uses the state-of-art waveforms and Bayesian parameter estimation for both frequency bands to assess the potential of multi-band GW observations of the stellar-mass BBHs starting from 3 years before the merger (LISA) up to the coalescence (ground-based detectors: upgraded LVK or 3G -- either ET or CE). Such observations will allow to determine all 17 parameters describing the binary with at least percent-level accuracy, possibly to identify a likely host galaxy and to alert about the expected merger days in advance significantly improving the chances of detection of any electromagnetic signature associated with it.

In \cite{Franciolini:2021xbq} the measurement accuracy by LISA, ET and LVK has been quantified for such PBH binary discriminators as eccentricity, spin and mass-spin correlation, as well as the horizon redshift for subsolar-mass mergers with negligible spins and eccentricity has been determined.

A better reconstruction of the black hole mass, rate and redshift distribution will help to distinguish between primordial and stellar scenarios, as well as the black hole environment and binary formation channel (merger from a cluster, stellar binary origin, AGNs...). This could be particularly the case for mergers with black hole masses in the pair-instability mass gaps that would be seen as short bursts in LVK detectors, not allowing for a precise mass reconstruction. LISA will significantly help to determine their still enigmatic (possibly primordial) origin.

\subsection{Quasi-monochromatic continuous waves from subsolar PBHs}

\begin{figure}[t!]
\includegraphics[width=0.95\textwidth]{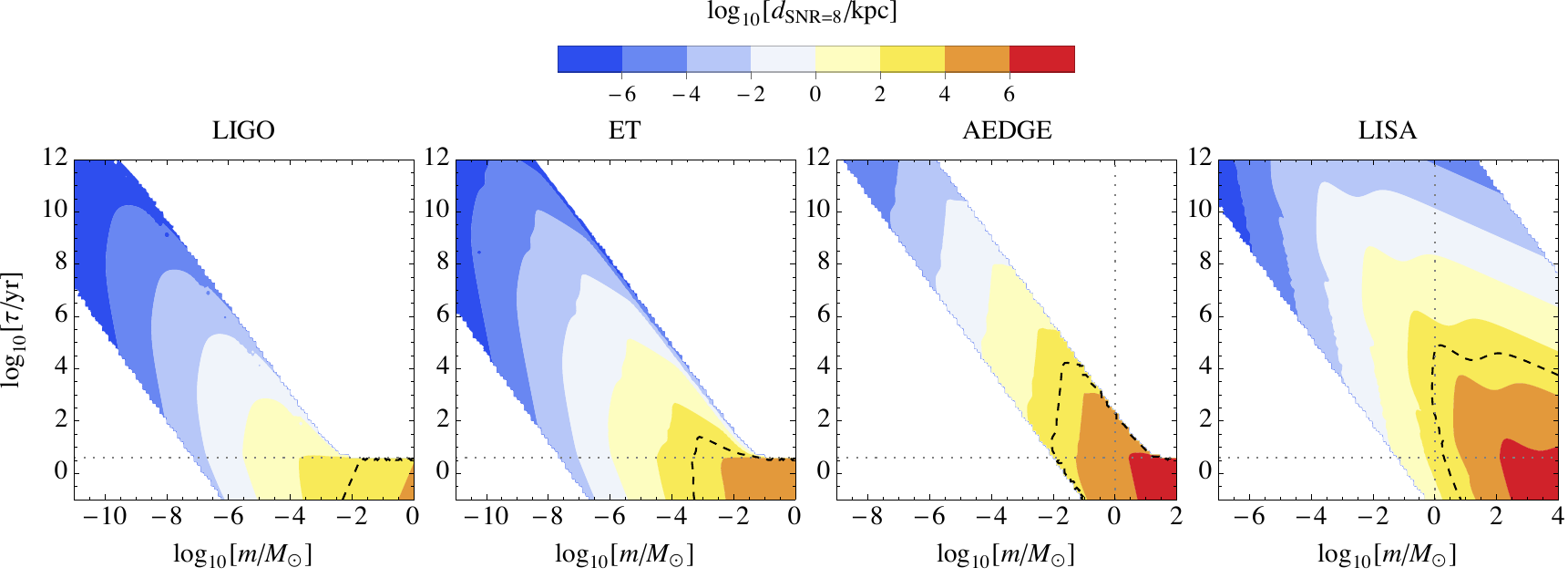}
\centering
\caption{Distance for which a circular binary of total mass $m$ and coalescence time $\tau$ leads to quasi-continuous gravitational waves detectable with a SNR $=8$ in different detectors.  The expectations for LISA are shown in the right panel.  Figure from Ref.~\cite{Pujolas:2021yaw}.
} \label{fig:cwLISA}
\end{figure}

It has been pointed out in~\cite{Pujolas:2021yaw,Barsanti:2021ydd,Guo:2022sdd} that LISA will probe the existence of galactic subsolar PBHs in binaries with very long coalescence times up to distances of tens of kiloparsecs, allowing to probe the galactic center or PBH clusters in the galactic dark matter halo.   This range is shown in the right panel of Fig.~\ref{fig:cwLISA} coming from~\cite{Pujolas:2021yaw} as a function of the binary mass and time before collapse.  The most interesting region is probably for $10^{-2} M_\odot < m < 1 M_\odot$ and collapse time $\tau $ up to $10^8$ years, for which the range is of order of tens of parsecs. 

Therefore, with the use of continuous-wave methods, LISA will be able to probe different mass and collapse time scales than ground-based instruments like Einstein telescope~\cite{DeLuca:2021hde}.   It will therefore be complementary to probe the existence of subsolar black holes or primordial origin in our galaxy, which if they are found, would lead to groundbreaking implications in cosmology.  

\subsection{Observing near-Earth asteroid-mass primordial black holes}

The LISA instrument will have the sensitivity to detect the gravitational influence of dark matter clumps or dark asteroid-mass compact objects like tiny primordial black holes passing nearby the detector.  This possibility was recently studied by Baum et al. in~\cite{Baum:2022duc} (see also the old work by Seto \& Cooray~\cite{Seto:2004zu}).   

With typical galactic velocities of hundreds kilometers per second, the signal duration is of order of hours or days, and would correspond to a burst at a frequency of order $10^{-6}-10^{-4}$ Hz, within the LISA sensitivity band.  Since the signal would come from a transient change in the acceleration on the three mirrors resulting from the gravitational force exerted by the compact object, the shape of the signal is expected to be different than for a gravitational-wave tensor fluctuation.  Therefore specific analysis tools are probably needed to search for them, but those could be used to detect near-Earth asteroids as well, which would be an interesting bi-product of such an analysis.  For instance, one expects one observable transit every 20 years~\cite{Baum:2022duc} for PBHs with mass of order $10^{14}$kg, if they constitute all the DM density.  The expected LISA range to asteroid-mass PBHs and the corresponding rate of expected events is shown in Fig.~\ref{fig:LISA-asteroidmass}. 

Observing tiny primordial black  holes in the solar system would be clearly revolutionary and LISA will probably be the first instrument to reach the required sensitivity to make such a great achievement. 

\begin{figure}
    \centering
    \includegraphics[width=0.6\textwidth,angle=-90]{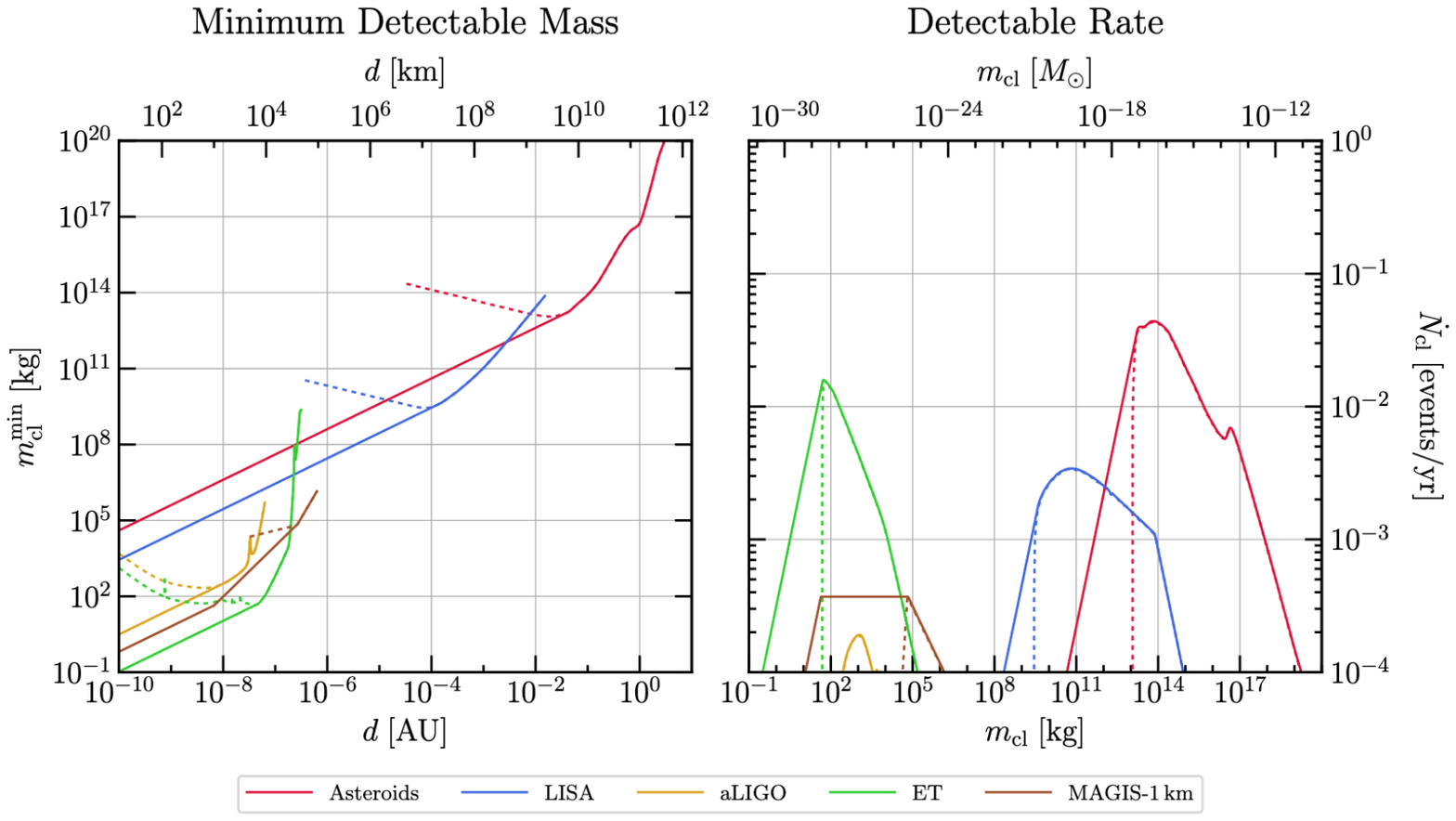}
    \caption{Left:  Expected mass of dark matter clumps or compact objects like PBHs $m_{\rm cl}$ in the inner solar system that could be detected at a distance $d$ by LISA and other types of gravitational-wave detectors, due to the transient acceleration of test masses (which for LISA corresponds to {the mirrors in the three space probes}).  Right:  Corresponding rate of expected observable events as a function of the object mass, if they comprise all the dark matter and assuming galactic velocities.  Figure from~\cite{Baum:2022duc}.   }
    \label{fig:LISA-asteroidmass}
\end{figure}

\subsection{Summary}

LISA will have the ability to probe primordial black holes of a very broad range of masses, based on different signals and GW production mechanisms:
\begin{itemize}
    \item $10^3-10^9 M_\odot$:  GWs from the merging of intermediate-mass and supermassive black holes, including at redshift $z \gtrsim 20 $, and their SGWB.
    \item $10-100 M_\odot$:   GWs from the merging of binaries with extreme-mass ratios and their SGWB.
    \item $0.01-1 M_\odot$:  Quasi-continuous GWs from sub-solar PBH in our galaxy.
    \item $10^{-15} - 10^{-8} M_\odot$:  SGWB from second order perturbations.
    \item $10^{-19} - 10^{-16} M_\odot$:  Newtonian gravitational force from PBHs in the inner solar system.
    \item $<10^{-20} M_\odot$: SGWB induced by Poisson PBH energy density fluctuations.
\end{itemize}

This includes the two best ways to almost unambiguously prove the existence of PBHs, i.e. detecting subsolar black holes (in binaries with extreme mass ratios) and black holes at higher redshift than the stellar formation scenario can account for.

As a consequence, LISA will be an instrument of premiere importance in order to probe the existence of PBHs, which will be complementary to other astrophysical and cosmological signals, as well as other types of GW detectors.

\section{Conclusions}
\label{sec:ccl}

Since the first detection of gravitational waves in 2015, primordial black holes (PBHs) have been the subject of an exponential growth of interest.   Not only could PBHs explain some of the LVK black hole mergers but they could also have important cosmological consequences in cosmology:  their existence would unveil new high-energy physics needed for their formation in the early Universe, they could significantly contribute to or even entirely constitute the dark matter, they may be related to baryogenesis, may have seeded supermassive black holes at the center of galaxies, and may as well have significantly modified the formation and growth of structures through the Universe's history.

With this work, we provide a review including the most recent scenarios of primordial black hole formation.  The most up-to-date formalisms to be used for end-to-end calculations of their gravitational-wave signatures are also reviewed.  A particular emphasis is placed on the observational perspectives in the context of the LISA mission.  The present astrophysical and cosmological limits are also briefly reviewed and discussed, as well as the possible clues for their existence.

Among the plethora  of formation mechanisms, we have discussed in particular the recent classes of models in which PBHs are formed from Gaussian or non-Gaussian inhomogeneities that could be produced from quantum fluctuations or quantum diffusion   of one or more scalar fields during inflation.  Wherever possible, we provided the concrete formula to be used to calculate the primordial power spectrum of curvature fluctuations in these models and their possible non-Gaussian statistics.  We went beyond the standard formalism of PBH formation by reviewing recent progress on the relation between the curvature perturbation and the density contrast, the model-dependent computation of the overdensity threshold leading to PBH formation, including thermal effects at the QCD transition that provide a good theoretical motivation for a peak in the PBHs distribution at the stellar-mass scale (possibly associated with LVK observations).  We also went beyond the standard computation of the present PBH mass function and spin distribution by including non-trivial effects related to accretion, which have been studied recently. 

PBHs can lead to different types of signatures in gravitational-wave observations and we have reviewed their calculation in a model-dependent way.   We provided the most recent prescriptions for the calculation of the PBH merging rates, for different binary formation mechanisms (early binaries, binaries formed in clusters, disrupted binaries).  These mergers would leave distinct GW signatures that are searched for with GW template-based methods.  The PBH mass function and the underlying formation model could be reconstructed from the rate, mass, spin and redshift distribution of black hole mergers with present and future gravitational-wave interferometers.  Recent analysis of LVK observations have revealed that there could be some statistical significance in favor of the PBHs hypothesis, depending on which astrophysical channels one compares them to.  Another source of GWs comes from the transient bursts produced during close hyperbolic encounters, for which we also provide the expected rate and typical shape.  Besides transient events, PBH binaries and hyperbolic encounters should also have produced a stochastic GW background covering the full frequency range accessible with observations of pulsar timing arrays, LISA and ground-based interferometers.  The general formalism to compute this background has been reviewed and  particular examples based on broad PBH mass functions including thermal effects have been given and discussed.  At second order in perturbation theory, one gets another source of gravitational-waves associated with PBHs, coming from the large density fluctuations that are at the origin of their formation or, for lighter PBHs which have already evaporated today, that are produced by their Poisson distribution.  We have reviewed the most recent prescriptions to calculate these SGWBs for any PBH model, and discussed their observability.  

The future LISA space-based interferometer will probe PBHs in several ways.  First, by searching for intermediate-mass PBH mergers.   Based on the expected noise power spectral density, we estimated the number and distribution of PBH merger events in some models.  LISA will also be able to detect mergers with very low mass ratios, for instance between an intermediate-mass PBH and a solar-mass PBH produced at the QCD transition.  Also, it will probe the SGWB from PBH binaries, which could be strongly boosted for wide PBH mass functions, compared to a peaked PBH mass function.  Therefore, LISA will be very useful to disentangle the different possible PBH mass functions and formation scenarios.  Searching for high-redshift mergers is another way to distinguish astrophysical and primordial origins of black holes and test the hypothesis that supermassive black hole seeds are primordial.   Finally, the broad gravitational-wave frequency range covered by LISA will allow to probe the stochastic GW background induced by scalar perturbations, corresponding to PBH scales ranging from $10^{-15} M_\odot$ to $10^{-8} M_\odot$, therefore covering the interesting asteroid-mass range where there is no significant limit on the PBH abundance.   By probing a scalar-induced background from Poisson fluctuations in even lighter, evaporating PBHs, that would have totally disappeared from the Universe today due to Hawking evaporation.  This would be another unique way to probe physics at much higher scales than can be done with particle accelerators.

Our work will soon be accompanied by a numerical code to compute all these gravitational-wave observations, for a wide variety of models.  This code is currently under development and testing, but some of our key figures were already produced using it.  This is a first promising perspective of this work.   Then, our review also sheds light on the physical processes that are still subject of large uncertainties, like the role of PBH accretion and clustering, or the PBH binary disruption.  We also discuss the fact that all signatures are highly model dependent.  

One important conclusion of our analysis of the recent literature on PBHs is that strong claims, in one or another direction, are certainly premature and generally rely on hypotheses that are very hard to test.  Thus, there is still a huge amount of work to be done before making precise predictions.  We therefore encourage others to pursue this direction and hope that our work will provide motivation and tools for researchers from other fields to join the growing PBH community.

%\textit{Three main items: Formation, signatures, detectability.   Emphasize the specificities:  latest calucalation of the PBH abudnance, non-Gaussian models, redshift evolution, thermal history effects, limitations of the constraints, LISA-specific observations.  Perspectives:  what remains to do, lines of research to explore. }

\newpage
\section{Table of notations}

\begin{table}[h]
    \centering
    \begin{tabular}{|c|l|}
    \hline
    $\zeta$ & Curvature fluctuation \\
    $\delta$ & Density fluctuation \\
    $ \mathcal P_{\zeta}$ & Primordial curvature power spectrum (dimensionless) \\
    $ \mathcal P_{\delta}$ & Primordial density power spectrum (dimensionless) \\
    $n_{\rm s}$ & Scalar spectral index of the power spectrum \\
    $P$ & Probability density function of curvature/density fluctuations \\
    $ \sigma $ & Variance of $P$ \\
    $ a$ & Scale factor \\
    $H$ & Hubble parameter \\
    $N$ & e-fold number \\
    $m_{\rm Pl}$  & Planck mass  \\
    $M_{\rm Pl}$ & Reduced Planck mass \\
    $m_{\rm PBH}$  & PBH mass \\
    $\mathcal M_{\rm c}$ & Chirp mass of a binary\\
    $M_{\rm tot}$ & Total binary mass  \\
    $m_1$ & Primary mass of a binary\\
    $m_2$ & Secondary mass of a binary\\
    $ M_{\rm H}$ & Horizon mass at horizon crossing \\
    $ t_{\rm H}$ & Horizon crossing time \\
    $\gamma$ & Ratio between PBH initial mass and Hubble mass at horizon crossing \\
    $\chi$ & PBH spin parameter  \\
    $f_{\rm PBH}$ & Total dark matter fraction made of PBHs  \\
    $\psi, \psi_1$ & Normalised PBH density distribution (per unit PBH mass interval) \\
    $f$ & Normalised PBH density distribution (per unit neperian logarithmic PBH mass interval) \\
    $\beta$ & PBH density distribution at formation (per neperian logarithmic mass interval)  \\
    $ \Omega_{\rm GW}$ & GW background amplitude \\
    $ \Omega_{\rm DM}$ & DM density parameter \\
    $ \rho_{\rm GW}$ & Energy density associated to GWs \\
    $ \delta_{\rm c}$ & Overdensity threshold for PBH formation \\
    $ \delta_{\rm m}$ & Matter perturbation amplitude \\
    $ \rho_{\rm PBH}$ & PBH energy density profile \\
    $ n_{\rm PBH}$ & PBH number density per comoving volume  \\
    $ w$ & Equation of state parameter \\
    $ \rho_{\rm c} $ & Critical energy density of a spatially flat Universe \\
    $ f_{\rm NL}$ & Non-Gaussianity parameter  \\
    $ R_{\rm EB/LB}$ & Merger rate in a comoving volume for early/late binaries \\
    & (per neperian logarithmic PBH mass interval)  \\
         &  \\
         \hline
    \end{tabular}
    \caption{List of symbols and notations used through the review, in more than one section.}
    \label{tab:my_label}
\end{table}

%\begin{comment}
\newpage
\section*{Acknowledgements}
E.B. and S.C. acknowledge support from the Belgian Francqui Foundation through a start-up grant and to the Belgian Fund for Research FNRS-F.R.S. through the \textit{Mandat d'impulsion scientifique} MIS/VA - F.4509.22 and an IISN grant.  
V.DL. is supported by funds provided by the Center for Particle Cosmology at the University of Pennsylvania.  
JME is supported by the European Union’s Horizon 2020 research and innovation program under the Marie Sklodowska-Curie grant agreement No. 847523 INTERACTIONS, and by VILLUM FONDEN (grant no. 53101 and 37766).
The work of G.F. partly received financial support provided under the European Union's H2020 ERC, Starting Grant agreement no.~DarkGRA--757480 and under the MIUR PRIN programme, and support from the Amaldi Research Center funded by the MIUR program ``Dipartimento di Eccellenza" (CUP:~B81I18001170001).
J.G.B., E.R.M., and S.K acknowledge support from the Spanish Research Project PID2021-123012NB-C43 [MICINN-FEDER], and the Centro de Excelencia Severo Ochoa Program CEX2020-001007-S at IFT, funded by MCIN/AEI/10.13039/501100011033. 
CJ is supported in part by the National Key Research and Development Program of China Grant No. 2021YFC2203004, and by the FNRS-FRIA Grant No. 1.E.070.19F and the FNRS \textit{Fonds de la Recherche Scientifique} Grant No. T.0198.19.
R.K.J. acknowledges financial support from the new faculty seed start-up grant of IISc, the Core Research Grant CRG/2018/002200 from the Science and Engineering Research Board, Department of Science and Technology, Government of India and the Infosys Foundation, Bangalore through the Infosys Young Investigator award. S.K. is supported by the Spanish Atracci\'on de Talento contract no. 2019-T1/TIC-13177 granted by Comunidad de Madrid, the I+D grant PID2020-118159GA-C42 funded by MCIN/AEI/10.13039/501100011033, the Consolidaci\'on Investigadora 2022 grant CNS2022-135211 and the i-LINK 2021 grant LINKA20416 of CSIC, and Japan Society for the Promotion of Science (JSPS) KAKENHI Grant no. 20H01899, 20H05853, and 23H00110. 
I.M. has received funding from the European Union’s Horizon2020 research and innovation programme under the Marie Skłodowska-Curie grant agreement No 754496. TP acknowledges financial support from the \textit{Foundation for Education and European Culture in Greece} as well as  from the \textit{Fondation CFM pour la Recherche in France} and from the \textit{Alexander S. Onassis Public Benefit Foundation} through the scholarship FZO 059-1/2018-2019. 
A.Ra. acknowledges funding from the Italian Ministry of University and Research (MUR) through the ``Dipartimenti di eccellenza'' project ``Science of the Universe''.
A.Ri. is supported by the Boninchi Foundation.
S.~RP is supported by the European Research Council under the European Union's Horizon 2020 research and innovation programme (Starting Grant agreement No 758792, project GEODESI). 
C.U. is supported by Ben Gurion University Kreitman Fellowship and the Israel Academy of Sciences and Humanities (IASH) \& Council for Higher Education (CHE) Excellence Fellowship Program for International Postdoctoral Researchers, also European Structural and Investment Funds and the Czech Ministry of Education, Youth and Sports (Project CoGraDS - CZ.02.1.01/0.0/0.0/15 003/0000437). 
D.W. is supported by UK STFC grants ST/S000550/1 and ST/W001225/1. E.B. is supported by the FNRS-IISN (under grant number 4.4501.19).

%\end{comment}

\newpage

%\section*{Bibliography}
%\medskip
%\printbibliography

%\bibliographystyle{plain}
%\bibliographystyle{ieeetr}
\bibliographystyle{unsrt}
\bibliography{main}
%\bibliography{main.bib}

\end{document}